\documentclass[a4paper,usenatbib,useAMS]{mnras}

\usepackage{graphicx}	
\usepackage{amsmath}	
\usepackage{amssymb}	
\usepackage{multicol}        
\usepackage{bm}		
\usepackage{pdflscape}	
\usepackage{dcolumn}   
\usepackage{grffile}
\usepackage{color}

\usepackage[T1]{fontenc}
\usepackage{ae,aecompl}

\usepackage{newtxtext}

\newcommand\Eqn[1]     {Eq.\,(\ref{#1})}
\newcommand\Eqns[2]    {Eqs\,(\ref{#1}) and~(\ref{#2})}
\newcommand\Eqnss[2]   {Eqs\,(\ref{#1})--(\ref{#2})}

\newcommand\eqns[2]    {Eqs\,(\ref{#1}) and~(\ref{#2})}

\newcommand\nn         {\nonumber}

\newcommand\mat[1]{\textsf{\bf {#1}}}

\def\mnras{{Mon.~ Not.~ R.~ Astron.~ Soc.~}}

\def\prd{{Phys.~ Rev.~ D.~}}

\def\apj{{Astrophys.~ J.~}}
\def\apjs{{Astrophys.~ J.~ Suppl.~}}
\def\apjl{{Astrophys.~ J.~ Lett.~}}

\def\nat{{Nature (London)~}}

\def\mnras{{MNRAS}}

\def\prd{{PRD}}

\def\apj{{ApJ}}
\def\apjs{{ApJS}}
\def\apjl{{ApJL}}
\def\aap{{A\&A}}
\def\nat{{Nature}}

\def\physrep{{Phys.~ Rep.~}}
\def\jcap{Journal of Cosmology and Astro-Particle Physics}
\def\araa{Anual Reviews of Astronomy \& Astrophys.}

\newcommand{\be}{\begin{equation}}
\newcommand{\ee}{\end{equation}}
\newcommand{\ba}{\begin{eqnarray}}
\newcommand{\ea}{\end{eqnarray}}

\def\pp1{{\prime}}
\def\pp2{{\prime\prime}}

\def\2D{{\rm 2D}}

\def\Vu{{V_{\mu}}}

\def\cyc{{\rm cyc}}

\def\bx{{\bf x}}
\def\br{{\bf r}}

\def\bk{{\bf k}}
\def\bq{{\bf q}}

\def\1Loop{{\rm 1Loop}}

\def\Msol{h^{-1}M_{\odot}}

\def\Mpc{\, h^{-1}{\rm Mpc}}

\def\Gpccube{\, h^{-3} \, {\rm Gpc}^3}
\def\kMpc{\, h \, {\rm Mpc}^{-1}}

\def\dx{{\rm d}^3{\bf x}}
\def\dk{{\rm d}^3{\bf k}}
\def\dq{{\rm d}^3{\bf q}}

\def\d{\delta}

\def\nbar{\bar{n}}

\def\fun#1#2{\lower3.6pt\vbox{\baselineskip0pt\lineskip.9pt
        \ialign{$\mathsurround=0pt#1\hfill##\hfil$\crcr#2\crcr\sim\crcr}}}

\title[Precision modelling of the matter power spectrum]
      {Precision modelling of the matter power spectrum in a Planck-like Universe}

      \author[Smith \& Angulo]{Robert E. Smith$^{1,2}$\thanks{r.e.smith@sussex.ac.uk}
        and Raul E. Angulo$^{3,4}$\thanks{reangulo@gmail.com} 
\\
$^{1}$Astronomy Centre, Department of Physics \& Astronomy, University of Sussex, Brighton, BN1 9RH, UK\\
$^{2}$Max-Planck-Institut f\"ur Astrophysik, Karl-Schwarzschild-Str. 1, 85748 Garching, Germany\\
$^{3}$Donostia International Physics Centre (DIPC), Paseo Manuel de Lardizabal 4., 20018 Donostia-San Sebastian, Spain.\\
$^{4}$IKERBASQUE, Basque Foundation for Science, E-48013 Bilbao, Spain
}

\date{\today}

\pubyear{\today}

\begin{document}

\label{firstpage}

\pagerange{\pageref{firstpage}--\pageref{lastpage}}

\maketitle

\begin{abstract}
  We use a suite of high-resolution $N$-body simulations and
state-of-the-art perturbation theory to improve the code {\tt
  halofit}, which predicts the nonlinear matter power spectrum. We
restrict attention to parameters in the vicinity of the Planck
Collaboration's best fit. On large-scales ($k\lesssim0.07\kMpc$), our
model evaluates the 2-loop calculation from the Multi-point Propagator
Theory of Bernardeau\,et\,al.\,(2012).  On smaller scales
($k\gtrsim0.7\kMpc$), we transition to a smoothing-spline-fit model,
that characterises the differences between the
Takahashi\,et\,al.\,(2012) recalibration of {\tt halofit2012} and our
simulations.  We use an additional suite of simulations to explore the
response of the power spectrum to variations in the cosmological
parameters. In particular, we examine: the time evolution of the dark
energy equation of state ($w_0$, $w_a$); the matter density
$\Omega_m$; the physical densities of CDM and baryons
$(\omega_c,\omega_b)$; and the primordial power spectrum amplitude
$A_s$, spectral index $n_s$, and its running $\alpha$. We construct
correction functions, which improve {\tt halofit}'s dependence on
cosmological parameters. Our newly calibrated model reproduces all of
our data with $\lesssim1\%$ precision. Including various systematic
errors, such as choice of $N$-body code, resolution, and through
inspection of the scaled second order derivatives, we estimate the
accuracy to be $\lesssim3\%$ over the hyper-cube:
$w_0\in\{-1.05,-0.95\}$, $w_a\in\{-0.4,0.4\}$, $\Omega_{\rm
  m,0}\in\{0.21,0.4\}$, $\omega_{\rm c}\in\{0.1,0.13\}$, $\omega_{\rm
  b}\in\{2.0,2.4\}$, $n_{\rm s}\in\{0.85,1.05\}$, $A_s\in\{1.72\times
10^{-9},2.58\times 10^{-9}\}$, $\alpha\in\{-0.2,0.2\}$ up to
$k=9.0\kMpc$ and out to $z=3$.  Outside of this range the model
reverts to {\tt halofit2012}. We release all power spectra data with
the C-code {\tt NGenHalofit} at: {\normalsize \tt
  https://CosmologyCode@bitbucket.org/ngenhalofitteam/ngenhalofitpublic.git}.

\end{abstract}



\section{Introduction}

The power spectrum of matter fluctuations contains a wealth of
information about the cosmological model and the initial distribution
of density perturbations in the early universe. Its accurate
measurement and evolution is therefore one of the main goals of modern
cosmology. In recent years a number of different methods have been
devised to extract this information: galaxy clustering
\citep[e.g.]{DavisPeebles1983,Feldmanetal1994}, cluster counts
\citep{Whiteetal1993,LimaHu2004}, shear-shear correlation functions
\citep[e.g.][]{Miralda-Escude1991,Kaiser1992}, correlations of
absorption features in the Lyman alpha forest \citep{Croftetal1998},
correlations in the 21cm emission from neutral hydrogen
\citep{Loebetal2004}, etc.  Each of these observables has a number of
problematic modelling issues, however one commonality is the need to
understand the matter density power spectrum in the nonlinear regime.
Currently, following this to high accuracy over a wide range of scales
can only be achieved using $N$-body methods. However, computation of
the nonlinear power spectrum for all of the cosmological models of
interest is currently prohibitively expensive.

On large-scales, before shell-crossing, one can use Eulerian
perturbative methods to follow the evolution of density and velocity
divergence perturbations in to the weakly nonlinear regime
\citep{Juszkiewicz1981,Vishniac1983,Goroffetal1986,Makinoetal1992,JainBertschinger1994}. Until
relatively recently such methods (described as Standard Perturbation
Theory SPT), were hindered by the fact that the so-called `loop
corrections' would result from the cancellation of large positive and
negative higher order terms to produce a small correction to the
linear spectrum. Extending this to higher orders would result in even
more fine cancellations. However, in the last decade significant
progress was made on this problem through the development of
renormalised perturbation theory \citep[][hereafter
  RPT]{CrocceScoccimarro2006a,CrocceScoccimarro2006b} and the
multi-point propagator approach \citep[][hereafter
  MPT]{Bernardeauetal2008}. In this framework certain diagrams in the
perturbative series could be summed to include an infinite number of
terms, leaving a sequence that involved the summation of positive
terms that were of decreasing importance on a given quasi-linear
scale. In this approach the power spectrum can be modelled at $z=0.0$
to subpercent accuracy on scales $k<0.2\kMpc$. Currently, work is
ongoing to develop an effective field theory approach to modelling the
nonlinear evolution of the cosmic fields
\citep{Carrascoetal2012}. This treats the coarse grained phase space
perturbatively, with the small scale smoothed components of the
phase-space generating an effective sound speed, and thus requiring
the modelling of a stress tensor. It has been claimed by
\citet{Carrascoetal2014} that this method can yield predictions
accurate to better than 1\% on scales $k<0.6\kMpc$. However, recent
work by \citet{Baldaufetal2015} suggests that, owing to the
scale-dependence of the effective-sound-speed parameter $c_{\rm s}$,
the gains are more likely to be limited to $k<0.3\kMpc$. Ultimately,
it likely that the perturbative approaches will be limited to the
scales associated with shell-crossing.

To probe deeper into the nonlinear regime various semi-analytic
methods have also been developed: \citet{Hamiltonetal1991} developed a
scaling ansatz for the integrated correlation function, that was
extended to the power spectrum by
\citet{PeacockDodds1994,PeacockDodds1996}.  In this approach it was
assumed that the power spectrum on a given scale was a function of the
linear spectrum on a remapped scale that was based on continuity
arguments.  At the turn of the Millennium the halo model was developed
by various authors \citep{Seljak2000,PeacockSmith2000,MaFry2000a}, in
which the large-scale contribution arises from the correlations
between different haloes, and the small-scale one from the correlation
of dark matter particles in the same halo.  Elements of these ideas
and the SPT were melted together in the {\tt halofit} code developed
in \citet{Smithetal2003}. This method was further improved by
\citet{Takahashietal2012} who recalibrated the fitting function and
introduced an explicit dependence on the dark energy equation of state
parameter $w_0$. The current claimed precision is $5\%$ for $k<1\kMpc$
for $z<10$ and 10\% for $1\le k\le 10 \kMpc$.

More recently, several empirical approaches have been developed,
largely inspired by techniques borrowed from the field of machine
learning.  An example of such an approach is the Neural Network model
of \citet{Agarwaletal2012,Agarwaletal2014}. They mapped a six
parameter cosmological parameter space using more than 6380
simulations and claim that their {\tt PkANN} code can generate power
spectra with better than 1\% precision on scales $k<0.9\kMpc$ for
redshifts $z\le2$. However, the simulations are, relatively speaking,
of low resolution and of small volume ($N=256^3$, $L=200\Mpc$) by
modern standards. The small boxes mean that they do not accurately
capture large-scale nonlinearities associated with the Baryon Acoustic
Oscillations \citep{Smithetal2007,CrocceScoccimarro2008}. Furthermore,
as was demonstrated in \citet{Heitmannetal2010} and more recently
\citet{Schneideretal2016}, such low-resolution runs are unlikely able
to capture the small-scale structure ($k>1\kMpc$) with the accuracy
required for lensing surveys.  This unfortunately reduces the
practical utility of {\tt PkANN}.

Another impressive development is the {\tt Coyote Universe} project,
which has led to the construction of various emulators and in
particular the {\tt CosmicEMU} code for predicting nonlinear matter
power spectra
\citep{Heitmannetal2009,Heitmannetal2010,Lawrenceetal2010,Heitmannetal2014}.
In \citet{Heitmannetal2010} it was claimed that the {\tt CosmicEMU}
code captured the matter power spectrum to better than 1\% precision
for $k<1\,{\rm Mpc}^{-1}$ ($k\lesssim 1.4\,\kMpc$). This was upgraded
in \citet{Heitmannetal2014} to include variations in the Hubble
parameter and an extension to smaller scales $k<8.6\,{\rm Mpc}$
($k\lesssim 12.3\,\kMpc$).

Lastly, another important contribution is the work of
\citet{Meadetal2015,Meadetal2016} who takes yet again a different
approach to the problem. In their work they assume that the halo model
is broadly correct at generating nonlinear power spectra. However
they argue that it is wrong in detail and introduce several
phenomenological modifications, which they argue accounts for missing
physics from the model: first, BAO suppression is introduced through a
Gaussian damping \`{a} la RPT; a graceful vanishing of the 1-Halo term
on large scales to guarantee linear theory; halo oblation -- to
account for the fact that not all haloes are spherical NFW profiles
\citep{Navarroetal1997}. These modifications introduce 2 new free
parameters.  Using the node points of the {\tt CosmicEMU} code to
calibrate these, they find that their {\tt HMCODE} can recover power
spectra at the level of a few percent for $k<10\kMpc$. The main
advantage of this approach is that it enables physically motivated
extrapolation beyond the constrictive grids of models required by the
machine learning codes. It also enables flexibility for the inclusion
of new physics, such as baryonic effects and modifications to the dark
matter model and gravity.

The aim of this paper is to examine a number of these tools and
confront them with a new series of relatively high-resolution $N$-body
runs that are centred around the Planck CMB mission's best fit
cosmological parameter set \citep{Planck2014XVI}. Furthermore, we aim
to build a power spectrum tool that enables accurate and precise
predictions of the power spectra in this region of parameter
space. The requirements of the method are that: it must use an
accurate Einstein-Boltzmann solver linear input power spectrum, such
as can be provided by {\tt CAMB} \citep{Lewisetal1999}; it must
evaluate a state-of-the-art perturbation theory scheme to generate
predictions on large-scales that are suitable for modelling evolution
of BAO; it must interpolate to state-of-the-art $N$-body simulations
on small scales; lastly, it must gracefully return to one of the lower
precision methods outside of the Planck parameter region; it must be
fast to evaluate and cover $k<10\kMpc$ and $z<5$. One further
requirement is that it must be able to describe a time evolving dark
energy $w$CDM parameter space and with the inclusion of a potential
running of the primordial power spectral index.

The paper breaks down as follows: In \S\ref{sec:background} we provide
an overview of key theoretical concepts, define the cosmological
framework and identify 8 cosmological parameters that we wish to
constrain from data. In \S\ref{sec:sims} we describe the suite of
cosmological simulations and provide an overview of their
generation. In \S\ref{sec:spectraI} we describe how we estimate the
power spectra and construct a composite fiducial spectrum from various
runs. We also validate the initial conditions. In \S\ref{sec:PT} we
make a comparison between our spectra and parameter-free predictions
from the 2-loop MPT. In \S\ref{sec:semi} we compare our fiducial runs
with the predictions from various semi-analytic and fitting function
methods. In \S\ref{sec:newmodel} we present our new semi-analytic
model and show how well it can model results from our fiducial set of
runs. In \S\ref{sec:cosmodep} we explore the dependence of the
nonlinear power spectrum on the cosmological parameters and in
\S\ref{sec:cosmonewmodel} we build the cosmology dependent corrections
for our new model and test them.  Finally, in \S\ref{sec:conclusions}
we summarise our findings, conclude and discuss future improvements to
the method.


\section{Theoretical background}\label{sec:background}


\subsection{The power spectrum}

The density field of matter at spatial position $\bx$ and at time $t$
is written as $\rho(\bx,t)$. We denote the spatial mean of this field
as $\bar{\rho}(t)$. We are mostly interested in the matter density
contrast (or overdensity field sometimes simply referred to as the
density field):
\be \d(\bx,t)\equiv \frac{\rho(\bx,t)-\bar{\rho}(t)}{\bar{\rho}(t)}\ ,\ee
where the above field is by definition mean zero, i.e.
$\left<\delta(\bx,t)\right>=0$, where angled brackets denote an
ensemble average process at fixed coordinate time. A complete
statistical description of the $\delta$--field can be obtained through
determining the $N$-point correlation functions
\citep{ScherrerBertschinger1991}: where for example
$\left<\delta(\bx)\delta(\bx+\br)\right>=\xi(\br)$ is the two-point
correlation function.  For a number of reasons we will prefer to work
in Fourier space, with the transform convention:
\ba
\delta(\bx) & = & \Vu\int \frac{\dk}{(2\pi)^3} \delta(\bk) \exp\left[-i\bk\cdot\bx\right] \ ; \\
\delta(\bk) & = & \frac{1}{\Vu}\int \dx \,\delta(\bx) \exp\left[i\bk\cdot\bx\right] \ ,
\ea
where $\Vu$ is a sufficiently large volume that the coherence length
of the correlators is $\ll \Vu^{1/3}$.  On assuming that the two-point
correlation function is statistically homogeneous
(i.e. $\left<\d(\bx)\d(\bx+\br)\right>=\left<\d({\bf
  0})\d(\br)\right>$), one can easily show that the Fourier space dual
of the correlation function is the power spectrum:
\be
\Vu^2\langle \d(\bk_1)\d(\bk_2)\rangle  =  (2\pi)^3\d^{\rm D}(\bk_1+\bk_2) P(\bk_1)\, \label{eq:defP}; 
\ee
where $\d^{\rm D}$ denotes the Dirac delta function and the power spectrum
is given by:
\be P(\bk) \equiv \int \dx\, \xi(\bx)\exp\left[i\bk\cdot\bx\right] \ .\ee
For the case of statistically isotropic fields, the power spectrum
depends only on the magnitude of $|\bk|$. For the case of a Gaussian
Random Field all of the statistical information is fully captured in
the power spectrum. This makes the power spectrum the lowest order
clustering statistic of interest for cosmology and it is the main
subject of interest for this paper. In real surveys it is usually not
measured directly but it can be extracted modulo reasonable modelling
assumptions.


\subsection{Cosmological model and fiducial parameters:}\label{ssec:params}

The various combinations of large-scale structure data
\citep{BOSS2017a}, weak lensing data
\citep{Kohlingeretal2017,DES2017a}, CMB data \citep{Planck2014XVI} and
Type Ia Supernovae data \citep{Bertouleetal2014}, have identified the
flat, time evolving, dark energy dominated cold dark matter model
(hereafter $w$CDM) as the cosmological model of interest. One of the
major challenges for modern cosmology is to accurately determine the
best fit parameters in this model. The flat $w$CDM model can be
characterised by 8 parameters:
\be \bm \theta = \{w_0,w_a,\Omega_{\rm DE},\Omega_ch^2,\Omega_bh^2,A_s,n_s,\alpha_s\} \ ,\ee
where $w_0$ and $w_a$ are the parameters that govern the time
evolution of the equation of state for dark energy, assuming that the
equation of state can be parameterised in the form:
\be 
w(a)=w_0+w_a(1-a) \ ; \label{eq:EOS}
\ee 
$\Omega_{\rm DE}$ denotes the present day energy density of dark
energy; $w_{\rm c}\equiv \Omega_{\rm c} h^2$ and $w_{\rm
  b}\equiv\Omega_{\rm b} h^2$ are the physical densities in cold dark
matter and baryons today -- $h$ being the dimensionless Hubble
parameter. Note that owing to the assumed flatness, other parameters
can be derived from the above set: the matter density parameter is
$\Omega_{\rm m}=1-\Omega_{\rm DE}$, and the Hubble parameter is
$h=\sqrt{(w_{\rm c}+w_{\rm b})/\Omega_{\rm m}}$.

The matter power spectrum is initialised by specifying the primordial
power spectrum of curvature perturbations and we make use of the
following form \citep{Komatsuetal2009,Planck2014XVI}:
\def\PR{{\mathcal P_{\mathcal R}}}
\be \Delta^2_{\mathcal R}(k)
= A_{\rm s} \left(\frac{k}{k_{\rm p}}\right)^{(n_{\rm s}-1)+\alpha_{\rm s} \log(k/k_{\rm p})/2} \ ,
\label{eq:PRmodel}\ee
where $A_{\rm s}$ is the primordial amplitude, $n_{\rm s}$ and
$\alpha_{\rm s}$ are the spectral index and the running of the
spectral index, all of which are determined at the pivot scale $k_{\rm
  p}$. The running of the spectral index can also be equivalently
written as:
\be \alpha_{\rm s} \equiv \left. \frac{dn_{\rm s}}{d\log k}\right|_{k=k_{\rm p}} \ ,\ee
and is interesting to include, since placing constraints on this may
help constrain inflationary models \citep{Vieiraetal2017}. Hence, the
matter power spectrum can be written in terms of primordial quantites
as \citep[for a discussion of the relation between the primordial
  curvature power spectrum and the matter power spectrum see][in
  prep.]{SmithSimon2019}:
\def\aearly{a_{\rm early}}
\def\aprime{a_{\rm prime}}
\be
P_{\rm m}(k,a) = \frac{8\pi^2}{25}\frac{a^2g^2(\aearly,a)}{\Omega_{\rm m}^2}\frac{c^4}{H_0^4}
\,T^{2}(k,a)\,k\Delta^2_{\mathcal R}(k) \ ,
\ee
where $g(\aearly,a)$ is the growth supression factor from an early
epoch $\aearly$ to $a$, $T(k,a)$ is the matter transfer function at
epoch $a$ (note this should become unity as $k\rightarrow0$) and
$c/H_0$ gives the Hubble scale today.

In this work we will examine the dependence of the nonlinear matter
power spectrum on these parameters and also develop a semi-empirical
approach that will improve the accuracy of current modelling. In what
follows we will assume as out fiducial model parameters that are close
to the best fit set obtained from the \citet{Planck2014XVI}:
$w_0=1.0$, $w_a=0.0$, $\Omega_{\rm DE}=0.693$, $\Omega_ch^2=0.119$,
$\Omega_bh^2=0.0222$, $A_s=2.145\times 10^{-9}$, $n_s=0.9611$,
$\alpha_s=0.0$.  Before continuing, we note that we are not aware of
any recent study that has explored the dependence of the nonlinear
power spectrum on $\alpha_s$ using $N$-body simulations. In what
follows we shall present results for $\alpha_s\ne0$.


\begin{table*}
\centering{
\begin{tabular}{|c|c|c|c|c|c|c|c|c|}
\hline 
Simulation & 
$w_0$ & $w_a$ & $\Omega_{DE}$ & $\omega_c$  &  
$\omega_b$  & $n_s$ & $A_s [\times 10^{-9}]$ &  $\alpha$  \\
\hline
{\tt Fiducial}   & -1.0 & 0.0 & 0.6914 & 0.11889  & 0.022161 & 0.9611  & 2.14818 & 0.0\\
\hline
{\tt V1}   & {\bf -1.1} & 0.0 & 0.6914 & 0.11889  & 0.022161 & 0.9611  & 2.14818 & 0.0\\
{\tt V2}   & {\bf -0.9} & 0.0 & 0.6914 & 0.11889  & 0.022161 & 0.9611  & 2.14818 & 0.0\\
\hline
{\tt V3}   & -1.0 & {\bf -0.2} & 0.6914 & 0.11889  & 0.022161 & 0.9611 & 2.14818 & 0.0\\
{\tt V4}   & -1.0 & {\bf 0.2}  & 0.6914 & 0.11889  & 0.022161 & 0.9611 & 2.14818 & 0.0\\
\hline
{\tt V5}   & -1.0 & 0.0 & {\bf 0.72597} & 0.11889  & 0.022161 & 0.9611 & 2.14818 & 0.0\\
{\tt V6}   & -1.0 & 0.0 & {\bf 0.65683} & 0.11889  & 0.022161 & 0.9611 & 2.14818 & 0.0\\
\hline
{\tt V7}   & -1.0 & 0.0 & 0.6914 & {\bf 0.124835}  & 0.022161 & 0.9611 & 2.14818 & 0.0\\
{\tt V8}   & -1.0 & 0.0 & 0.6914 & {\bf 0.112945}  & 0.022161 & 0.9611 & 2.14818 & 0.0\\
\hline
{\tt V9}   & -1.0 & 0.0 & 0.6914 & 0.11889  & {\bf 0.0232691} & 0.9611 & 2.14818 & 0.0\\
{\tt V10}   & -1.0 & 0.0 & 0.6914 & 0.11889  & {\bf 0.021053}  & 0.9611 & 2.14818 & 0.0\\
\hline
{\tt V11}   & -1.0 & 0.0 & 0.6914 & 0.11889  & 0.022161  & {\bf 1.00916} & 2.14818 & 0.0\\
{\tt V12}   & -1.0 & 0.0 & 0.6914 & 0.11889  & 0.022161  & {\bf 0.913045} & 2.14818 & 0.0\\
\hline
{\tt V13}   & -1.0 & 0.0 & 0.6914 & 0.11889  & 0.022161  & 0.9611 & {\bf 2.363} & 0.0\\
{\tt V13}   & -1.0 & 0.0 & 0.6914 & 0.11889  & 0.022161  & 0.9611 & {\bf 1.93336} & 0.0\\
\hline
{\tt V15}   & -1.0 & 0.0 & 0.6914 & 0.11889  & 0.022161  & 0.9611 & 2.14818 & {\bf 0.01}\\
{\tt V16}   & -1.0 & 0.0 & 0.6914 & 0.11889  & 0.022161  & 0.9611 & 2.14818 & {\bf -0.01}\\
\hline
\end{tabular}}
  \caption{{\tt D\"ammerung} cosmological parameters -- Columns are:
    (1) and (2) denote the equation of state parameter for the dark
    energy $P_{w}=w\rho_{\rm w}=w_0+(1-a)w_a$; (3) density parameter
    for dark energy; (4) and (5) physical densities of CDM and baryons
    (where $\omega_c\equiv\Omega_ch^2$ and $\omega_b\equiv
    \Omega_bh^2$; (6), (7) and (8) the spectral index, amplitude and
    running of the primordial curvature power spectrum. Note that
    since we are assuming flatness, the following parameters are
    derived quantities: the matter density is obtained through
    $\Omega_{\rm m}=1-\Omega_{\rm DE}$; the Hubble parameter is obtained via
    $h=\sqrt{(\omega_c+\omega_b)/(1-\Omega_{\rm DE})}$.
    \label{tab:cospar}}

\vspace{0.5cm}

\begin{tabular}{|c|c|c|c|c|c|c|c|c|c|c|}
\hline 
Simulation & $N_{\rm part}$ & $L \ [\Mpc]$ &  $l_{\rm soft}\ [\Mpc]$ & $p_{\rm mass}\ [10^{9}\Msol]$
& ${\rm PMGRID}$ & $N_{\rm ensemble}$ & $z_{\rm IC}$ & ${\rm \#\, snapshots}$ \\
\hline
{\tt F1 (Big Box)} & $2048^3$ & 3000.0 & 0.05 & 269.0 & $2048^3$ & 1 & 49.0 & 63\\
\hline
{\tt F1}--{\tt F10} & $2048^3$ & 500.0 & 0.008 & 1.246 & $2048^3$ & 10 & 49.0 & 63\\
\hline
{\tt V1} & $2048^3$ & 500.0 & 0.008 & 1.246 & $2048^3$ & 1 & 49.0   & 63\\
{\tt V2} & $2048^3$ & 500.0 & 0.008 & 1.246 & $2048^3$ & 1 & 49.0   & 63\\
\hline
{\tt V3} & $2048^3$ & 500.0 & 0.008 & 1.246 & $2048^3$ & 1 & 49.0   & 63\\
{\tt V4} & $2048^3$ & 500.0 & 0.008 & 1.246 & $2048^3$ & 1 & 49.0   & 63\\
\hline
{\tt V5} & $2048^3$ & 500.0 & 0.008 & 1.107 & $2048^3$ & 1 & 49.0   & 63\\
{\tt V6} & $2048^3$ & 500.0 & 0.008 & 1.386 & $2048^3$ & 1 & 49.0   & 63\\
\hline
{\tt V7} & $2048^3$ & 500.0 & 0.008 & 1.246 & $2048^3$ & 1 & 49.0   & 63\\
{\tt V8} & $2048^3$ & 500.0 & 0.008 & 1.246 & $2048^3$ & 1 & 49.0   & 63\\
\hline
{\tt V9} & $2048^3$ & 500.0 & 0.008 & 1.246 & $2048^3$ & 1 & 49.0   & 63\\
{\tt V10}& $2048^3$ & 500.0 & 0.008 & 1.246 & $2048^3$ & 1 & 49.0   & 63\\
\hline
{\tt V11}& $2048^3$ & 500.0 & 0.008 & 1.246 & $2048^3$ & 1 & 49.0   & 63\\
{\tt V12}& $2048^3$ & 500.0 & 0.008 & 1.246 & $2048^3$ & 1 & 49.0   & 63\\
\hline
{\tt V13}& $2048^3$ & 500.0 & 0.008 & 1.246 & $2048^3$ & 1 & 49.0   & 63\\
{\tt V14}& $2048^3$ & 500.0 & 0.008 & 1.246 & $2048^3$ & 1 & 49.0   & 63\\
\hline
{\tt V15}& $2048^3$ & 500.0 & 0.008 & 1.246 & $2048^3$ & 1 & 49.0   & 63\\
{\tt V16}& $2048^3$ & 500.0 & 0.008 & 1.246 & $2048^3$ & 1 & 49.0   & 63\\
\hline
\end{tabular}
  \caption{The {\tt D\"ammerung} simulation
    parameters \label{tab:gadget} and current run status. Important
    run parameters were set as follows: {\tt ErrTolIntAccuracy}=0.015,
    {\tt MaxRMSDisplacementFac}=0.2; {\tt MaxSizeTimestep}=0.02; {\tt
      MinSizeTimestep}=0.00; {\tt ErrTolTheta}=0.5; {\tt
      ErrTolForceAcc}=0.005; {\tt RCUT=4.5} ; {\tt ASMTH}=1.25} \label{tab:SimPar}
\end{table*}


\section{The D\"aemmerung Simulations}\label{sec:sims}

We have generated a series of $N$-body simulations to explore cosmic
structure formation in the nonlinear regime.

\subsection{Overview of fiducial runs}

For our fiducial simulations we have generated two types of run: small
box runs that have high resolution and a large volume run to capture
large-scale nonlinearities. Each of the simulations was generated as
follows: first, we adopted a fiducial cosmological model and for this
we chose cosmological parameters that were consistent with the Planck
CMB analysis best-fit \citep{Planck2014XVI}. The exact values that we
used were presented in \S\ref{ssec:params} and are also repeated in
the first row of Table~\ref{tab:cospar}. We then ran the
Einstein-Boltzmann solver code {\tt CAMB} to generate the linear
theory matter power spectrum at $z=0$ for the fiducial model.  This
was used as the input linear power spectrum for our upgrade of the
publicly available {\tt 2LPT}\footnote{\tt
  http://cosmo.nyu.edu/roman/2LPT/} C-code developed in
\citet{Scoccimarroetal2012} -- our upgrade makes various modifications
to the original code, in particular the use of {\tt FFTW3} MPI
parallel Fourier Transform libraries and the code has been tested for
particle loads up to $N=4096^3$. The linear power spectrum was
rescaled back to $z=49$ using the appropriate linear growth factor
(for further details as to how we calculate this for all our models
see \S\ref{ssec:growth}).  The $N$-body particles were distributed
onto a cubical lattice of size $N=2048^3$ and the particles were then
displaced off their lattice points using the 2LPT recipe.

The $N$-body simulations were then evolved under gravity in an
expanding universe framework using the OpenMPI, parallel Tree--PM code
{\tt Gadget-3} developed by \citet{Springel2005} and
\citet{Anguloetal2012} and used for the generation of the Millennium
XXL Virgo run. The upgraded features of the code meant that halo and
sub-halo catalogues along with various statistical measures, including
the matter power spectra, were calculated `on-the-fly'.  Each of the
small-box runs was performed with $N=2048^3$ dark matter particles, in
a comoving box of size $L=500\Mpc$, yielding a mass per particle of
$m_{\rm p}=1.2\times10^9\Msol$. The large-box runs also followed
$N=2048^3$ dark matter particles, but in a comoving box of size
$L=3000\Mpc$, yielding a mass per particle of $m_{\rm
  p}=2.69\times10^{11}\Msol$.

We output 60 snapshots between $z=49$ and $z=0$, with a hybrid
linear-logarithmic output spacing that matched the Millennium Run I
simulation \citep{Springeletal2005}. The simulations were run on the
SuperMUC machine at the Leibniz Rechnum Zentrum in Garching and also
the MPG Hydra cluster in Garching. The full particle data storage per
run was of the order 20 TB. Various properties of interest for the runs
are presented in Table~\ref{tab:SimPar}. For the small-box runs we
adopted this particle mass resolution and box-size for the reason that
this would lead to converged 1\% precision results on the power
spectrum on scales $\gtrsim 1\kMpc$ \citep{Schneideretal2016} and that
it would also enable accurate tracking of sub-structures that are
required for semi-analytic galaxy formation modelling.

For completeness we have also included in Table~\ref{tab:SimPar} a
list of the {\tt Gadget-3} code parameter settings that we used.  As
was shown in \citet{Reedetal2013} and \citet{Smithetal2014} a careful
choice of these parameters is required to keep numerical errors below
the percent level.  Here, in order to increase integration accuracy,
we have chosen a relatively small-timestep, which was set through the
parameter $\eta={\tt ErrTolIntAccuracy}=0.015$, and where $\Delta t
\propto \eta^{1/2}$. We have used this for all runs and a typical
small box run required several thousand timesteps to complete.


\begin{figure*}
\centerline{
  \includegraphics[angle=0,width=8.4cm]{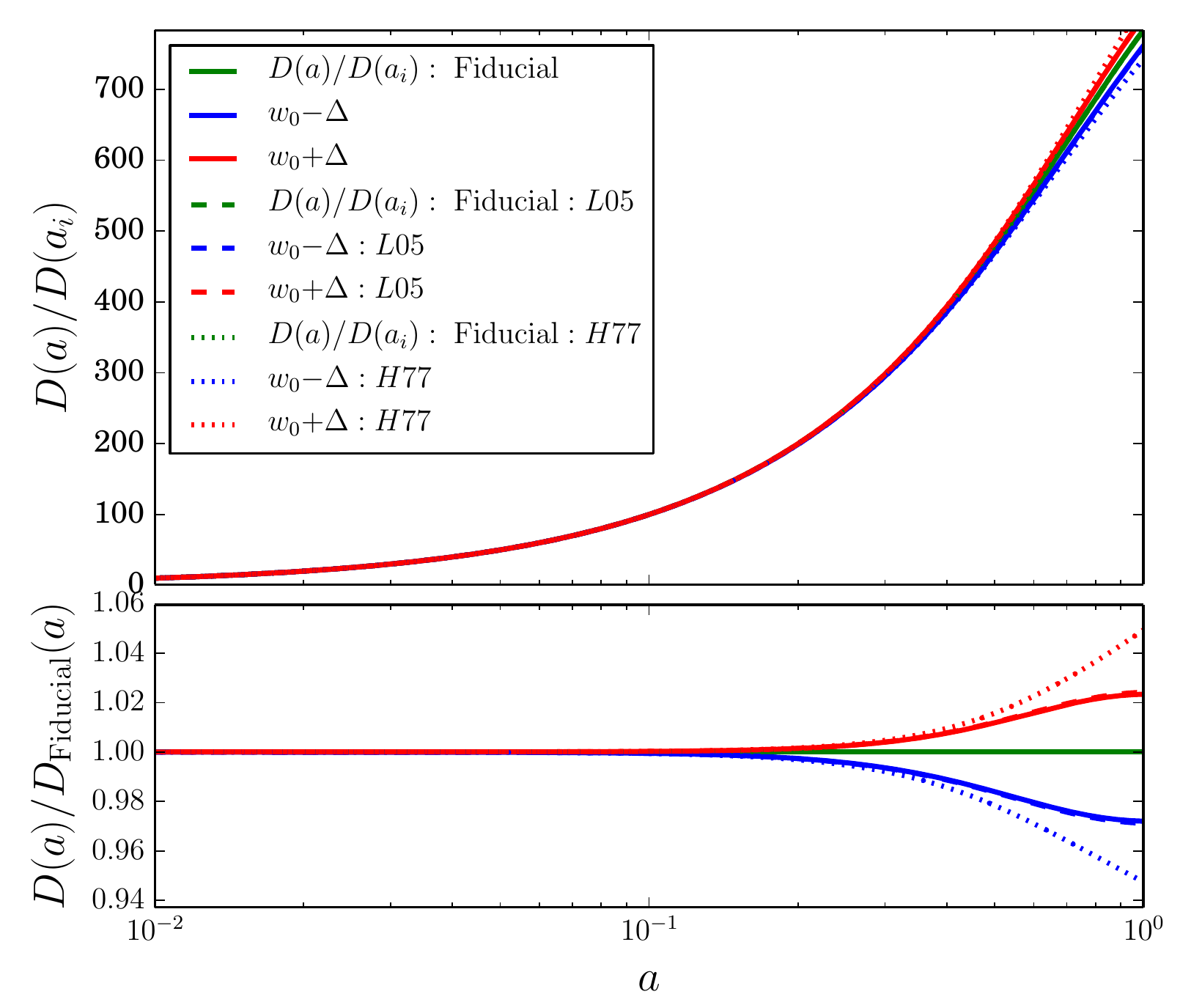}\hspace{0.3cm} 
  \includegraphics[angle=0,width=8.4cm]{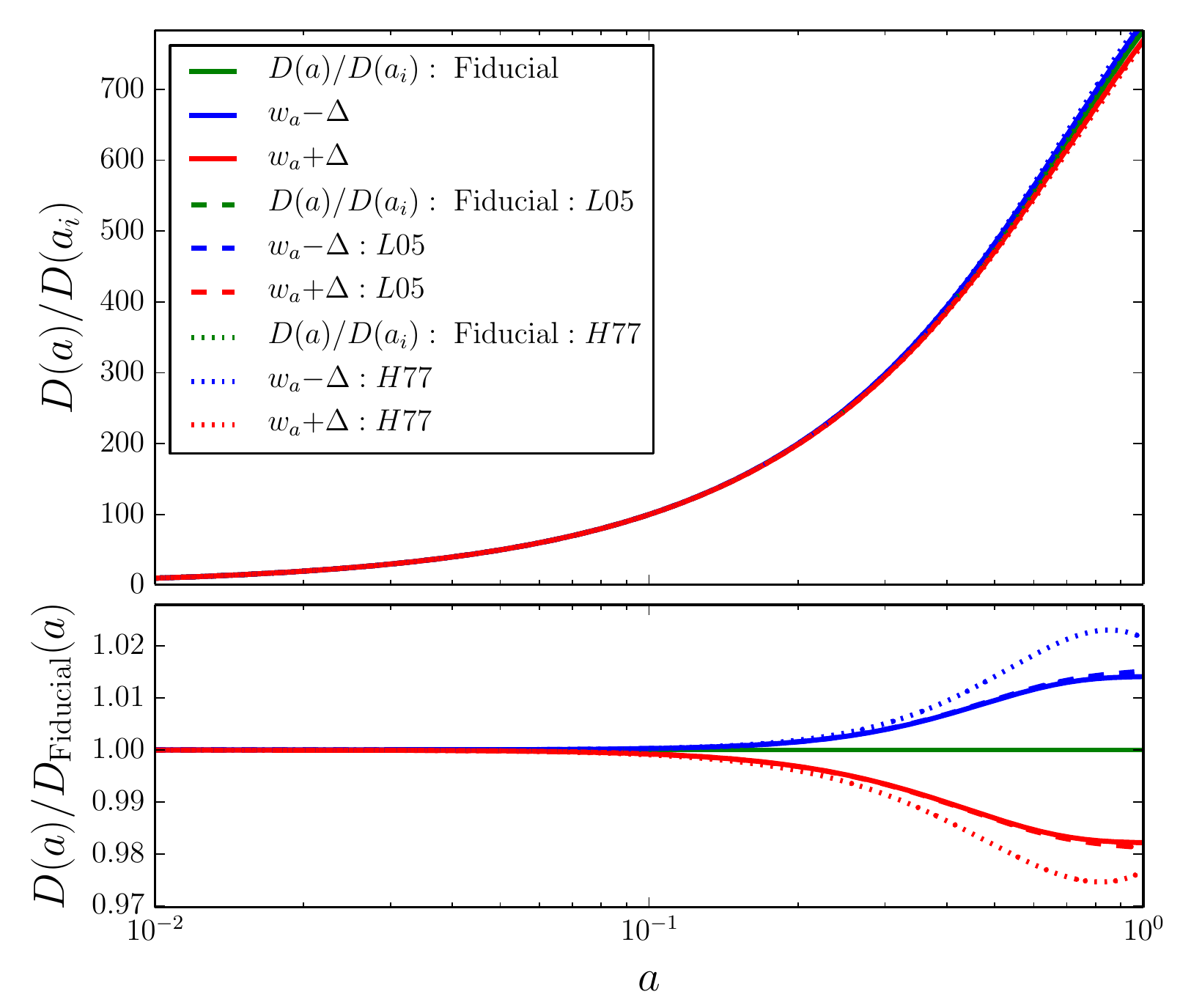} }
\caption{\small{Evolution of the linear growth factor as a function of
    expansion factor. The left and right sub-figures show the results
    for variations in $w_0$ and $w_a$, respectively. The top panels
    show the growth normalised to the initial time and the lower
    panels show the ratio of the variations in the cosmological model
    with respect to the fiducial model. In all panels the solid lines
    denote the 4th order Runge-Kutta solution of \Eqn{eq:g2}, the
    dashed lines denote the approximate expression from
    \citet{Linder2005} and the dotted lines denote the
    \citet{Heath1977} expression, which is only exact for presureless
    fluids. The green lines denote the fiducial model, the red lines
    show $w_0-0.9$ (let panel) and $w_a=-0.2$ (right panel), and the
    blue lines denote $w_0=-1.1$ (let panel) and $w_a=0.2$ (right
    panel).}
\label{fig:growth}}
\end{figure*}


\subsection{Overview of cosmology variations}

In order to explore the dependence of the nonlinear structure
formation on the cosmological parameters we have generated a further
set of 16 small-box simulations. Rather than sample our 8-dimensional
parameter space for maximum coverage, as has been done for example in
the Coyote Universe Project, where a Latin hypercube approach was
adopted, we instead focus on the idea of generating a Taylor expansion
model around our fiducial point, but relative to some preexisting
theoretical model. For our 8 parameters this can be done to good
accuracy by running an extra two simulations for each parameter: one
that represents a small positive increase in the parameter and another
that gives the response for a negative change.  The exact values of
the cosmological parameters that we have used for each variation are
listed in Table~\ref{tab:cospar}.

The procedure for generating each of the variational simulations was
exactly as described for the case of the small-box fiducial runs. In
order to minimise large-scale cosmic variance between runs we have
matched the Fourier mode phase distribution of each run to those of
the first fiducial small-box run.

As stated, all of the variation simulations were performed using the
standard {\tt Gadget-3} code, with the exception of the runs that
explore variations in the time evolution of the dark energy equation
of state parameter $w(a)=w_0+(1-a)w_a$. In order to perform these runs
we made the following modifications to the {\tt Gadget-3} code. First,
we made the global replacement of the parameter ${\tt
  OmegaLambda}\rightarrow {\tt OmegaDE}$ throughout the code.  Second,
we introduced two new free parameters {\tt w0} and {\tt wa} into the
structure {\tt global\_data\_all\_processors} and the snapshot header
structure {\tt io\_header}, ensuring that the total byte size remained
the same. Third, we made the following global replacement for the
Hubble parameter, contained in the {\tt darkenergy.c} file accessed
through {\tt double INLINE\_FUNC hubble\_function(double a)}]:
\ba & & H^2(a) = H^2_0\left[\frac{}{}\Omega_{\rm m,0} a^{-3} + \Omega_{{\Lambda},0}
  + \Omega_{k,0}a^{-2} 
\right]\ \nn \\
\Longrightarrow \hspace{-0.6cm} && 
H^2(a) = H^2_0\left[\frac{}{}\Omega_{\rm m,0}a^{-3}+\Omega_{\rm DE,0} f(a)a^{-3}  + \Omega_{k,0}a^{-2} \right] 
\ea
and where we have defined
\be
f(a)\equiv a^{-3(w_0+w_a)}\exp\left[3w_a\left(a-1\right)\right] \ ,
\ee
and $\Omega_{k,0}=(1-\sum_i\Omega_{\rm i,0})$ is the curvature density
parameter.
Finally, some additional small adjustments were also necessary to the
following parts of the code {\tt io.c, begrun.c and read\_ic.c}, which
was mainly for I/O of the new parameters.


\subsection{Linear growth factor}\label{ssec:growth}

In generating the initial particle distributions using the 2LPT
algorithm we need to evolve back the $z=0$ linear theory power
spectrum (from {\tt CAMB}) to the start redshift, which for all our
runs was $z=49$. We do this by computing the linear growth factor and
for all of the models that we consider this can be done as follows.

At early times the matter density \mbox{$\delta_{\rm m}=f_{\rm b}\delta_{\rm
  b}+f_{\rm c}\delta_{\rm c}$}, can be expressed as:
\be \delta_{\rm m}(a,\bx)=\frac{D(a)}{D(a_i)}\delta_{\rm m}(a_i,\bx)\ ,\ee
where $D(a)$ gives the time evolution of the density perturbation in
the growing mode and $a_i$ gives the normalised expansion factor at
the initial time. Under the assumption that CDM and baryon
fluctuations are unbiased with respect to each other \citep[see][for a
  discussion]{SomogyiSmith2010}, this function can be obtained by
solving the second order, ordinary differential equation (ODE) that
results for a single collisionless fluid \citep{LinderJenkins2003}:
\be
D''(a)+\Gamma_1(a)D'(a)+\Gamma_2(a)D(a)  =0  \label{eq:g2}
\ee
where $'\equiv d/da$ indicates derivatives with respect to the expansion factor
and for the case of our time evolving dark energy model given by
\Eqn{eq:EOS}, the time dependent coefficients are:
\ba 
\Gamma_1(a) & = & \frac{3}{2a}\left[1-\frac{w_0+(1-a)w_a}{1+X(a)}\right] \ ; \\
\Gamma_2(a) & = & -\frac{3}{2a^2}\left[\frac{X(a)}{1+X(a)}\right] \ ; \\
X(a) & = & \frac{\Omega_{\rm m,0}}{1-\Omega_{\rm m,0}}f(a) \ . 
\ea
In Appendix~\ref{app:one} we show how one can use a 4th order
Runge-Kutta algorithm to solve the above differential equation to
obtain the growing solution $D(t)$ and as a by product the logarithmic
growth rate $f(a)\equiv d\log D/d\log a$. The appendix shows that the
relative error in the solution is $10^{-5}$ for all times of interest.

Figure~\ref{fig:growth} compares the time evolution of the linear
growth factor for the four dark energy models listed in
Table~\ref{tab:cospar} with that of our fiducial model.  The left
panel shows the results for the two constant $w_0$ models and the
right panel the same but for the $w_a\ne0$ models. All of the growth
functions have been normalised to have the same amplitude at the
initial time $a_i$. The solid line denotes the results from the 4th
order Runge-Kutta method.  We see that a 10\% variation in $w_0$ will
lead to growth variations of $\pm2.5\%$ and that variations in
$w_a=\pm0.2$ lead to variations of the order $\pm1.5\%$, and roughly
double that for their impact on the power spectrum.

The figure also shows the growth functions that you would get if you
were to assume that the solution of \citet{Heath1977} would hold for
the case of pressured fluids (for more details and discussion see
Appendix~\ref{app:one}). As the figure clearly shows, the growth
functions from the Heath approach are inaccurate at the level of
several percent for the dark energy models of interest and so should
not be used for model building and predictions where high accuracy is
required.

Lastly, the figure also shows the result of evaluating the approximate
expression (dashed lines in Fig.~\ref{fig:growth}):
\be D(a) = a g(a) = a \exp\left[\int_{a_i}^{a} \frac{da}{a}\left(\Omega_{\rm
    m}^{\gamma}(a)-1\right)\right] \ , \label{eq:growlinder}\ee
where $\gamma=0.55+0.05\left[1-w(a=0.5)\right]$
\citep{Linder2005}. This provides an excellent description (of the
order $\sim$0.1\% precision) of the variations in the growth for the
dark energy models considered. However, since this also involves a
numerical integral we recommend the reader to code up the Runge-Kutta
solution, since it is more general and flexible.


\section{The measured power spectra}\label{sec:spectraI}

\subsection{Estimating the power spectrum}

For a given realisation of the density field, an estimator for the
power in a given Fourier mode is:
\be
\hat{P}(\bk_1,\bk_2)  = \Vu\d(\bk_1)\d(\bk_2)\delta^{K}_{\bk_1+\bk_2=\bf 0} \ ,
\label{eq:Pkest1}\ee
where $V_{\mu}=L^3$ is the volume of the simulation. However, the
noise in such an estimate is very large and in order to overcome this
one must sum over a set of Fourier wavemodes in a thin $k$-shell.
Introducing the binning function $\tilde\Pi_k(\bq) = 1$ for $|\bq|\in
[k-\Delta k/2, k+\Delta k/2]$ and zero otherwise, the power spectrum
estimate averaged over a bin of width, $\Delta k$, is given by
\ba
&& \hspace{-0.6cm}\hat{P}(k) = \int d^3 \bq_1\int d^3 \bq_2\, \delta^D\!(\bq_1 + \bq_2)\frac{\hat{P}(\bq_1,\bq_2)
  \tilde\Pi_k(\bq_1)\tilde\Pi_k(\bq_2)}{V_P(k)} \ \nn \\
&& \  
=\int d^3 \bq_1 \frac{\hat{P}(\bq_1,-\bq_1)
  \tilde\Pi_k(\bq_1)\tilde\Pi_k(-\bq_1)}{V_P(k)}  \nn \\
&& \ 
=\frac{\Vu}{V_P(k)} \int d^3 \bq_1 \left|\d(\bk_1)\right|^2 
\tilde\Pi_k(q_1) \ ,\label{eq:Pkest2}
\ea 
where
\ba V_P(k) & \equiv & \!\! \int d^3 \bq_1\int d^3 \bq_2 \d_D(\bq_1 + \bq_2)
\tilde\Pi_k(\bq_1)\tilde\Pi_k(\bq_2) \nn \\
& = & \!\!\int d^3\bq_1
\tilde\Pi_k(\bq_1)\tilde\Pi_k(-\bq_1) \nn \\
& = & \!\!\int d^3\bq_1\tilde\Pi_k(\bq_1) = 4 \pi k^2 \Delta k
\left[1+\frac{1}{12}\left(\frac{\Delta k}{k}\right)^2\right] \  \ea
is the volume of a spherical shell satisfying these conditions.  A
practical implementation of the above estimator for $N$-body
simulations is described in \citet{Smithetal2003} and
\citet{Jing2005}.

The computation of the power spectrum that we employ is embedded in
the {\tt Gadget-3} code itself and so makes efficient use of the
built-in domain decomposition routines and also the fast MPI parallel
FFTs as implemented by {\tt FFTW} \citep{FFTW}. In brief, the
simulation particles are distributed on to a cubical lattice. The
particles are then assigned to the FFT mesh using the `cloud-in-cell'
(CIC) mass assignment scheme (note that we reuse the PM force grid for
this so the grid size is specified by {\tt PMGRID}). This gives
$\delta_W^{\rm d}(\bx)$ the discrete density field convolved with the
CIC window. This we Fourier transform using the FFT algorithm to
obtain $\delta_W^{\rm d}(\bk)$. The CIC scheme is then deconvolved
using \citep{Jing2005}:
\be \delta^{\rm d}(\bk) = \frac{\delta_W^{\rm d}(\bk)}{W_{\rm
    CIC}(\bk)} \ee
where
\be W_{\rm CIC}(\bk)=\prod_{i\in \{x,y,z\}}\left\{\rm Sinc\left(\frac{\pi
  k_i}{2k_{\rm Ny}}\right)\right\}^2\ , \ee
where ${\rm Sinc}(x)\equiv\sin x/x$. The power spectrum of the
point-sampled field is then estimated using \Eqn{eq:Pkest2}:
\be \hat{P}^{\rm d}(k) = \frac{\Vu}{N_k}\sum_{i=1}^{N_k} \left|\delta^{\rm d}(\bk_i)\right|^2 \ ,\ee
where $N_k=V_{P}(k)/(2\pi)^3/L^3$ is the number of Fourier modes in a
$k$-space shell.

Following \citet{Peebles1980} the above estimate of the true power
spectrum is biased by the addition of the variance from the point
sampling procedure. However, this can be corrected for using:
\be P^d(k)=P^c(k)+\frac{1}{\nbar} \ , \label{eq:autospectrum}\ee
where $P^c$ is the power spectrum of the underlying continuous
field.

In what follows, we use FFT grids with $N_{\rm grid}=2048$ cells per
dimension, and this sets the minimum and maximum spatial frequencies
for a given power spectrum to: $k_{\rm min}=2\pi/L$ and $k_{\rm
  Ny}=\pi N_{\rm grid}/L$. In practice, the power on length scales
$k>k_{\rm Ny}$ will get `aliased' to larger spatial scales
\citep{Jing2005}, and so we take the rule of thumb $k_{\rm max}=k_{\rm
  Ny}/2$. Thus for the $L=500\Mpc$ runs the low-$k$ and high-$k$
cut-offs for the power spectra are $k^{500}_{\rm fun}=0.012\kMpc$ and $k^{500}_{\rm
  max}=6.4\kMpc$, respectively. For the $L=3000\Mpc$ box run the
cut-offs are: $k^{3000}_{\rm fun}=0.0021\kMpc$ and $k^{3000}_{\rm max}=1.072\kMpc$.


\begin{figure*}
\centerline{
  \includegraphics[angle=0,width=7.2cm]{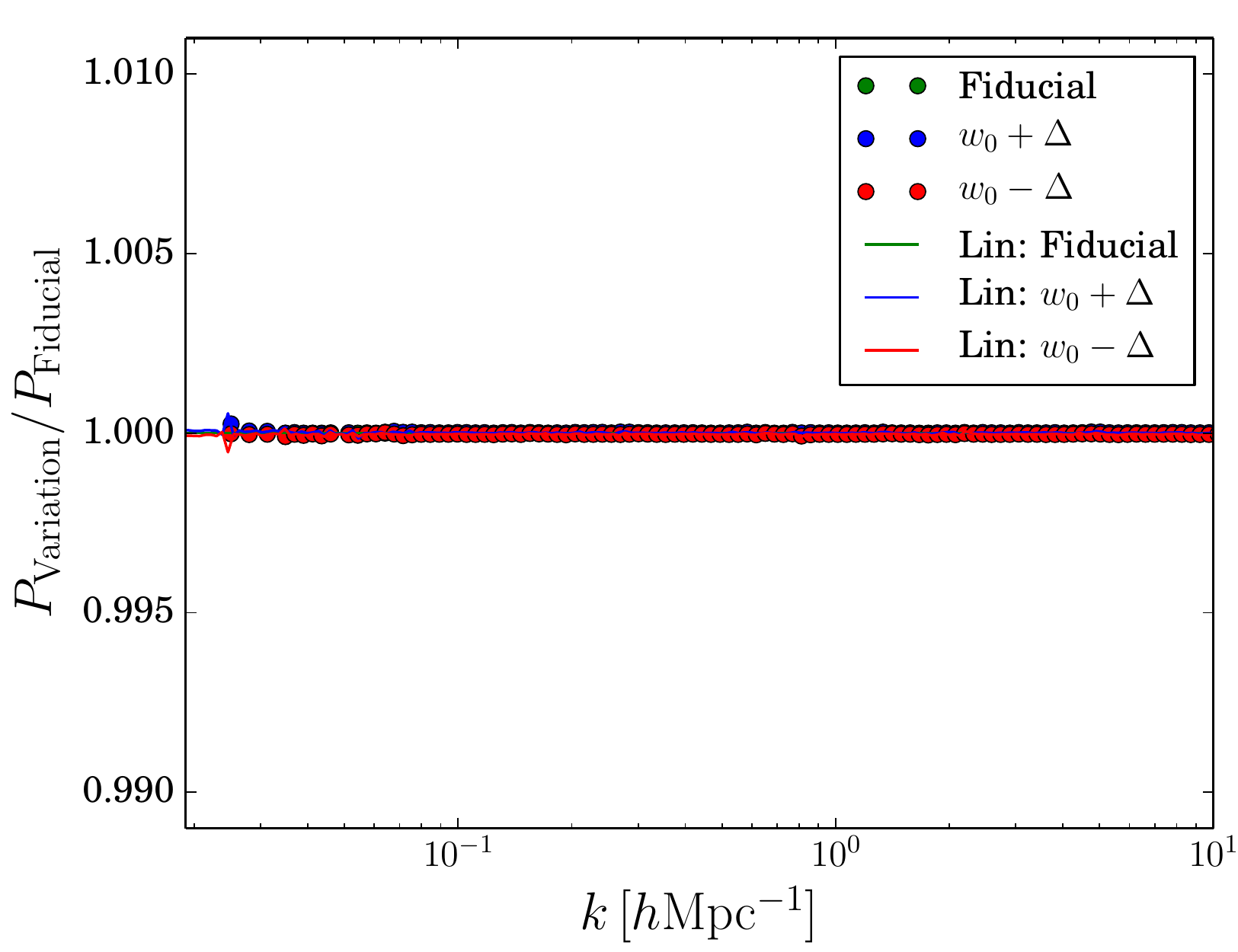}\hspace{0.5cm}
  \includegraphics[angle=0,width=7.2cm]{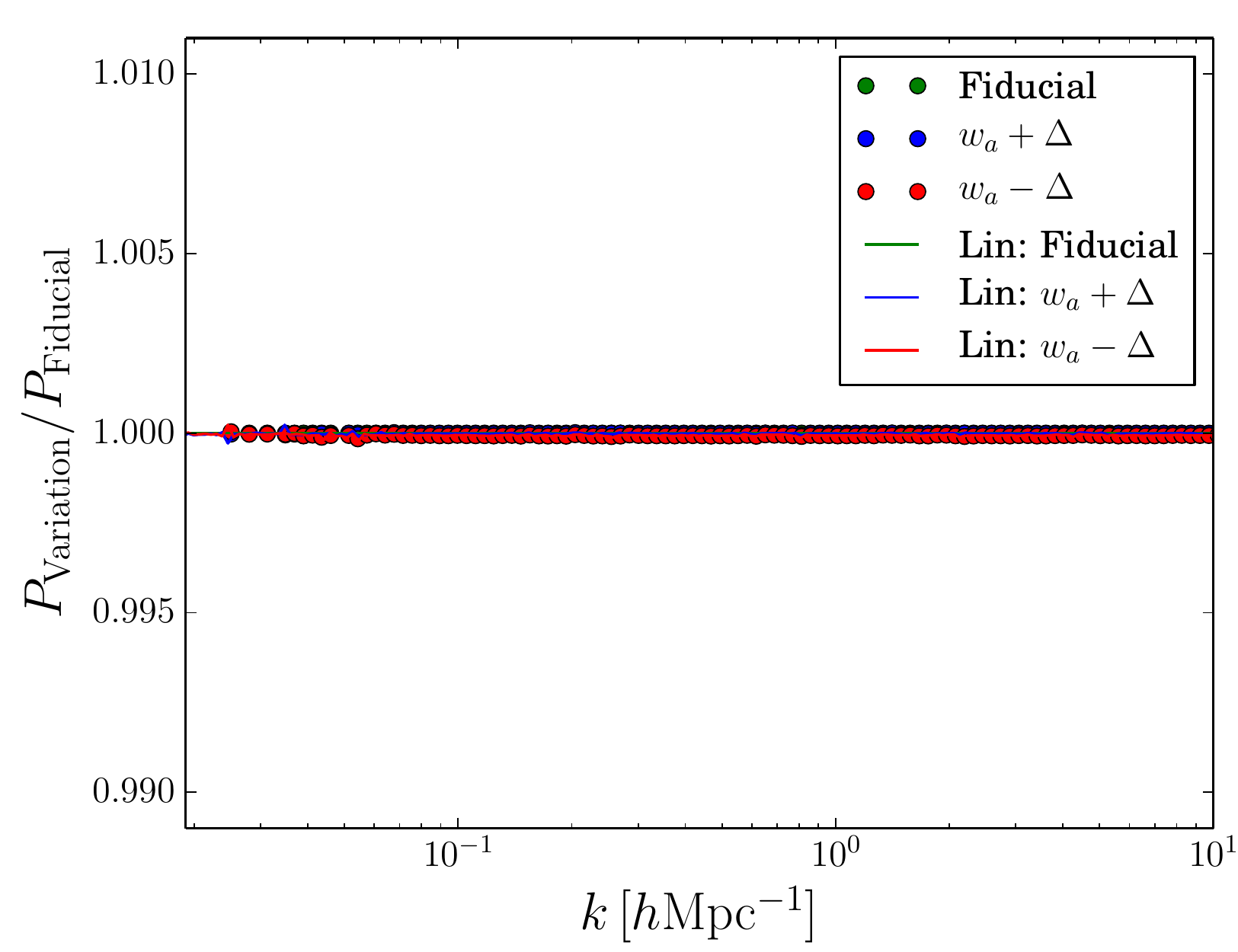}}
\centerline{
  \includegraphics[angle=0,width=7.2cm]{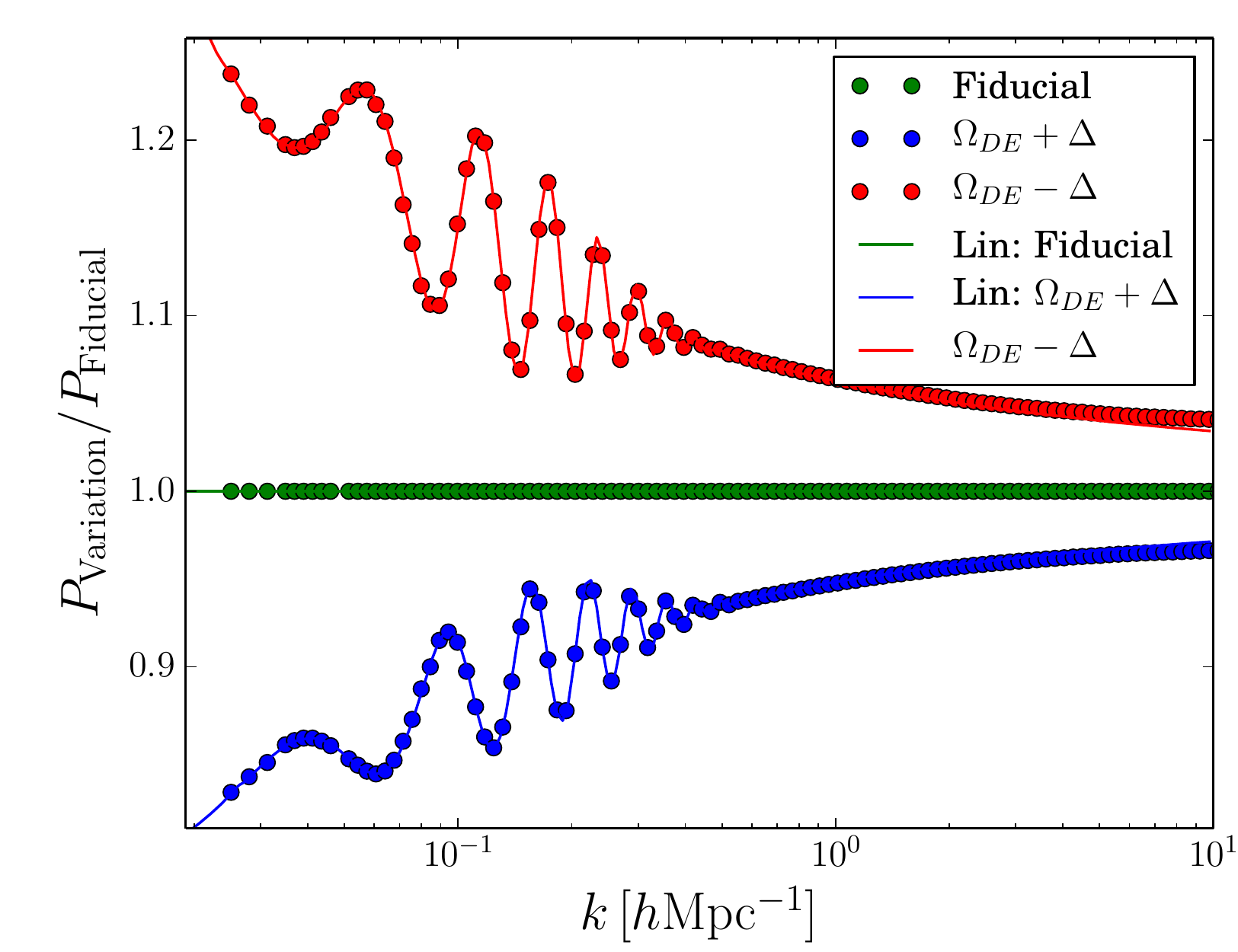}\hspace{0.5cm}
  \includegraphics[angle=0,width=7.2cm]{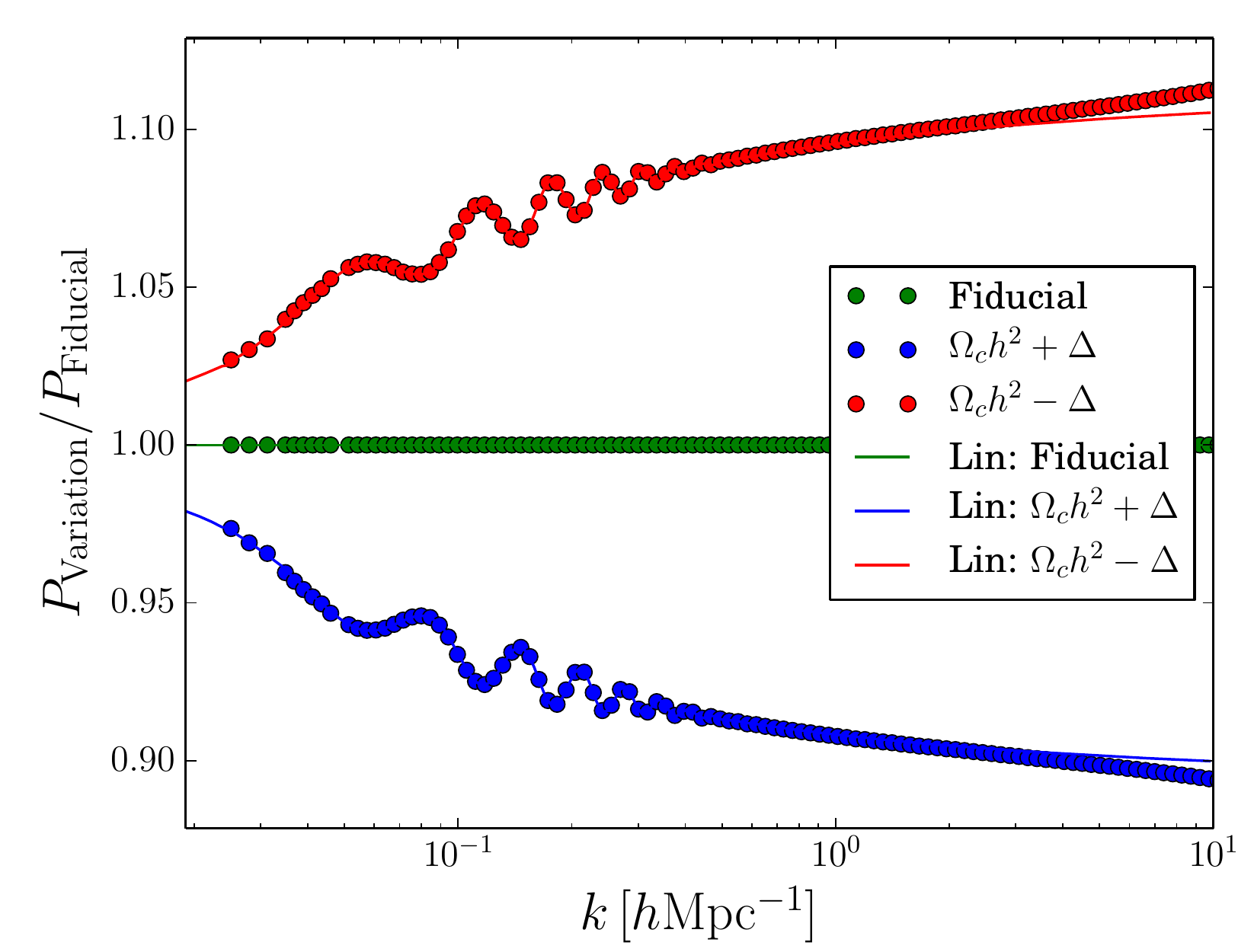}}
\centerline{
  \includegraphics[angle=0,width=7.2cm]{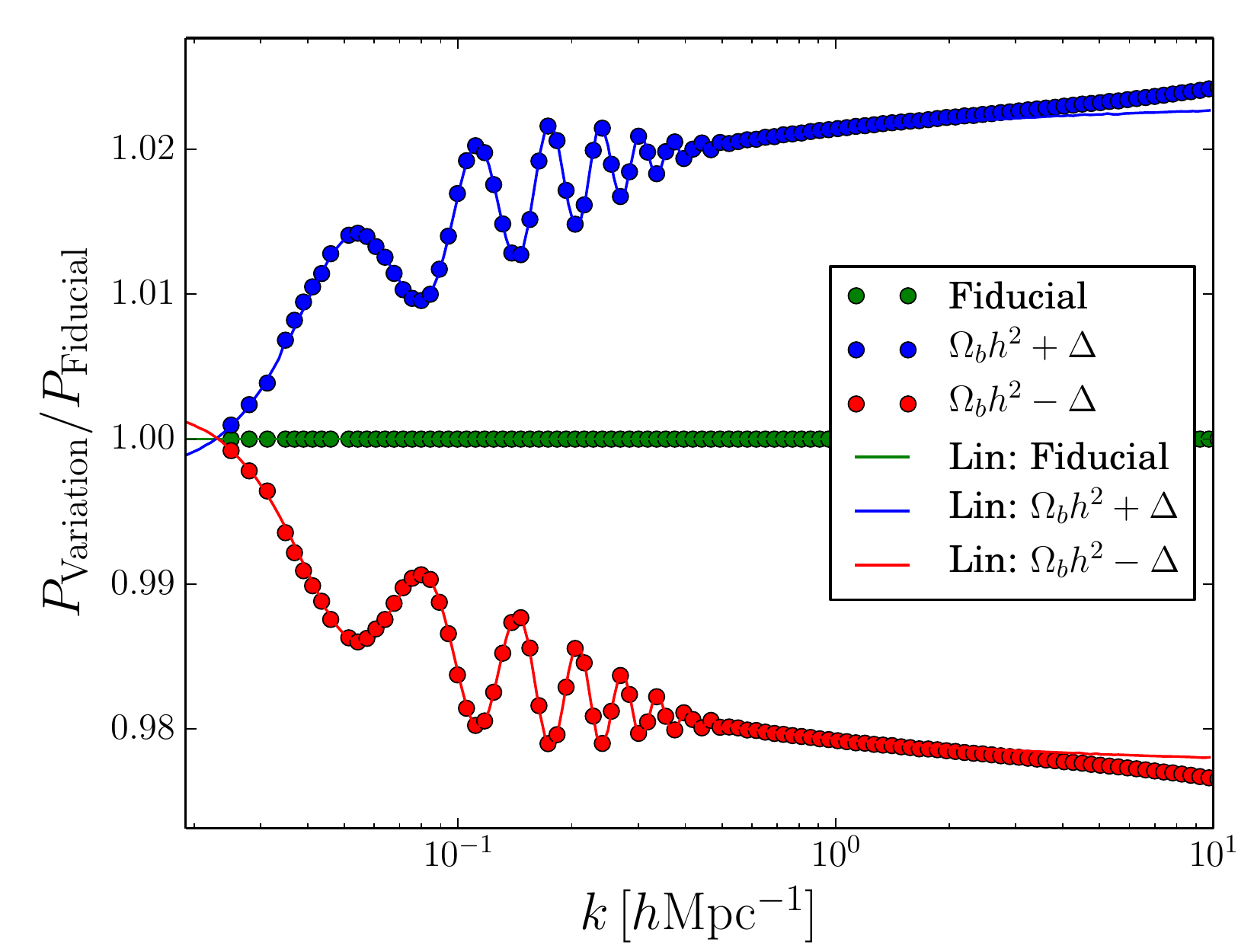}\hspace{0.5cm}
  \includegraphics[angle=0,width=7.2cm]{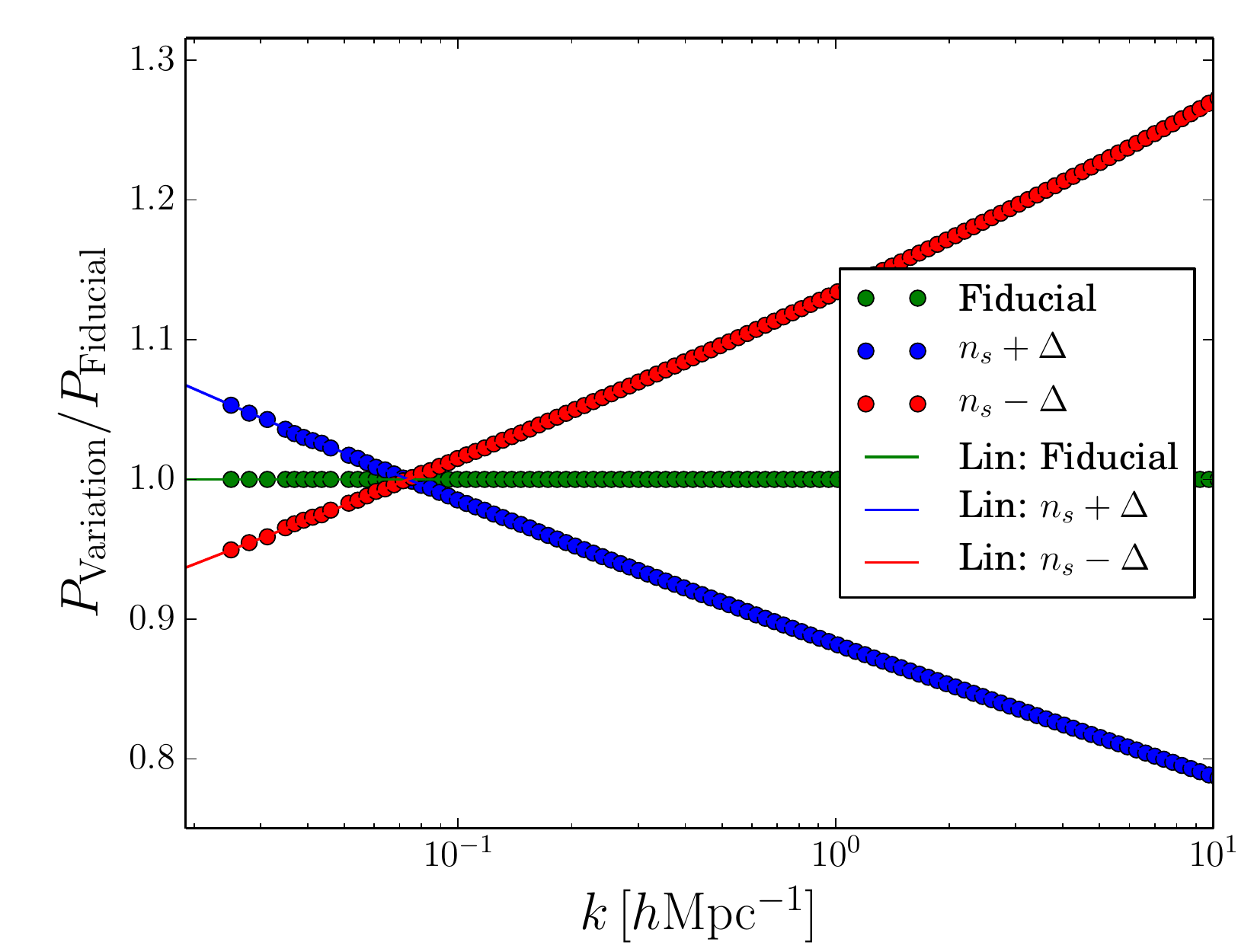}}
\centerline{
  \includegraphics[angle=0,width=7.2cm]{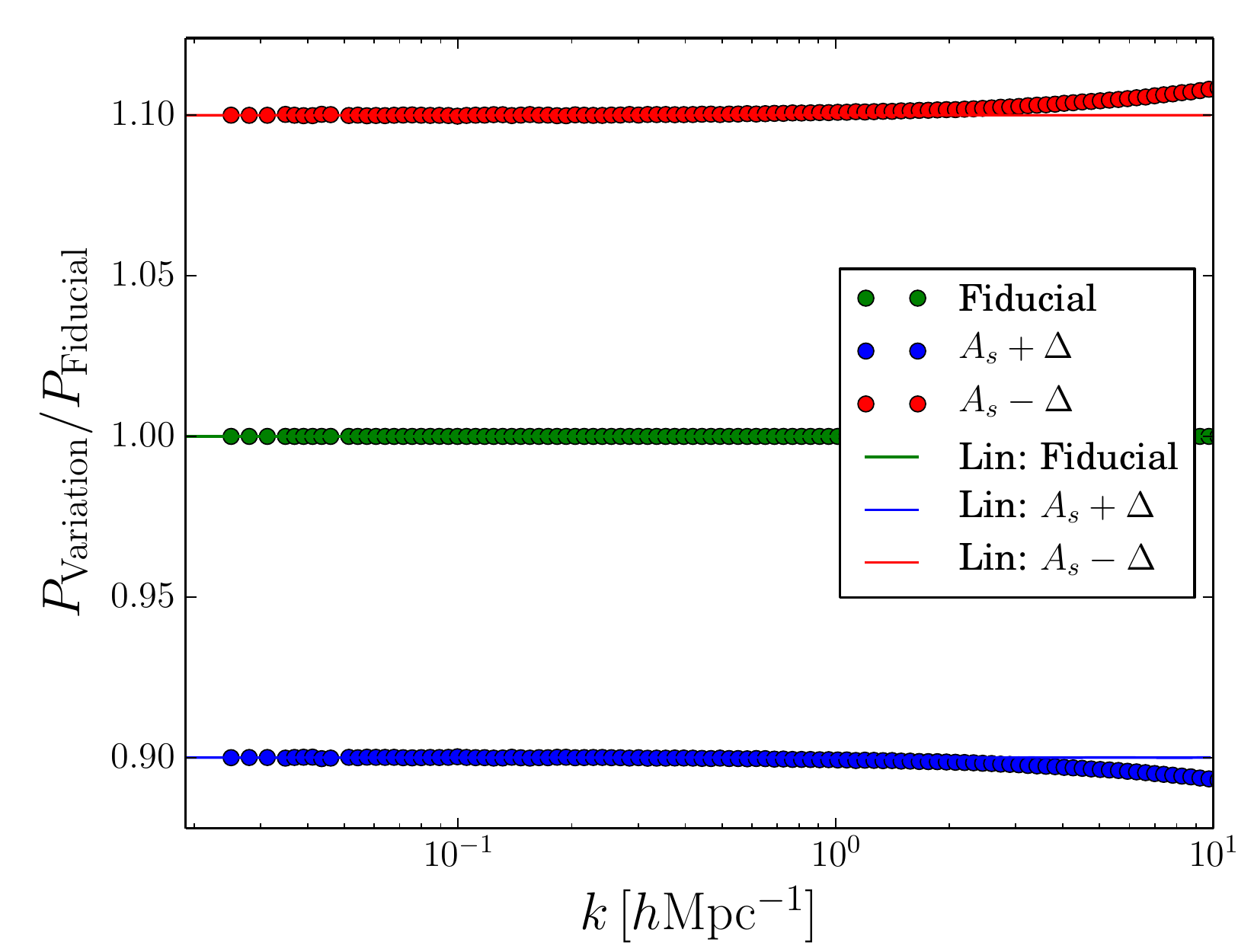}\hspace{0.5cm}
  \includegraphics[angle=0,width=7.2cm]{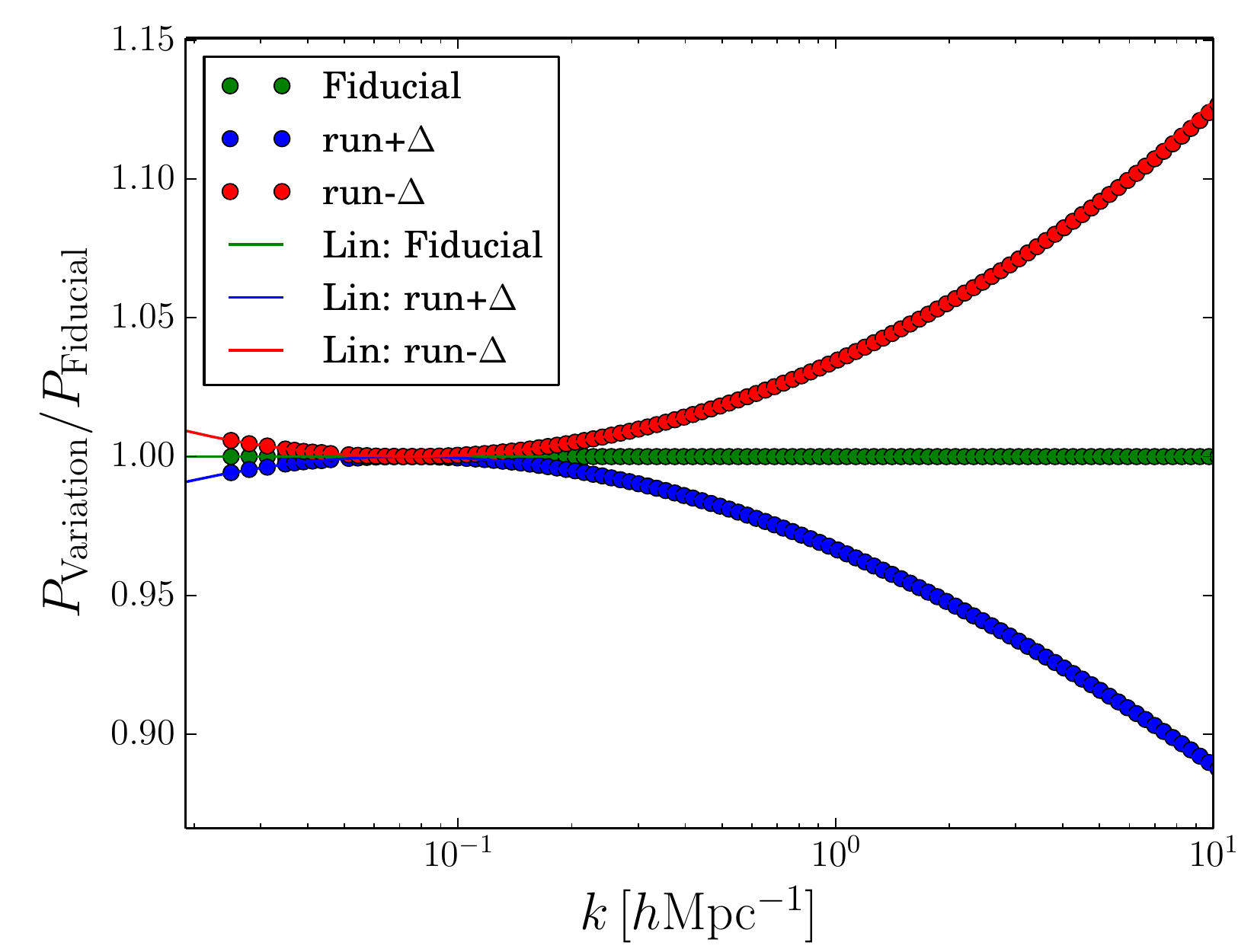}}
\caption{\small{Ratio of the power spectra of the {\tt 2LPT} initial
    conditions for the variational runs with the corresponding power
    spectrum from the fiducial run. All spectra were measured at
    $z=30$. The points show the measurements from the simulations and
    the lines show the linear theory from {\tt CAMB}. The red and blue
    points denote the positive and negative variation in the particular
    parameter from the fiducial, which is shown in green. From left to
    right, the top row shows the variations in $\{w_0,\,
    w_a,\,\Omega_{\rm DE}\}$, the middle row, $\{w_c=\Omega_{\rm
      c}h^2,\, w_b=\Omega_{\rm b}h^2,\,n_s\}$, the bottom row
    $\{A_s,\,\alpha_s\}$.}
\label{fig:PkIC}}
\end{figure*}


\subsection{Validation of the initial conditions}

In Figure~\ref{fig:PkIC} we present the ratios of the initial power
spectra for the variational runs with respect to the fiducial model as
generated by our {\tt 2LPT} code and evolved using {\tt Gadget-3} to
$z=30$. Each of the eight panels shows the results for variations in
one of the cosmological parameters, with all of the others frozen. The
positive/negative variation in each of the cosmological parameters is
denoted by the red/blue points. The solid red and blue lines denote
the predictions from linear theory as obtained from the {\tt CAMB}
code. One can see that taking the ratio with matched phase initial
condition means that the cosmic variance has been cancelled on
large-scales. We also note that the measured data points and linear
predictions agree to very high accuracy, validating the initial density
spectra for each simulation.


\begin{figure*}
\centerline{
  \includegraphics[width=6.2cm]{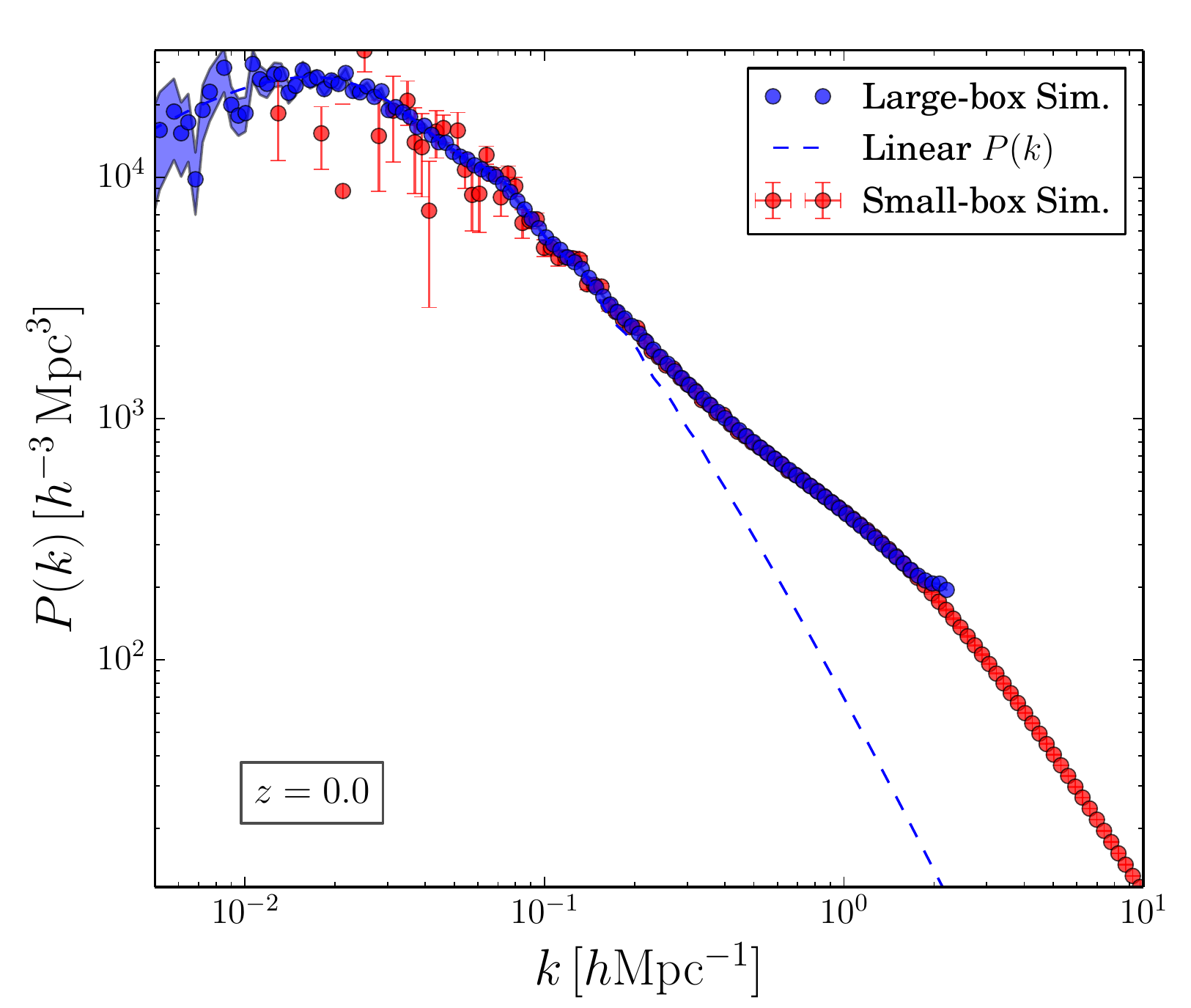}\hspace{0.3cm}
  \includegraphics[width=6.2cm]{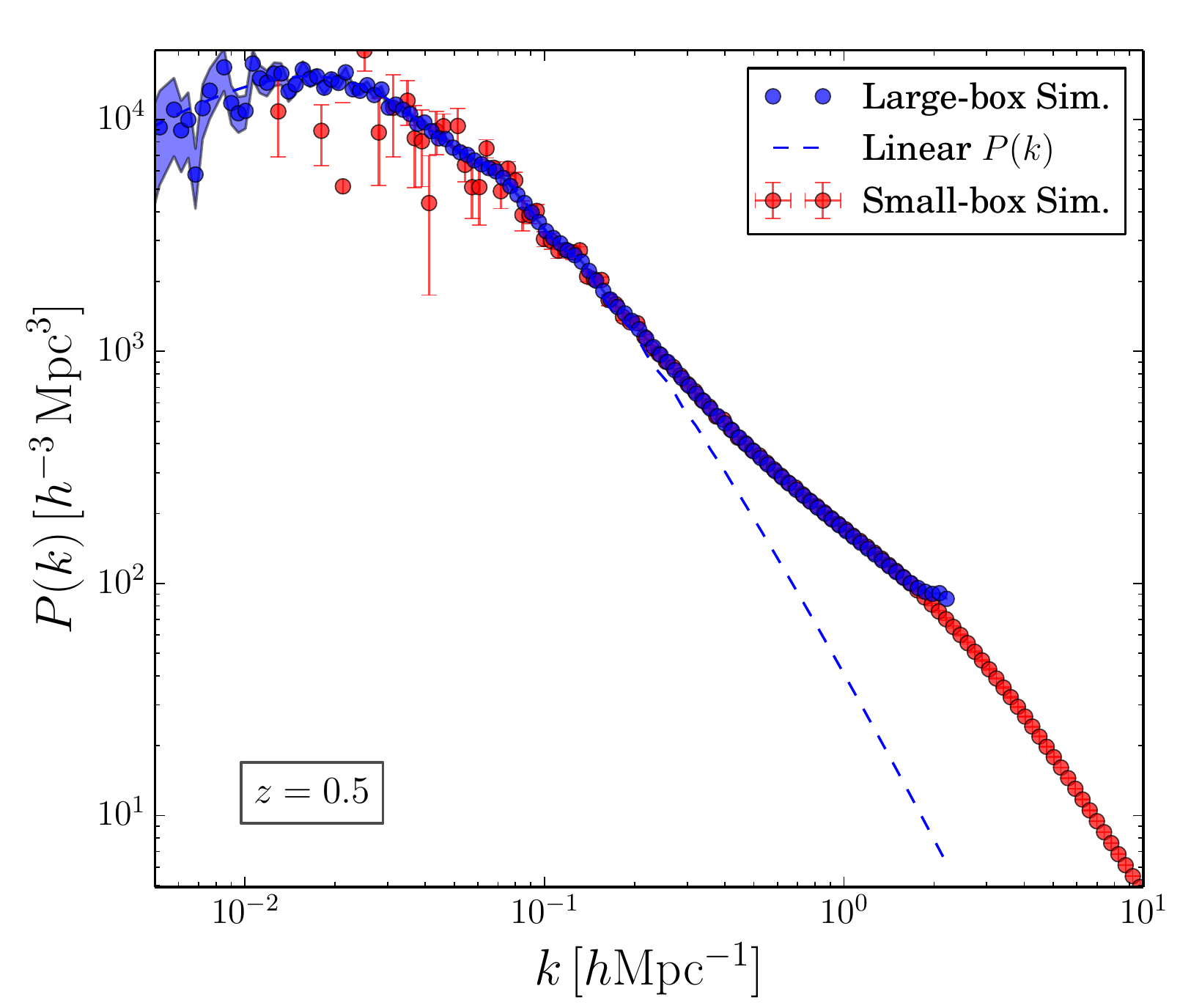}}
\centerline{
  \includegraphics[width=6.2cm]{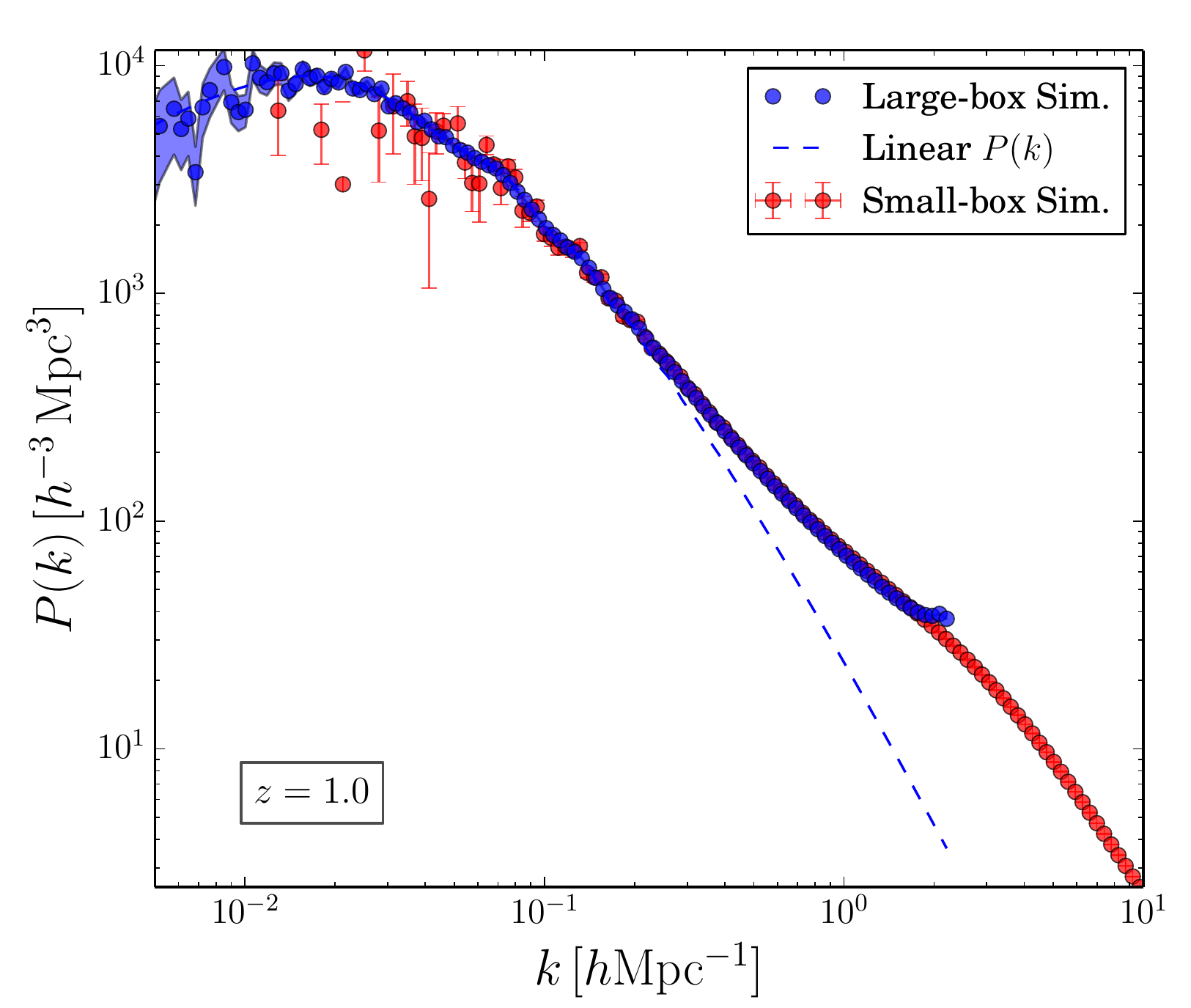}\hspace{0.3cm}
  \includegraphics[width=6.2cm]{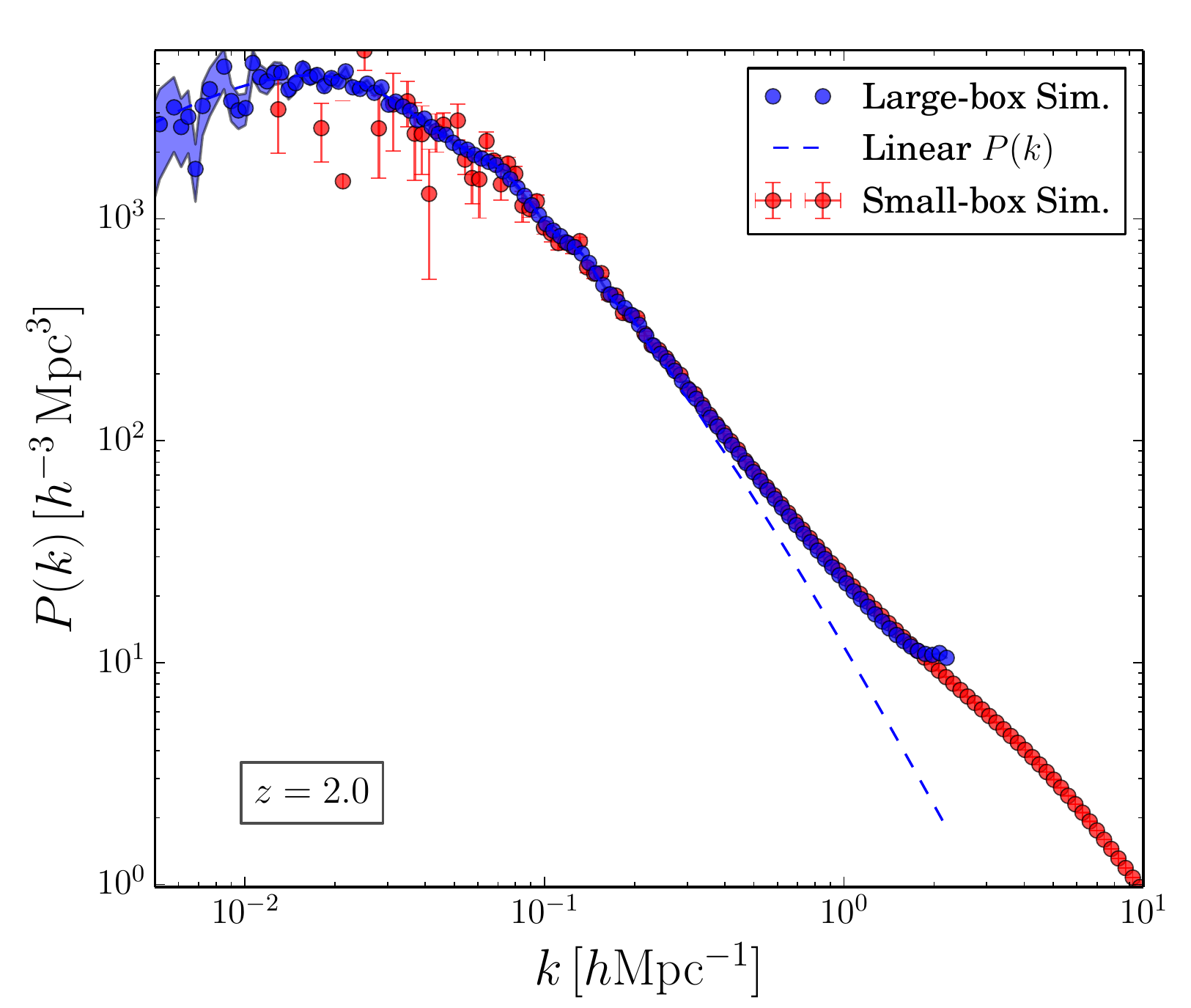}}
\caption{\small{Evolution of the nonlinear matter power spectra for
    the Fiducial Planck-like model as a function of spatial
    wavenumber. The top left, top-right, bottom left, bottom right
    sub-figures show the spectra for $z=0$, $z=0.5$, $z=1.0$, and
    $z=2.0$, respectively.  In all panels, the red points with error
    bars show the mean power spectrum and its 1$\sigma$ errors for the
    small-box runs ($L=500\Mpc$). The blue points show the results for
    the large-box run ($L=3000\Mpc$). The shaded blue region gives the
    predictions for the Gaussian error on the blue points.  The dash
    line shows the linear theory prediction according to {\tt CAMB}.}
\label{fig:PkFidRaw}}
\centerline{
  \includegraphics[width=6.2cm]{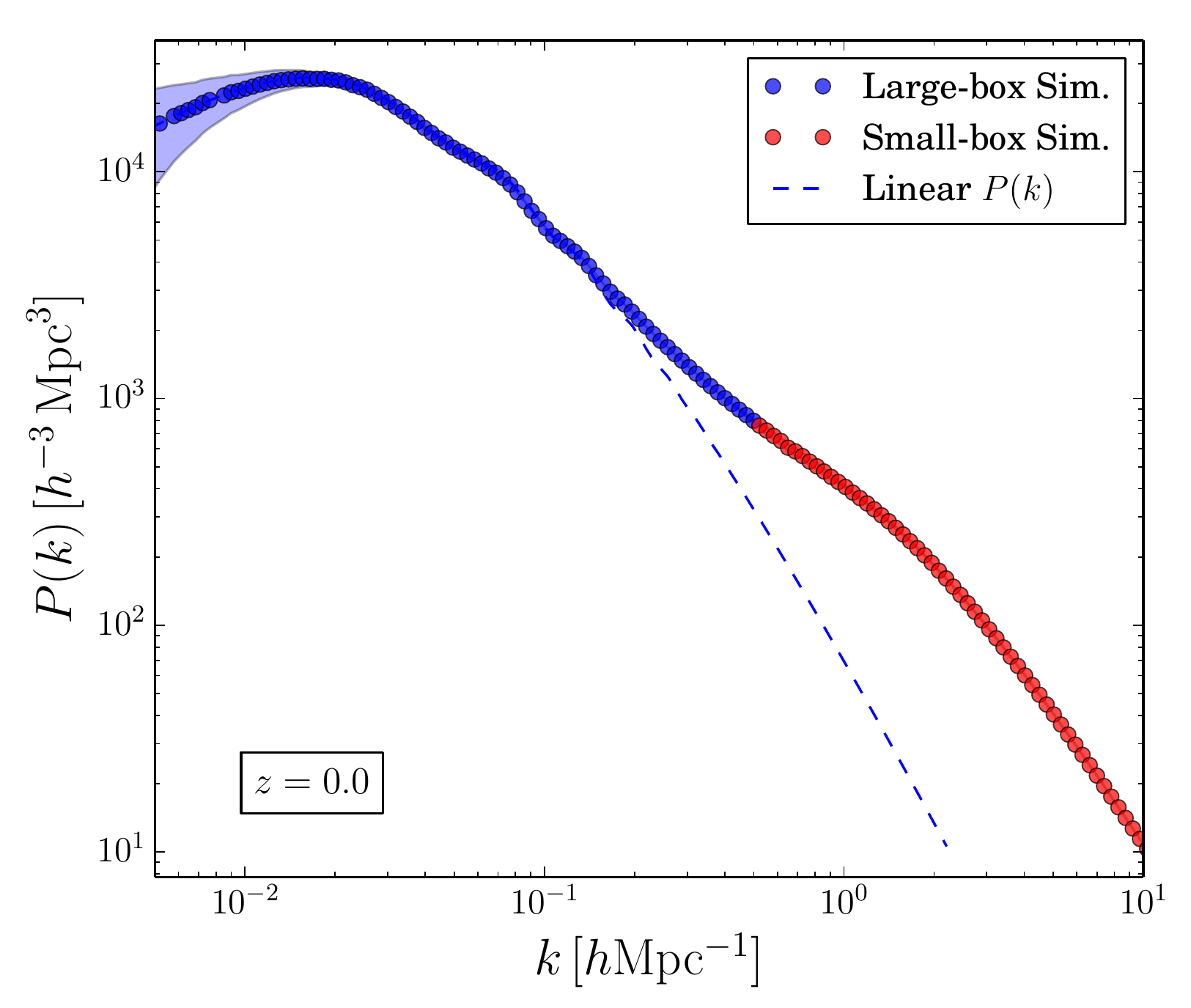}\hspace{0.3cm}
  \includegraphics[width=6.2cm]{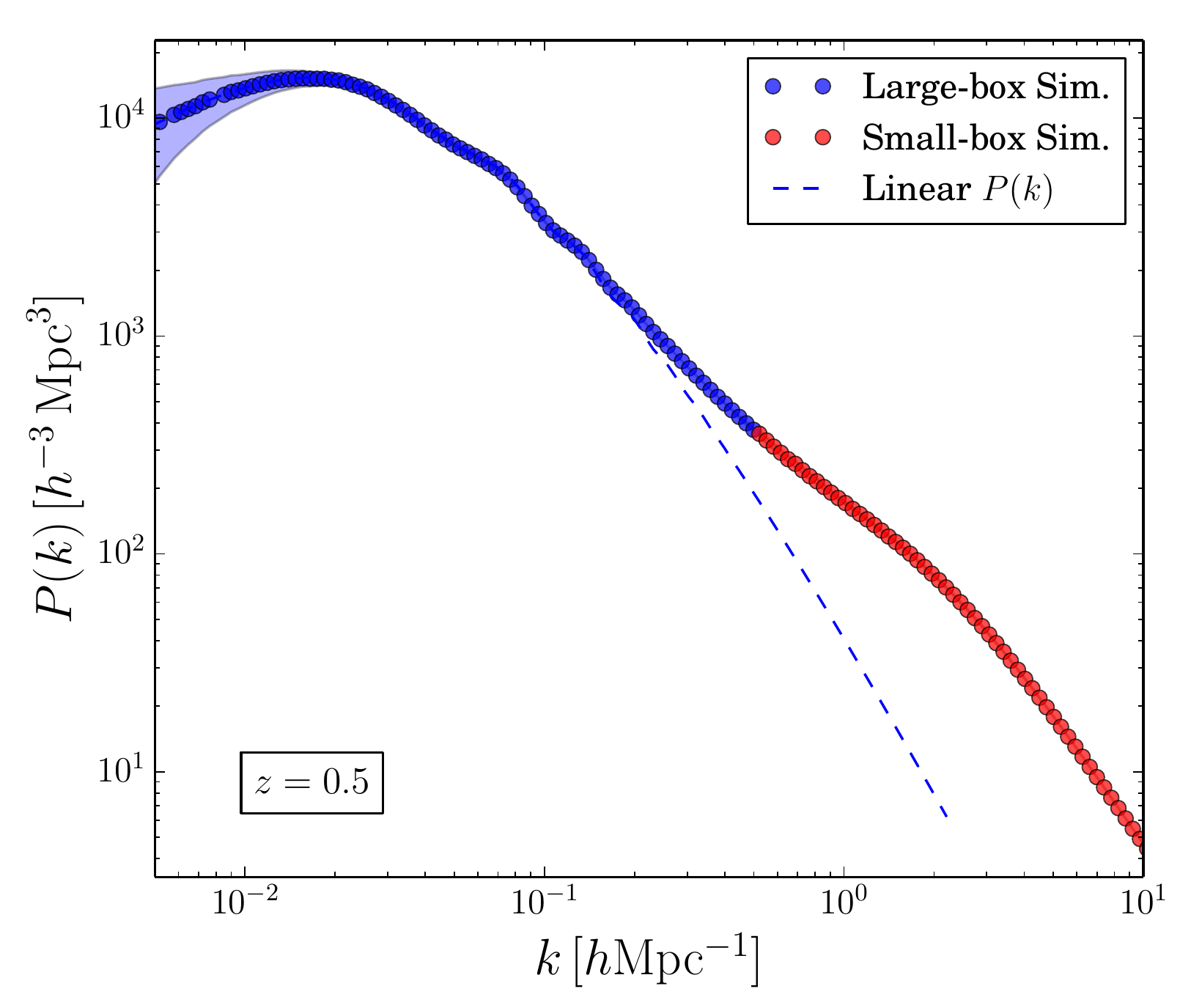}}
\centerline{
  \includegraphics[width=6.2cm]{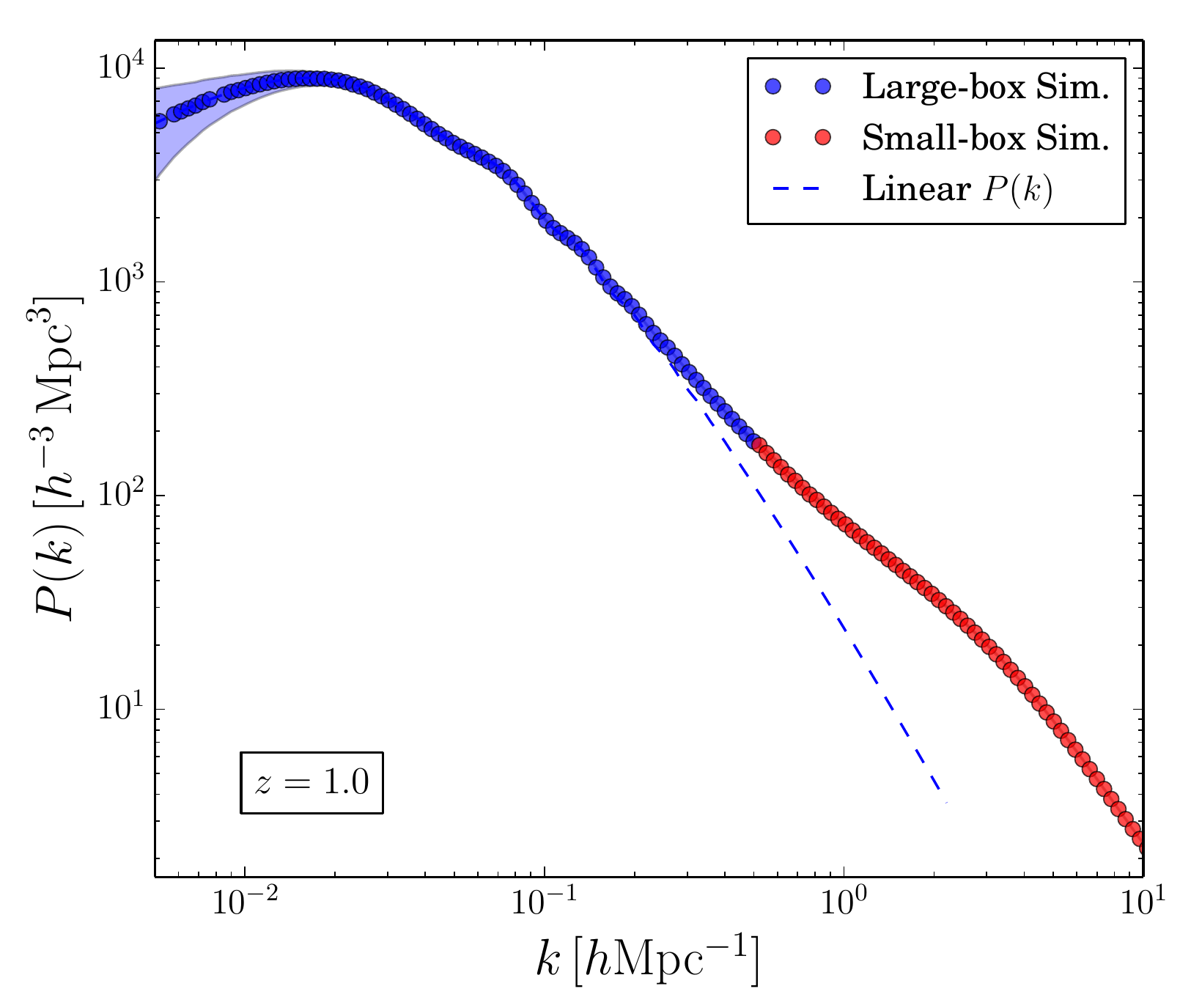}\hspace{0.3cm}
  \includegraphics[width=6.2cm]{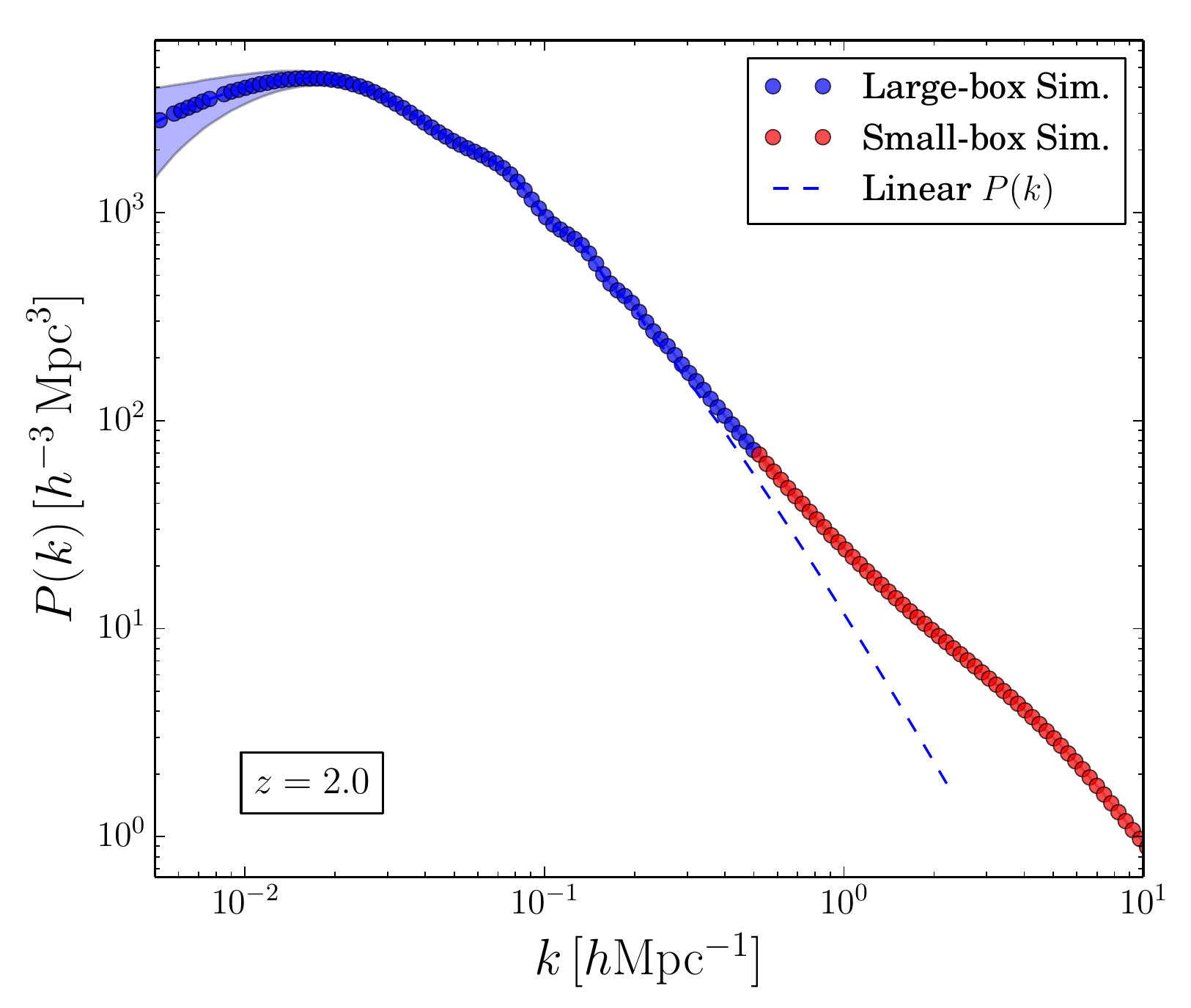}}
\caption{\small{Same as Figure.~\ref{fig:PkFidRaw} only showing the
    composite spectrum generated from the combination of the big-box
    and small-box runs.}
\label{fig:PkFidSuperComp}}
\end{figure*}


\subsection{Evolution of the raw fiducial power spectra}

In Figure~\ref{fig:PkFidRaw} we show the evolution of the raw matter
power spectra measured over several redshifts.  In each panel the red
points denote the mean of the small-box runs and the error bars are
the standard deviation as determined from the 10 realisations. The
blue points denote the results from the large-box run and the shaded
blue region shows the predicted 1--$\sigma$ error bar that results
from assuming the density field is Gaussianly distributed \citep[see
  for example][]{Scoccimarroetal1999c,Smith2009}:
\be \sigma^2_{P}(k) =
\frac{2}{N_{k}}\left[P(k)+\frac{1}{\nbar}\right]^2
\ , \label{eq:Gaussian}\ee
where $N_k$ is the number of Fourier modes in a given $k$-space shell
$N_k\approx\Vu 4 \pi k^2 \Delta k\left[1+\left(\Delta
  k/k\right)^2/12\right]/(2\pi)^3$ and where $\Delta k$ is the spacing
of the $k$-space shell. The dashed blue lines denote linear theory
predictions.

We see that there is very good agreement between the large-box and
small-box runs on large and quasi-linear scales. There is an increase
in the power associated with the big-box relative to the small-box
runs at around $k\sim2\kMpc$. This can be attributed to the effects of
aliasing. These effects can be mitigated by only considering scales
$k<k_{\rm Ny}/2$, where $k_{\rm Ny}=\pi N_{\rm g}/L$, with $N_{\rm g}$
being the size of the FFT grid.


\subsection{Construction of the composite fiducial spectra}

Rather than using the raw spectra in what follows we construct a super
composite of the large- and small-box fiducial runs. We do this by
selecting a partition scale $k_{\rm p}$ and then we select all of the
modes from the Big-box run with $k<k_{\rm p}$ and all of the data that
have $k>k_{\rm p}$ from the small-box runs.  We expect a small
discontinuity at the partition scale owing to the fact that the
large-box runs are more than a factor of 200 times lower resolution
than the small-box runs, which means that on small scales we expect
the large-box runs to be slightly lower in amplitude. For all spectra
we take $k_{\rm p}=0.6\,\kMpc$.

In Figure~\ref{fig:PkFidSuperComp} we show the result of the
construction of the composite spectrum. We have colour coded the
points with $k<k_{\rm p}$ using blue and those with $k\ge k_{\rm p}$
with red.  In this plot we have also removed some of the large-scale
cosmic variance by rescaling the spectrum of each realisation in the
following way:
\ba P^{\rm NoCV}(k_i,a) \!\! & = &  \!\!
W(k_i|k') \hat{P}^{\rm Sim}(k_i,a)  \left[\frac{P^{\rm Lin}(k_i,a)}{D^2(a) P^{\rm Sim}(k,a_i)}\right] \nn \\
& & +\left[1-W(k_i|k') \right] \hat{P}^{\rm Sim}(k_i) \ ,\ea
where we take $k'=0.05\kMpc$. Note that the shaded region gives the
1$\sigma$ error region computed before any rescaling takes place.  The
end result is that the composite spectrum smoothly covers more than
three orders of magnitude and cuts off at $k\sim 10\kMpc$.


\section{Comparison with analytic perturbation theory}\label{sec:PT}

\subsection{Perturbative methods}

As the Universe expands small primordial matter over-densities
aggregate through gravitational instability. When averaged over
sufficiently large enough scales, and on scales smaller than the
horizon, the evolution of these fluctuations can be modelled using the
Newtonian fluid equations expressed in expanding coordinates. As the
system evolves, nonlinear mode coupling takes place and the presence
of large-scale wave-modes modulates the growth of structure on all
scales \citep{Peebles1980,Bernardeauetal2002}.

As mentioned earlier, in recent years there has been significant
progress in developing methods to improve the range of applicability
of nonlinear perturbation theory. Of particular note are the RPT and
MPT approaches
\citep{CrocceScoccimarro2006a,CrocceScoccimarro2006b,Bernardeauetal2008}.
Subsequent research has focused on various extensions of these
schemes. We note that currently much interest surrounds the
construction of an Effective Field Theory for large-scale structure
\citep{Carrascoetal2012}. However, we will not explore this here,
since as has been shown in \citet{Baldaufetal2015}, the additional
complexity and need for a, possibly scale-dependent, free-parameter
means that this technique is not generally applicable `right out of
the box'. Moreover, even when fully calibrated this approach would
offer improvements over MPT only on scales where, in this work, we
will look to $N$-body simulations for the correct answer. In what
follows, we will therefore focus on implementing and testing the MPT
method against our runs.


\begin{figure*}
\centerline{
\includegraphics[width=8.6cm]{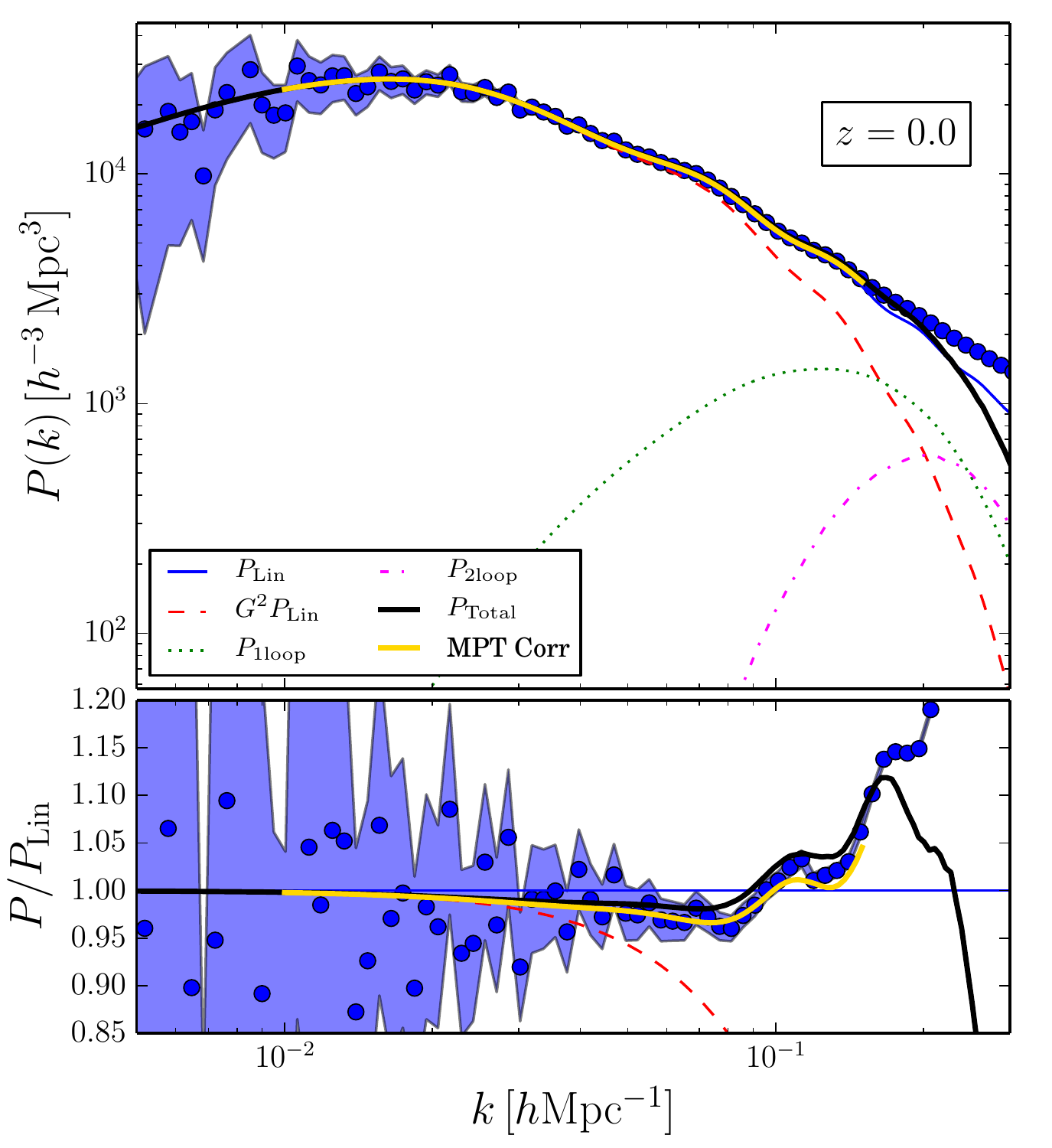}\hspace{0.2cm}
\includegraphics[width=8.6cm]{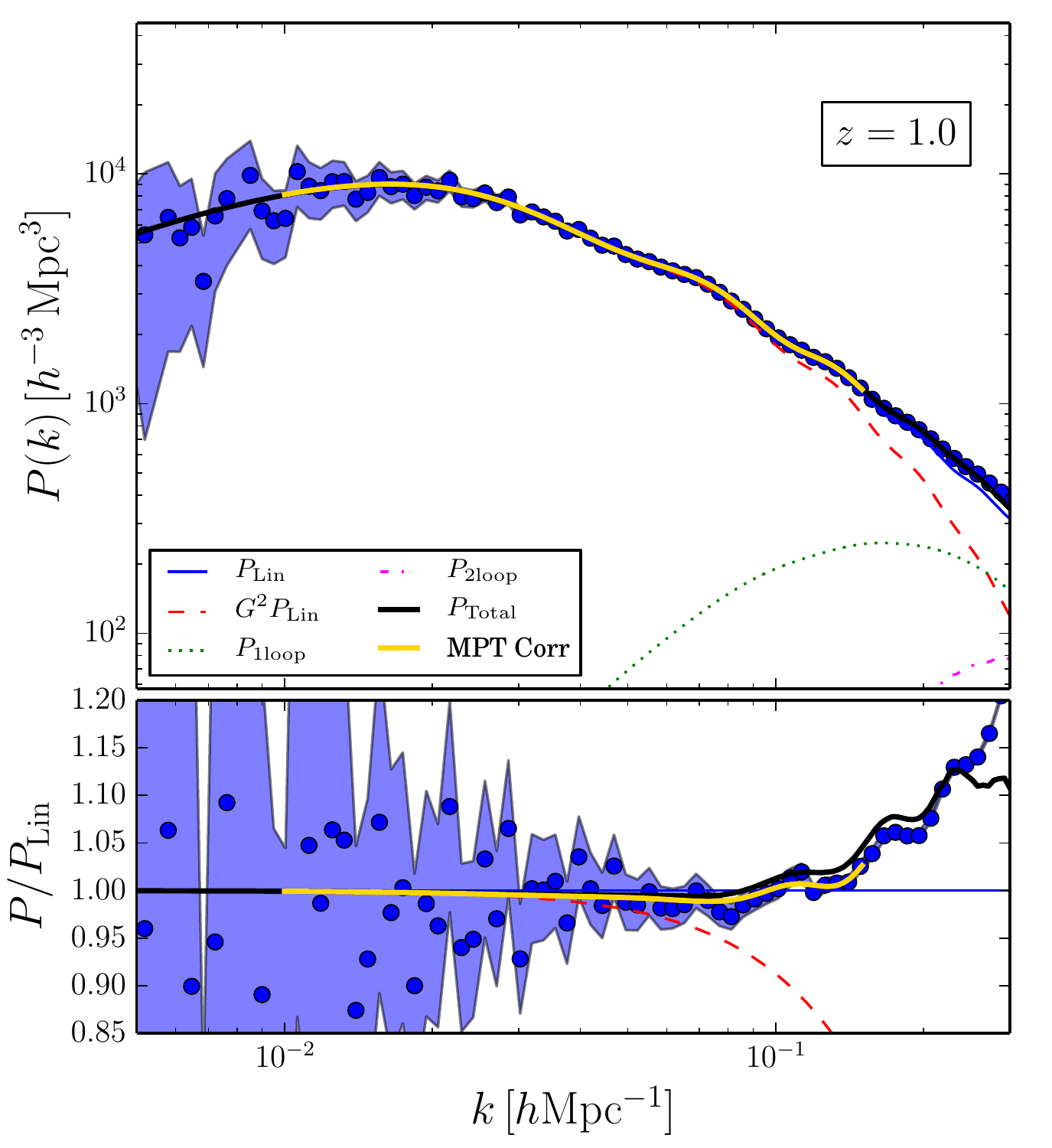}}
\caption{\small{Nonlinear power spectrum on large-scales as a function
    of wavenumber. {\bf Left figure, top panel}: the absolute power at
    $z=0$. The solid blue points present the results from our
    large-box simulation ($L=3000\Mpc$, $N=2048^3$) of our Fiducial
    model. The light shaded band shows the Gaussian prediction of the
    1-$\sigma$ errors on the measured data given by
    \Eqn{eq:Gaussian}. The thick solid line shows the predictions from
    MPT theory. The thin blue line shows linear theory and the dashed,
    dotted and dot-dashed lines show the individual contributions to
    the MPT predictions from the 1-, 2- and 3-point propagators,
    respectively. The thick solid yellow/gold line presents the
    correction to the MPTbreeze recipe described in \S\ref{ssec:MPTCorr}.
    {\bf Left figure, bottom panel}: The ratio of the measurements and
    MPT theory with respect to the linear theory predictions. {\bf
      Right figure}: the same as left figure, but for $z=1$.}
\label{fig:MPTPk}}
\end{figure*}


\subsection{The power spectrum in the Multi-point Propagator Theory}

In this theory the matter power spectrum can be expressed as an
infinite sum over nonlinear $n$-point propagators:
\ba
P_{\rm MPT}(k,z) \! \! \! & = &\! \!\! \sum_{n\ge 1} n!
\int \frac{\dq_1}{(2\pi)^3} \ldots \frac{\dq_n}{(2\pi)^3} \,
\delta^{\rm D}(\bk-\bq_{1}\dots-\bq_n) \nn \\
& & \hspace{-0.5cm}\times \left[\Gamma^{(n)}(\bq_1,\ldots,\bq_n;z)\right]^2
P_0(q_1) \ldots P_0(q_n) \ ,
\label{eq:gamexpansion}
\ea
where $P_0$ denotes the initial matter power spectrum determined at
some initial time. The $\Gamma^{(n)}$ are the `multi-point
propagators', which can loosely be understood as the memory that the
final field at a given point retains from multiple connections to the
initial field. More formally, they can be defined:
\ba
\frac{1}{n!}\, 
\left<\frac{\delta^n \Psi_{a}(\bk,z)}{\delta\phi_{b_{1}}(\bq_{1},z_i)
  \dots\delta\phi_{b_{n}}(\bq_{n},z_i)}\right>
&  & \nn \\
& & \hspace{-5cm}\equiv\frac{}{} \delta^{\rm D}(\bk-\bq_{1}-\dots-\bq_n)
\ \Gamma^{(n)}_{ab_{1}\dots b_{n}}\left(\bq_{1},\dots,\bq_{n},z \right),
\label{GammaAllDef}
\ea
where $\Psi_a(\bk,z)=\{\delta(\bk,z),\theta(\bk,z)\}$ is a doublet
which denotes the late time density and velocity divergence fields and
where $\phi_a(\bk,z_i) \equiv \Psi_a(\bk,z_i)$ denotes the doublet at
some initial time and when the initial fields are set to start in the
growing mode we have that $\phi_a(\bk,z_i)=u_a\delta_0(\bk)$ with
$u_a=\{1,1\}$.  The operation on the left-hand-side of
\Eqn{GammaAllDef} means take the $n$th functional derivative of the
final state with respect to $n$ initial states. The angle brackets
mean compute the expectation of the resultant expression. Lastly the
above definition for the $n$-point propagators can be connected to the
term $\Gamma^{(n)}$ in \Eqn{eq:gamexpansion} through the expression:
\be
\Gamma^{(n)}(\bq_1,\ldots,\bq_t)=\Gamma^{(n)}_{1 c_1 \ldots c_t}(\bq_1,\ldots,\bq_n)
\,u_{c_1} \ldots u_{c_t} ,
\ee
where repeated indices are summed over.

As can be understood from inspection of \Eqn{eq:gamexpansion} one
major advantage of the MPT expansion over the standard perturbation
theory scheme is that all of the terms in the expansion are clearly
positive. Further, as the order $n$ is increased the contributions to
the power spectrum appear to contribute to increasingly small scales.
Hence, on large-scales, one can confidently truncate the sequence at a
finite number of `loops'.


\subsection{MPT recipe {\em \`{a} la} MPTbreeze:}

In this work we will desire to use the MPT approach to describe the
power spectrum in the weakly nonlinear regime, and in order to do this
we follow \cite{Crocceetal2012} and truncate the propagator expansion
after the first three terms in Eq.~(\ref{eq:gamexpansion}).  In the
diagrammatic language this corresponds to renormalised versions of
tree level, one and two loop corrections, respectively. Explicitly,
this is:
\ba P_{\rm MPT}(k,z) & \approx & \left[\Gamma^{(1)}(k;z,z_i)\right]^2 P_0(k,z_i) \nn \\
& & \hspace{-2cm}+
2 \int \frac{d^3\bq}{(2\pi)^3} \left[\Gamma^{(2)}(\bk-\bq,\bq;z,z_i)\right]^2
P_0(|\bk-\bq|,z_i) P_0(q,z_i) \nn \\
& & \hspace{-2cm}+ 6 \int \frac{d^3\bq_1}{(2\pi)^3} \int \frac{d^3\bq_2}{(2\pi)^3} 
\, \left[\Gamma^{(3)}(\bk-\bq_{1}-\bq_2,\bq_1,\bq_2;z,z_i)\right]^2 \nn \\
& & \hspace{-2cm}\times P_0(|\bk-\bq_{1}-\bq_2|,z_i) P_0(q_1,z_i) P_0(q_2,z_i)\ , \label{eq:PMPT1}
\ea
and where the 1-, 2- and 3-point propagators are given by:
\ba
\Gamma^{(1)} \! \!& = & \! \!D(z,z_i)F^{(\rm s)}_1(\bq_1)e^{\left[f(|\bq_1|)D^2(z,z_i)\right]} \label{eq:gam1}\ ;\\
\Gamma^{(2)} \! \!& = & \! \!D^2(z,z_i)
F_2^{(\rm s)}(\bq_1,\bq_2)e^{\left[f(|\bq_1+\bq_2|)D^2(z,z_i)\right]} \label{eq:gam2}\ ;\\
\Gamma^{(3)} \! \!& = & \! \!D^3(z,z_i)F^{(\rm s)}_3(\bq_1,\bq_2,\bq_3)
e^{\left[f(|\bq_1+\bq_2+\bq_3|)D^2(z,z_i)\right]} \label{eq:gam3}
\ea
where for brevity we have suppressed the arguments of the
$\Gamma^{(n)}$ functions. The symmetrised gravitational mode coupling
kernels $F_n^{(\rm s)}$ up to third order and the function $f(q)$ are
given in Appendix~\ref{app:MPT}.

Considering \Eqn{eq:PMPT1}, the first term on the right-hand side is
directly proportional to the linear power spectrum and the square of
the 1-point propagator, which is a direct indicator of the `memory' of
the initial conditions on a particular scale to that same scale at late
times. The second term is the `one-loop' correction, this term can be
simplified as follows: firstly, a quick inspection of the $F^{(\rm
  s)}_2(\bq_1,\bq_2)$ kernel indicates that it depends only on the
magnitudes of the two vector arguments and the cosine of the angle
between them. Hence, on choosing the $\bk$ vector to denote the
$z$-axis of a spherical polar coordinate system one can immediately
integrate out the azimuthal angle. Secondly, the exponential term
depends on the sum of the two vector arguments, and by momentum
conservation we have that the sum always results in $k$, hence it may
be factored out of the integrals. This leaves us with:
\ba P_{\rm MPT}^{(1\ell)}(k,z) & = &
\frac{4\pi D^4(z,z_i)}{(2\pi)^3} \exp\left[2f(k)D^2(z,z_i)\right]\nn \\
& & \hspace{-2cm}\times \int_0^{\infty} dq q^2 \int_{-1}^{1} d\mu
\left[\tilde{F}_2^{(\rm s)}(q,k,\mu)\right]^2P_0(k\psi(y,\mu)) P_0(q)\ ,
\label{eq:oneloop}
\ea 
where 
\be
\tilde{F}_2^{(\rm s)} = \frac{5}{7}+\frac{1}{2}\left[\frac{\mu-y}{\psi(y,\mu)}\right]
\left[\frac{y}{\psi(y,\mu)}+\frac{\psi(y,\mu)}{y}\right]+\frac{2}{7}
\left[\frac{\mu-y}{\psi(y,\mu)}\right]^2 \ ,
\ee
with $\psi(y,\mu)\equiv\left(1+y^2-2y\mu\right)^{1/2}$ and $y\equiv
q/k$.  We are thus left with an integration in 2-D which can be
rapidly evaluated with a standard Gaussian quadrature routine and we
employ repeated use of the {\tt GSL} standard library routine {\tt
  gsl\_integration\_qag}.

Consider the third term on the right-hand-side of \Eqn{eq:PMPT1}, this
is termed the `two-loop' correction and in order evaluate this we
first substitute $\Gamma^{(3)}$ in. Again, we notice that exponential
term can be extracted. Next, we follow \citet{Crocceetal2012} and
reduce the dimensionality of the integration by one integral through
noting the following: without loss of generality we can fix the $\bk$
to lie along the $z$-axis. Next we can restrict $\bq_1$ to lie in the
$x-z$ plane. Finally, $\bq_2$ must be unrestricted. With these
choices, the relevant vectors in spherical polar coordinates as:
\ba
  \bk   &=& k \,\, (0,0,1) \nonumber \\
  \bq_1 &=& q_1(\sin \theta_1, 0, \cos \theta_1) \nonumber \\
  \bq_2 &=& q_2 (\sin \phi_2 \sin \theta_2, \cos \phi_2 \sin \theta_2, \cos \theta_2) . \nonumber
\ea
On integrating out the redundant azimuthal angle $\phi_1$ of the
$\bq_1$ vector we are left with the following 5-D integral
\ba
  P_{\rm MPT}^{(2\ell)}(k,z) & = & \frac{12\pi D^6(z,z_i)}{(2\pi)^6}
  \exp\left[2f(k)D^2(z,z_i)\right] \\
  && \hspace{-1.5cm}\times \int_{0}^{\infty} dq_1q_1^2 \int_{0}^{\infty} dq_2q_2^2
  \int_{-1}^1 d\mu_1 \int_{-1}^1 d\mu_2
  \int_{0}^{2\pi} d\phi_2 \, \nonumber \\
  && \hspace{-1.5cm} \times \left[\frac{}{}F_{3}^{(\rm s)}(\bk-\bq_{1}-\bq_2,\bq_1,\bq_2;z)\right]^2 \nn \\
  && \hspace{-1.5cm} \times \frac{}{}P_0(|\bk-\bq_{1}-\bq_2|)
  P_0(q_1) P_0(q_2) \label{eq:twoloop} \ ,
\ea
were $\mu_1 = \cos \theta_1$ and $\mu_2 = \cos \theta_2$.  This
integral can be efficiently computed using Monte Carlo integration
techniques and we employ the {\tt Vegas} algorithm supplied by the
{\tt CUBA-4.2} package \citep{Hahn2016}. Note that in the numerical
implementation of the integrals \eqns{eq:oneloop}{eq:twoloop} we
perform the radial integrals over the restricted domain:
$q_i\in\left[k_{\rm min},k_{\rm max}\right]$ where the lower bound is
fixed at $k_{\rm min}=0.001 \kMpc$ and the upper limit varies
according to $k_{\rm max}={\rm max}\left[20k,2\pi\right]$.

Figure~\ref{fig:MPTPk} shows the 2-loop calculation of the MPT power
spectrum of \Eqn{eq:PMPT1} evaluated using the above procedure and for
our Planck-like fiducial model. The left panel shows $z=0$ and the
right $z=1$. The upper panel of each figure shows the absolute power
and one can see that the sum of the MPT propagators adds signal to the
spectrum at increasingly higher wavemodes. It is interesting to note
that at $z=0$, relative to linear theory, there is a 2--3\% suppression
of power on very large scales ($k\sim0.07\kMpc$), followed by an
amplification that starts around $k\sim0.1\kMpc$. In addition, while
the 1-point propagator shows tens of percent difference from linear
theory at $k\sim0.1\kMpc$, the sum of the 1- and 2-point propagators
gives an amplitude that is coincidentally only different by a percent
or so.

Figure~\ref{fig:MPTPk} also compares the theoretical predictions with
the composite power spectrum blue points.  The shaded blue region
shows the 1--$\sigma$ error region, obtained assuming that the density
field is a Gaussian random field (c.f. \Eqn{eq:Gaussian}). The lower
panels in the figure show the ratio of the spectra with respect to the
linear theory. We see that the MPT calculation describes the results
of the simulation up to $k\lesssim0.15\kMpc$ with relatively good
accuracy. There are however some small but notable differences, in
particular, we see that for the $z=0$ data at $k\gtrsim0.075\kMpc$ the
MPT predictions slightly over-predict the data by $\gtrsim1\%$. For
the case of the $z=1$ data this discrepancy is pushed to slightly
higher wavenumbers. We therefore explored whether we might correct for
this.


\subsection{An {\em ad hoc} correction to MPT}\label{ssec:MPTCorr}

We found that the MPT predictions and the data could be brought into
better agreement by introducing the following {\em ad hoc} correction.
Considering again the propagator expansion of the power spectrum
\Eqn{eq:gamexpansion}, we note that if we were to slightly increase
the amount of decorrelation of the initial conditions with the final
conditions, then this would reduce the predictions on the relevant
scales. We find that this can most easily be achieved by recomputing
$f(q)$ that appears in \Eqn{eq:gam1} with a nonlinear matter power
spectrum model such as {\tt halofit2012}, which we might call $f_{\rm
  NL}$ i.e. in \Eqn{eq:fk} we make the replacement $P_0(q)\rightarrow
P_{\rm halofit}(q)$ (see \Eqn{eq:fkNL}).

Unfortunately, since the resumed 1-point propagator multiplies all of
the higher order propagators (see \Eqns{eq:gam2}{eq:gam3}), this
alteration has the effect of considerably damping all of the loop
terms. We obviate this by only using $f_{\rm NL}(q)$ in the
computation of the 1-point propagator:
\ba 
&& \hspace{-0.5cm} \Gamma^{(1)}=D(z,z_i)F^{(\rm s)}_1(\bq_1)e^{\left[f(|\bq_1|)D^2(z,z_i)\right]} \nn \\
&& \hspace{0.1cm} \Longrightarrow \Gamma^{(1)}_{\rm NL}=D(z,z_i)F^{(\rm s)}_1(\bq_1) e^{\left[f_{\rm NL}(|\bq_1|)D^2(z,z_i) \right]}
\ea
and leaving $\Gamma^{(2)}$ and $\Gamma^{(3)}$ unchanged. In
implementing this approach we had to pay special attention to the
limits of the integral for $f_{\rm NL}(q)$ since we found that the
amount of damping was sensitive to the upper limit. After some trial
and error we found that adopting the upper limit $k=1.0\kMpc$ produced
acceptable results. A higher value for this cut-off would lead to too
much damping.

In Figure~\ref{fig:MPTPk} we indicate our correction to the MPT power
spectrum implementation by the solid yellow/gold lines. Up to
$k=0.1\kMpc$ we see that for both of the redshifts considered this
recipe leads to improved predictions.  We shall therefore adopt this
corrected MPT formulation as the means for generating the nonlinear
matter power spectrum on scales $k\lesssim0.1\kMpc$. Before
continuing, we also point out that our large-box simulation has a
volume of $27\Gpccube$, and since typical surveys cover a smaller
volume we expect that the modelling errors on these large-scales would
fall below sample variance errors. 


\begin{figure*}
\centerline{
  \includegraphics[width=8.5cm]{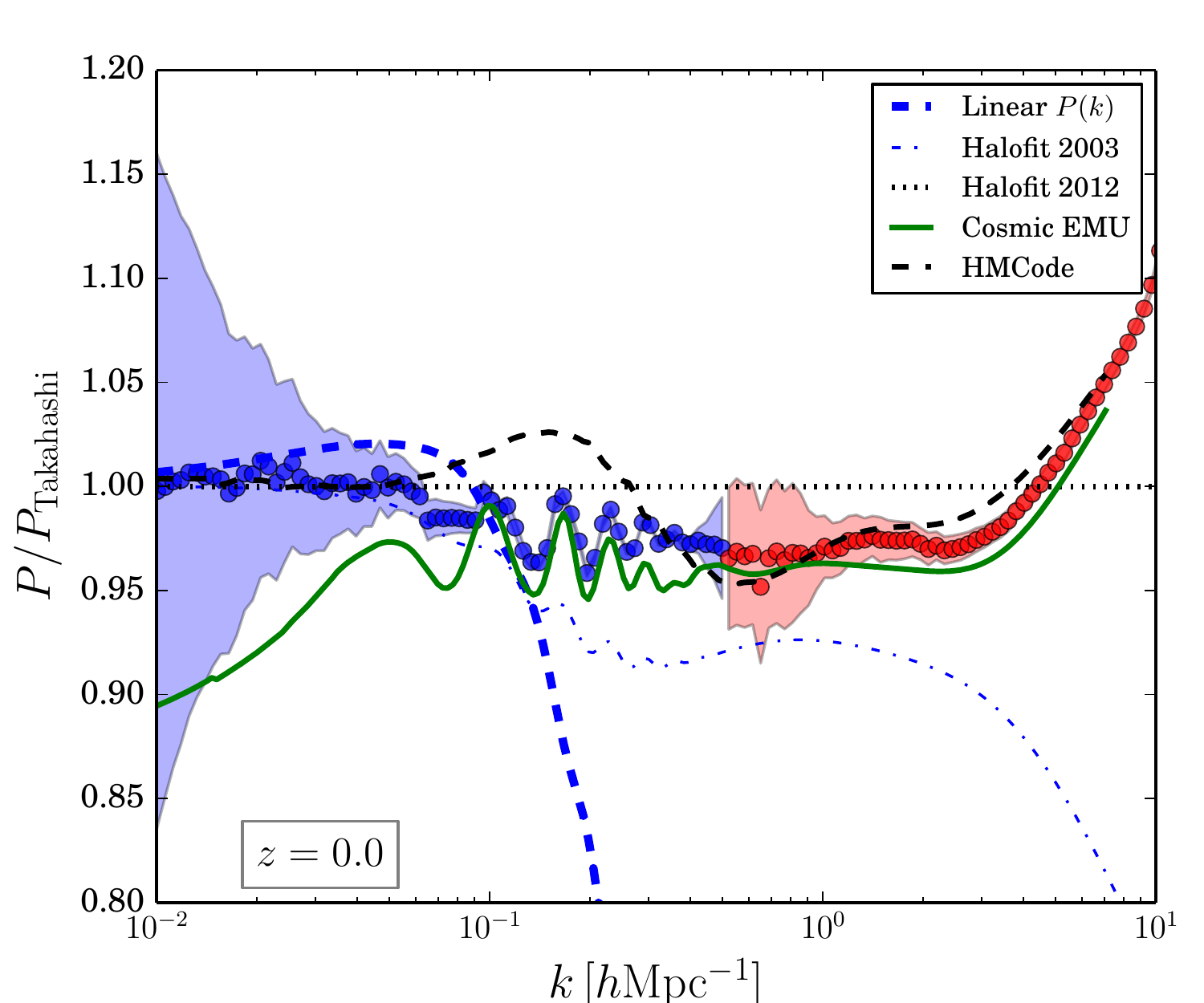}\hspace{0.3cm}
  \includegraphics[width=8.5cm]{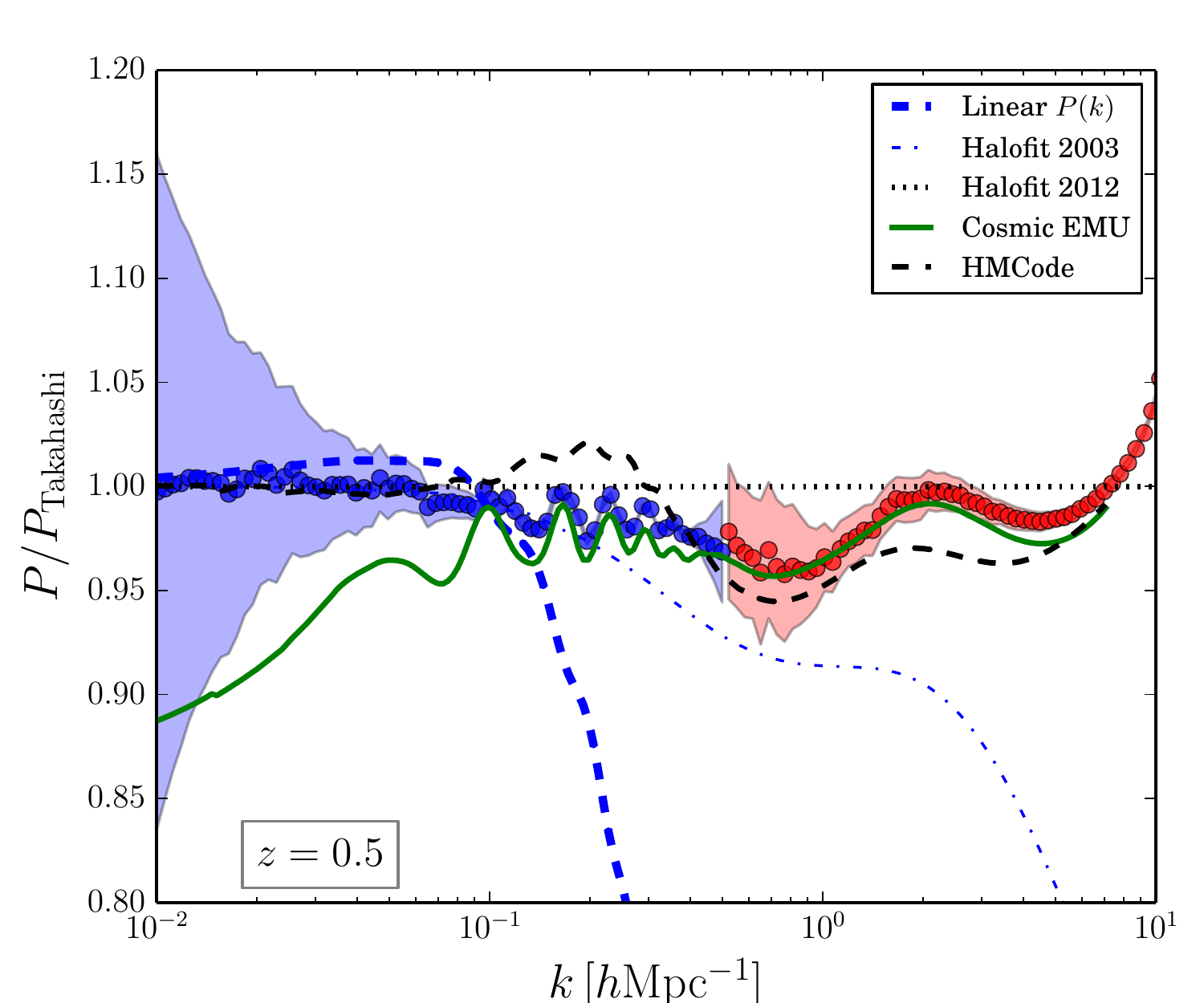}}
\centerline{
  \includegraphics[width=8.5cm]{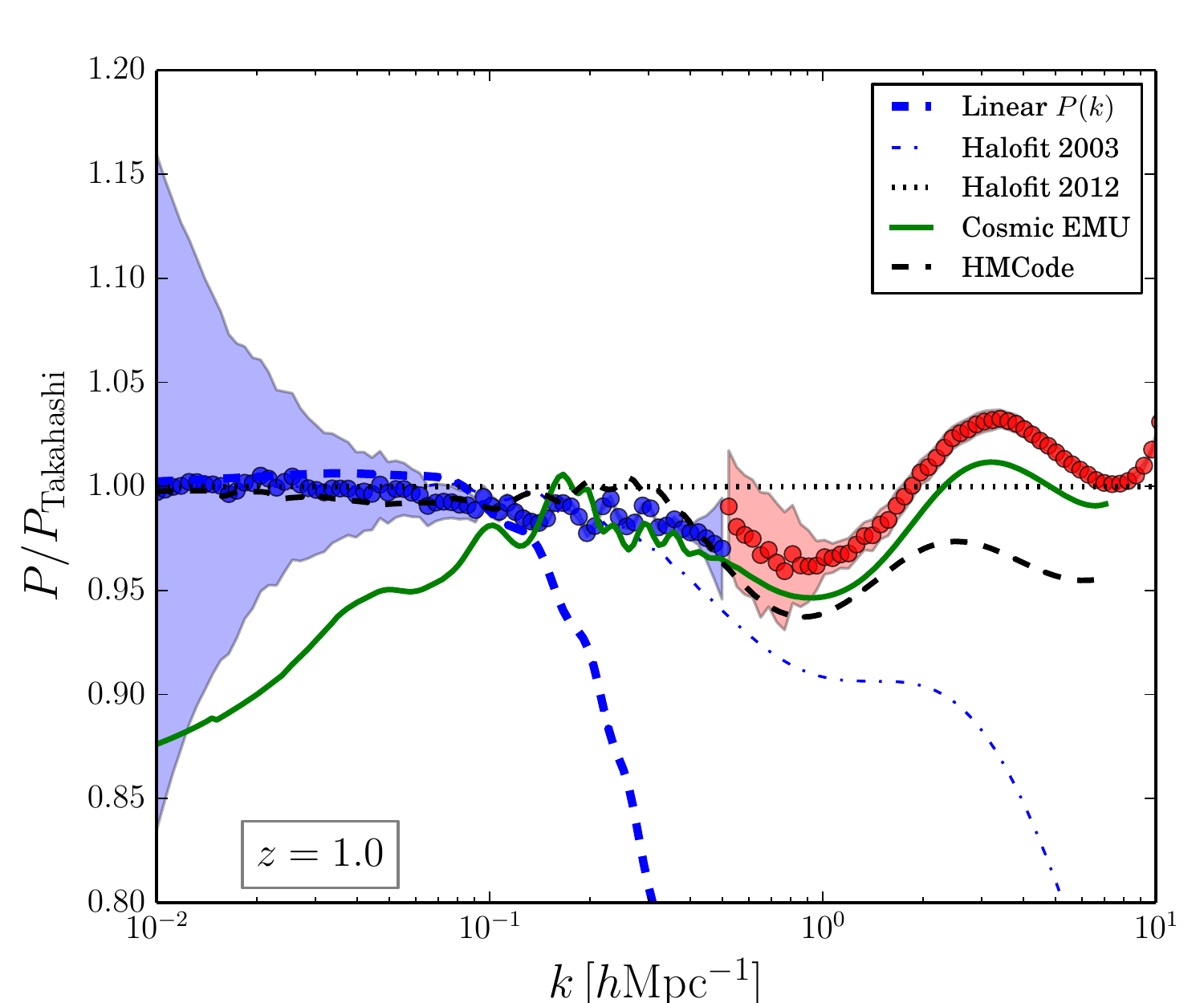}\hspace{0.3cm}
  \includegraphics[width=8.5cm]{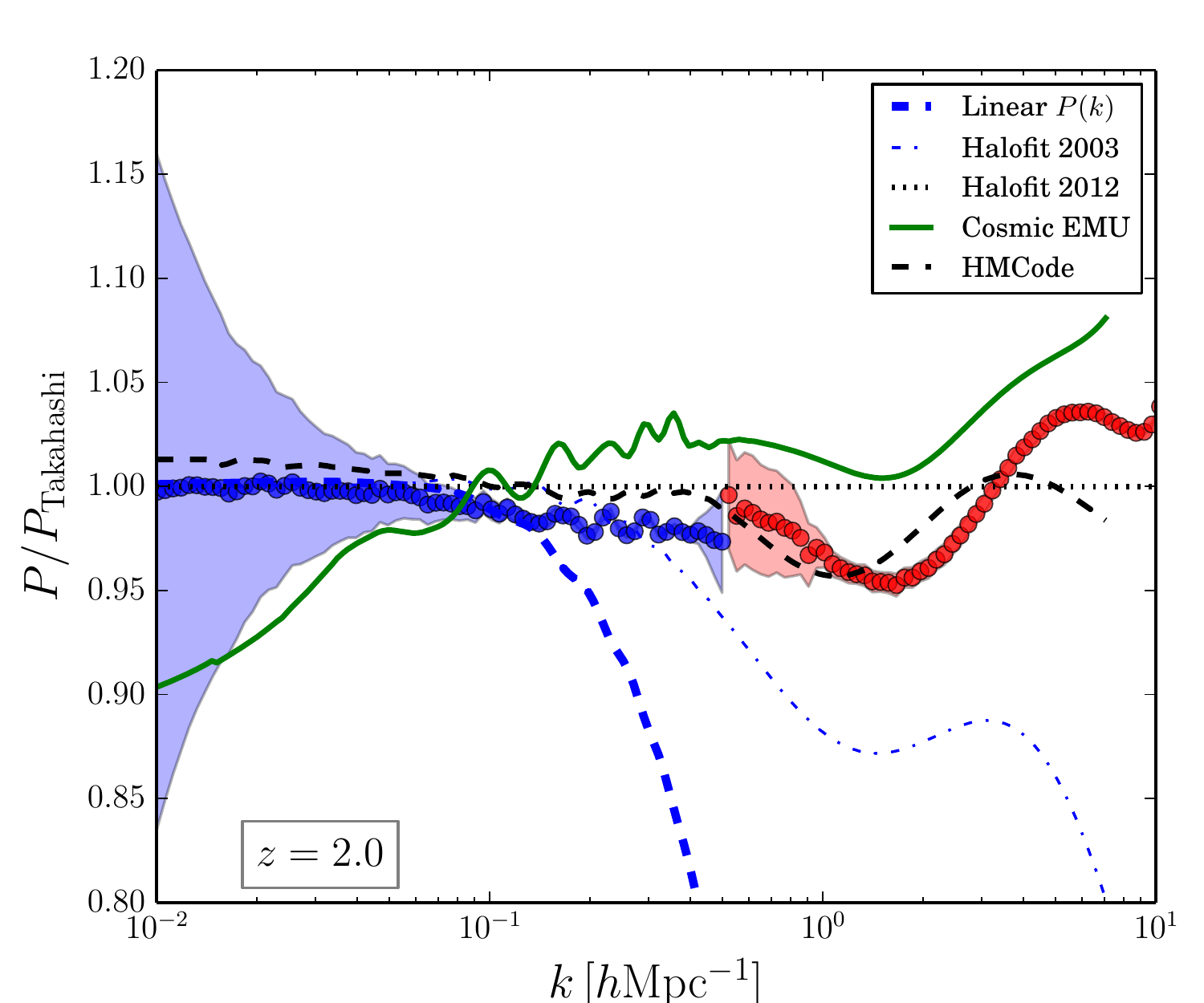}}
\caption{\small{Evolution of the ratio of the measured nonlinear
    matter power spectra with the predictions from the {\tt halofit2012}
    model of \citet{Takahashietal2012} as a function of spatial
    wavenumber. The top left, top-right, bottom left, bottom right
    sub-figures show the spectra for $z=0$, $z=0.5$, $z=1.0$, and
    $z=2.0$, respectively. In all panels, the red points with error
    bars show the mean and 1$\sigma$ errors for the power spectra
    measured from the Planck-2013-like fiducial model simulations in
    boxes of side $L=500\Mpc$.  The solid blue line presents our {\tt
      NGenHalofit} model, the dotted line shows the updated {\tt
      halofit2012} model of \citet{Takahashietal2012}, the dot-dash line
    shows the original {\tt halofit} model of \citet{Smithetal2003},
    and the dash line shows the linear theory.}
\label{fig:PkFidRatio}}
\end{figure*}


\section{Comparison with semi-analytic methods}\label{sec:semi}

We now compare the power spectra measured from our $N$-body runs with
various semi-analytic and emulator methods.


\begin{figure*}
\centerline{
\includegraphics[width=8.5cm]{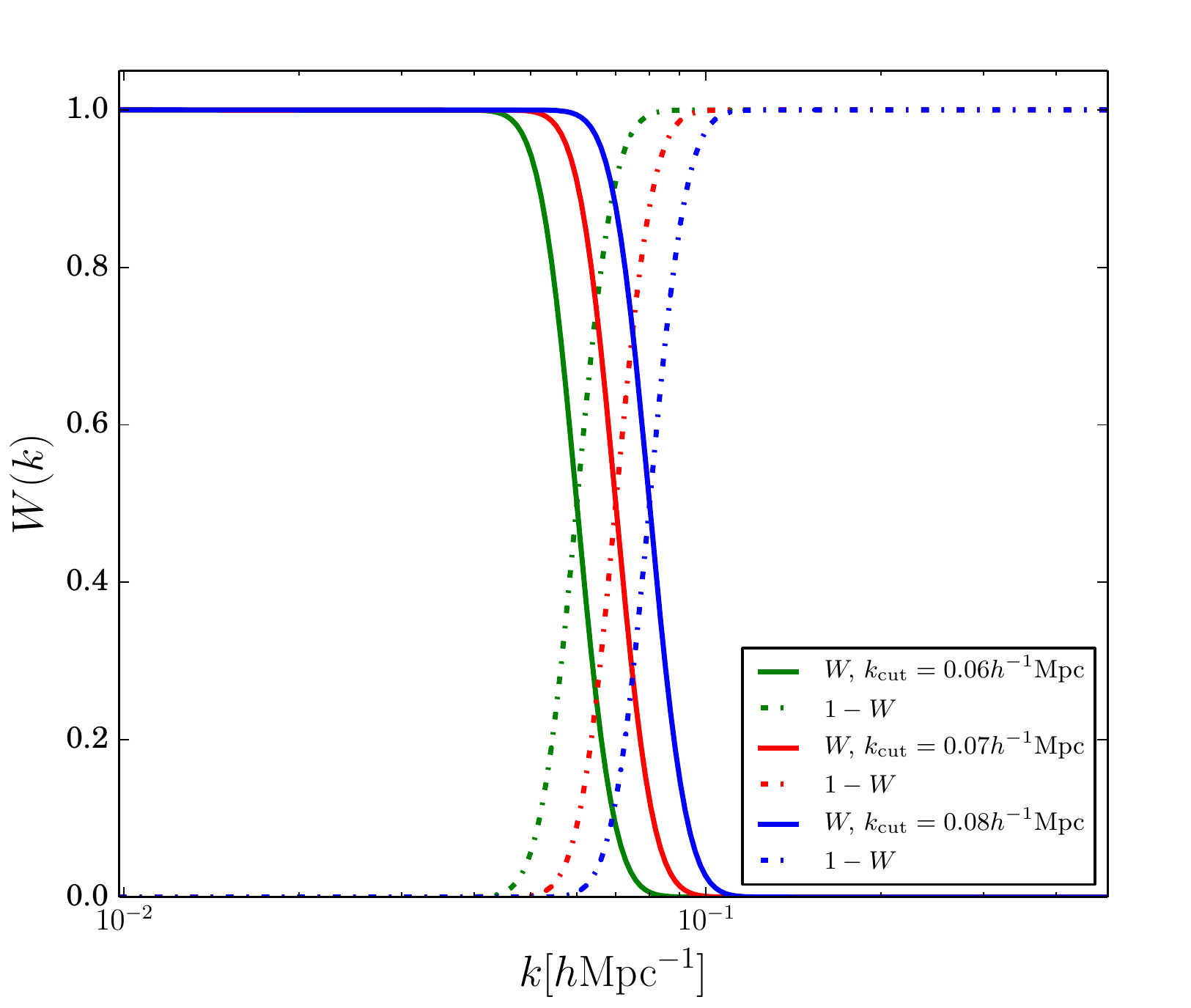}
\includegraphics[width=8.5cm]{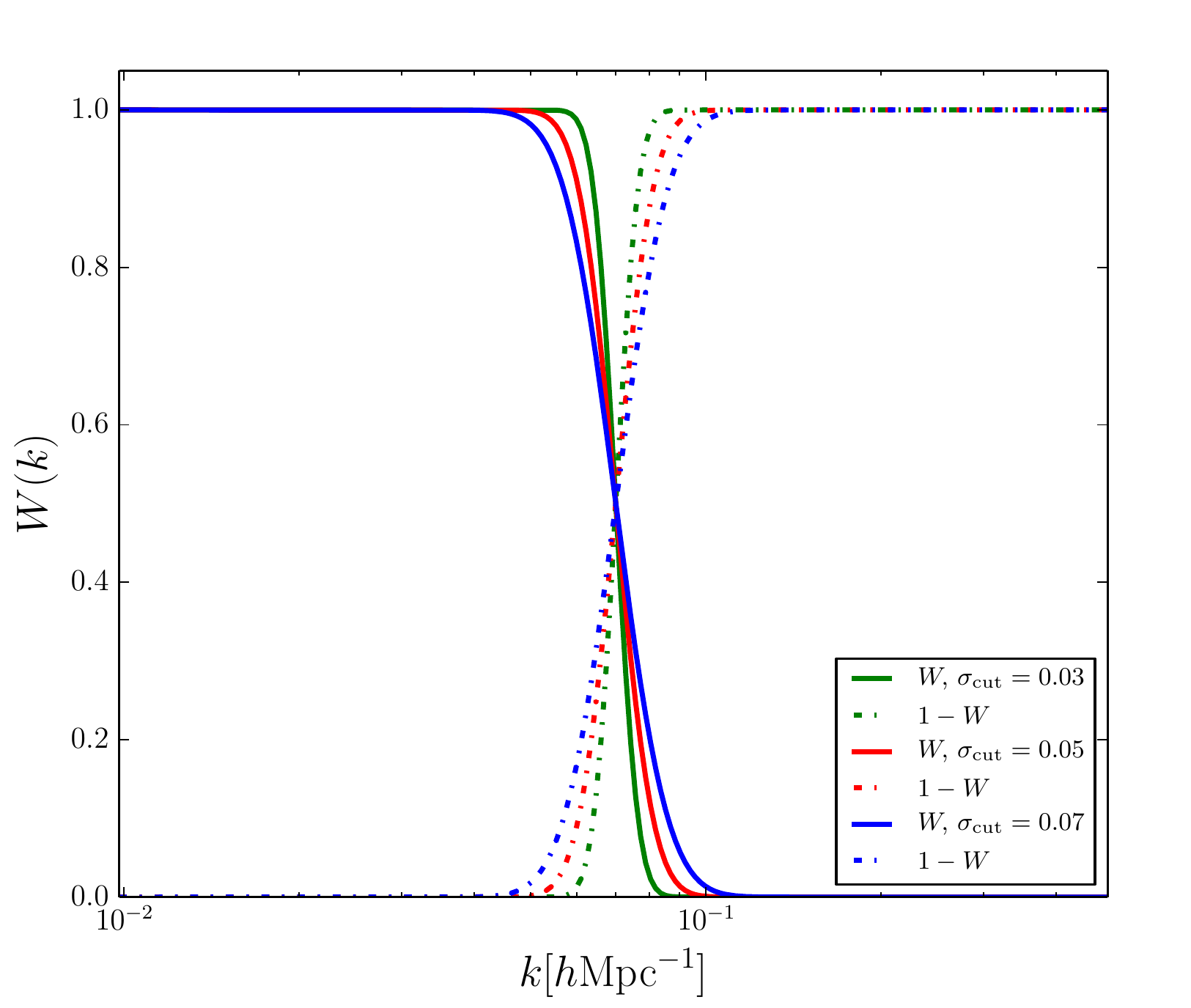}}
\caption{\small{Transition functions as a function of wavenumber. {\bf
      Left panel:} variation of transition functions $W$ and $1-W$
    with the choice of the cut-scale parameter $k_{\rm cut}$ centred
    on our fiducial choice. {\bf Right panel}: same as right panel,
    but this time showing variation with respect to the width
    parameter $\sigma_{\rm cut}$. In both panels, the solid lines
    denote $W$ and the dotted ones give $1-W$.}
\label{fig:filter1}}
\centerline{
\includegraphics[width=8.5cm]{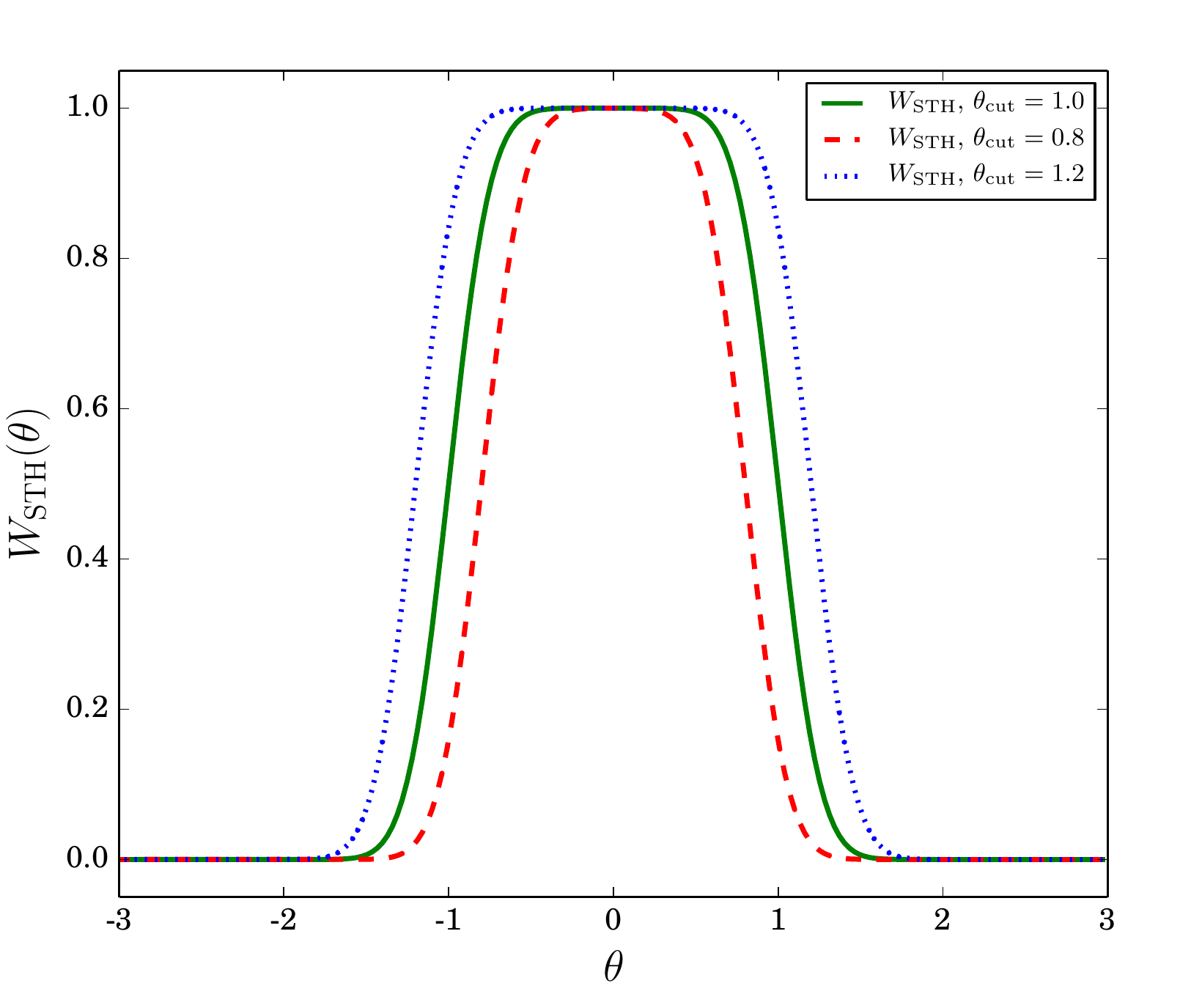}
\includegraphics[width=8.5cm]{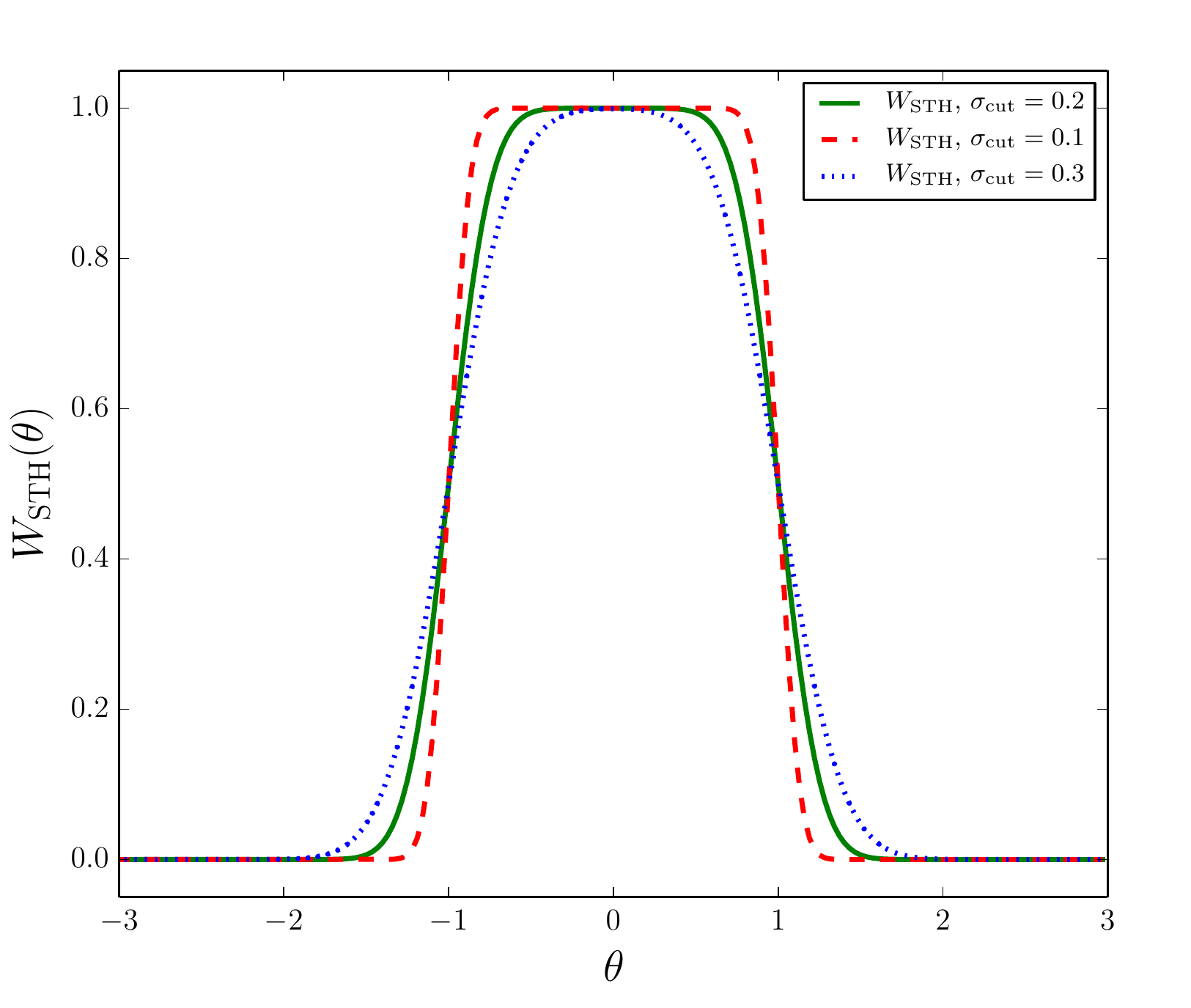}}
\caption{\small{Transition functions as a function of wavenumber. {\bf
      Left panel:} variation of transition functions $W$ and $1-W$
    with the choice of the cut-scale parameter $k_{\rm cut}$ centred
    on our fiducial choice. {\bf Right panel}: same as right panel,
    but this time showing variation with respect to the width
    parameter $\sigma_{\rm cut}$. In both panels, the solid lines
    denote $W$ and the dotted ones give $1-W$.}
\label{fig:filter2}}
\end{figure*}


\subsection{Comparison with {\tt halofit} and {\tt halofit2012}}

The perturbative methods such as SPT and MPT are unable to describe
the evolution of structure once shell-crossing takes place. This we
shall refer to as the deeply nonlinear regime. In order to understand
how structures collapse and evolve on these scales we need to make use
of fully non-perturbative schemes such as $N$-body simulations and
study the phenomenology of the structures formed.

The {\tt halofit} prescription for modelling the nonlinear matter
power spectrum, originally developed in \citet{Smithetal2003} has the
following form for the power spectrum:
\be P_{\rm halofit}(k,z|{\bm\theta}) = P_{\rm L
  }(k,z|{\bm\theta}){\mathcal G}(k,z|\bm\theta) + P_{\rm
  H}(k,z|\bm\theta) \ee
where ${\mathcal G}(k,z|\bm\theta)$ represents the quasi-linear
suppression and amplification of nonlinear structures that are loosely
associated with the phenomenological effects arising from a 2-halo
like term and $P_{\rm H}(k,z|\bm\theta)$ is a 1-halo shot-noise like
term (borrowing loosely from the language of the halo
model). Parameterised analytic forms for these two functions were
devised in \citet{Smithetal2003} and the best fit parameters were
established through fitting to the root mean square difference between
the model and a suite of power spectra from $N$-body simulations on
small scales and 1-loop SPT on large scales.  This approach was
recently upgraded in the work of \citet{Takahashietal2012} who
recalibrated the fitting parameters against a series of improved
simulations and extended the method to include a time evolving dark
energy equation of state.

Figure~\ref{fig:PkFidRatio} shows the ratio of the composite power
spectra derived from our fiducial runs, with the updated {\tt halofit}
model of \citet{Takahashietal2012}. In all of the panels the blue and
red points with error bars denote the composite spectrum from the
simulations. The blue dashed line shows the linear theory prediction,
which is clearly a poor fit to the measured data for most of the
scales of interest. The original {\tt halofit} algorithm is given by
the blue dot-dashed line. This appears to underestimate the true
nonlinear power on scales $k>0.2\kMpc$ at $z=0$ by roughly 10\%. For
$k>3\kMpc$ this discrepancy rises sharply.

On the other hand, the recalibrated {\tt halofit2012} model of
\citet{Takahashietal2012} provides a good description of the data at
the 5\% level on all scales. This rises slightly at $k>8\kMpc$,
however this discrepancy is likely owing to the fact that our data
approaches the Nyquist frequency of the FFT mesh at $k\sim12\kMpc$,
which leads to a small increase in the power. However, it fails to
capture the BAO oscillations on large scales, and also seems to
overpredict the amount of structure on quasi-linear scales by
$\lesssim5\%$. This is entirely consistent with the claimed accuracy
of the fitting function.


\subsection{Comparison with {\tt CosmicEMU}}

We now examine how the predictions from the {\tt CosmicEMU} model
\citep{Heitmannetal2014} compare with the estimates from our
simulations. We obtained the latest version of the code (Version 3),
which included updated constraints from the ``Mira-Titan Universe''
runs \citep[for details see][]{Lawrenceetal2017}.  Since the code
returns $P(k)$ in ${\rm Mpc^3}$ and $k$ in units of ${\rm Mpc}^{-1}$,
we multiply the returned power spectra by $h^3$ and the returned
wavemodes by $h$ to obtain our standard units.  In
Figure~\ref{fig:PkFidRatio} the results from the {\tt CosmicEMU} code
are presented as the solid green lines. We see that on quasi-linear
and nonlinear scales ($k>0.1\kMpc$) the emulator does an excellent job
to predict the nonlinear power at high precision. At $z=0$ the
differences are at the level of a few percent. At $z=0.5$ and $1.0$
the results are in even better agreement. At $z=2$ the results begin
to disagree at roughly $\lesssim 5\%$.
 
On large scales, however, the model appears to be less accurate.  For
$k<0.1\kMpc$ we find that the predictions underestimate the true power
by between 5\% and 10\%. We speculate that this owes to the fact that
the {\tt CosmicEMU} code does not use an externally computed linear
theory power spectrum -- something that the other methods do by design
-- but instead interpolates over a set of pre-generated linear spectra
(the 37+1 models used to calibrate the fitting function) to make
predictions.


\subsection{Comparison with {\tt HMCode}}

In Figure~\ref{fig:PkFidRatio} we also compare our fiducial runs with
the predictions from the {\tt HMCode} of \citet{Meadetal2015} (denoted
as the black dashed lines in all panels). We use the latest version of
{\tt CAMB} to evaluate this model for our fiducial cosmological
model. As can be seen from the figure, this semi-analytic model
provides perhaps the best description of our data overall. On large
scales it exactly recovers the linear theory. On small scales it is
calibrated to the {\tt CosmicEMU} model, which as we noted above,
provides a precise match to our data on these scales. Its main
shortcoming appears to be the precise modelling of the quasi-nonlinear
regime, in particular the nonlinear processing of the baryon acoustic
oscillations in the interval $k\in[0.07\kMpc,0.5\kMpc]$, especially
for the $z=0$ data.

Note that we do not consider the {\tt PkANN} code of
\citet{Agarwaletal2014}, since as discussed earlier the quite
restrictive set of scales of applicability ($k<1\kMpc$) means that
this neural network approach, whilst very promising is of limited use
for the applications of interest.

\vspace{0.2cm}

Before moving on we also note that in Figure~\ref{fig:PkFidRatio} as
one goes to higher redshift, a small systematic offset between the
spectra from the small- and large-scale box runs emerges. Our
investigations of the Millennium Run and Millennium-XXL simulations
has shown similar results \citep[see][]{SmithSimon2019}. This led us
to the conclusion that our simulations suffer from mass-resolution
effects \citep[for detailed studies
  see][]{Heitmannetal2010,Schneideretal2016}. Based on the difference
beween the small and large-box runs at the joining scale we therefore
estimate that there will be a systematic error in our fiducial spectra
of the order $\sim3\%$. This means that some of the differences that
we have noted above may be entirely driven by resolution effects. We
also note that we investigated a more elaborate joining scale
criterion that was set by finding the $k$-mode at which the size of
the sample variance error-bar equalled the error due to Poisson
shot-noise. However, this did not yield better results than the
empirical joining scale that we adopted through inspection of the
power spectral ratios -- one obvious problem with this more
sophisticated approach was the need to know {\em a priori} the
nonlinear power spectrum.


\section{A new angle on synthesising nonlinear power spectra}\label{sec:newmodel}

Based on the insufficiencies of the previous methods we now describe a
new approach to improving the accuracy of the current methods for a
wide range of cosmological models, but especially in the vicinity of
the peak of the posterior for the Planck data.


\subsection{The {\tt NGenHalofit} method}

To start, we assert that on large scales $k<k_{\rm cut}$ the power
spectrum can, to high accuracy, be described by a 2-loop MPT
calculation and that on smaller scales $k\ge k_{\rm cut}$ it can be
well represented by a nonlinear fitting function (we will use the
corrected MPT expressions described at the end of
\S\ref{sec:PT}). Thus our new scheme must enable us to interpolate
between these two regimes and we do this using,
\ba P_{\rm NGenHalofit}(k,z|\bm\theta) & = &
W(k|k_{\rm cut}) P_{\rm MPT}(k,z|\bm\theta) \nn \\
& & +\left[1-W(k|k_{\rm cut})\right] P_{\rm true}(k,z|\bm\theta) \ ,\ea
where $P_{\rm true}(k)$ is the `true' nonlinear matter power spectrum
given the cosmological model and redshift, and is free from noise.
For the interpolation function we adopt a scaled error function form,
\be W(k|k_{\rm cut})=\frac{1}{2}\left(1-{\rm Erf}
\left[\frac{\log_{10}(k/k_{\rm cut})}{\sqrt{2}\sigma_{\rm
      cut}}\right]\right)\ ,\ee
where $k_{\rm cut}$ controls the scale at which the filter $W(k|k_{\rm
  cut})$ acts as a low-pass filter and $1-W(k|k_{\rm cut})$ acts as a
high-pass filter. The strength of the transition is controlled by the
parameter $\sigma_{\rm cut}$. Figure~\ref{fig:filter1} shows how the
transition functions $W$ and $1-W$ behave for various choices for
$k_{\rm cut}$ and $\sigma_{\rm cut}$. After experimenting with various
values we found $k_{\rm cut}=0.07\kMpc$ and $\sigma_{\rm cut}=0.05$
gave the desired transition scale and speed.

In order to proceed further all we need is the true nonlinear model
power spectrum! Unfortunately we do not have access to this, however
what we do have access to is discrete realisations of this under the
assumption that the simulations provide a proxy for the ``truth''. We
also have analytic models that are accurate to $<10\%$. Let us denote
these as $P_{\rm model}(k)$. One can thus rewrite our desired true
spectrum as:
\ba 
P_{\rm true}(k,z|\bm\theta)
& = & P_{\rm model}(k,z|\bm\theta)\frac{P_{\rm true}(k,z|\bm\theta)}{P_{\rm model}(k,z|\bm\theta)} \nn \\
& = & P_{\rm model}(k,z|\bm\theta) y(k,z|\bm\theta) \ ,
\ea
where we have defined a function $y(k,z|\bm\theta)$ that characterises
the deviations from the known analytic model and $|y-1|<0.1$. Let us
next perform a Taylor expansion of the deviation function
$y(k,z|\bm\theta)$ with respect to the cosmological parameters
$\bm\theta$, about some fiducial model $\bm\theta_0$, whereupon
\ba 
\frac{P_{\rm true}(k,z|\bm\theta)}{P_{\rm model}(k,z|\bm\theta)}
& = & y(k,z|\bm\theta_0)
+\sum_{i}\left.\frac{\partial y(k,z|\bm\theta_0)}{\partial\theta_i}\right|_{\bm\theta_0}\Delta\theta_i \nn \\
  && \hspace{-2cm}
+\frac{1}{2}\sum_{ij}\left.\frac{\partial^2 y(k,z|\bm\theta_0)}
{\partial\theta_i\partial\theta_j}\right|_{\bm\theta_0}
  \Delta\theta_i\Delta\theta_j
  +\dots  \ .
\ea
If we now factor out the first term from the square brackets on the
right-hand-side, then we see that the true power spectrum can be
written as:
\ba 
\frac{P_{\rm true}(k,z|\bm\theta)}{P_{\rm model}(k,z|\bm\theta)}
\!\!& = &\!\!  y(k,z|\bm\theta_0)
\left[1+
  \sum_i{\mathcal R}_i^{(1)}(k,z|{\bm\theta_0})\Delta\theta_i \right. \nn \\
  & & \left. +
  \frac{1}{2}\sum_{i,j}{\mathcal R}_{ij}^{(2)}(k,z|{\bm\theta_0})\Delta\theta_i\Delta\theta_j+\dots
\right]
\ea
where we have defined the 1st and 2nd order, power spectral ratio,
logarithmic derivative functions as:
\ba
   {\mathcal R}^{(1)}_i(k,z|{\bm\theta}) & \equiv &
   \frac{\partial \log
     y(k,z|\bm\theta)}{\partial\theta_i} \ ;\\
        {\mathcal R}^{(2)}_{ij}(k,z|{\bm\theta}) & \equiv &
        \frac{1}{y(k,z|\bm\theta)}\frac{ \partial^2 
          y(k,z|\bm\theta)}{\partial\theta_i\partial\theta_j} \ .
        \ea
Finally, if we back substitute for $y(k,z|\bm\theta_0)$ in the above
expression, then on keeping all terms up to quadratic order we arrive
at the approximate formula:
\ba
\frac{P_{\rm true}(k,z|\bm\theta)}{P_{\rm model}(k,z|\bm\theta)}
\!\!& \approx &\!\!  \frac{P_{\rm true}(k,z|\bm\theta_0) }{P_{\rm model}(k,z|\bm\theta_0)}
\left[1+ \sum_i{\mathcal R}_i^{(1)}(k,z|{\bm\theta}_0)\Delta\theta_i\right.\nn \\
& & \hspace{-1cm}\left. +
\frac{1}{2}  \sum_{i,j}{\mathcal R}_{ij}^{(2)}(k,z|{\bm\theta}_0)\Delta\theta_i\Delta\theta_j
\right] \ \label{eq:PNgen} .
\ea
Several points are worth noting: first, if we are exactly evaluating
the fiducial model, then $\Delta\theta_i=0$ and the bracketed terms
vanish, leaving only $y(k,z|\bm\theta_0)$ to recalibrate the $P_{\rm
  model}$. This can be obtained directly from a set of simulations of
the same cosmological model as the fiducial point. Second, the more
accurately our assumed model describes the simulated data, the more
rapidly our series of derivative functions will converge to zero,
i.e. ${\mathcal R}^{(1)}\rightarrow0$ and ${\mathcal
  R}^{(2)}\rightarrow0$. Third, if $|{\mathcal R}^{(1)}|<0.1$ then a
10\% calibration would produce power spectrum predictions that are
$\lesssim1\%$ accurate for parameter differences
$|\Delta\theta|=\sqrt{\theta_1^2+\dots+\theta_N^2} <0.1$.  Fourth, in
this work we assume that $P_{\rm true}$ can be obtained from $N$-body
simulations and on large scales MPT. In the future, it is hoped that
if and when new improved analytic models and simulations are
available, these ingredients should be easy to interchange, thus
providing a rapid upgrade path for this approach.

One can implement this method at several levels. The zeroth order
implementation would simply be to drop all of the response functions
and so recalibrate our model at the fiducial point. The 1st order
implementation would be to include ${\mathcal R}^{(1)}$. To do this
one needs to generate the vector of first order partial derivatives of
the simulated power spectrum with respect to each of the cosmological
parameters at the fiducial point. The 2nd order would be to build the
Hessian of the power spectrum with respect to the parameters. In this
work we will aim to achieve the 1st order correction, but also a
hybrid between 1st and 2nd order corrections, since we can populate
the diagonal entry of the Hessian using our current data.


\subsection{Practical issues}\label{ssec:practical}

In what follows we shall take $P_{\rm model}\rightarrow P_{\rm
  Halofit}$ as the approximate model, using the recalibrated model of
\citet{Takahashietal2012}. The recalibration of the approximate model
and the nonlinear response functions can be computed from a small set
of $N$-body simulations, which we will describe in the following
section. However, before we do this there are a few obstacles to
overcome in order for the above approach to be practicable.

\vspace{0.2cm}

\noindent $\bullet$ {\bf Parameter space coverage:} One of the above
virtues, can also prove to be a drawback to this approach: if one
considers large deviations from the fiducial point $\bm\theta_0$, such
that $\Delta\theta_i\gtrsim 1$, then the Taylor expansion will break
down.  Hence, if one desires to perform cosmological parameter
estimation using an uninformative prior, then the accuracy of the
theoretical model will vary across the parameter space.  This gives
rise to the pitfall: a blind application of this method may lead to
artificially good constraints on the posterior in the regions around
the fiducial model where the model is good (see for example the
discussion in Marian et al. 2018, in prep. for more details about this
effect). We overcome this by making the following modification to the
procedure: we replace the second term in the bracket on the
right-hand-side of \Eqn{eq:PNgen} with:
\be
\sum_i{\mathcal R}_i^{(1)}(k,z|{\bm\theta_0})\Delta\theta_i \rightarrow
W_{\rm STH}(\bm\theta)\sum_i{\mathcal R}_i^{(1)}(k,z|{\bm\theta_0})\Delta\theta_i\ ,
\ee
where 
\be W_{\rm STH}(\bm\theta|\bm\theta_{\rm cut},\bm\sigma^{\theta}_{{\rm cut}})\equiv 
\prod_{j=1}^{N_{\rm par}}W_{\rm STH}(\Delta\theta_j|\theta_{{\rm cut},j},\sigma^{\theta}_{{\rm cut},j}) \ee
and where $N_{\rm par}$ is the number of cosmological parameters and the
function $W_{\rm STH}(\Delta\theta_j|\theta_{{\rm
    cut},j},\sigma^{\theta}_{{\rm cut},j})$ behaves as a smoothed
top-hat function centred on the fiducial parameter choice. A similar
transition function multiplies the term involving ${\mathcal
  R}_i^{(2)}$. In what follows we use the form:
\ba
W_{\rm STH}(\theta) \!\!\! & = & \!\!\!
-\frac{1}{2} 
\left\{
{\rm Erf}\left[\frac{\theta-\theta_{\rm cut}}{\sqrt{2}\sigma^{\theta}_{\rm cut}}\right] 
 +{\rm Erf}\left[\frac{-\theta-\theta_{\rm cut}}{\sqrt{2}\sigma^{\theta}_{\rm cut}}\right]
\right\} \ .
\ea 
For models with $|\Delta\theta|>\theta_{\rm cut}$, this procedure will
guarantee a smooth transition (modulated by $\sigma^{\theta}_{\rm
  cut}$) to the original theoretical model $P_{\rm
  model}(k,a|{\bm\theta})$, but recalibrated at the fiducial point
${\bm\theta_{0}}$.  It therefore will enable $P_{\rm
  model}(k,a|{\bm\theta})$ to cover the full parameter space beyond
where the Taylor expansion breaks down. Within the vicinity of the
fiducial model the theory will become significantly more accurate.

Figure~\ref{fig:filter2} demonstrates how the smoothed top-hat
function behaves for an arbitrary parameter for various choices of
$\theta_{\rm cut}$ and $\sigma^{\theta}_{\rm cut}$. We discuss how we 
choose the values of $\theta_{\rm cut}$ in \S\ref{ssec:precision}.


\vspace{0.2cm}

\noindent $\bullet$ {\bf Smooth functions:} As written in
\Eqn{eq:PNgen}, one needs to possess a {\em smooth analytic} model for
the function $P_{\rm true}(k,z|\bm\theta_0)$ -- i.e. the true
nonlinear power spectrum at the Fiducial point in our parameter
space. Secondly, one also needs the same thing for the response
functions ${\mathcal R}_i^{(1)}(k,z|{\bm\theta_0})$ -- where one
function obtains for every cosmological parameter that we consider --
and similarly for ${\mathcal R}_i^{(2)}(k,z|{\bm\theta_0})$, but this
time giving $N(N+1)/2$ functions.

Considering $P_{\rm true}(k,z|\bm\theta_0)$, let us assume that an
accurate $N$-body simulation can measure this function over a set of
scales $(k_{\rm min}<k<k_{\rm max})$ and over a range of expansion
factors $(a_{\rm min}<a<a_{\rm max})$, but that we only have this
information at a discrete set of lattice points:
\be
{\mat{P}_{\rm true}}=\left(\begin{array}{ccc}
  P_{\rm true}(a_M,k_1) & \dots & P_{\rm true}(a_M,k_N) \\
  \vdots & \vdots & \vdots \\
  P_{\rm true}(a_1,k_1) & \dots & P_{\rm true}(a_1,k_N) 
\end{array}\right)\ ,
\ee
where we have suppressed the dependence on the cosmological parameters
$\bm\theta_0$. At this point one might consider using a bicubic
interpolation scheme between the lattice points. However, there is a
small degree of ambiguity as to how one should do this since the
growth between two timesteps is not linear -- for instance on large
scales we know that $P\propto D^2(a)$. One way to obviate such
problems is to rescale the elements of the above matrix by the factor
$P_{\rm model}(k,z|\bm\theta_0)$:
\be
{\mat{Y}}=\left(\begin{array}{ccc}
  y(a_N,k_1) & \dots & y(a_M,k_N) \\
  \vdots & \vdots & \vdots \\
  y(a_1,k_1) & \dots & y(a_1,k_N)
\end{array}\right) \ .
\ee
This ratio has the useful property that for spatial- and time-scales
$k\in\{(a_1,k_1),\dots,\,(a_l,k_i)\}$ that are large enough and early
enough for linear theory to be accurate, $\mat{Y}_{ij}\rightarrow
1$. Nevertheless, provided the model is good, the
$y(a_i,k_j)\approx1+\epsilon$, where we will assume $\epsilon \lesssim
0.1$. In the absence of errors in the determination of the
$y(a_i,k_j)$, one would then simply compute a bi-cubic spline
interpolation of the data to obtain the model for $y(a,k)$.

\vspace{0.2cm}

\noindent $\bullet$ {\bf Noisy data}: Our next concern is that the
elements of the matrix $\mat{Y}$ have errors associated with them,
since we do not obtain the true $y(a_i,k_j)$ directly, but only an
estimate $\hat{y}(a_i,k_j)$ from an ensemble of simulations. The
errors on large scales are dominated by cosmic variance and those on
small scales by shot-noise. We may account for this by using `Basis
splines' or `B-spline' functions. These differ from interpolating
splines in that the resulting curve is not required to pass through
each data point. Instead, one constructs a sequence of cubic
polynomials that are piece-wise connected together at a set of
carefully chosen node points, where the piece-wise polynomials are
connected at the node points in such a way that the resultant function
is continuous at the nodes. The free parameters that govern each
piece-wise polynomial between two nodes are obtained in such a way to
reproduce the data in a least squares sense -- hence this is also
called a smoothing spline \citep[for a detailed discussion of
  B-splines see][]{deBoor1978}. In the {\tt NGenHalofit} code we use
the routines provided in the {\tt GSL}
libraries \footnote{www.gnu.org\/software\/gsl\/manual\/html\_node\/Basis-Splines.html},
in particular {\tt gsl\_bspline\_$\star$}. 

Owing to our desire to accurately capture the baryon acoustic
oscillation features, we perform two separate B-spline fits to the
data.  The first was for the spectra over scales
$0.05\kMpc<k<0.4\kMpc$, with ${\tt NCOEFFS=20}$ and therefore giving
${\tt NBREAK=NCOEFFS-2}=18$ logarithmically spaced spectra node
points. This had the desired flexibility to capture rapid features,
whilst still being coarse enough to filter out noise. The second
B-spline was for the data over scales $0.4\kMpc<k<10\kMpc$, with ${\tt
  NCOEFFS=10}$ and therefore giving ${\tt NBREAK=NCOEFFS-2}=8$
logarithmically spaced spectra node points. The B-spline approach
required us to specify the variance of the data points and for the
small-box fiducial runs we used the errors from the ensemble, whereas
for the large-box run we assumed these to follow the Gaussian error
model of \Eqn{eq:Gaussian}.

Owing to the fact that we are not aware of a publicly available
bicubic B-spline routine, we overcome this problem by making use of
the B-spline functions to model each row of the matrix $\mat{Y}$. We
then recompute the values $y(a_i,k_j)\rightarrow y_{\rm
  B-spline}(a_i,k_i)$ and so form the matrix: $\mat{Y}\rightarrow
\mat{Y}_{\rm B-spline}$.  For this smoothed matrix we are now able to
use a bi-cubic spline routine to interpolate between the elements of
the $(a_i,k_j)$ matrix. We use the bi-cubic spline routine as
implemented in the {\tt GSL} libraries as routines {\tt
  gsl\_spline2d\_$\star$}.

Finally, the approach described above for modelling $y(a,k)$ can be
implemented identically for the functions ${\mathcal
  R}_i^{(1)}(k,z|{\bm\theta_0})$ and ${\mathcal
  R}_i^{(2)}(k,z|{\bm\theta_0})$ etc, only one needs to generate
arrays of such spline functions.


\begin{figure*}
\centerline{
  \includegraphics[width=8.5cm]{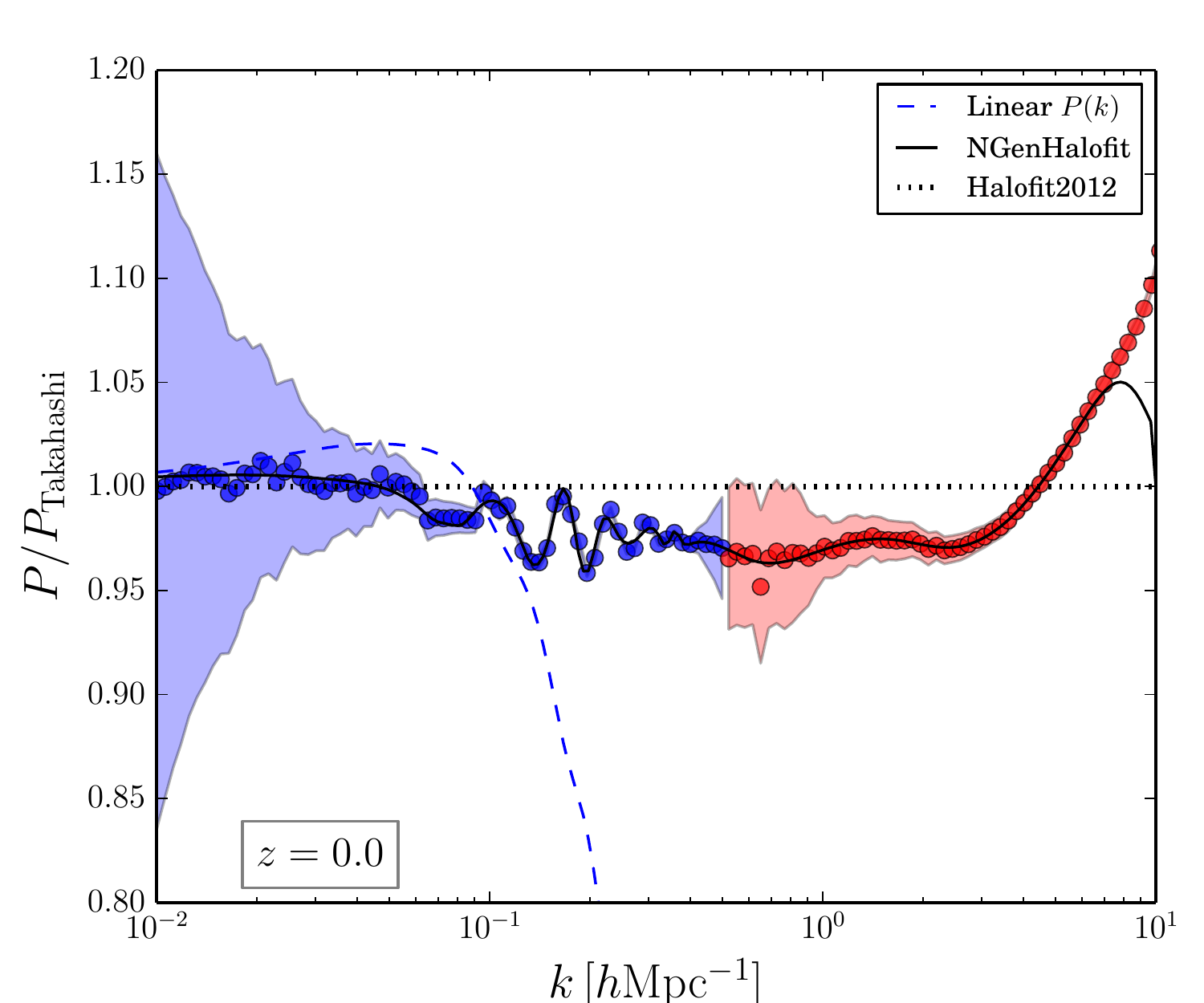}\hspace{0.3cm}
  \includegraphics[width=8.5cm]{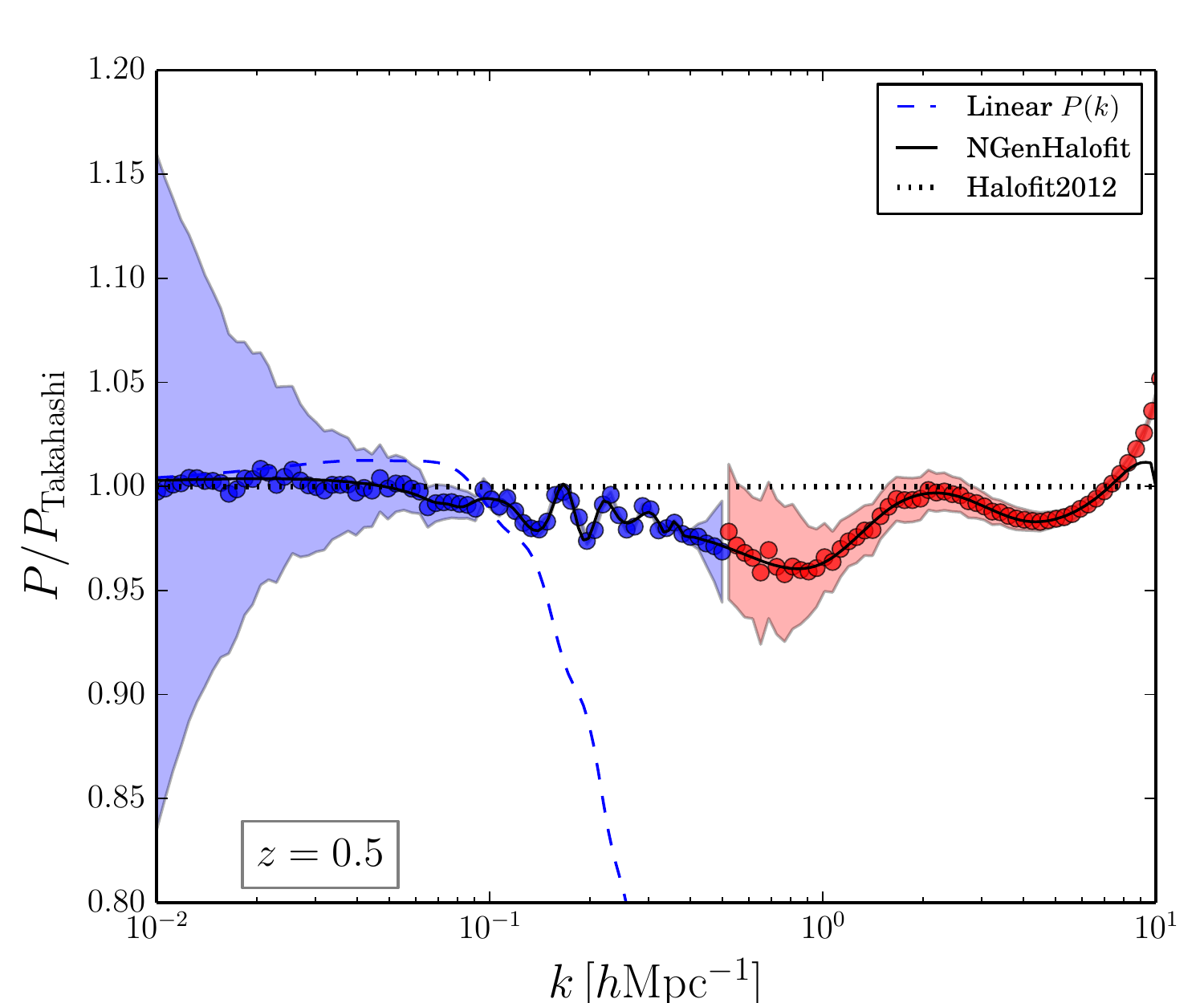}}
\centerline{
  \includegraphics[width=8.5cm]{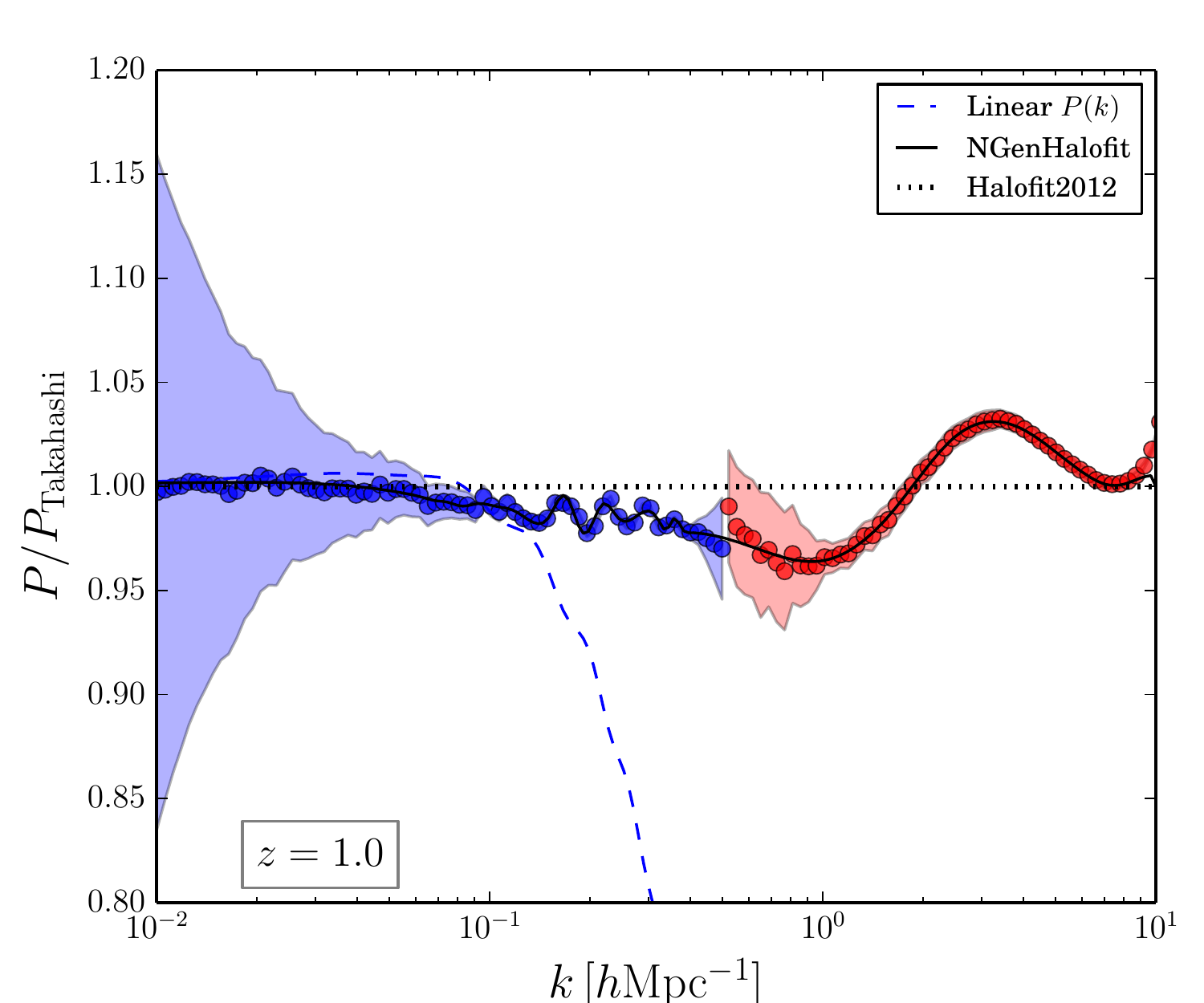}\hspace{0.3cm}
  \includegraphics[width=8.5cm]{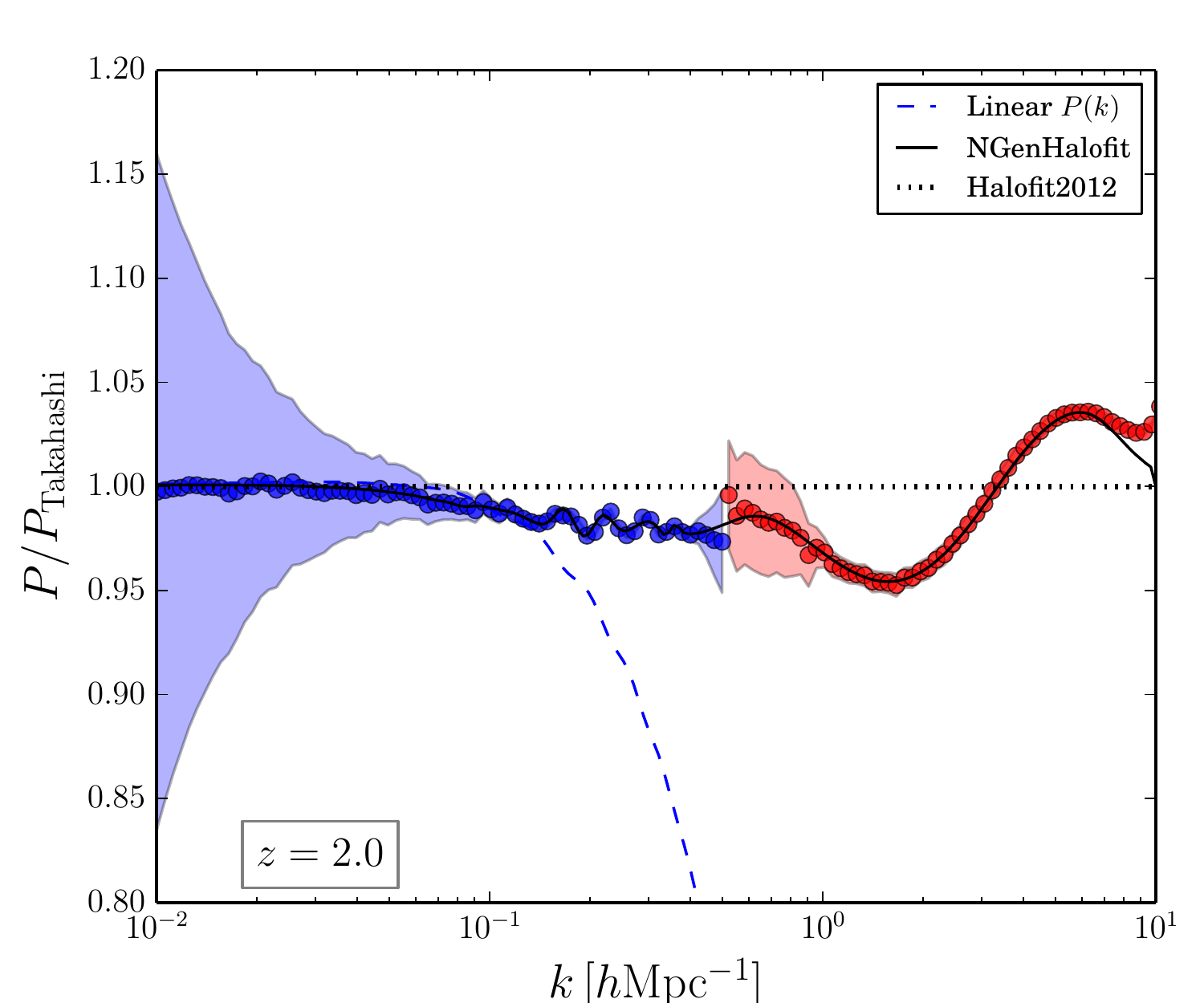}}
\caption{\small{Evolution of the ratio of the measured nonlinear
    matter power spectra with the predictions from the {\tt
      halofit2012} model of \citet{Takahashietal2012} as a function of
    spatial wavenumber. The top left, top-right, bottom left, bottom
    right sub-figures show the spectra for $z=0$, $z=0.5$, $z=1.0$,
    and $z=2.0$, respectively. In all panels, the red points with
    error bars show the mean and 1$\sigma$ errors for the power
    spectra measured from the Planck-2013-like fiducial model
    simulations in boxes of side $L=500\Mpc$.  The solid blue line
    presents our {\tt NGenHalofit} model, the dotted line shows the
    updated {\tt halofit2012} model of \citet{Takahashietal2012}, the
    dot-dash line shows the original {\tt halofit} model of
    \citet{Smithetal2003}, and the dash line shows the linear theory.}
\label{fig:PkFid}}
\end{figure*}


\subsection{The new model compared to the fiducial data}\label{ssec:fiducial}

Figure~\ref{fig:PkFid} is similar to Fig.~\ref{fig:PkFidRatio} and
shows the evolution of the ratio of the nonlinear matter power
spectrum with respect to the updated {\tt halofit2012} model developed by
\citet{Takahashietal2012}. However, in this version of the figure we
now compare the measured results with the predictions from our new
{\tt NGenHalofit} code. We see that the data and model predictions are
in excellent agreement to high precision on all scales. This good
agreement is not surprising, since our zeroth order correction to
{\tt halofit2012}, is to renormalise to match our fiducial simulations
through the factor $y(k,z|\bm\theta_0)$.

There are however, two key points to note: first, the solid lines in
the figure have been evaluated for an arbitrary set of $k$-modes,
meaning that our interpolation scheme is working correctly. Second,
one can see that the smoothing spline reproduces most of the features
in the data, especially the nonlinear processing of the baryon
acoustic oscillations. On the other hand, it has enough restrictions
that it is not reproducing all of the noise fluctuations seen in the
figure, especially around the joining scale between the spectra from
the large- and small-box runs.



\section{Cosmology dependence}\label{sec:cosmodep}

In this section we examine the cosmological dependence of the
nonlinear matter power spectra as a function of scale.


\begin{figure*}
\centerline{
\includegraphics[width=6.5cm,angle=0]{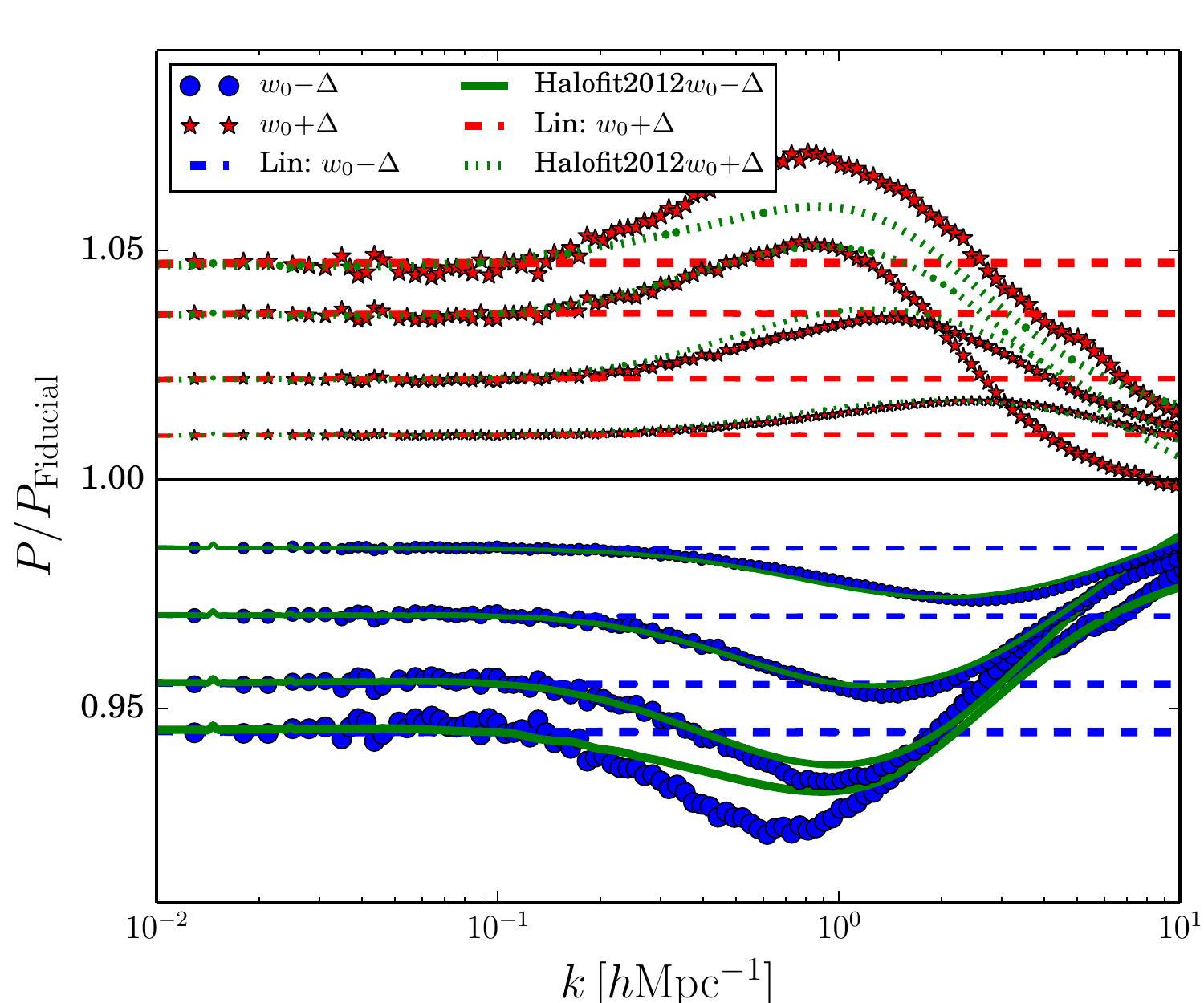}\hspace{0.3cm}
\includegraphics[width=6.5cm,angle=0]{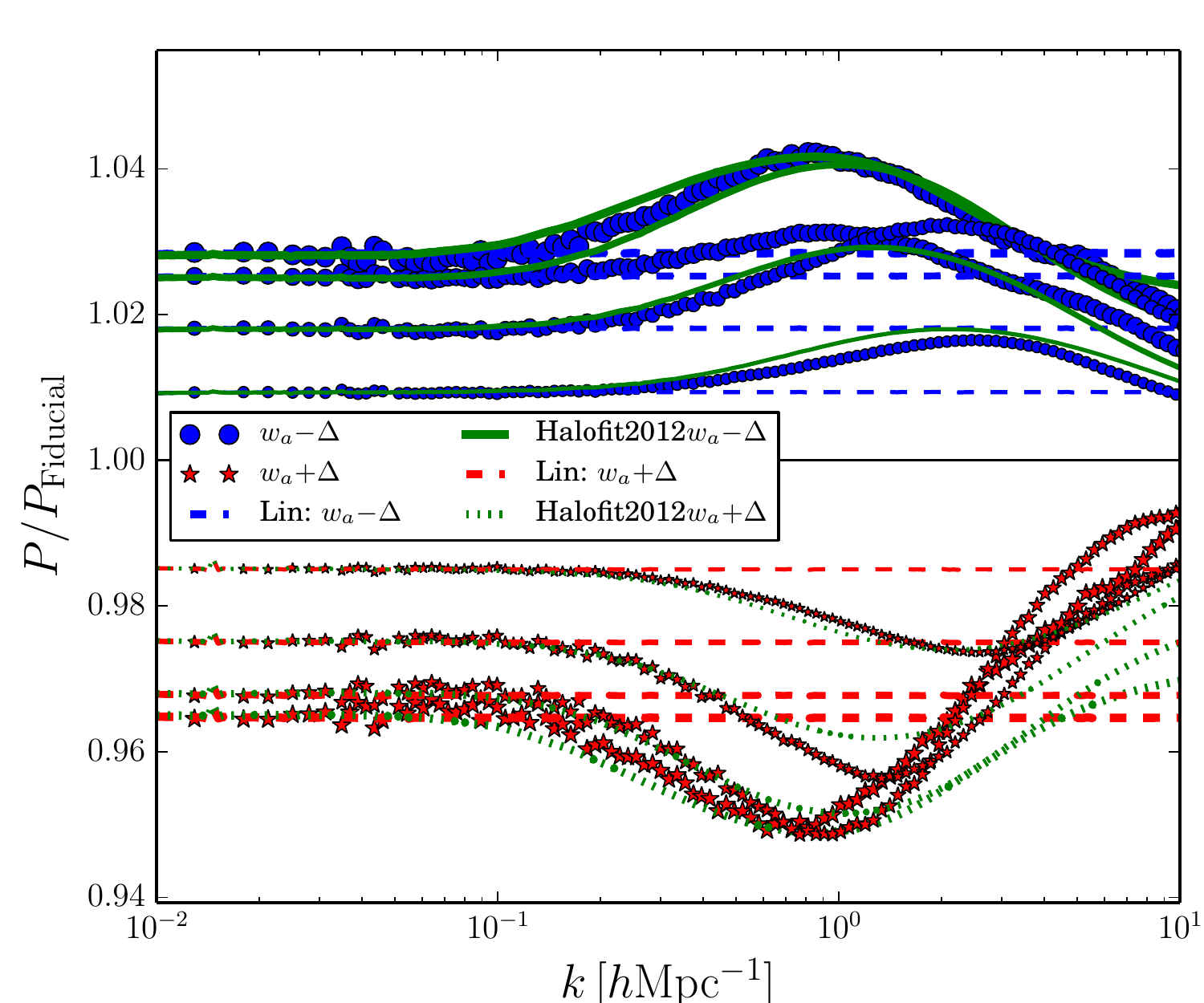}}
\vspace{0.2cm}
\centerline{
\includegraphics[width=6.5cm,angle=0]{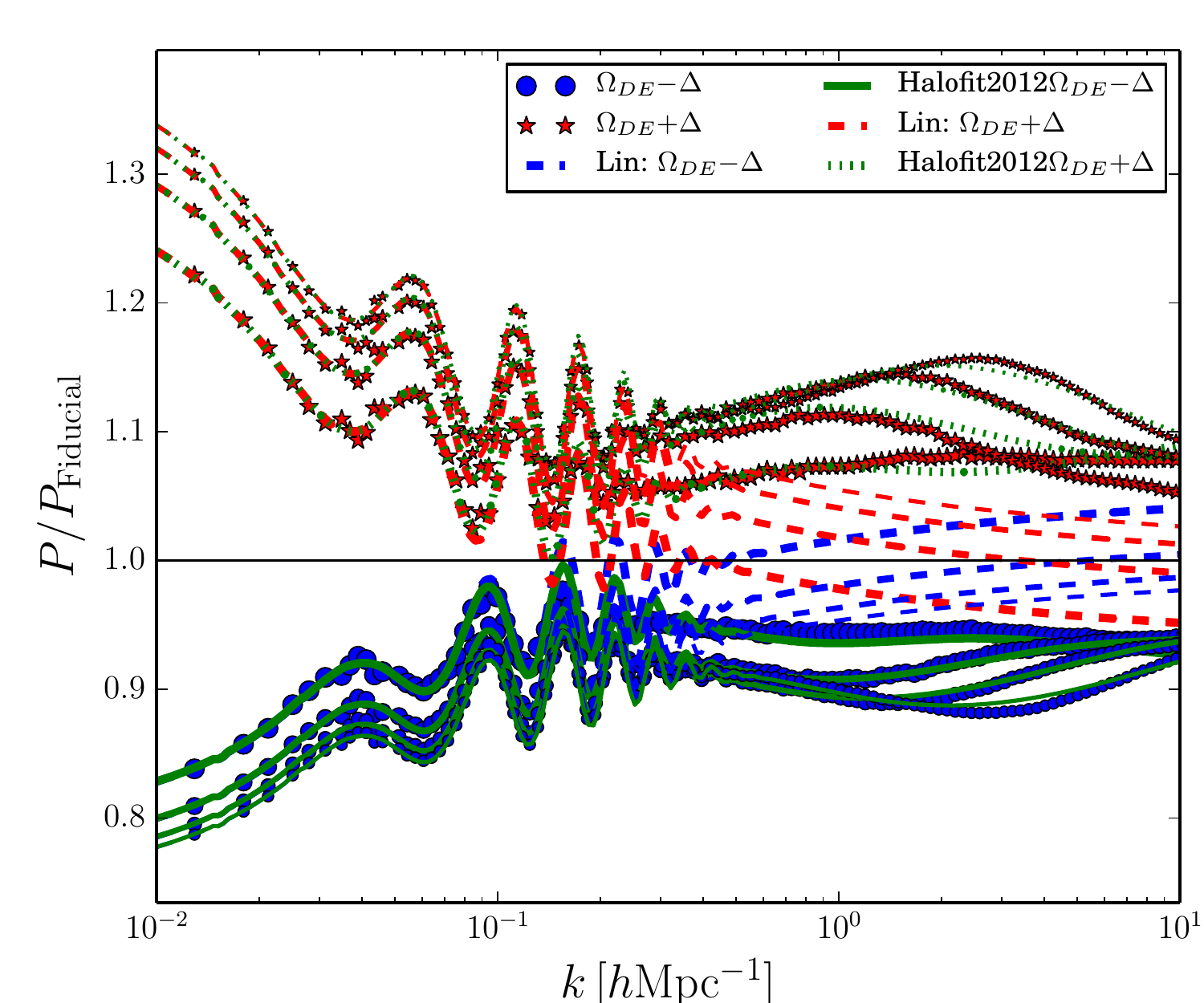}\hspace{0.3cm}
\includegraphics[width=6.5cm,angle=0]{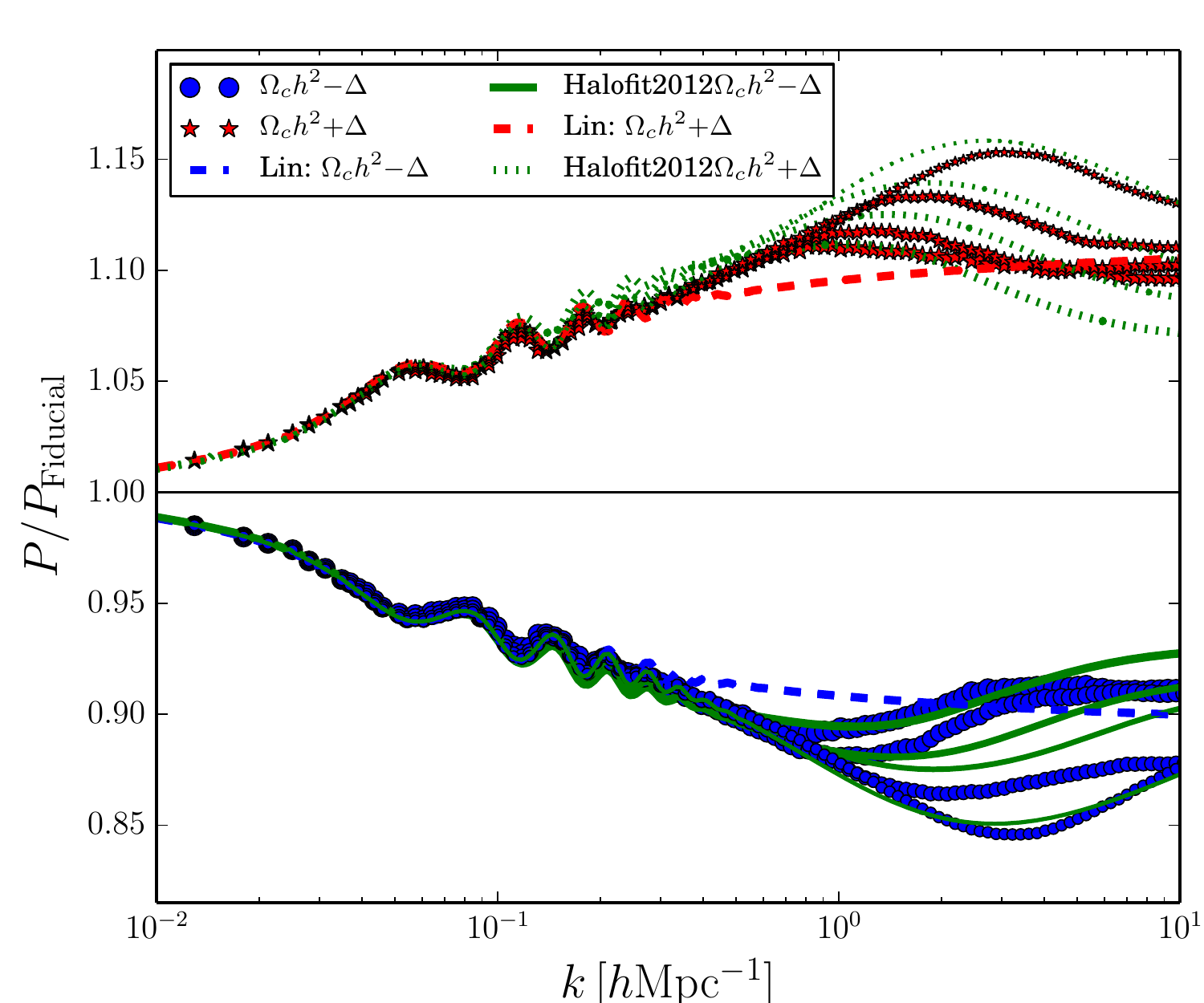}}
\vspace{0.2cm}
\centerline{
\includegraphics[width=6.5cm,angle=0]{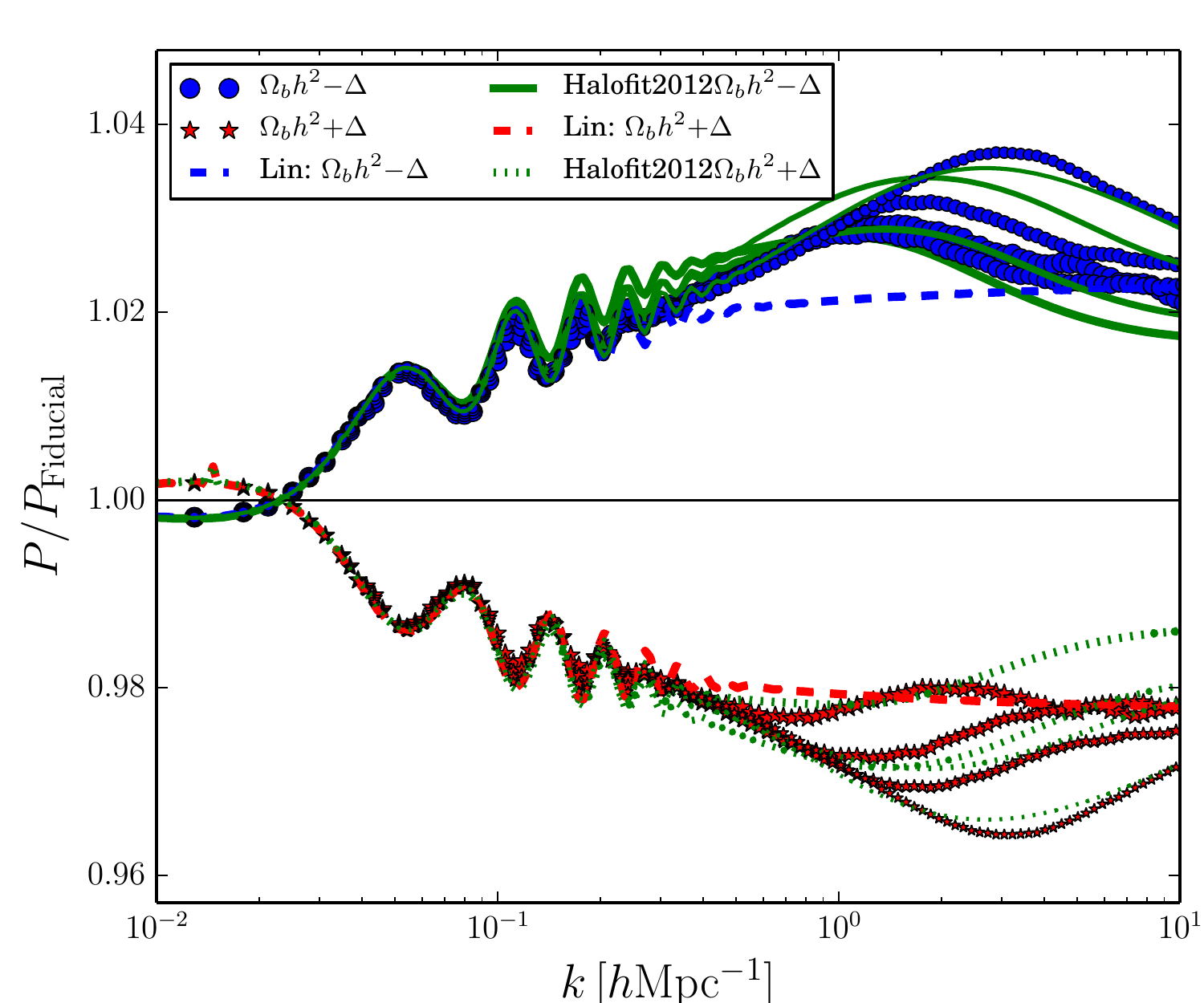}\hspace{0.3cm}
\includegraphics[width=6.5cm,angle=0]{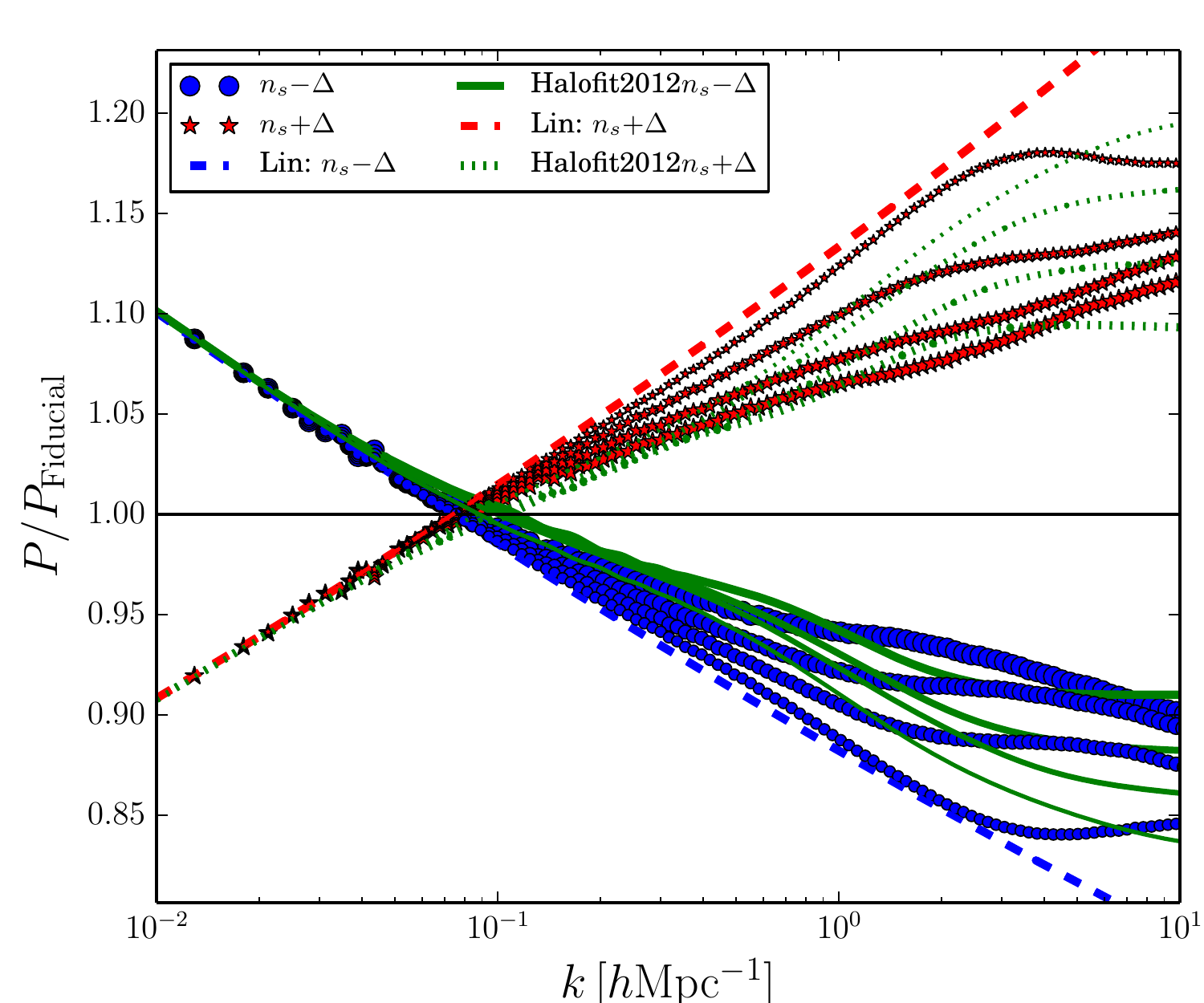}}
\vspace{0.2cm}
\centerline{
\includegraphics[width=6.5cm,angle=0]{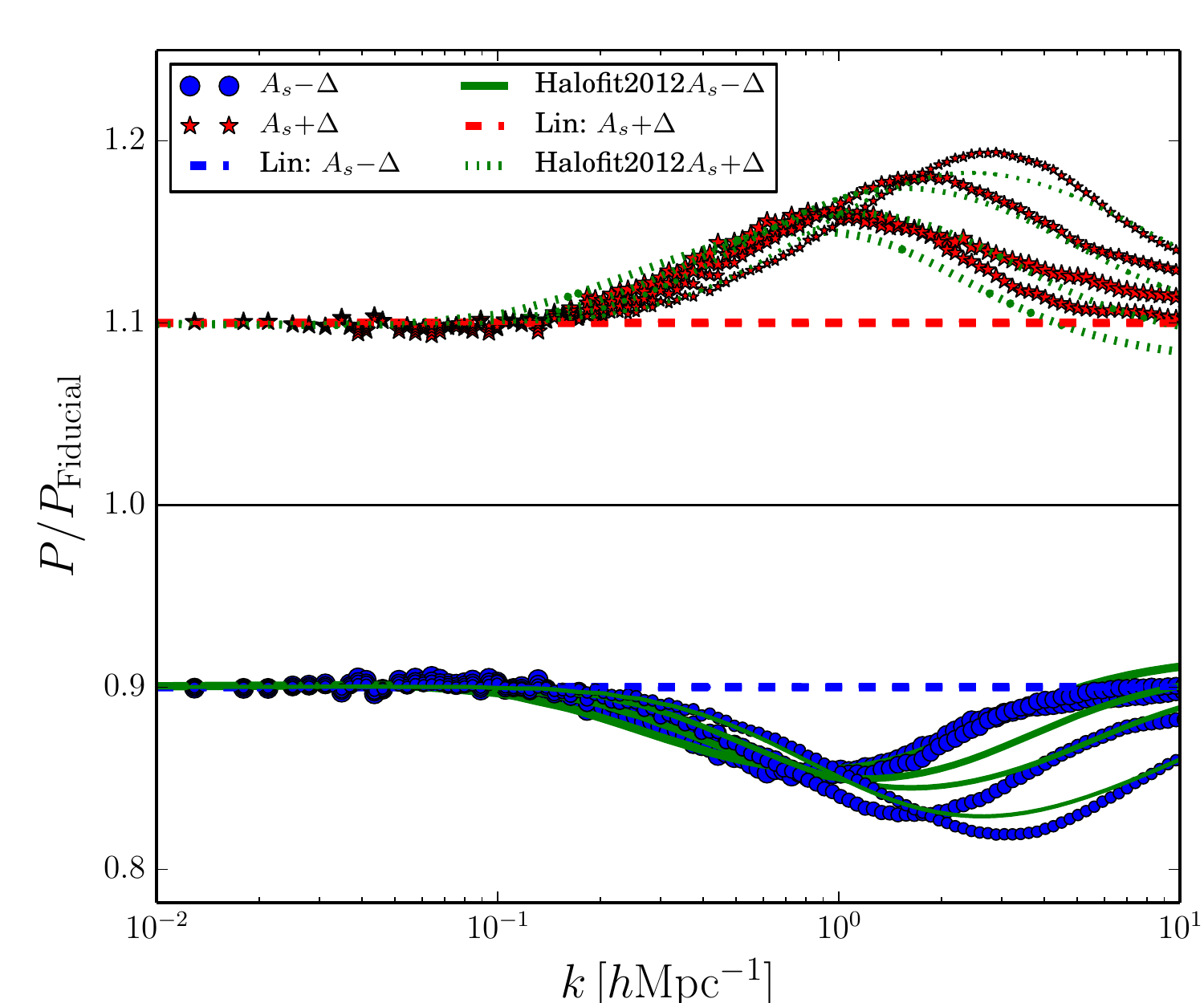}\hspace{0.3cm}
\includegraphics[width=6.5cm,angle=0]{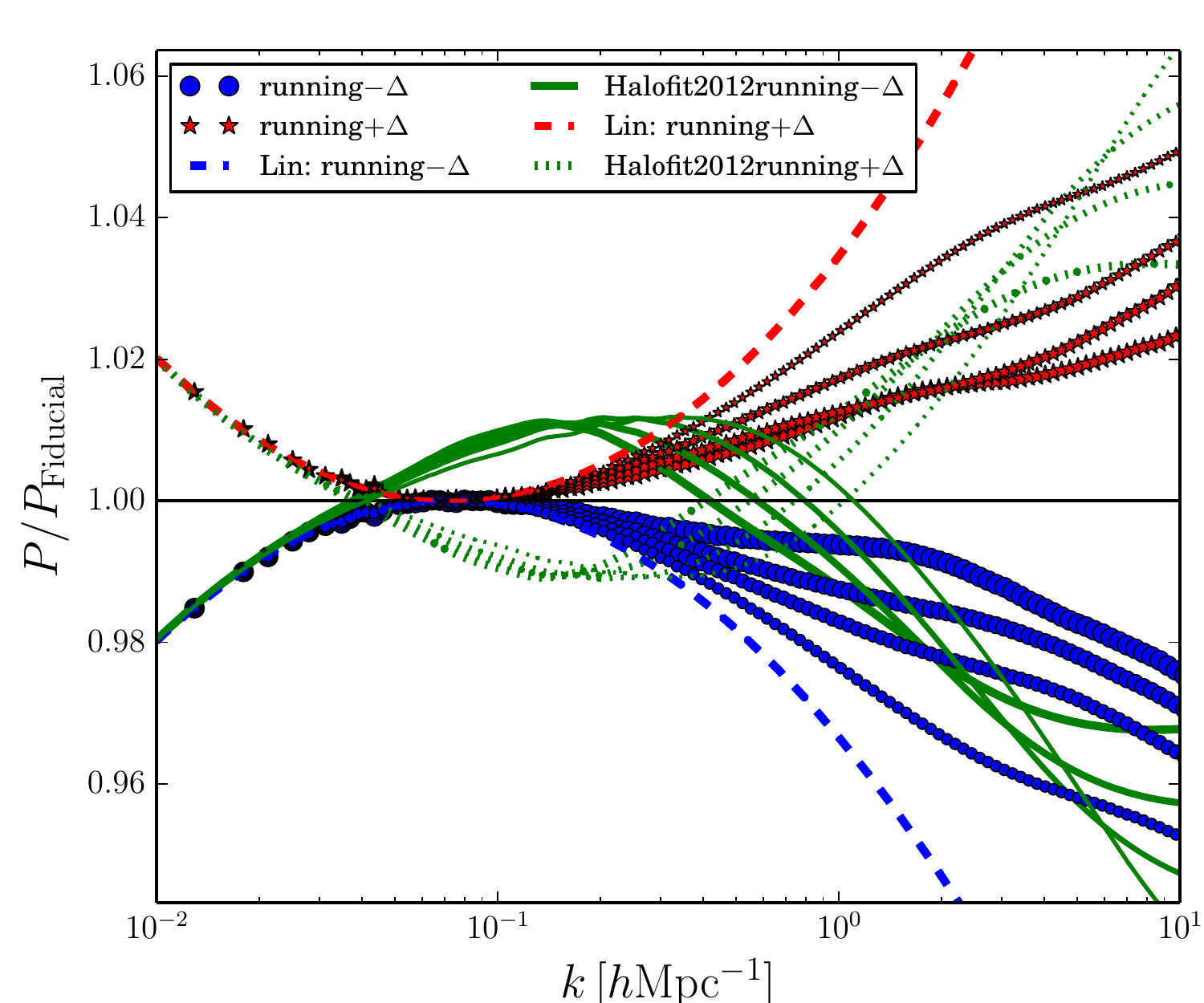}}
\caption{\small{Dependence of the power spectrum on variations in the
    cosmological parameters. All plots show the ratio of the
    variational models with respect to the fiducial model. Each panel
    shows the variations for a single parameter as a function of
    scale. The red and blue points show positive and negative
    variations, respectively and the point size increases with
    decreasing redshift, with $z\in\{2,1,0.5,0.0\}$. The dashed lines
    show the results for linear theory. }
\label{fig:PkCosmoRatios}}
\end{figure*}


\subsection{Spectral corrections}\label{ssec:corr}

For some of our smaller box runs we found that there were some small
$\sim1\%$ deviations from linear theory on very large-scales. In
Appendix~\ref{app:one} we performed an investigation to understand the
origin of this error.  In turned out to be due to two effects.

First, the code {\tt Gadget-3} does not dump snapshots exactly at the
listed outputs. There can be small but significant deviations from the
requested expansion factor -- even for the case of the same
cosmological model simulated with different initial conditions.  Owing
to that, we evolved all of the snapshots to exactly the same set of
expansion factors. This was done by assigning a reference list of
snapshot expansion factors, and for this we used the fiducial run1
small box set. For each simulation we then computed the linear growth
factor to the exact output redshift of the snapshot and also the
growth factor to the desired reference expansion factor. The corrected
spectrum can then be obtained through linear extrapolation:
\be P^{(1)}_{\rm Sim}(k,a_*) = \frac{D^2(a_*)}{D^2(a_{\rm sim})} P^{(0)}_{\rm Sim}(k,a_{\rm sim}) \ ,\ee
where $a_*$ and $a_{\rm sim}$ are the desired reference and actual
simulated expansion factors, respectively. Note that here we are
assuming that the nonlinear spectrum does not evolve appreciably
between $a_{\rm sim}$ and $a_{*}$. This is reasonable, since we are
considering deviations of the order $\lesssim1\%$ between $a_{\rm
  sim}$ and $a_{*}$.

Second, as the Appendix \ref{app:one} shows, if one explores how the
power spectral modes with $k<0.04\kMpc$ evolve compared to linear
theory, then in most cases the results are $\lesssim 0.6\%$
agreement. However, for some cases, the differences could be as high
as $1.5\%$ (in the appendix we speculate about the cause of this
effect). We deal with this by applying a small correction to all of
our variation spectra to guarantee that the large-scale modes
$k<0.03\kMpc$ match linear theory exactly. We estimate the correction
through reference to the ratio of the variation to the fiducial model,
through:
\be
C({\bm\theta},a) = \frac{1}{N_<} \sum^{N_{<}}_i
\frac{[P_{\rm Lin}(k_i,a|{\bm \theta})/P_{\rm Lin}(k_i,a|{\bm \theta_0})]}
{[P^{(1)}_{\rm Sim}(k_i,a|{\bm \theta})/P^{(1)}_{\rm Sim}(k_i,a|{\bm \theta_0}]} \ .
\label{eq:corr}\ee
where $N_<$ are all the modes less than $k=0.03\kMpc$. Thus our second
correction can be written:
\be P^{(2)}_{\rm Sim}(k,a|{\bm\theta}) =
C({\bm\theta},a) P^{(1)}_{\rm Sim}(k,a|{\bm\theta}) \ ,\ee
Clearly, this correction can only affect the runs with
$\bm\theta\ne\bm\theta_0$, since for the fiducial runs
$C({\bm\theta_0},a)\equiv1$. No such correction is applied to the
fiducial runs. On large-scales this is of no consequence, since we
make use of the large-box run, for which the large-scale power very
accurately reproduces linear theory. However, on small scales, where
the small boxes are used, this may lead to a small ($\lesssim0.5\%$)
discontinuity between the large- and small-box solutions at the
joining scale (The $z=2$ panel of Fig.~\ref{fig:PkFid} shows some
evidence of this).


\subsection{Power spectral ratios}

In Figure~\ref{fig:PkCosmoRatios} we show the ratios of the measured
nonlinear power spectra for the variations in the cosmological models
described in Table~\ref{tab:cospar} with the fiducial power spectrum,
as a function of scale and for several epochs. While the absolute
value of the power for any given simulation is very noisy on
large-scales, owing to the fact that we have used the same phase
realisation, the ratio with respect to the corresponding fiducial run
cancels out most of the cosmic variance on large-scales, leaving a
relatively noiseless quantity\footnote{See \citet{Smithetal2007} where
  this technique was applied to reduce cosmic variance errors in halo
  clustering.}. Thus, after implementing the corrections described
above, we see that in all of the panels the large-scale ratios
accurately match linear theory predictions. Whilst the calibrations
were made for scales $k<0.03\kMpc$ we clearly see good agreement with
the linear theory ratios up to scales of the order $k\sim0.1\kMpc$.
On smaller scales, nonlinear evolution drives the measured ratios away
from the linear theory prediction. 

For the case of the dark energy variations we see that modifications
to $w(a)$ primarily affect the linear growth rate on
large-scales. However, as we see they also lead to increased/decreased
nonlinear power as $w(a)$ becomes more/less negative \citep[see
  also][]{LinderJenkins2003}. Considering the case of a running power
spectral index, we see that for $\alpha$ positive/negative the power
is boosted/suppressed on all scales compared to the fiducial model,
with the exception of the pivot scale of the primordial power
spectrum. For a variation $\alpha=\pm0.01$ the nonlinear power can be
boosted by several percent, with the correction increasing with
increasing redshift.

It is interesting to note that the linear theory ratios reach deeper
into the nonlinear regime than the absolute value of the linear theory
-- compare with Fig.~\ref{fig:PkFid}, where 2-3\% departures on
scales of $k\sim0.05\kMpc$ are already present for the absolute value.

In the plots we also show how well the {\tt halofit2012} mode does at
matching the cosmological dependence of the suite of D\"ammerung runs
(depicted as the thin solid and dashed green lines in the panels). For
the parameters $\{w_0,w_a,\Omega_{\rm m},\Omega_{\rm c}h^2,A_s\}$ the
model does excellently, with deviations being less than a few
percent. However, for the parameters $\{\Omega_{\rm
  b}h^2,n_s,\alpha\}$, the prescription does less well with errors
being of 5\% or more, with the case of the running of the primordial
power spectrum being especially bad.


\begin{figure*}
\centerline{
\includegraphics[width=6.5cm,angle=0]{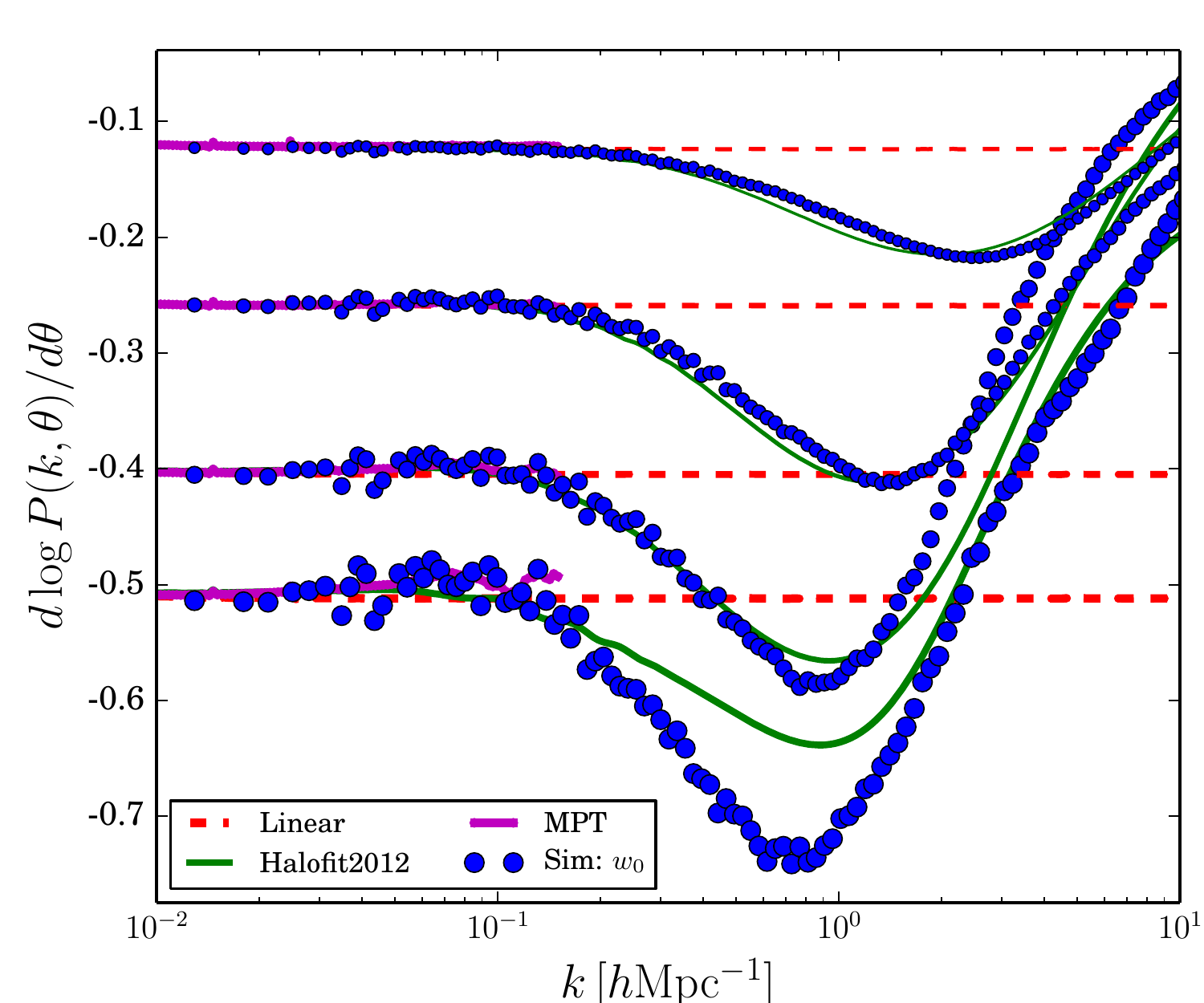}\hspace{0.3cm}
\includegraphics[width=6.5cm,angle=0]{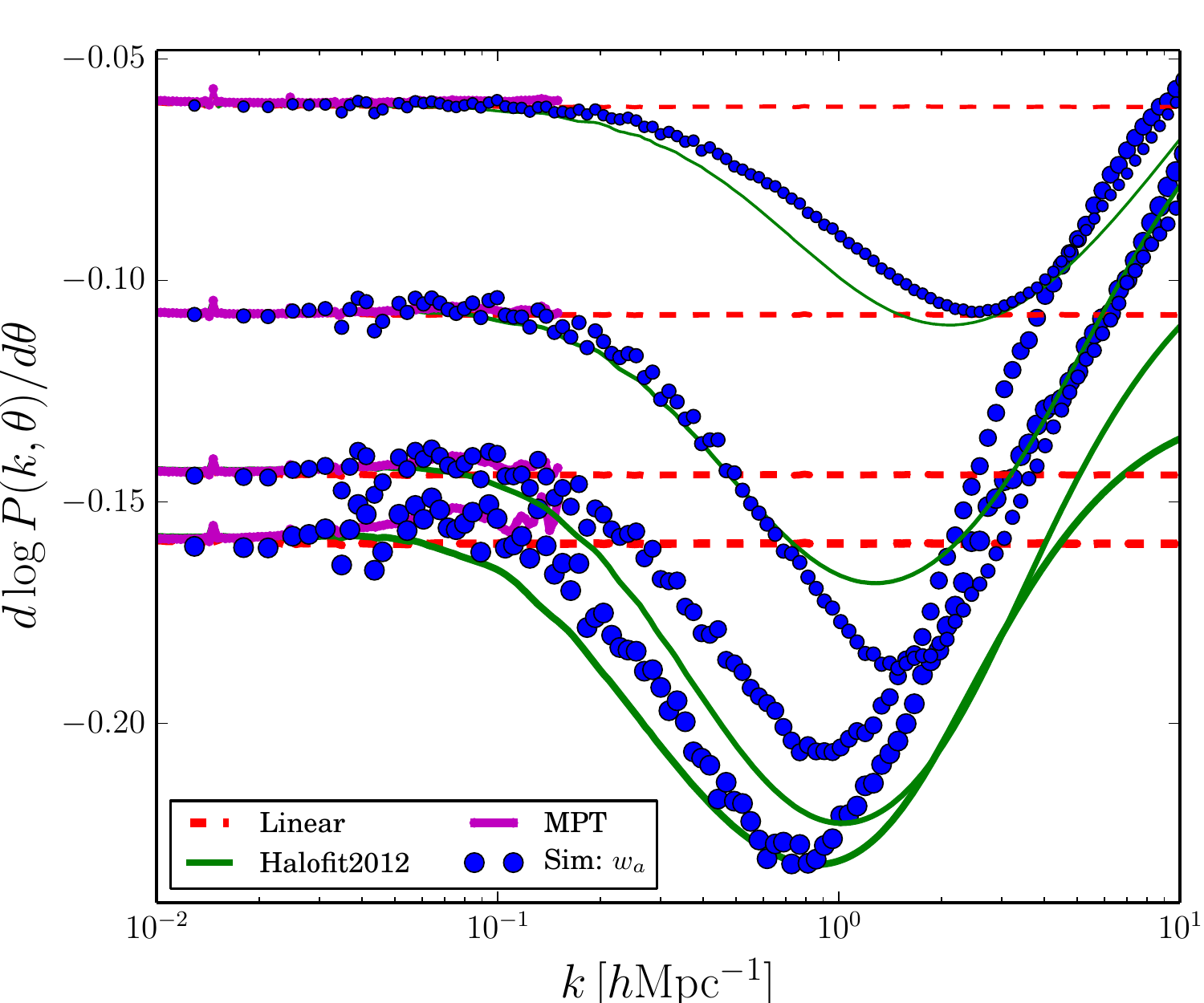}}
\vspace{0.2cm}
\centerline{
\includegraphics[width=6.5cm,angle=0]{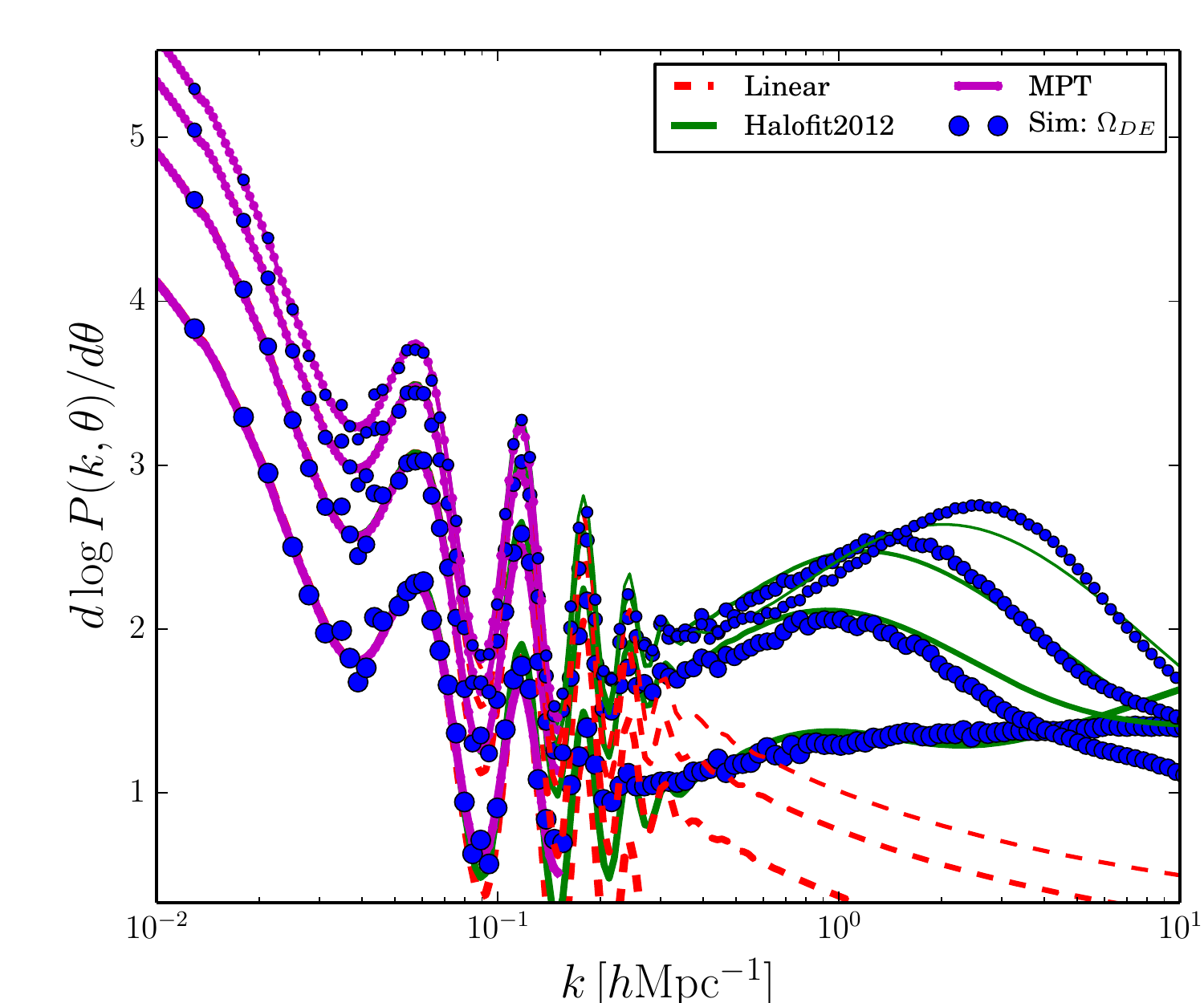}\hspace{0.3cm}
\includegraphics[width=6.5cm,angle=0]{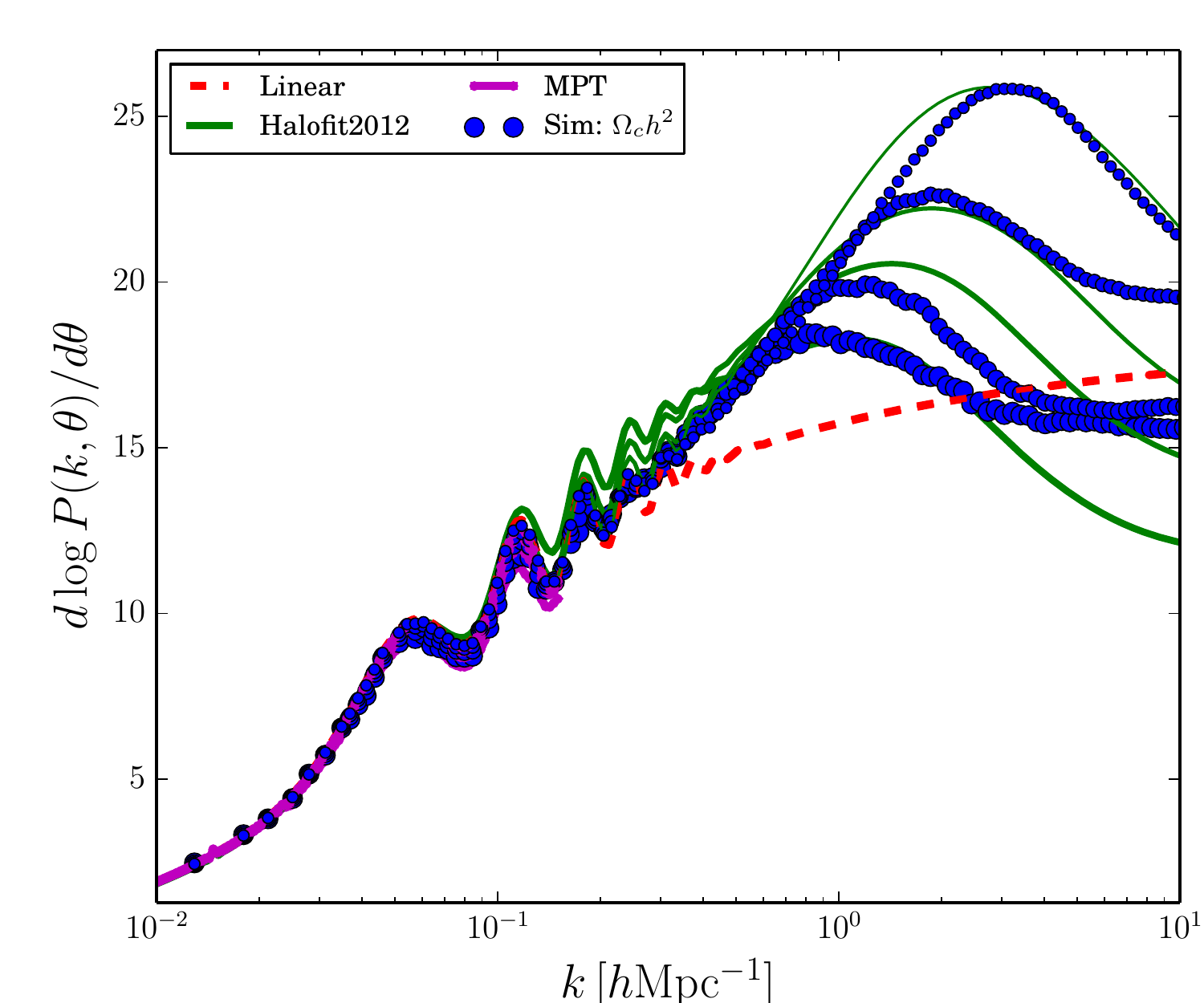}}
\vspace{0.2cm}
\centerline{
\includegraphics[width=6.5cm,angle=0]{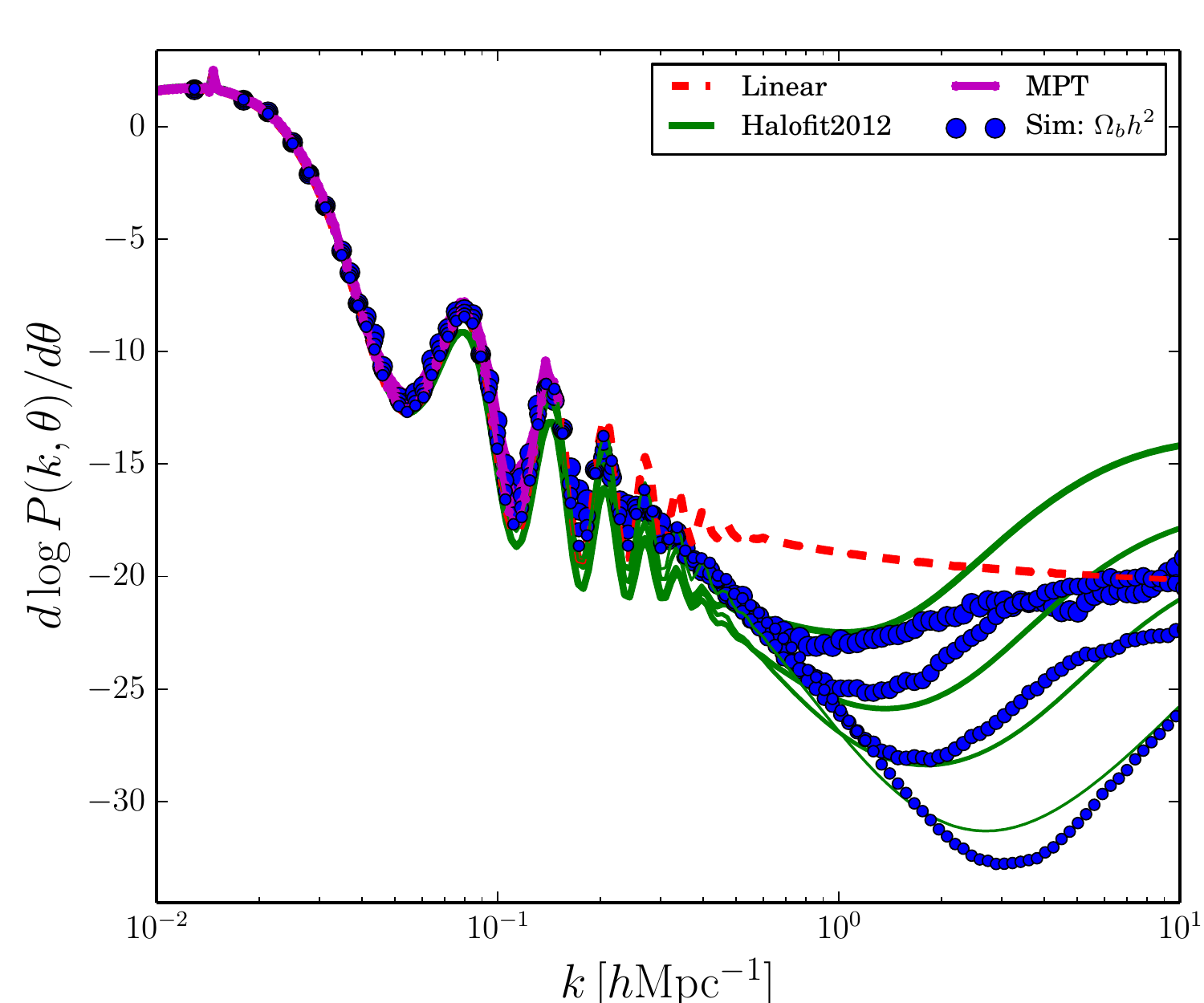}\hspace{0.3cm}
\includegraphics[width=6.5cm,angle=0]{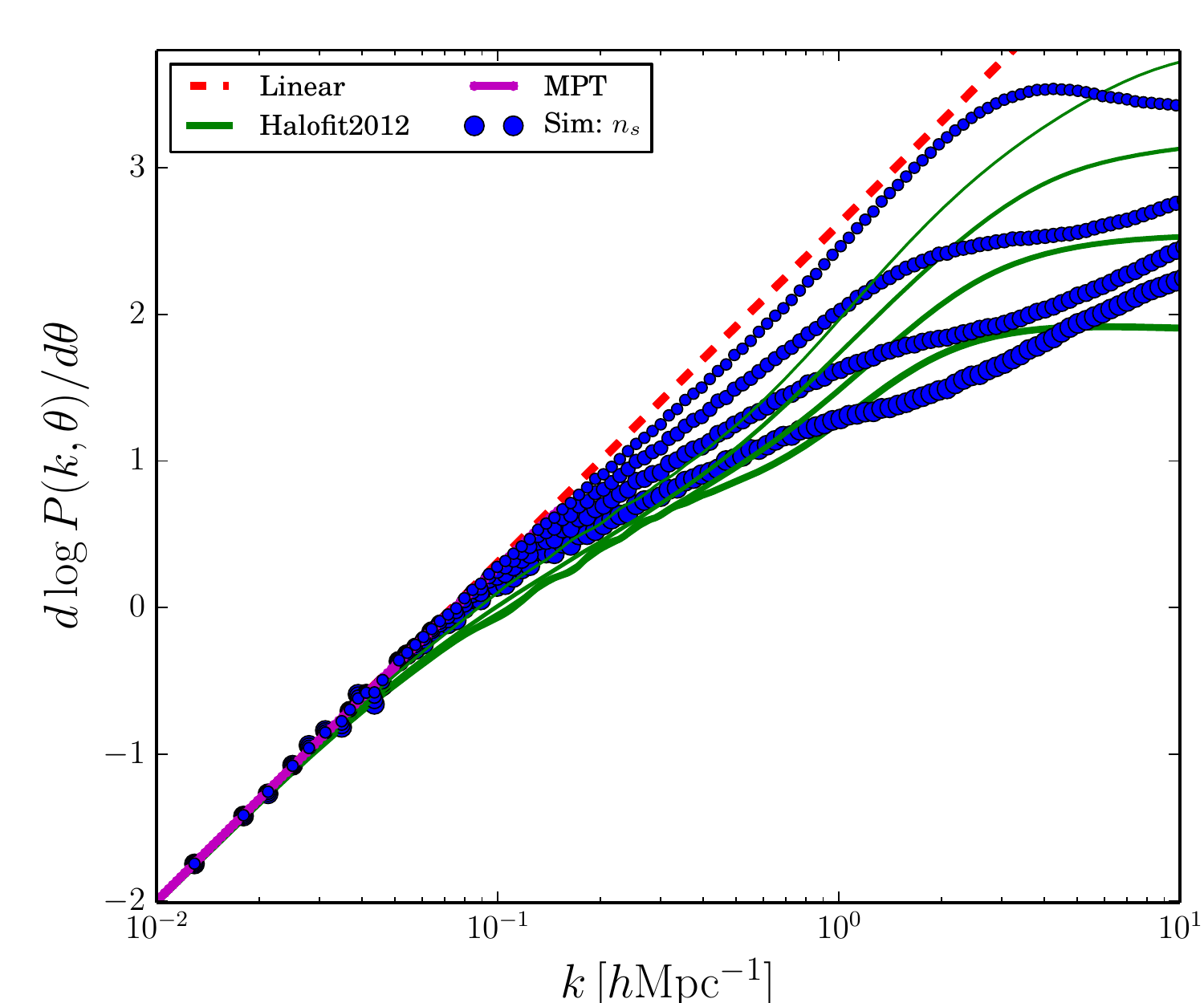}}
\vspace{0.2cm}
\centerline{
\includegraphics[width=6.5cm,angle=0]{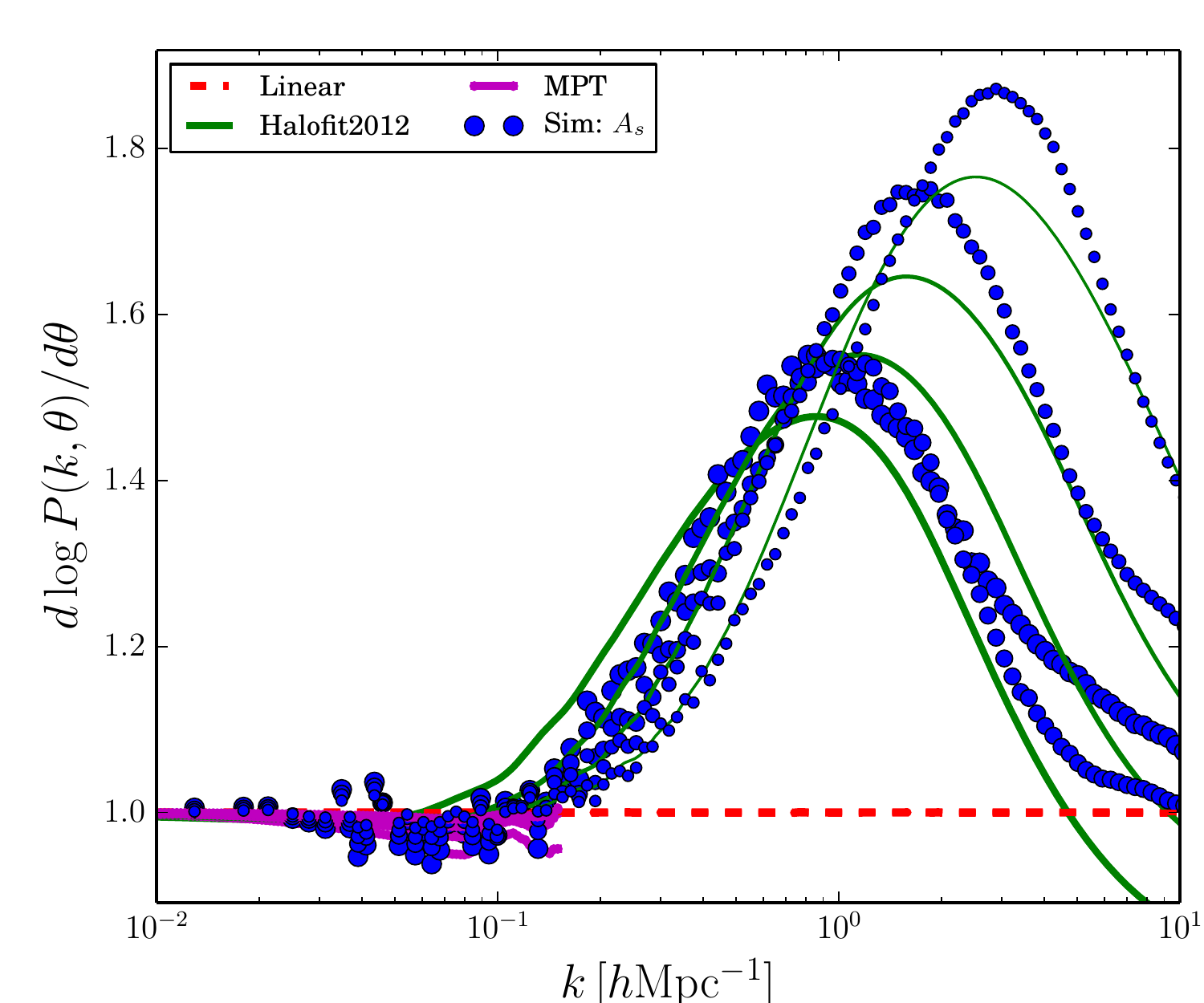}\hspace{0.3cm}
\includegraphics[width=6.5cm,angle=0]{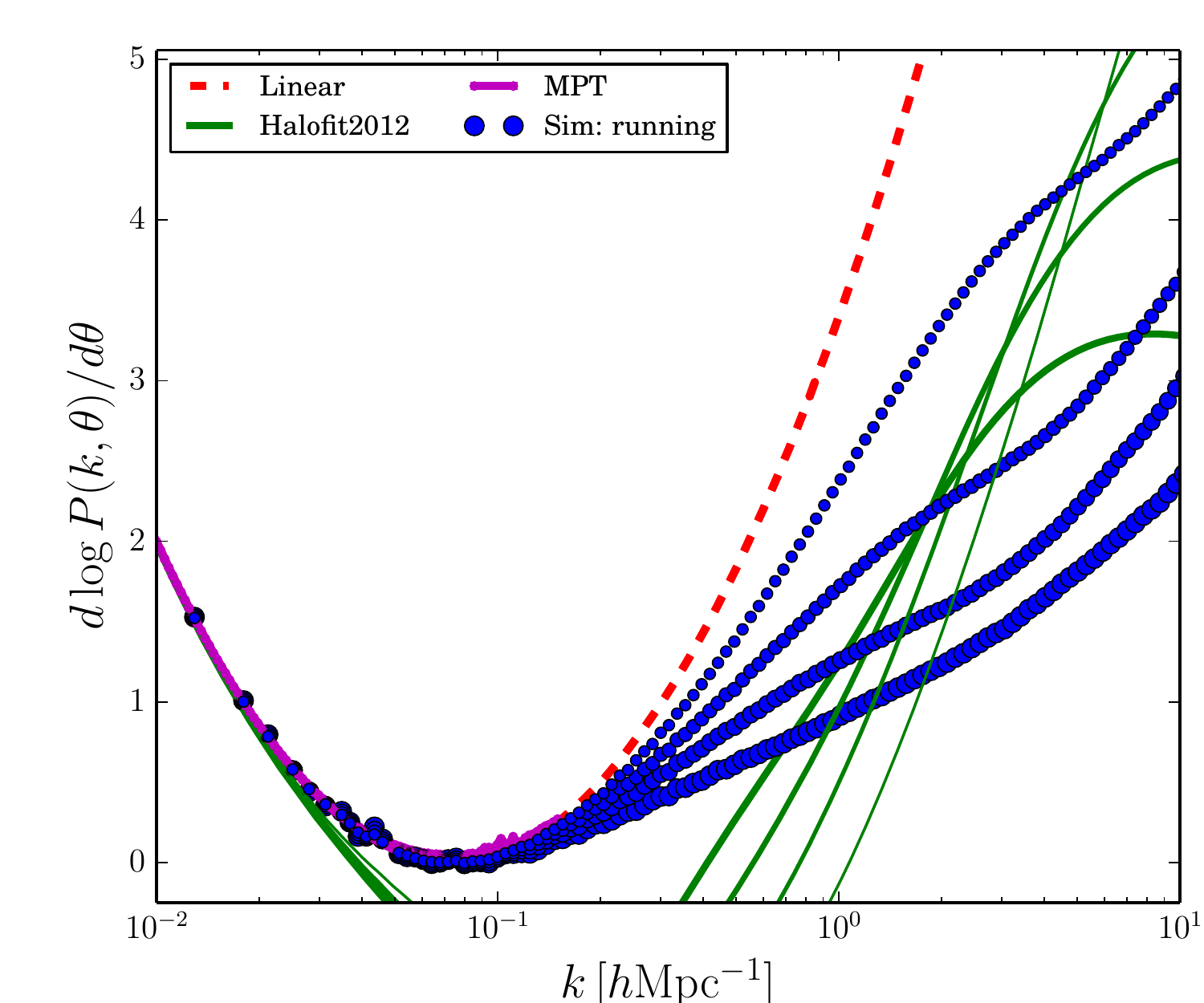}}
\caption{\small{Logarithmic derivative of the nonlinear matter power
    spectrum with respect 8 cosmological parameters considered in this
    paper as a function of scale.  The blue points show the results
    from the D\"ammerung simulations. The dashed lines show the
    results for linear theory and the solid green lines show the
    prediction from the updated version of {\tt halofit2012} from
    \citet{Takahashietal2012}. The point size and line thickness
    increases with decreasing redshift, with $z\in\{2,1,0.5,0.0\}$.}
\label{fig:PkDeriv1}}
\end{figure*}


\begin{figure*}
\centerline{
\includegraphics[width=6.5cm,angle=0]{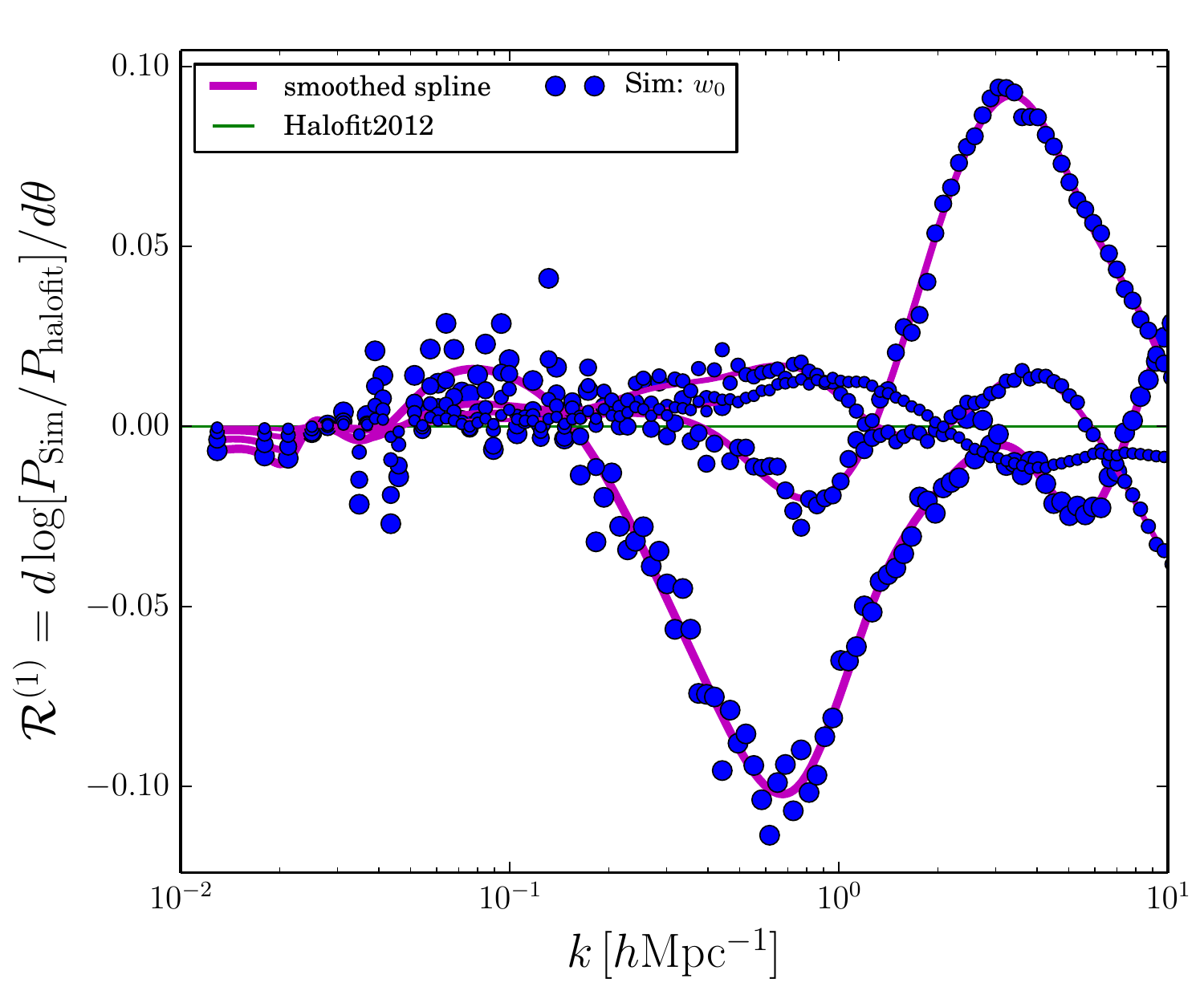}\hspace{0.3cm}
\includegraphics[width=6.5cm,angle=0]{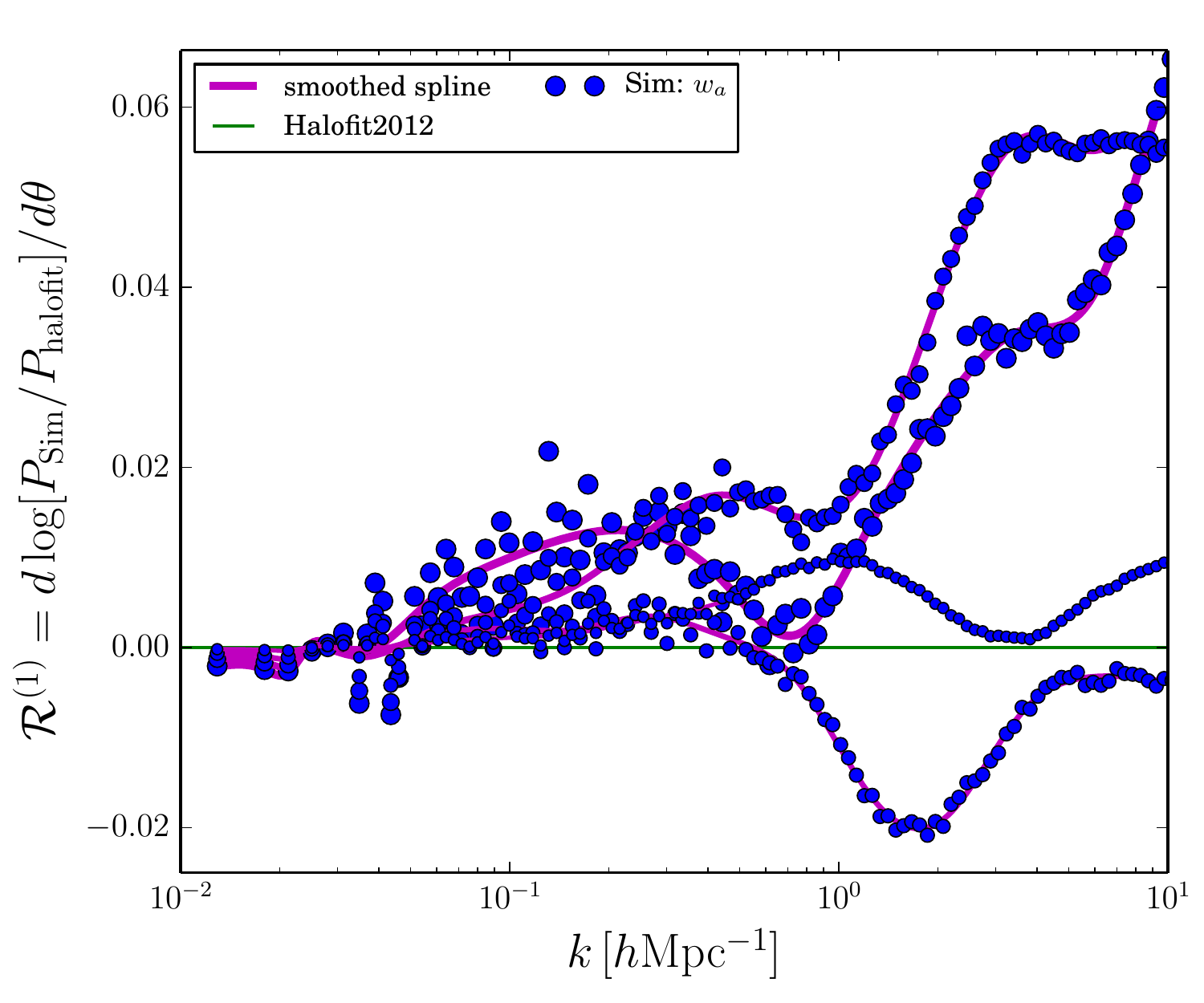}}
\vspace{0.2cm}
\centerline{
\includegraphics[width=6.5cm,angle=0]{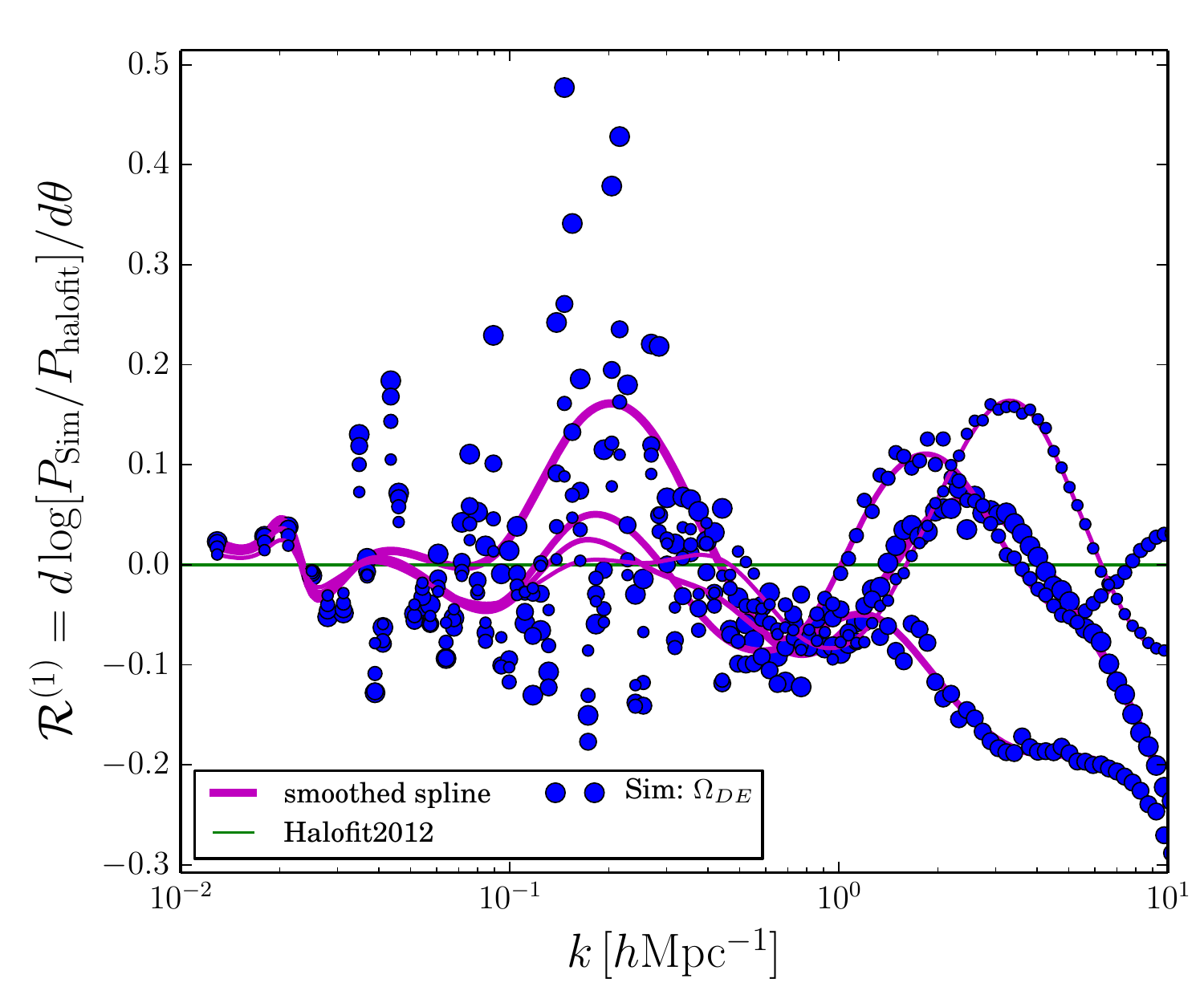}\hspace{0.3cm}
\includegraphics[width=6.5cm,angle=0]{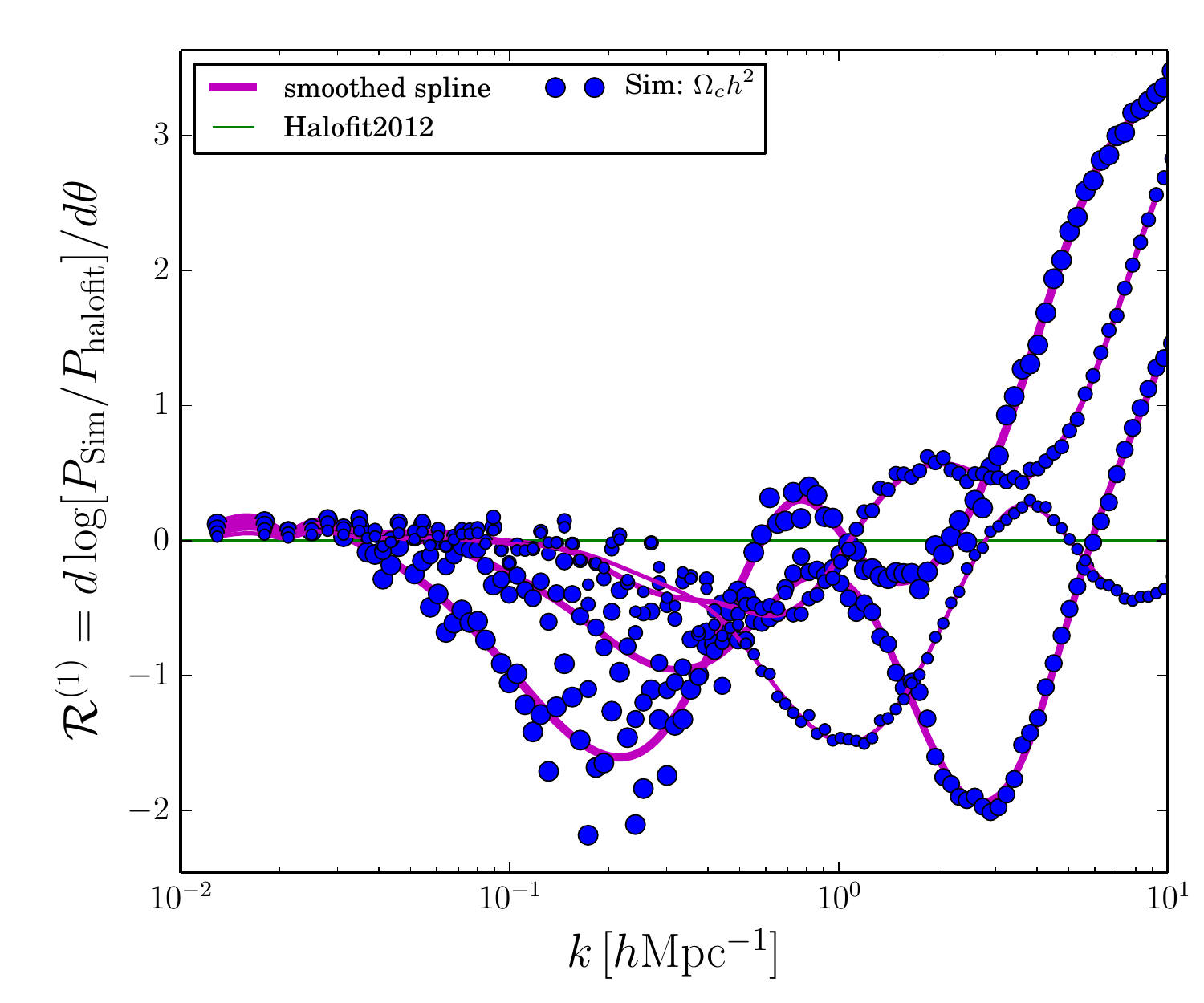}}
\vspace{0.2cm}
\centerline{
\includegraphics[width=6.5cm,angle=0]{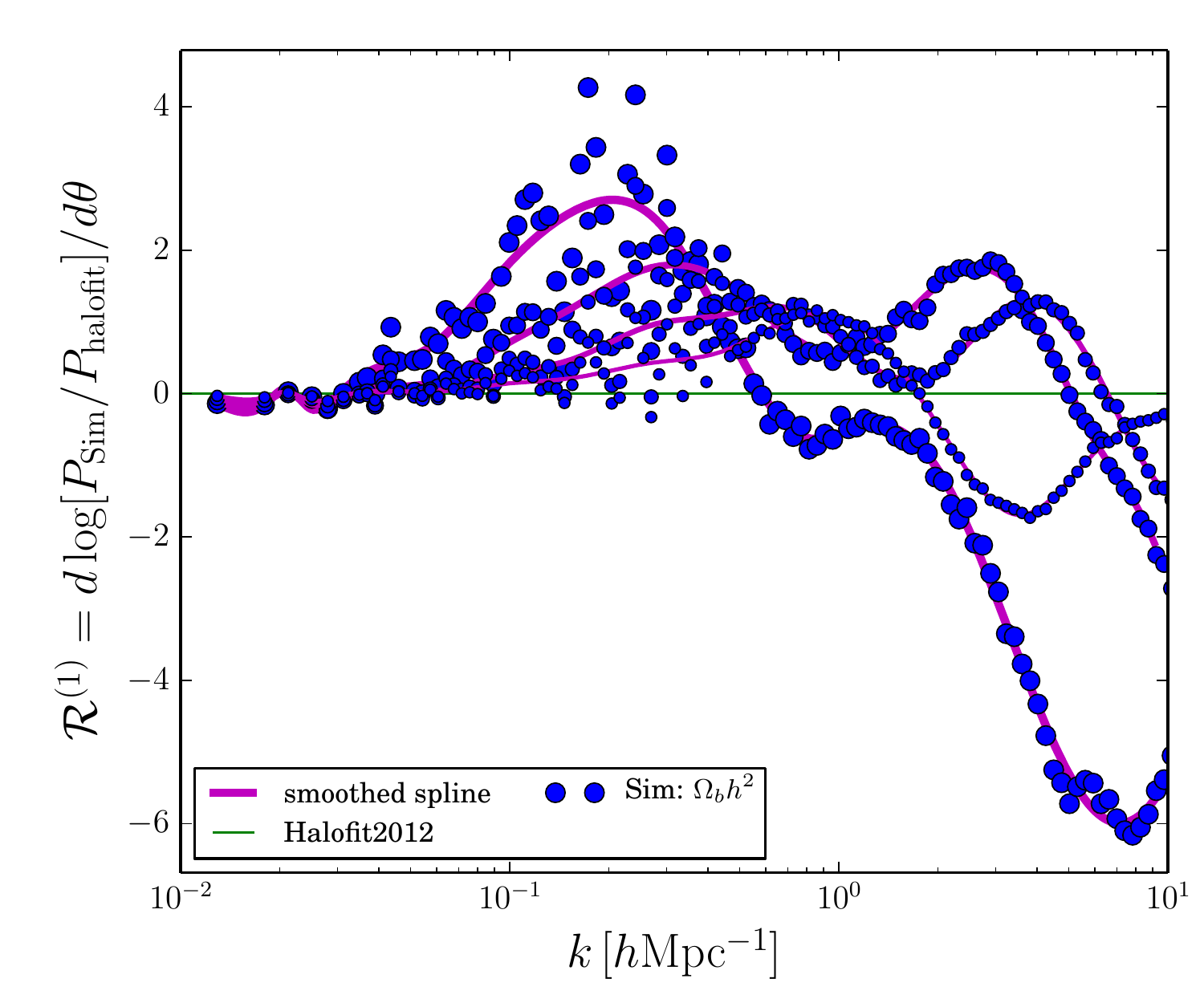}\hspace{0.3cm}
\includegraphics[width=6.5cm,angle=0]{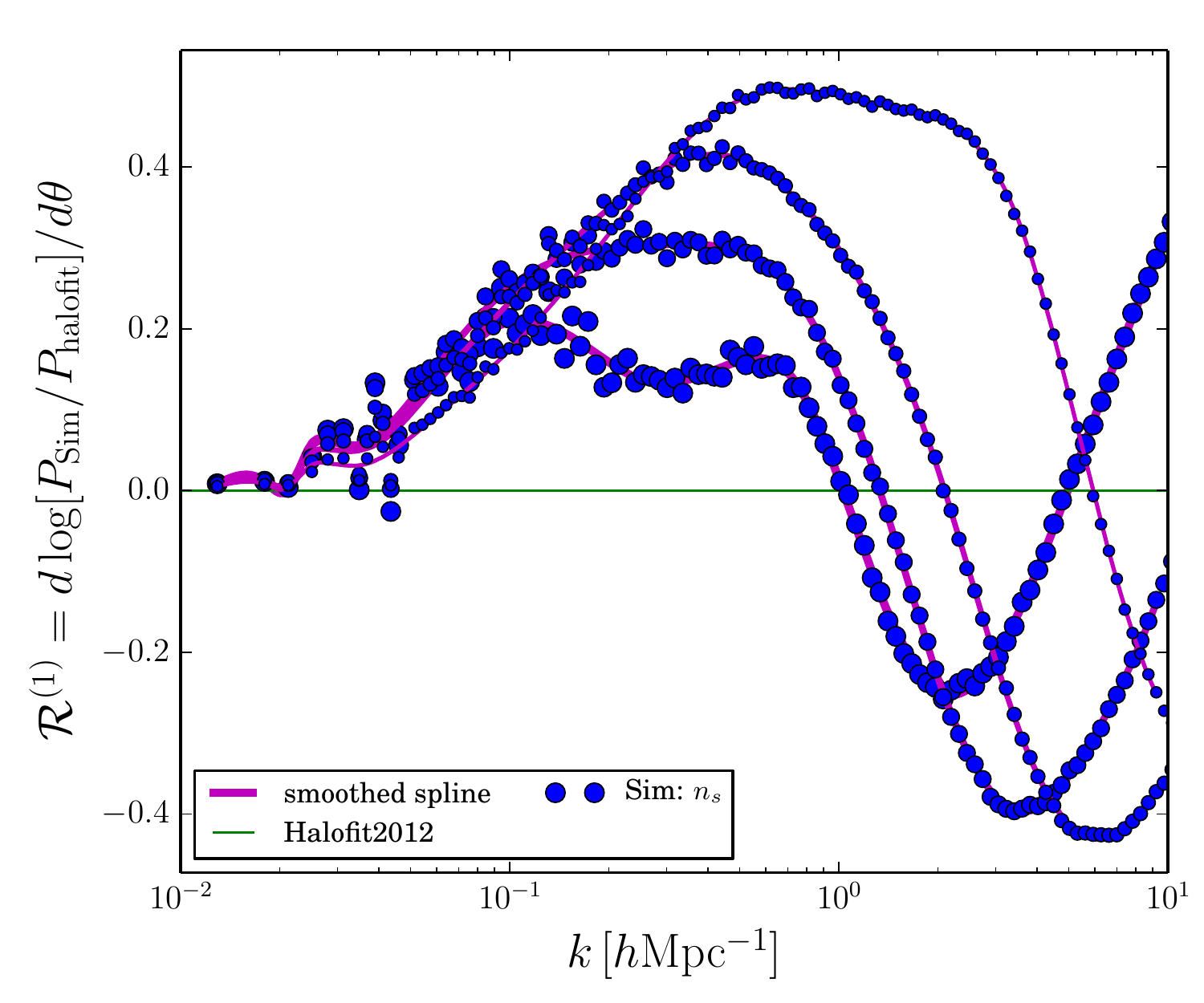}}
\vspace{0.2cm}
\centerline{
\includegraphics[width=6.5cm,angle=0]{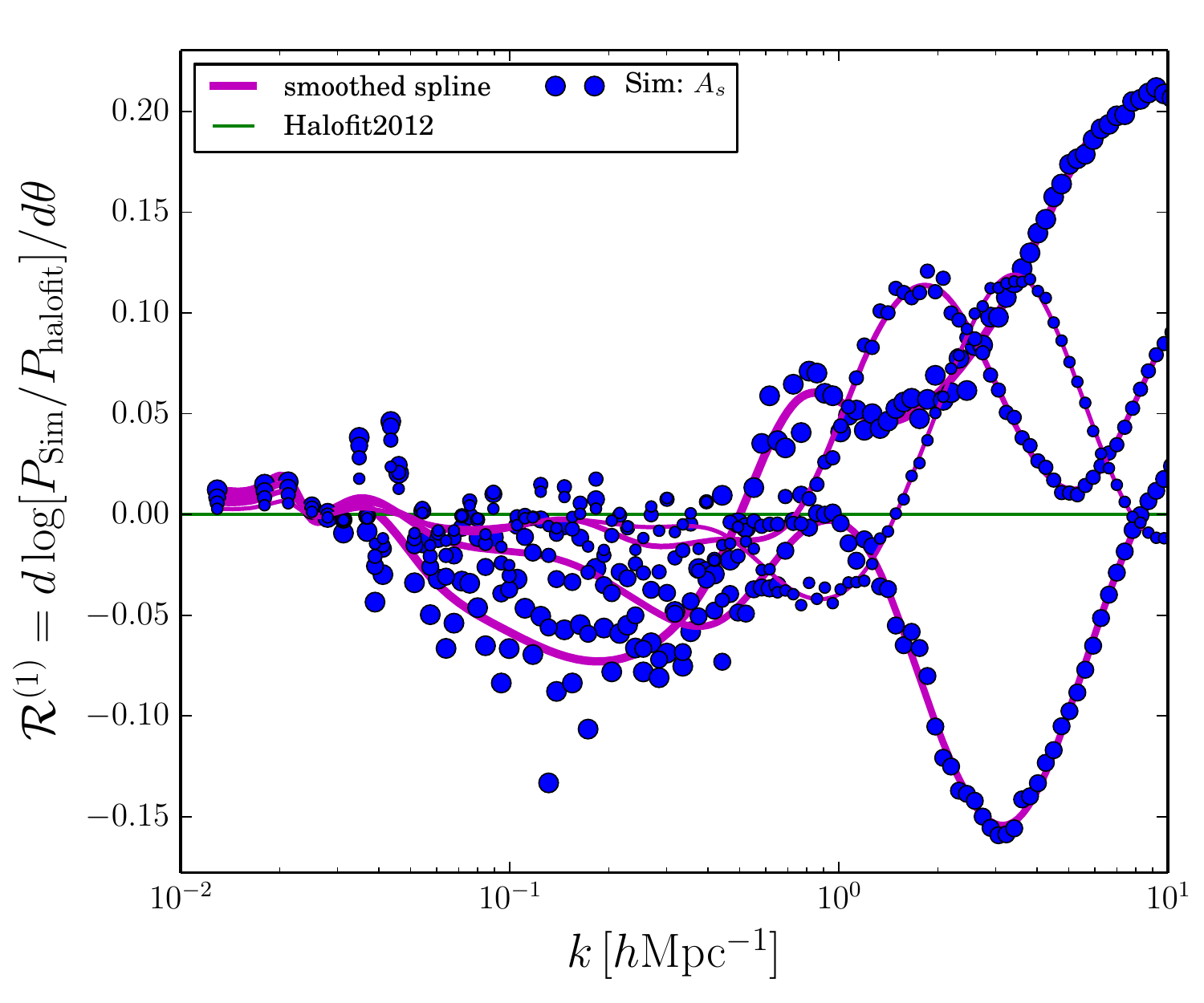}\hspace{0.3cm}
\includegraphics[width=6.5cm,angle=0]{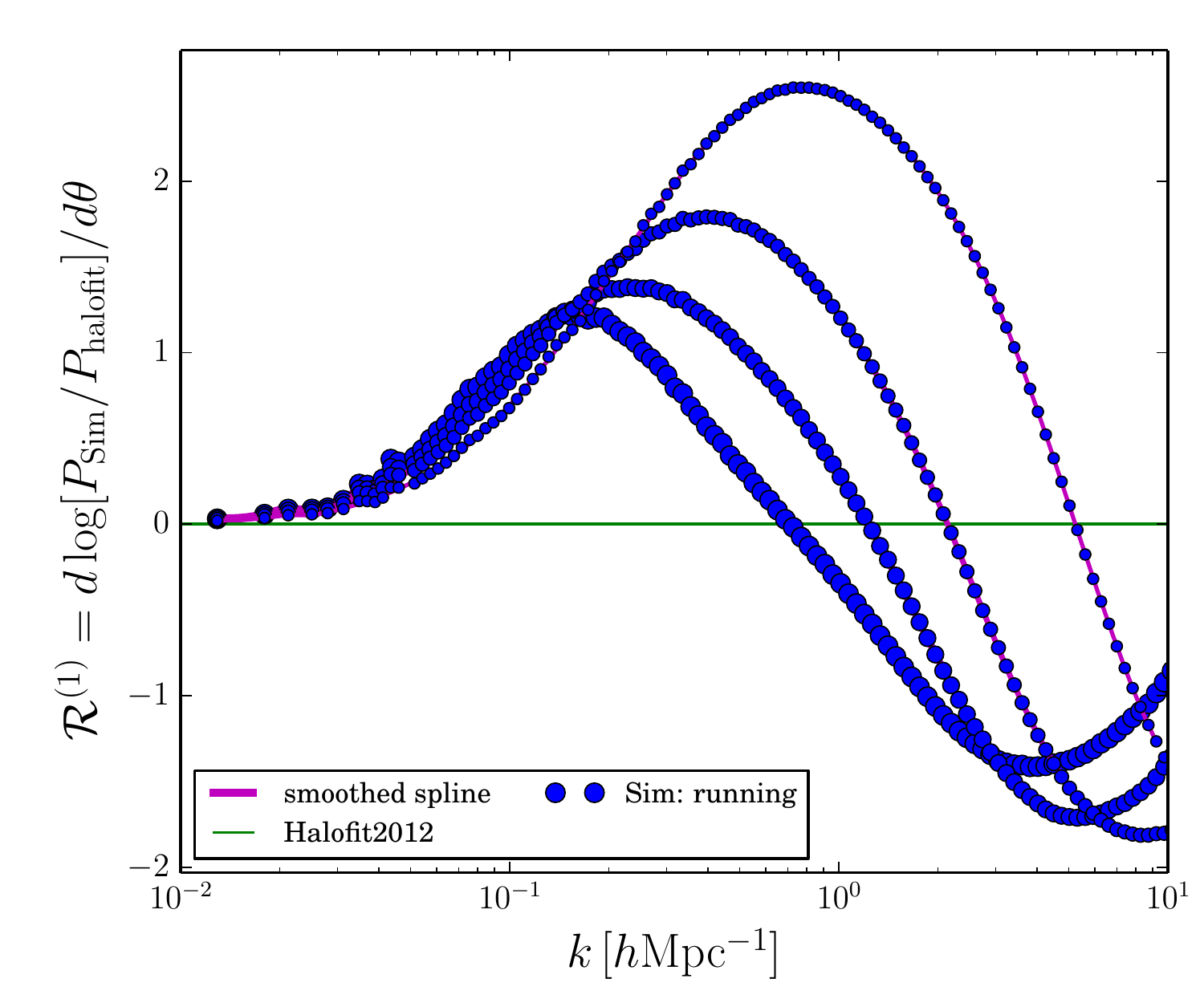}}
\caption{\small{Similar to Fig.~\ref{fig:PkDeriv1}, except this time
    the matter power spectra have been rescaled by the predictions
    from the {\tt halofit2012} model before computing the derivative
    with respect to the cosmological parameters. The magenta solid
    lines denote the result of applying a smoothing spline function to
    the measured scaled derivatives. Once again, increasing line
    thickness and point size corresponds to decreasing redshift with
    $z\in\{2,1,0.5,0.0\}$.}
\label{fig:PkDerivScaled}}
\end{figure*}


\begin{figure*}
\centerline{
\includegraphics[width=6.5cm,angle=0]{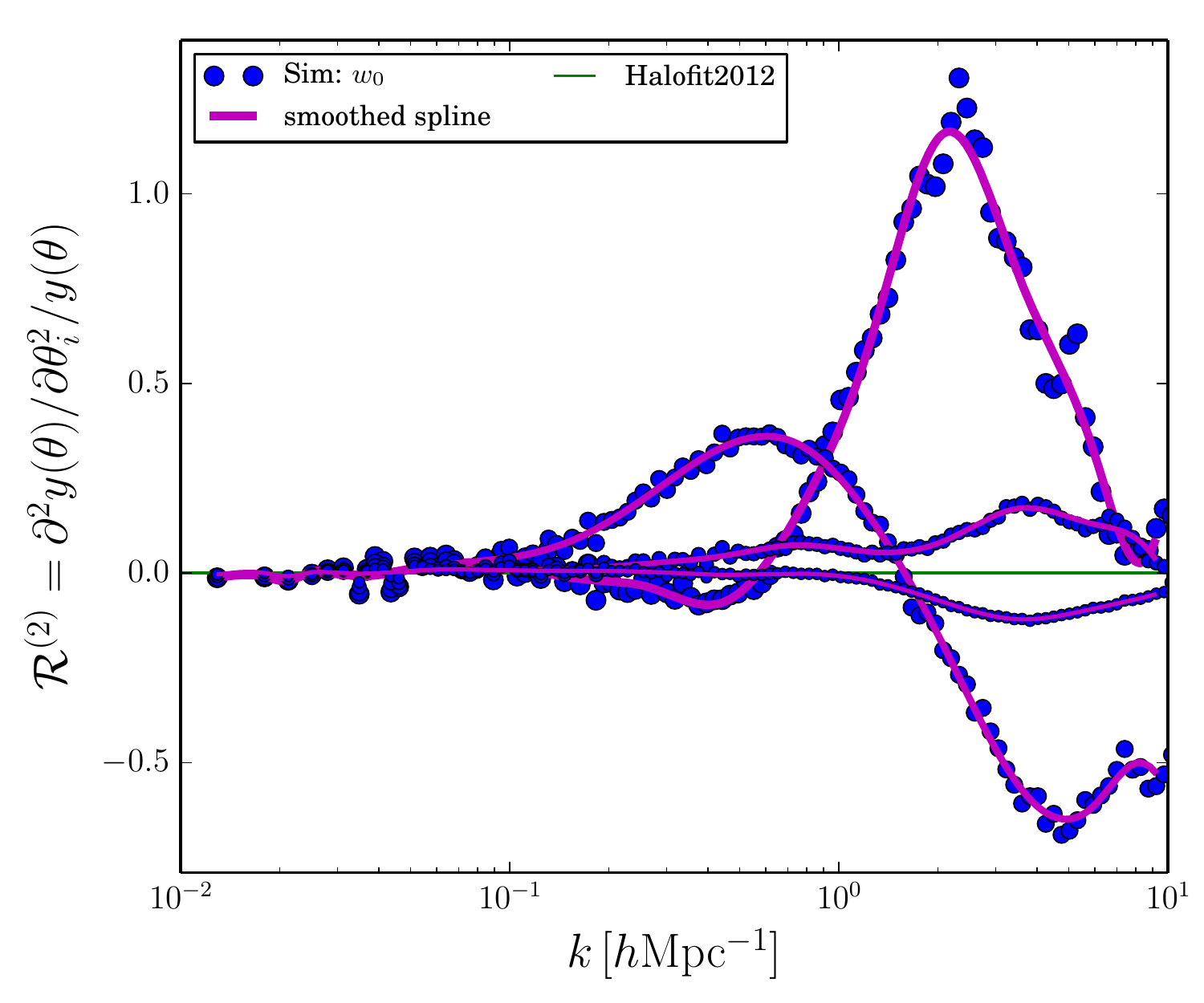}\hspace{0.3cm}
\includegraphics[width=6.5cm,angle=0]{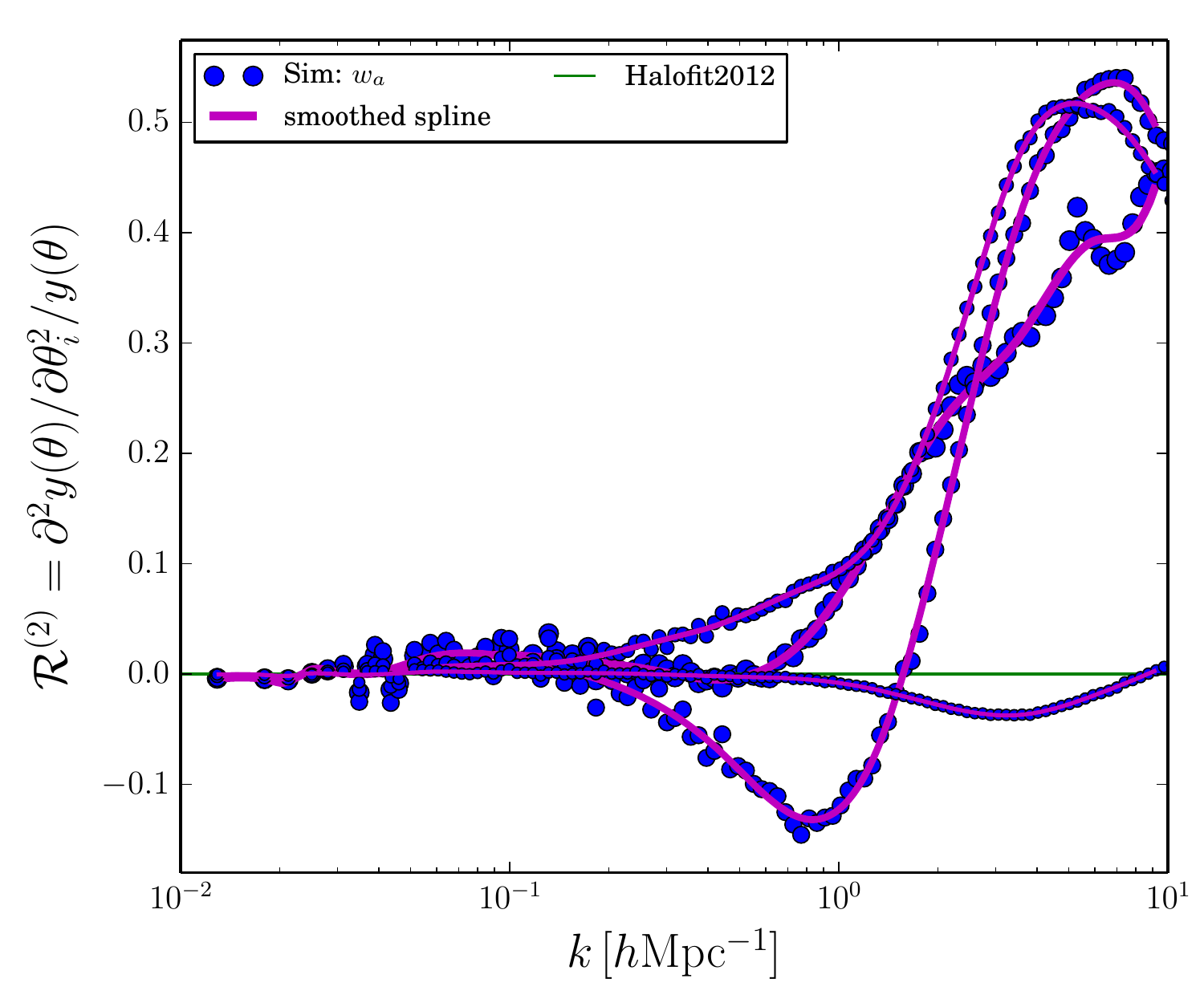}}
\vspace{0.2cm}
\centerline{
\includegraphics[width=6.5cm,angle=0]{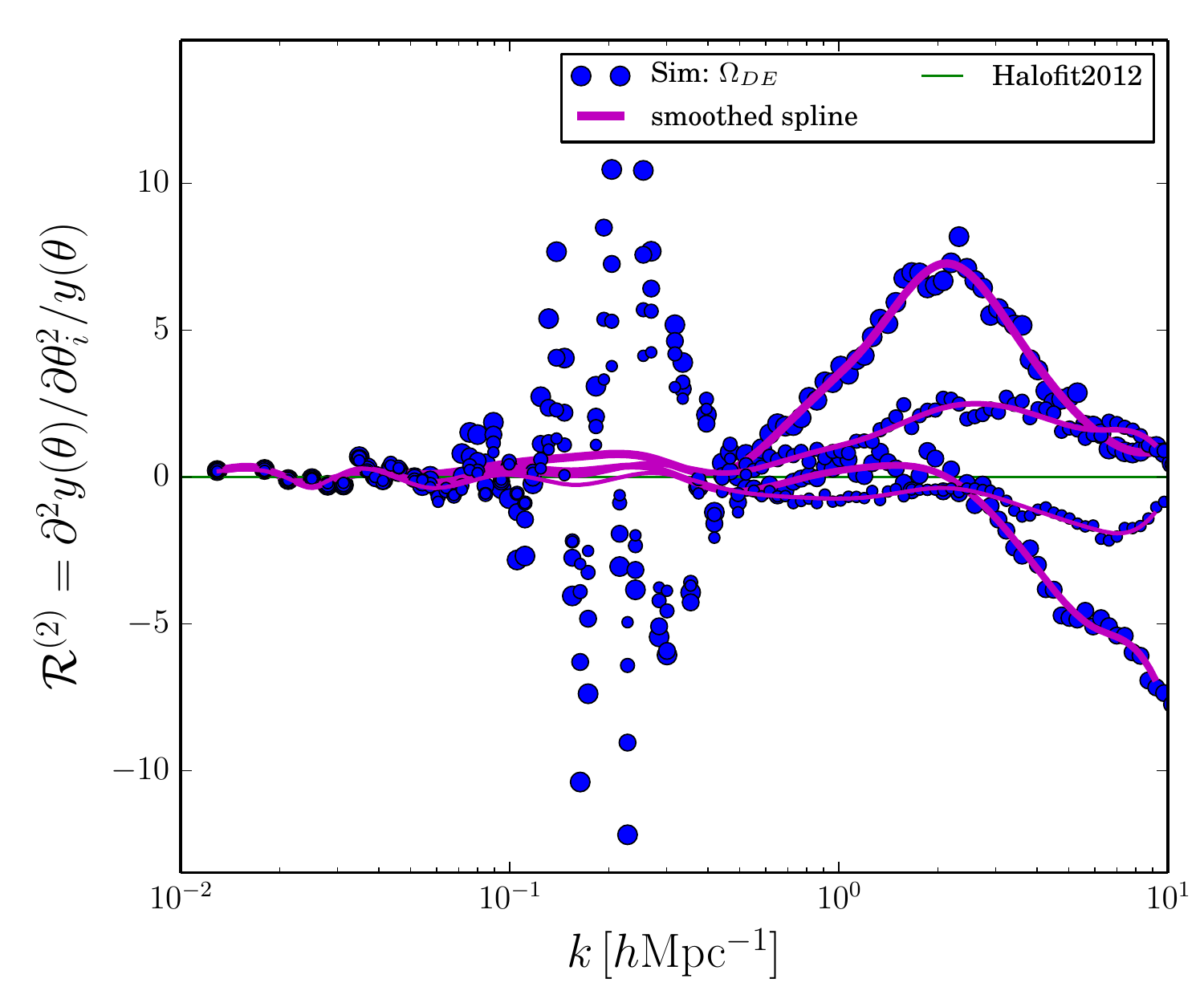}\hspace{0.3cm}
\includegraphics[width=6.5cm,angle=0]{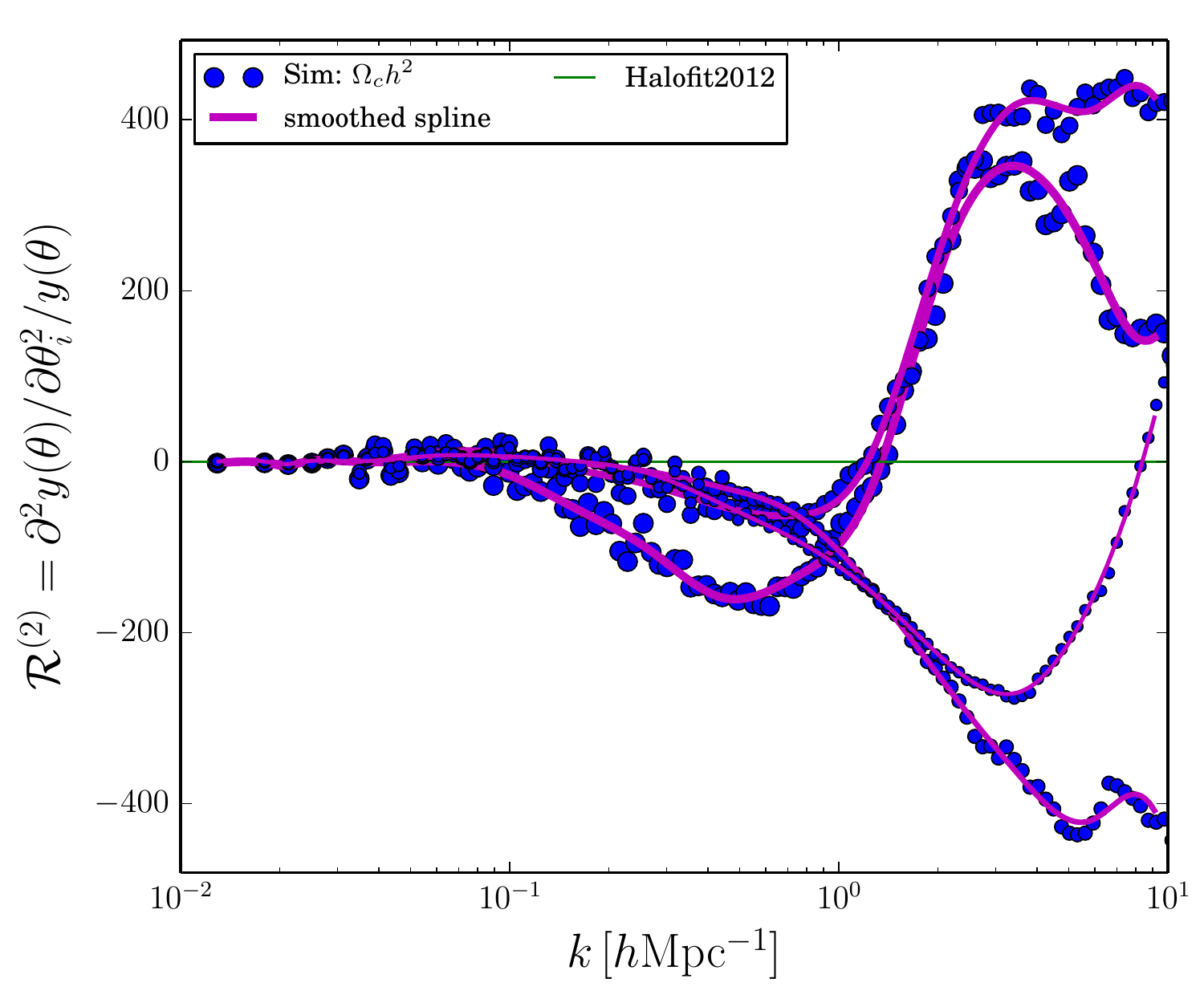}}
\vspace{0.2cm}
\centerline{
\includegraphics[width=6.5cm,angle=0]{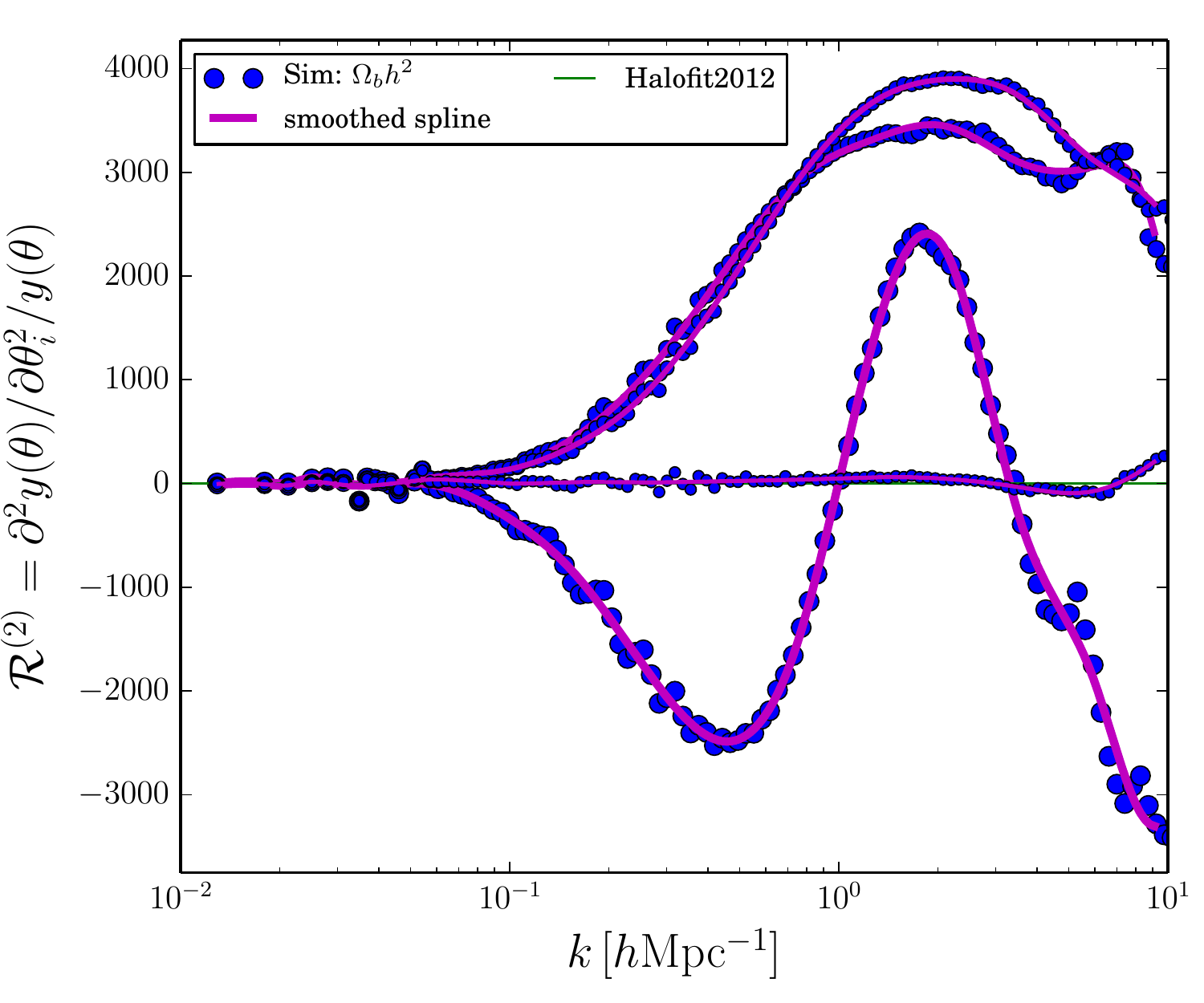}\hspace{0.3cm}
\includegraphics[width=6.5cm,angle=0]{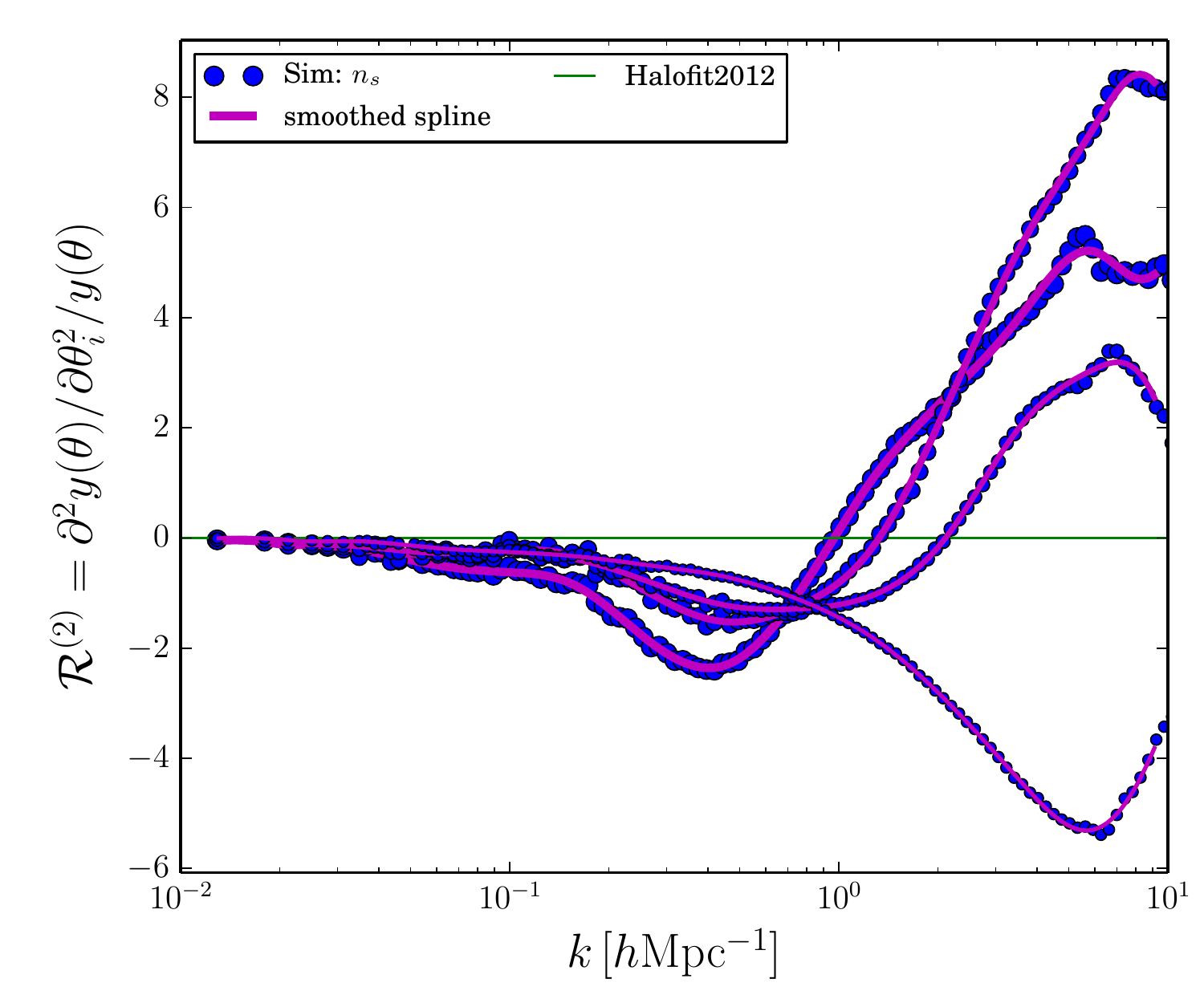}}
\vspace{0.2cm}
\centerline{
\includegraphics[width=6.5cm,angle=0]{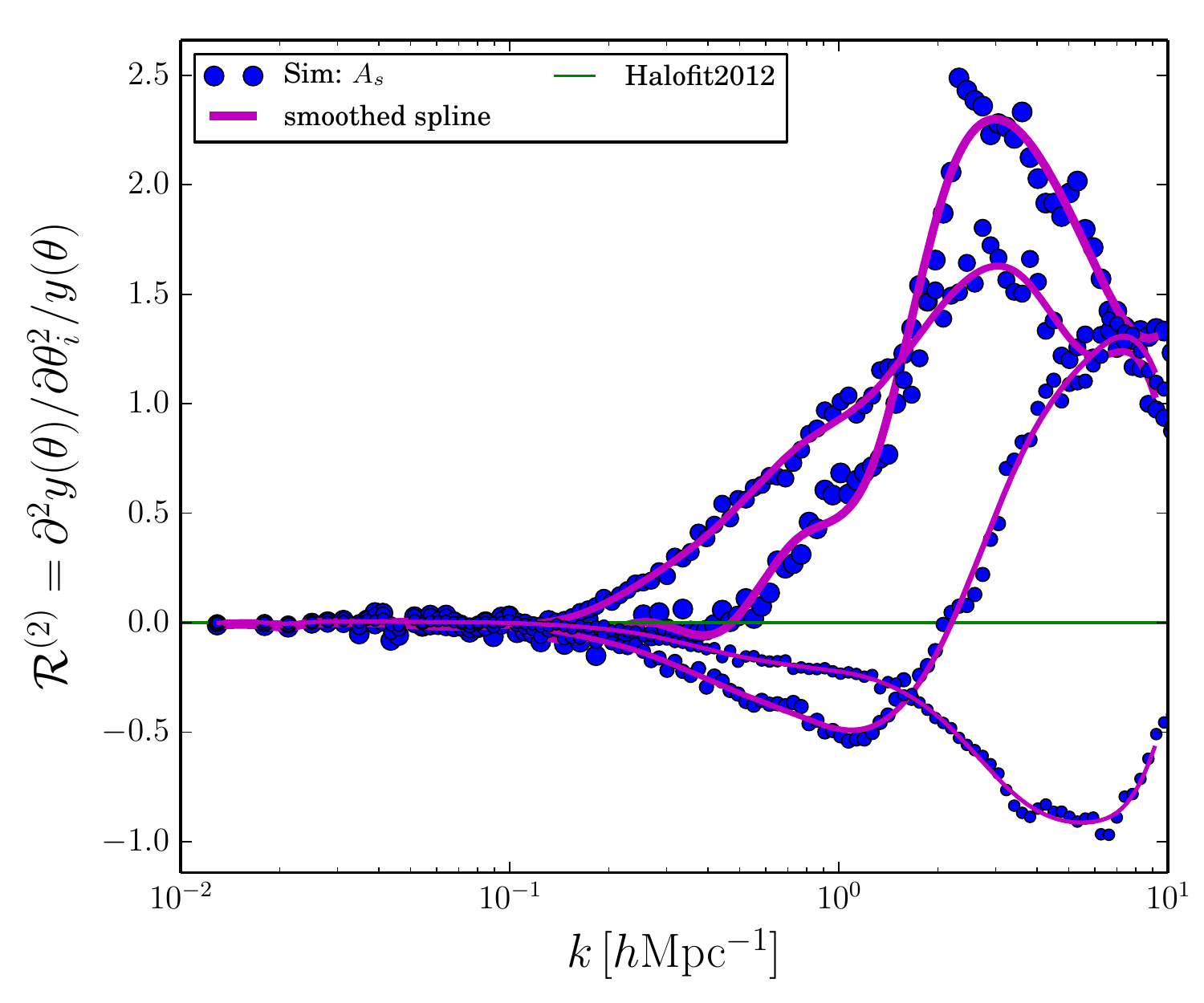}\hspace{0.3cm}
\includegraphics[width=6.5cm,angle=0]{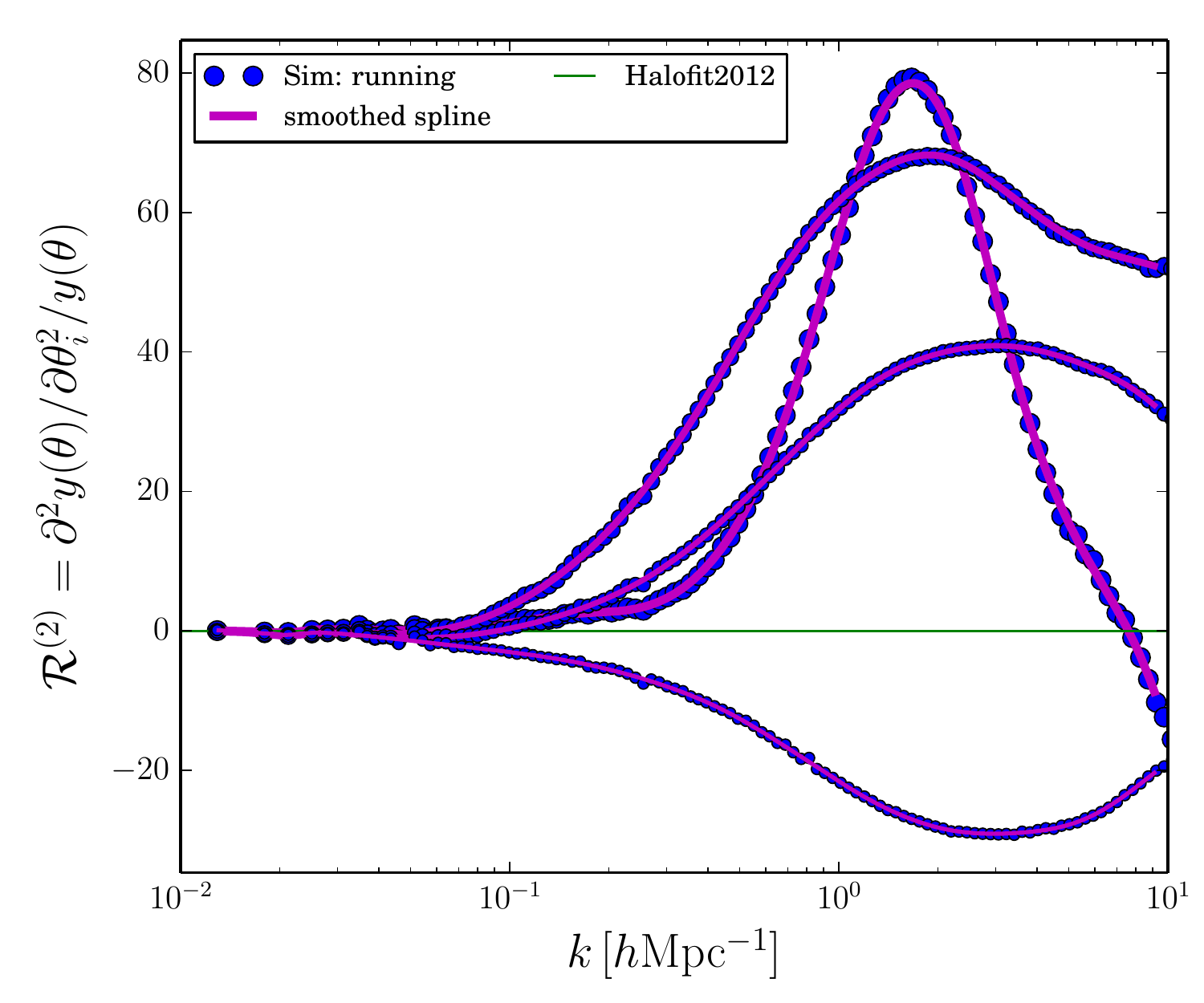}}
\caption{\small{Same as Fig.~\ref{fig:PkDerivScaled}, except here we show
    the 2nd order derivatives of the power spectra scaled by the {\tt
      halofit2012} predictions with respect to the cosmological
    parameters.}
\label{fig:PkDerivScaled2}}
\end{figure*}


\subsection{Derivatives with respect to cosmology}

We next turn our attention to the derivatives of the power spectra
with respect to the parameter variations. We construct the nonlinear
derivatives at the fiducial point $\bm\theta_0$ in parameter space
using the double sided derivative technique from
\citet[][]{Smithetal2014}, and for a given cosmological parameter
variation this means (suppressing the $k$ and $a$ dependence):
\be 
\widehat{\frac{\partial P(\bm\theta)}{\partial\theta_i}} \approx 
\frac{\widehat{P}(\theta_{0,i}+\Delta\theta_i)-\widehat{P}(\theta_{0,i}-\Delta\theta_i)}
     {2\Delta\theta_i} + {\mathcal O}(\Delta\theta_i^2)\ ,
\ee
where the estimate of the logarithmic derivative is given by:
\be \widehat{\frac{\partial\log P(\bm\theta)}{\partial\theta_i}} =
\frac{1}{\widehat{P}(\bm\theta_0)}\widehat{\frac{\partial P(\bm\theta)}{\partial\theta_i}}
\ee
Note that for the same set of simulations we can also construct the
second order derivative:
\ba \widehat{\frac{\partial^2 P({\bm\theta})}{\partial\theta_i^2}} & \approx &
\frac{
  \widehat{P}(\theta_{0,i}+\Delta\theta_i)+
  \widehat{P}(\theta_{0,i}-\Delta\theta_i)-2\widehat{P}(\theta_{0,i})}
     {(\Delta\theta_i)^2}  \nn \\
     & & + {\mathcal O}(\Delta\theta_i^2)\ ,
\ea
Unfortunately, we do not have enough simulations to fully populate the
Hessian matrix $\partial^2
P(\theta_i)/\partial\theta_i\partial\theta_j$ (see
\S\ref{sec:conclusions} for how one can obtain this).

Figure~\ref{fig:PkDeriv1} shows the logarithmic derivatives for the 8
cosmological parameter variations listed in
Table~\ref{tab:cospar}. The blue points denote the measurements at a
set of redshifts, with the point size decreasing with increasing
redshift. In the each panel we also show the predictions from the
linear theory as the red dashed lines and also the predictions from
{\tt halofit2012} as the green solid lines.  We see that on large
scales the measured logarithmic derivatives exactly agree with both
the linear and nonlinear predictions. This is not too surprising, owing
to corrections we have implemented (see \S\ref{ssec:corr}). However,
as pointed out in the previous section, while our corrections were
made using modes with $k<0.3\kMpc$, we see that the agreement between
linear theory and the measurements remains very good all the way up to
$k\sim0.1\kMpc$. On smaller scales, there are significant deviations.

We also show the predictions for the 2-loop MPT theory calculation
evaluated up to $k=0.15\kMpc$ as the solid magenta line. We see that
the MPT model accurately describes the first order derivatives on
these scales. Note that we pushed the calculation to $k=0.3\kMpc$ and
found strong deviations from the measured results, leading us to
believe that the 2-loop answer should be used with care for
$k>0.15\kMpc$. In our current approach we transition between MPT and
our corrected model on scales $k\sim0.07\kMpc$. Nevertheless, one can
see that the 2-loop calculation is more accurate than using linear
theory, in particular for capturing the cosmology dependence of the
nonlinear processing of the BAO features.

We also show the predictions for {\tt halofit2012} as the green solid
lines. Whilst on the whole it provides a much better description of the
measured derivatives, there are some noticeable deviations, and in
particular for the parameters $\{\Omega_{\rm b}h^2,n_s,\alpha\}$, the
model is quite poor.


\subsection{Modelling the scaled 1st order and 2nd order derivatives}

In Figures~\ref{fig:PkDerivScaled} and \ref{fig:PkDerivScaled2} we
show the first and second order derivative of the power spectra scaled
by the predictions from {\tt halofit2012} with respect to the
cosmological parameters, respectively. If {\tt halofit2012} provided a
perfect description of the nonlinear evolution, then all of the scaled
derivatives would be zero. We see that on large scales
($k\lesssim0.02\kMpc$), indeed the measured derivatives all converge
to this value or scatter about it.  For the cases of the parameters
$\{w_0,w_a,\Omega_{\rm DE},n_s,A_s\}$ and for $k<10.0\kMpc$ we have
${\mathcal R}^{(1)}_i=d\log y/d\theta_i<0.2$. This suggests that {\tt
  halofit2012} provides a very good description of the cosmology
dependence of the nonlinear power spectrum with respect to these
parameters.

On the other hand, for the parameters $\{\Omega_{\rm c}h^2,\Omega_{\rm
  b}h^2,\alpha\}$ we see that the scaled derivatives have values
$\lesssim4$ for the scales that we probe. One can also see in
Fig.~\ref{fig:PkDerivScaled2} that the scaled 2nd order derivatives
for these parameters also have large values. This implies that {\tt
  halofit2012} does not describe the cosmological dependence of the
nonlinear power spectrum for these parameters very well. However, we
note that, for example, a 10\% variation in $\Omega_{\rm c}h^2$ away
from the fiducial value would imply at most a $\lesssim$40\% error in
the {\tt halofit} prediction. Thus if we can calibrate the derivatives then
we should produce a method that is significantly more accurate.

Following the approach described in \S\ref{ssec:practical} we have
modelled the measured scaled derivatives using the B-spline approach.
The results of this for each parameter are presented in
Figures~\ref{fig:PkDerivScaled} and \ref{fig:PkDerivScaled2} as the
solid magenta lines. Clearly, this smoothed spline function approach
captures the trends seen in the data. These functions provide us with
the smooth ${\mathcal R}_i^{(1)}(k,z|{\bm\theta}_0)$ and ${\mathcal
  R}_{ii}^{(2)}(k,z|{\bm\theta}_0)$ components for \Eqn{eq:PNgen} on
quasi-linear and nonlinear scales.  We remind the reader that on large
scales the nonlinear predictions are exactly those of the MPT theory.


\section{Cosmology dependence of the new model}\label{sec:cosmonewmodel}

\subsection{Comparison with variational runs}

\begin{figure*}
\centerline{
\includegraphics[width=6.5cm,angle=0]{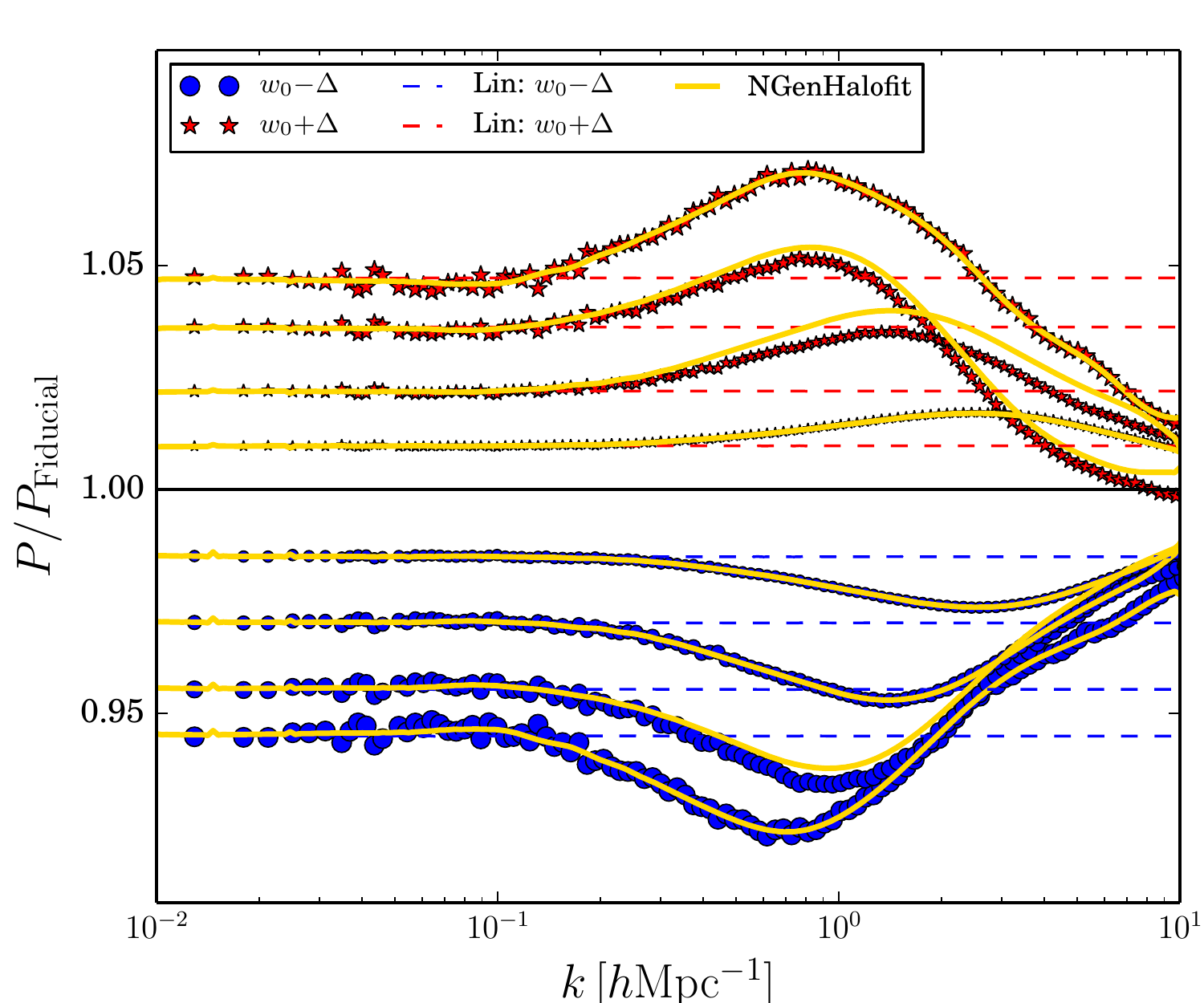}\hspace{0.3cm}
\includegraphics[width=6.5cm,angle=0]{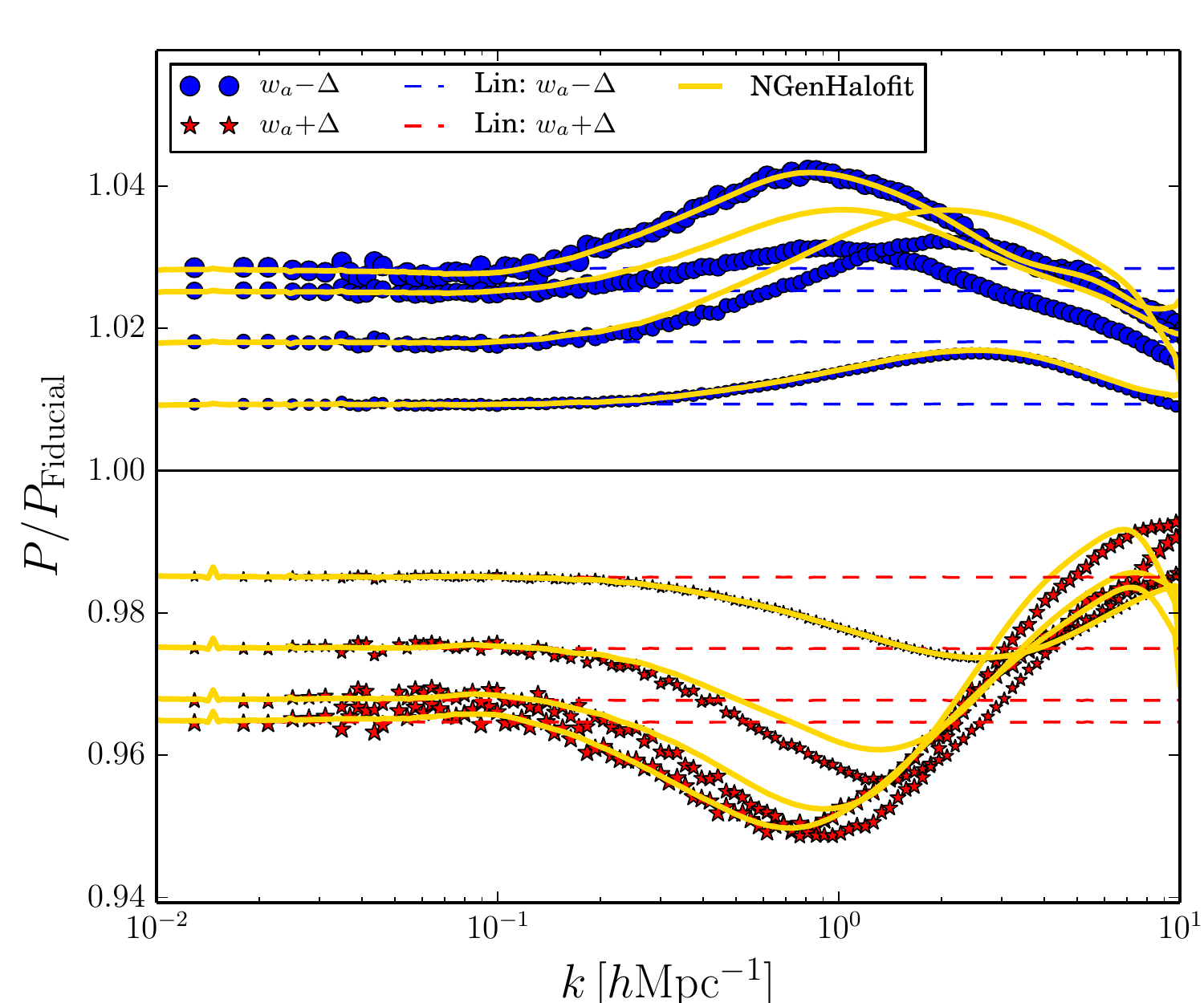}}
\vspace{0.2cm}
\centerline{
\includegraphics[width=6.5cm,angle=0]{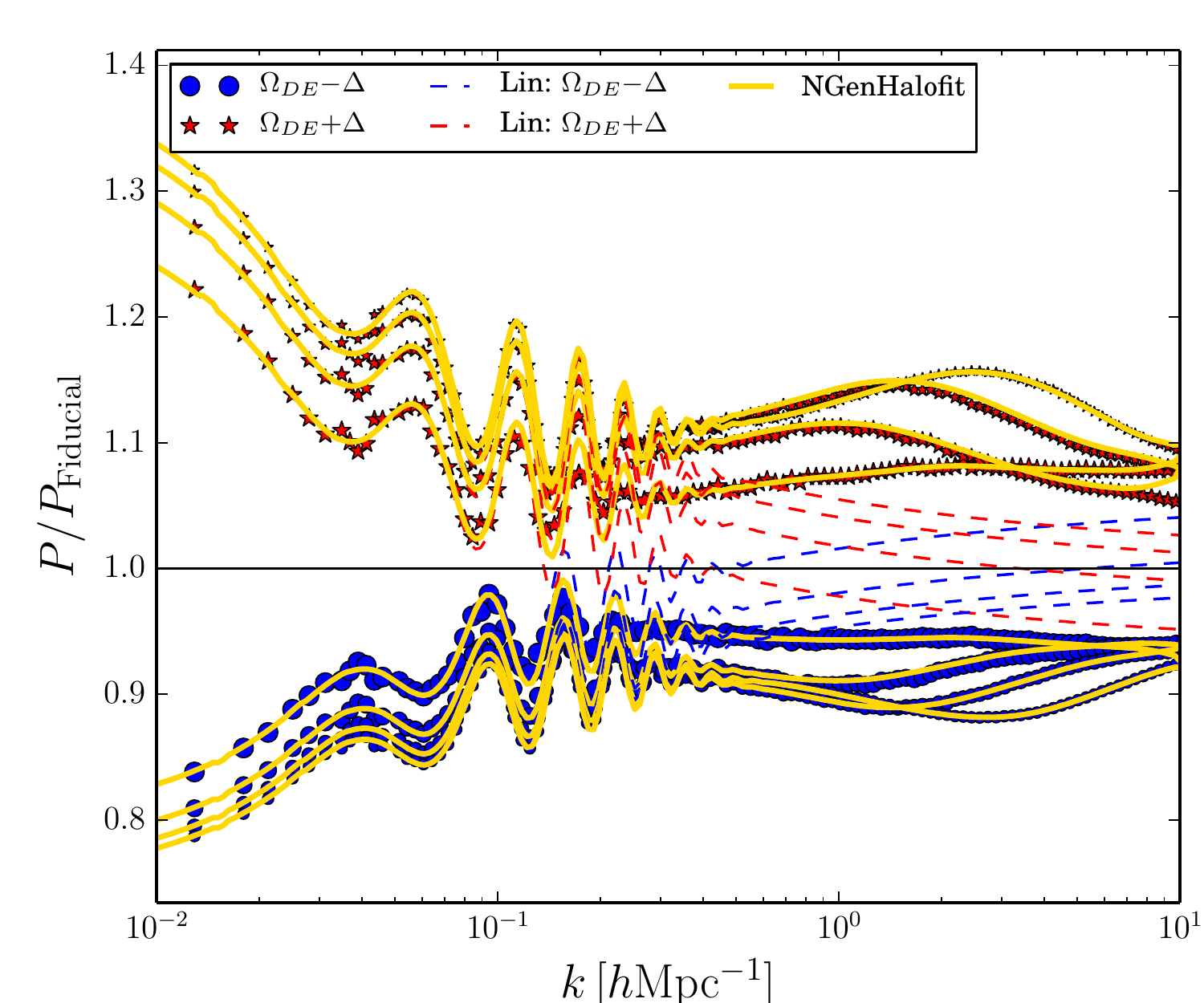}\hspace{0.3cm}
\includegraphics[width=6.5cm,angle=0]{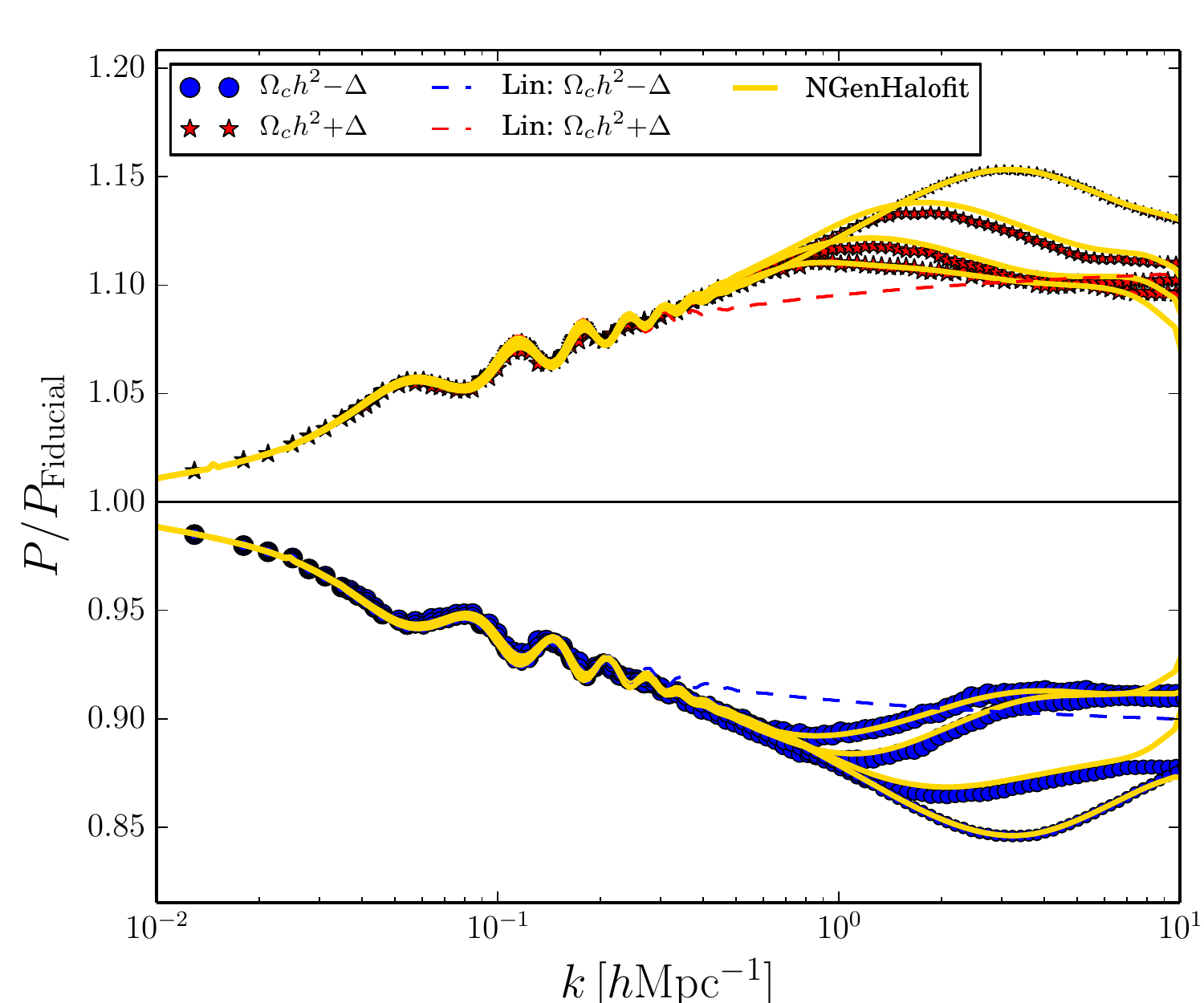}}
\vspace{0.2cm}
\centerline{
\includegraphics[width=6.5cm,angle=0]{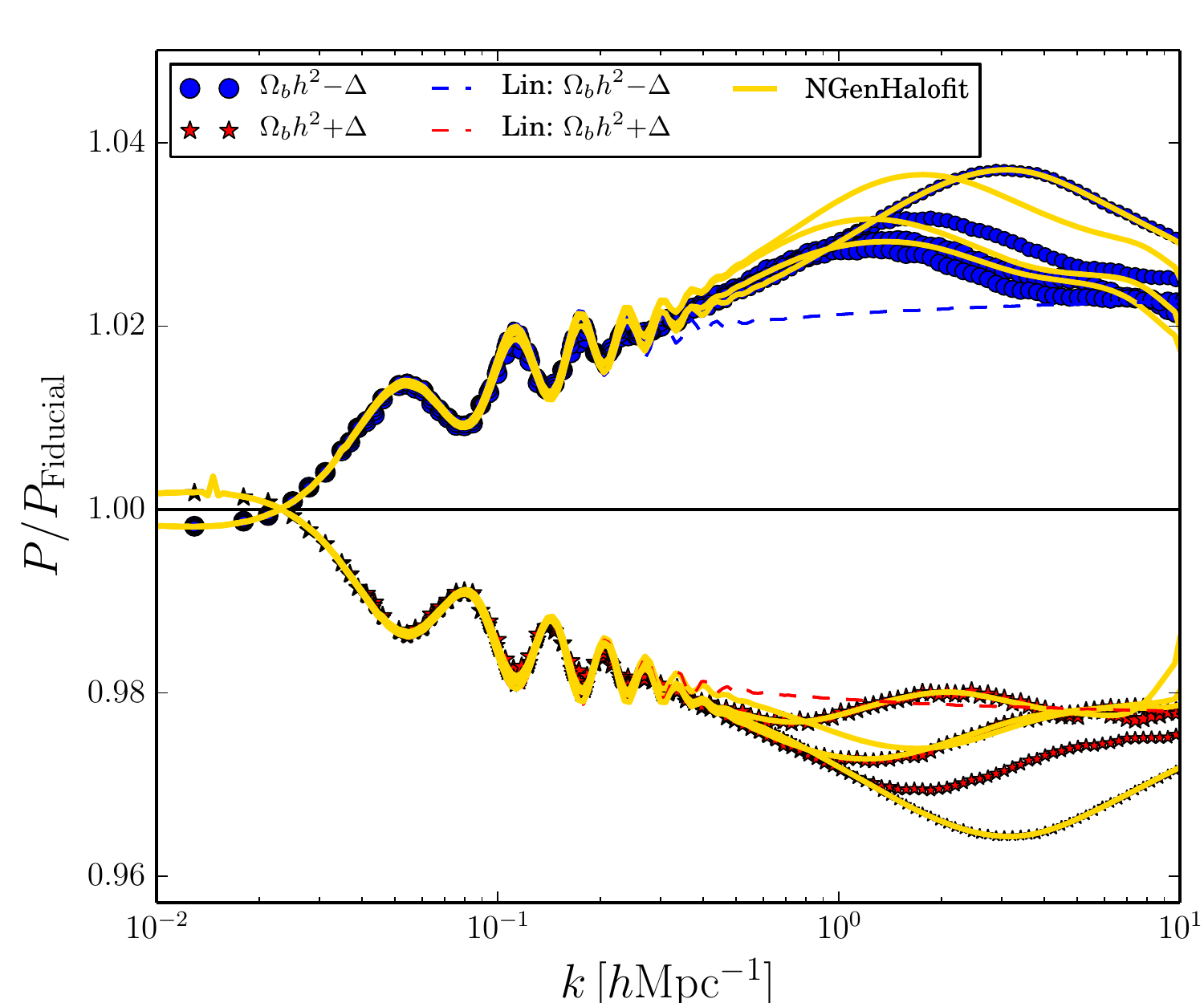}\hspace{0.3cm}
\includegraphics[width=6.5cm,angle=0]{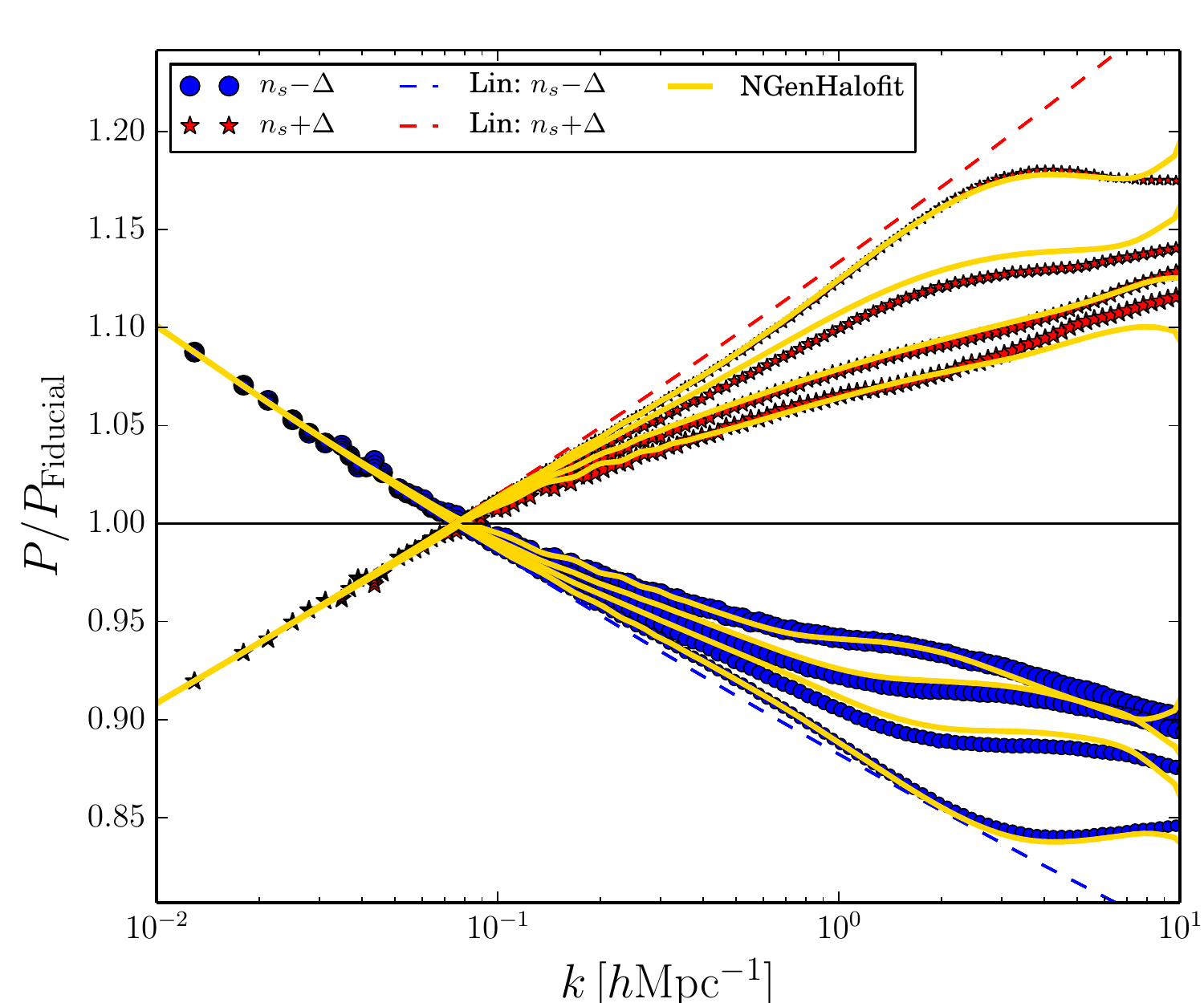}}
\vspace{0.2cm}
\centerline{
\includegraphics[width=6.5cm,angle=0]{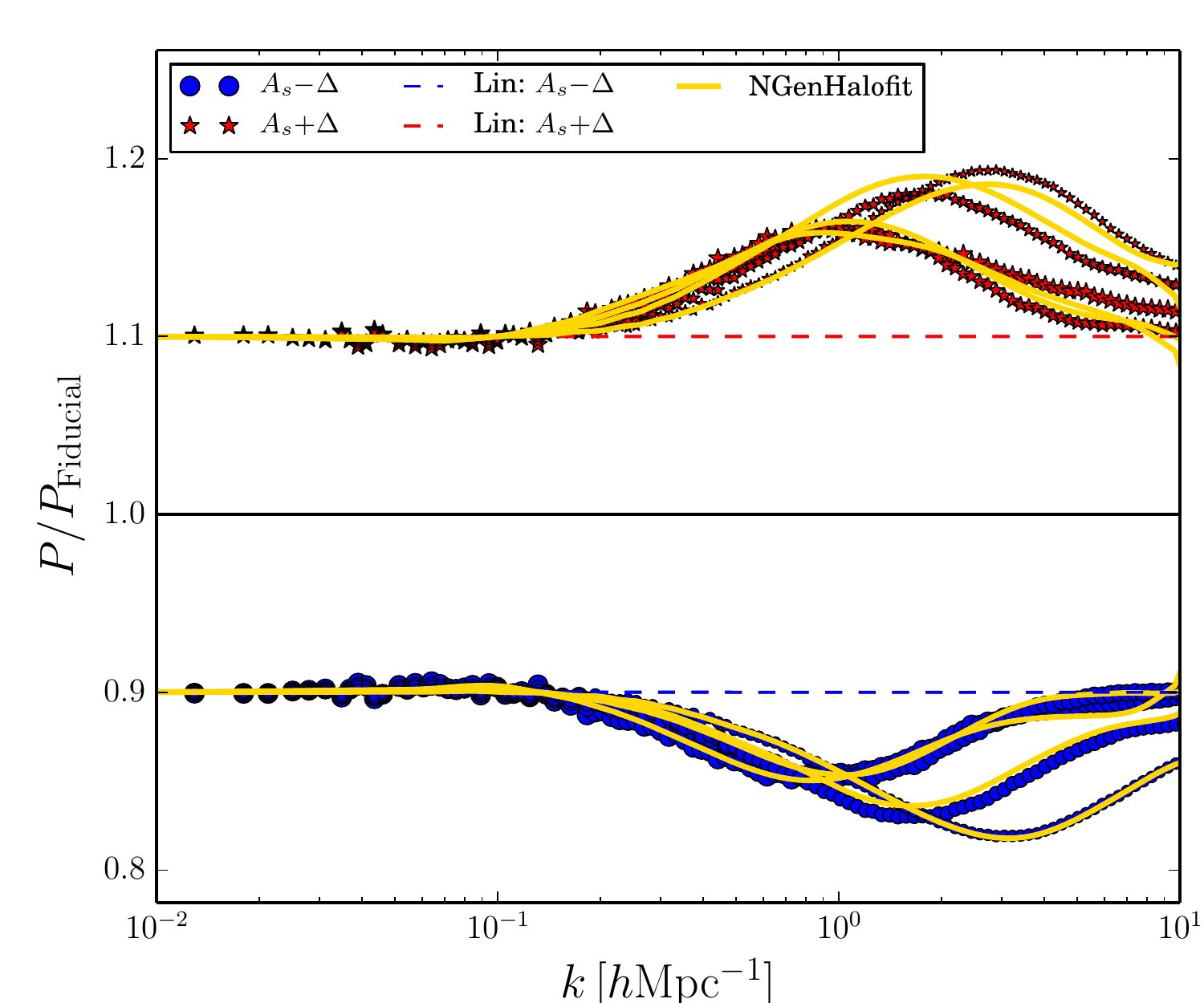}\hspace{0.3cm}
\includegraphics[width=6.5cm,angle=0]{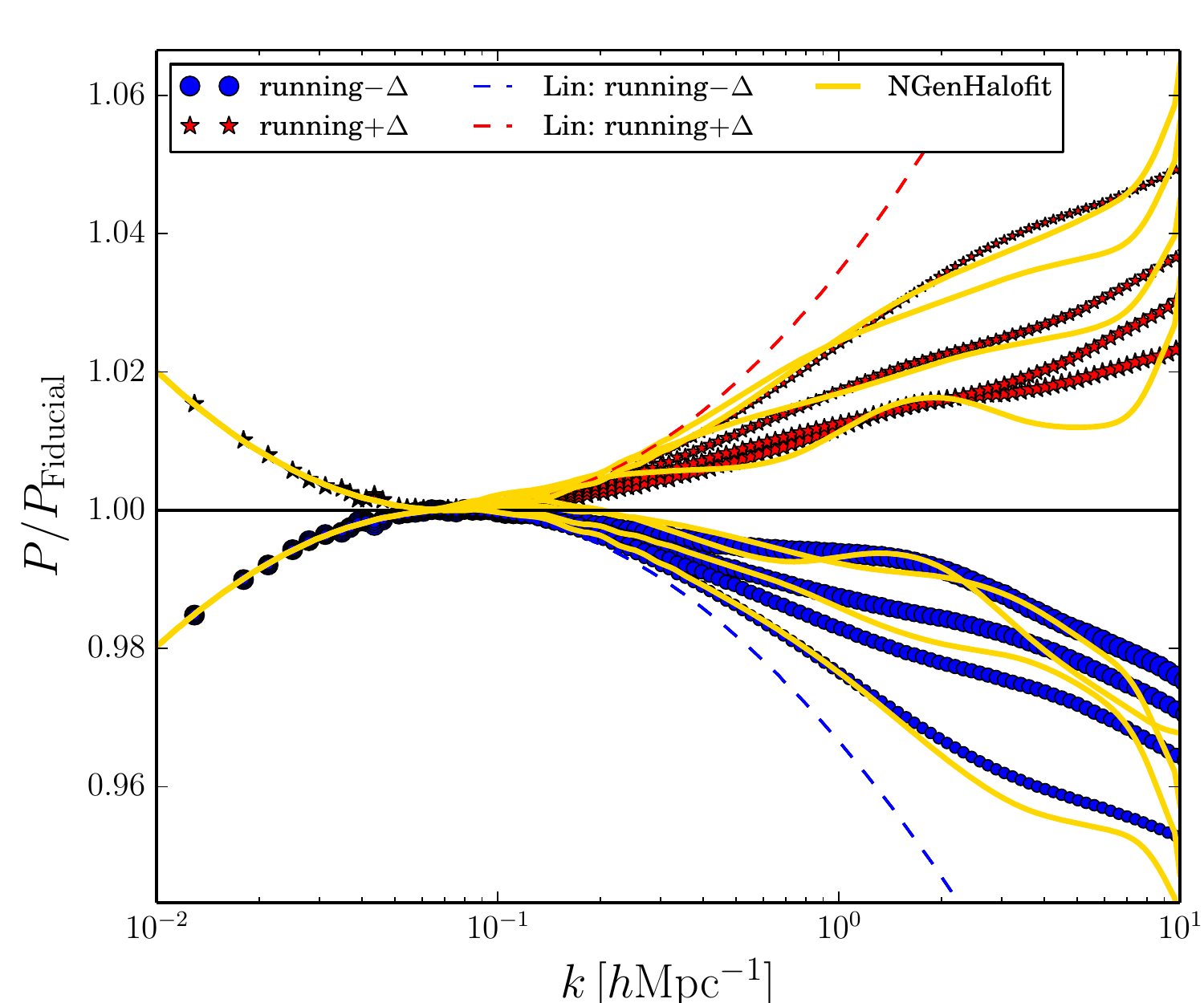}}
\caption{\small{Dependence of the power spectrum on variations in the
    cosmological parameters. All plots show the ratio of the
    variational models with respect to the fiducial model. Each panel
    shows the variations for a single parameter as a function of
    scale. The red and blue points show positive and negative
    variations, respectively and the point size increases with
    decreasing redshift, with $z\in\{2,1,0.5,0.0\}$. The dashed lines show the results for linear
    theory. }
\label{fig:PkCosmoRatios2}}
\end{figure*}

We now turn to the cosmology dependence of {\tt NGenHalofit}.
Figure~\ref{fig:PkCosmoRatios2} shows again the ratio of the
variational runs with respect to the power spectra from run 1 of the
fiducial model. However, this time we show the ability of {\tt
  NGenHalofit} to predict the 16 ratios for the 8 parameters -- this
is shown as the set of solid black lines in each panel.

At first this may not sound like a very stringent test, since we used
the variational results to construct the derivatives. However, there is
no guarantee that this would enable each of the extreme variations to
be accurately predicted, since we are not interpolating. For the case
of the parameters $\{w_0,\Omega_{\rm DE},\Omega_{\rm c}h^2,n_s,A_s\}$
the fits are virtually perfect. For the remaining parameters one can
see that there are some small deviations between the model and the
data. This stems from the fact that the linear response appears not to
be sufficient to fully capture the variations in these cosmological
models and one would need to add in the second order response function
to improve this.

Nevertheless, on comparing these results with the predictions from
{\tt halofit2012}, we see that in all cases the predictions from {\tt
  NGenHalofit} are better. In particular, for the running of the
primordial spectral index, the new approach has significantly improved
the predictions and appears to be highly accurate for $k<3\kMpc$.

Before moving on, although the ratio of the variation runs with
respect to the fiducial model appear to be not too badly described by
{\tt halofit2012}, we point out that the ratio is insensitive to
overall calibration errors, since:
\ba \frac{P_{\rm true}(k,a|{\bm\theta})}{P_{\rm true}(k,a|{\bm\theta}_0)} & = &
  1+
  \sum_i{\mathcal R}_i^{(1)}(k,z|{\bm\theta_0})\Delta\theta_i  \nn \\
  & & +
  \frac{1}{2}\sum_{i,j}{\mathcal R}_{ij}^{(2)}(k,z|{\bm\theta_0})
  \Delta\theta_i\Delta\theta_j+\dots
\ea

The combination of our exact 2-loop MPT modelling on large scales
$(k<0.07\kMpc)$, our precise modelling of the absolute value of the
fiducial power spectra, as demonstrated in Fig.~\ref{fig:PkFid}, and
our again high precision modelling of ${\mathcal R}^{(1)}_i$ and
${\mathcal R}^{(2)}_{ii}$ as demonstrated in
Fig.~\ref{fig:PkCosmoRatios2} mean that {\tt NGenHalofit} provides,
overall a very accurate description of the power spectra from the
D\"ammerung run $N$-body data.


\begin{figure*}
\centerline{
\includegraphics[width=6.5cm,angle=0]{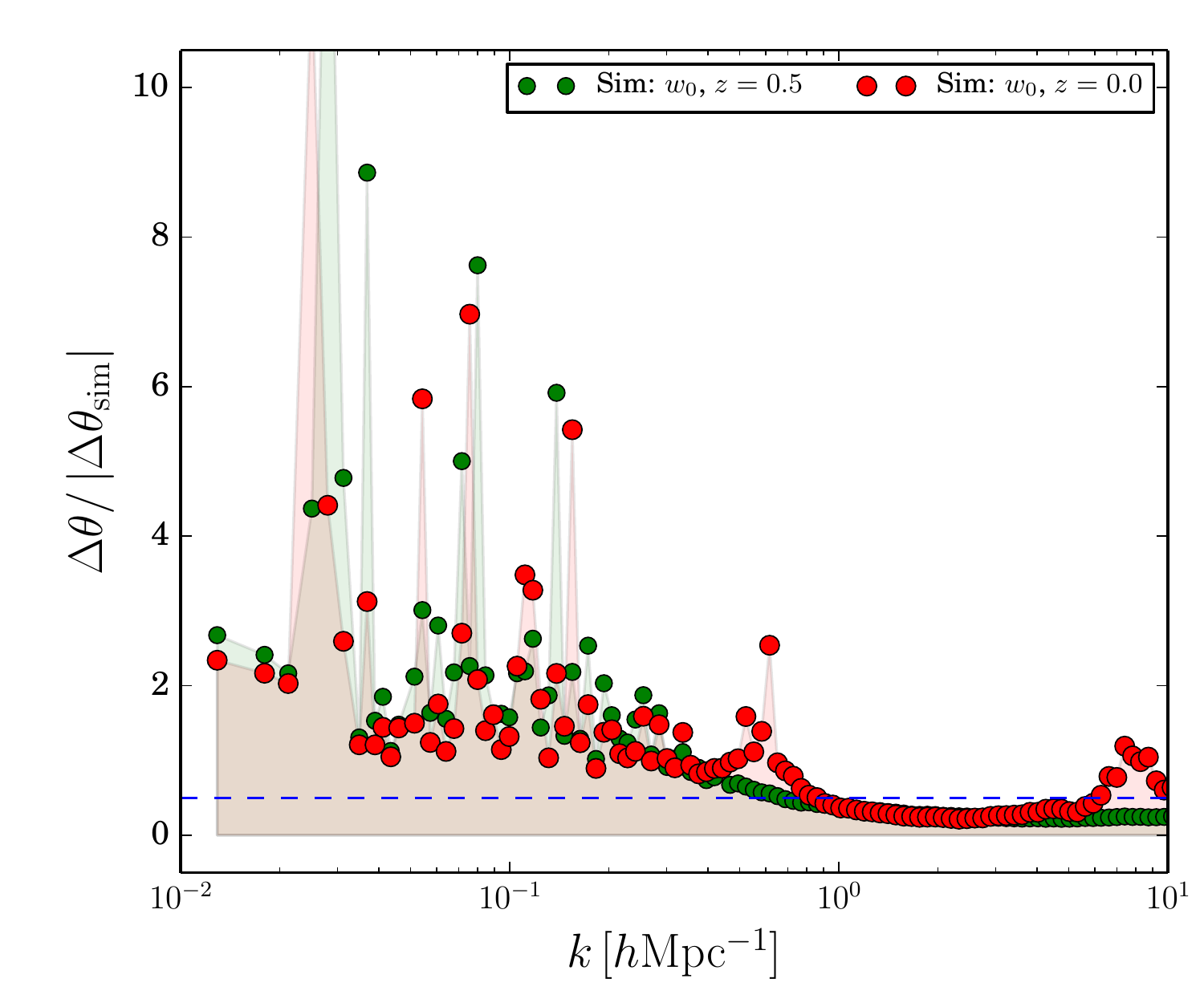}\hspace{0.3cm}
\includegraphics[width=6.5cm,angle=0]{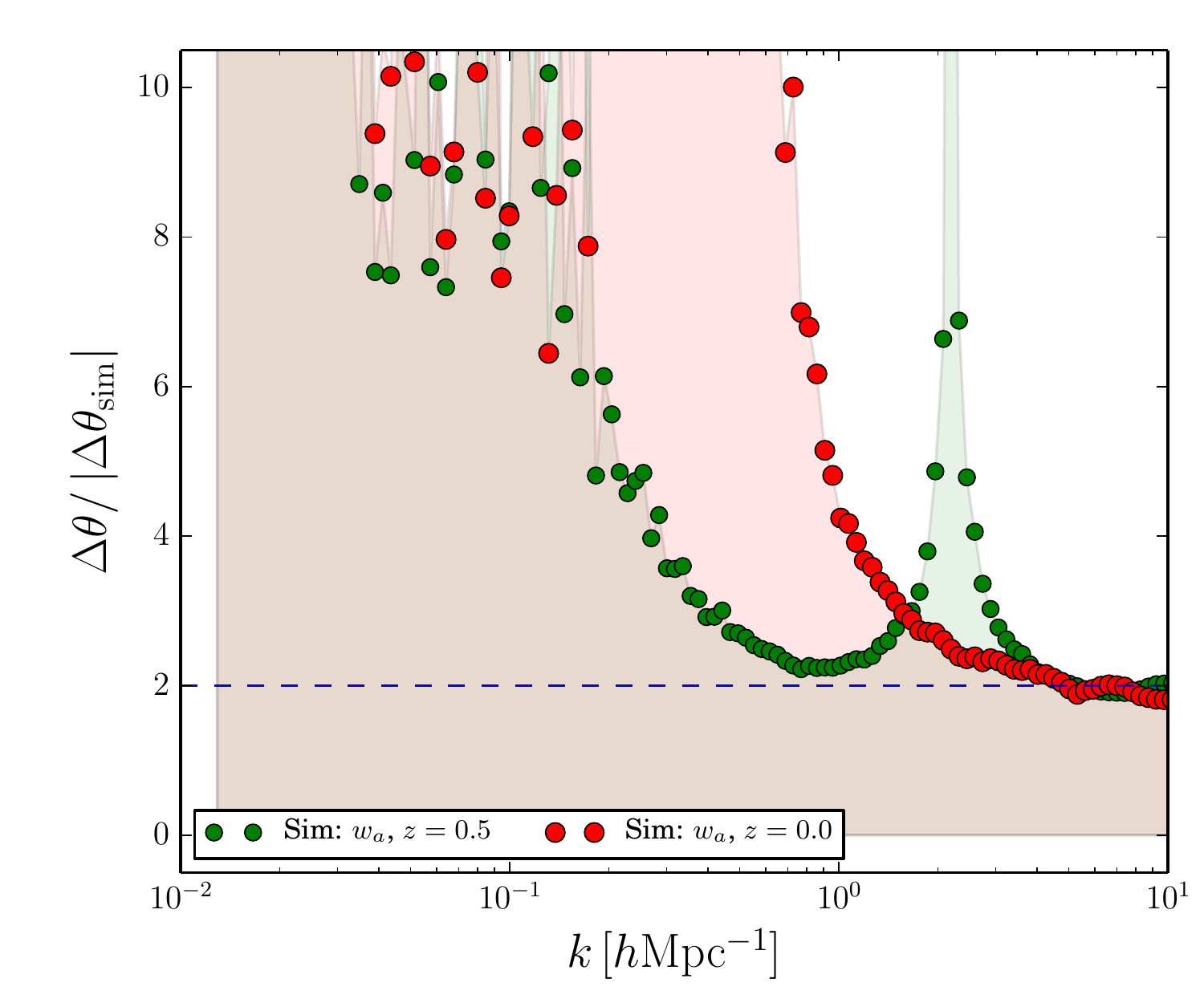}}
\vspace{0.2cm}
\centerline{
\includegraphics[width=6.5cm,angle=0]{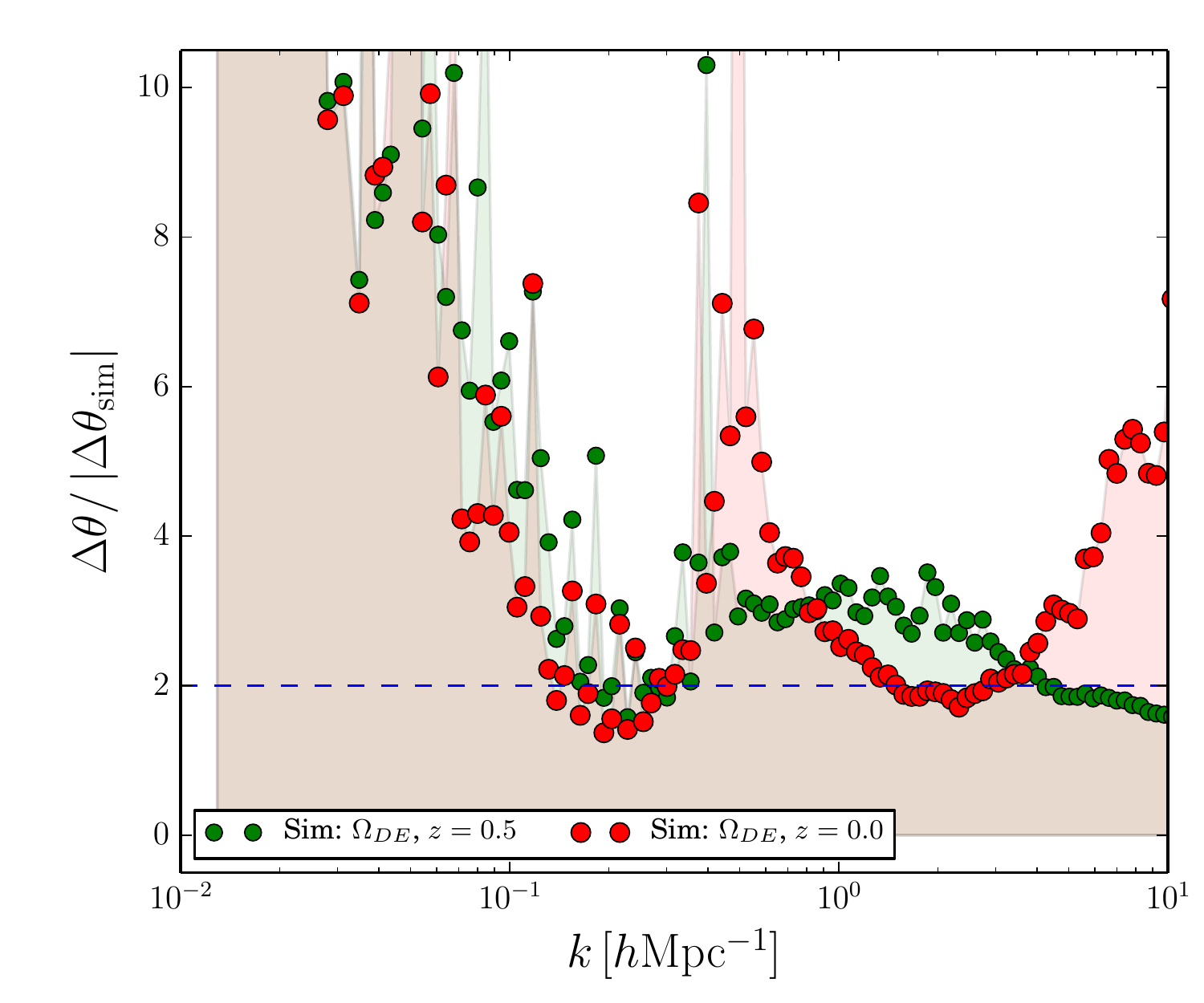}\hspace{0.3cm}
\includegraphics[width=6.5cm,angle=0]{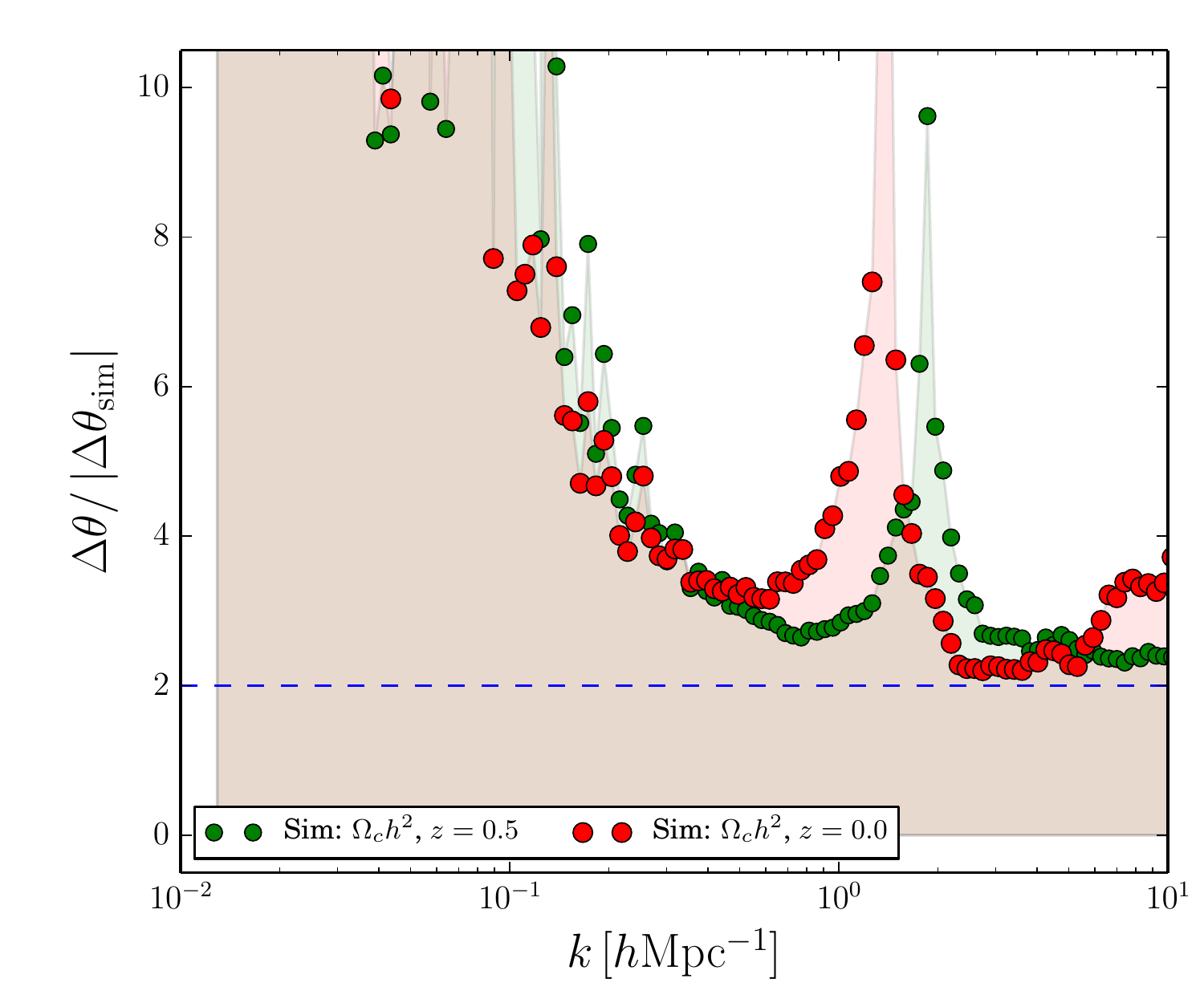}}
\vspace{0.2cm}
\centerline{
\includegraphics[width=6.5cm,angle=0]{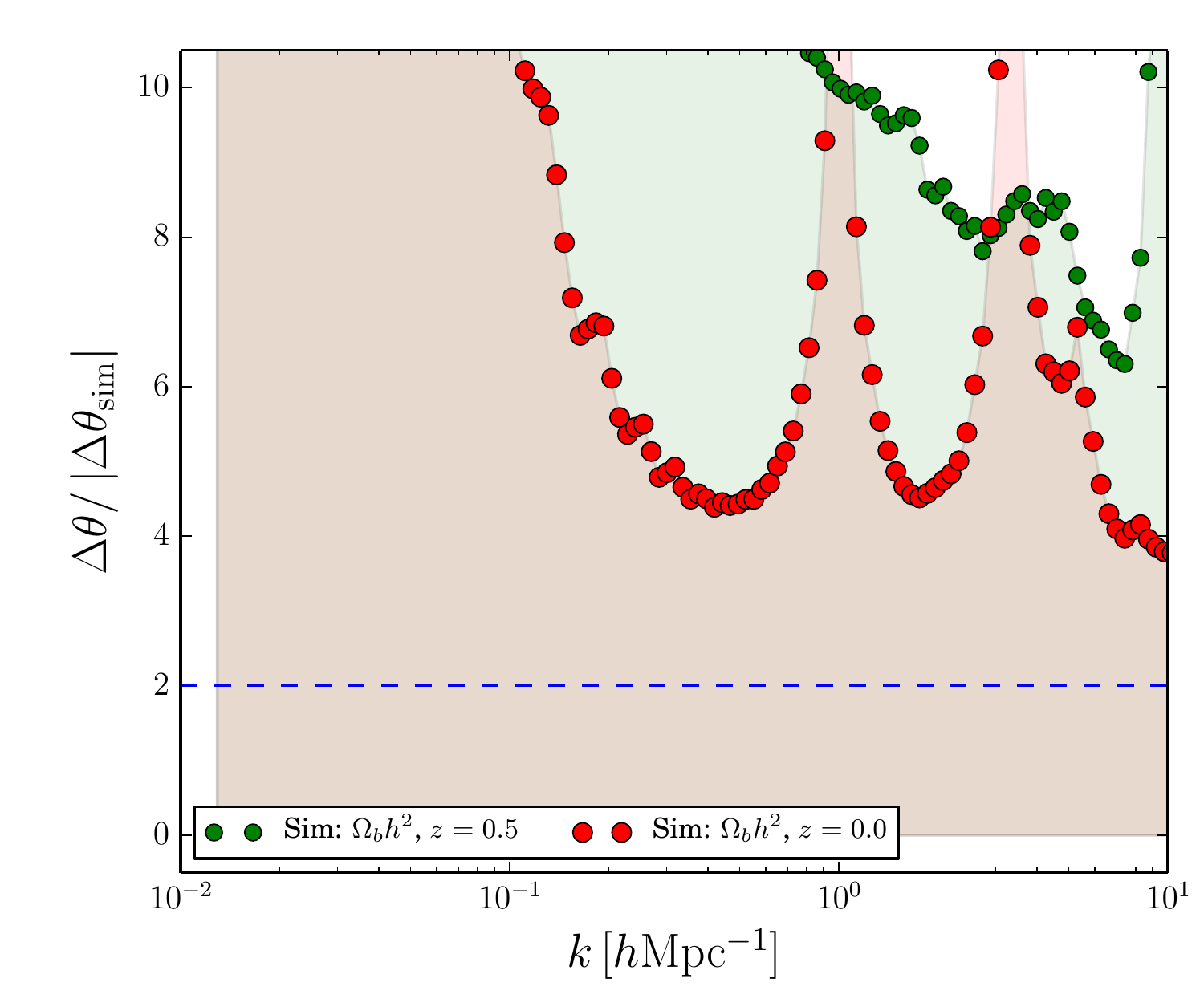}\hspace{0.3cm}
\includegraphics[width=6.5cm,angle=0]{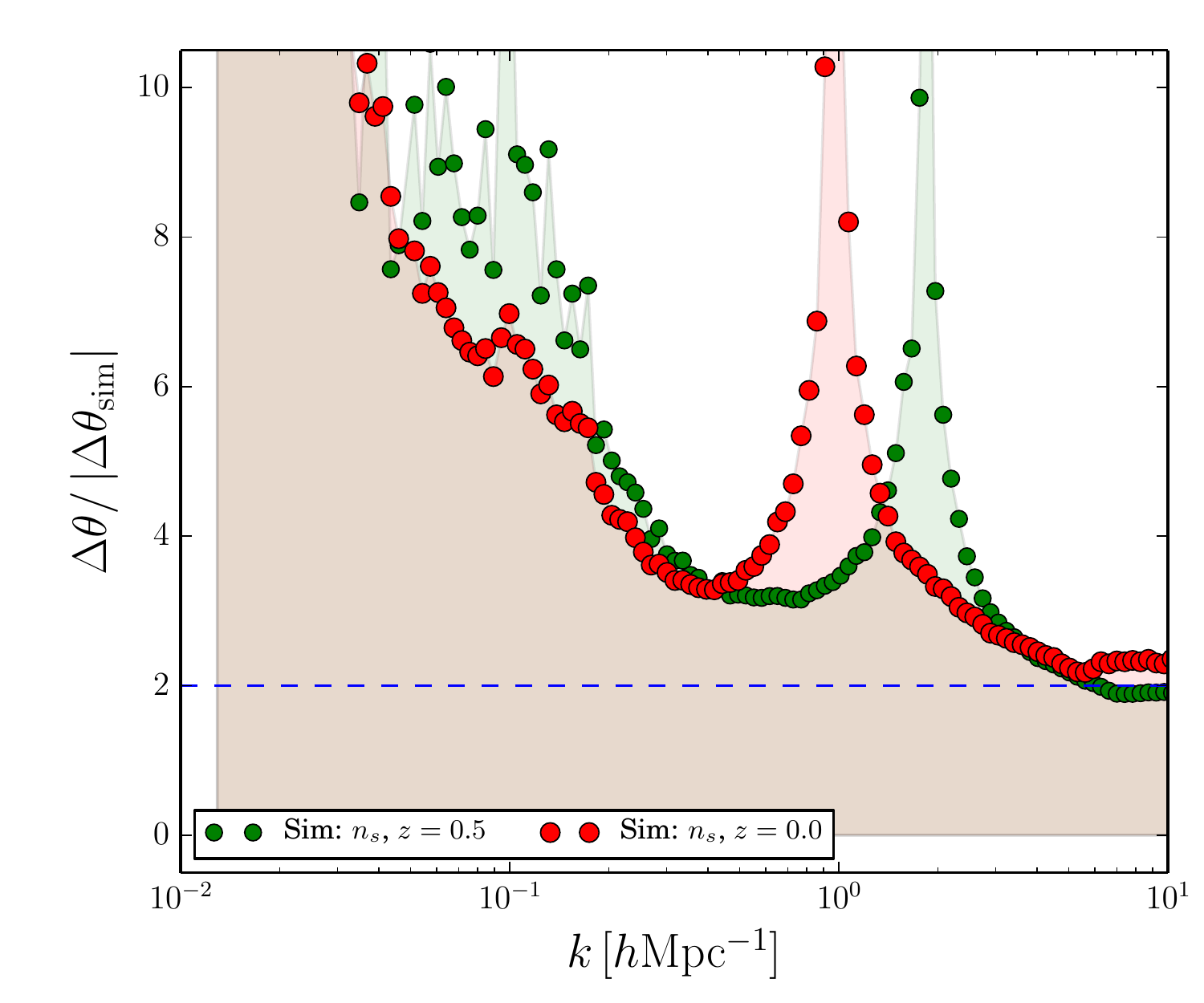}}
\vspace{0.2cm}
\centerline{
\includegraphics[width=6.5cm,angle=0]{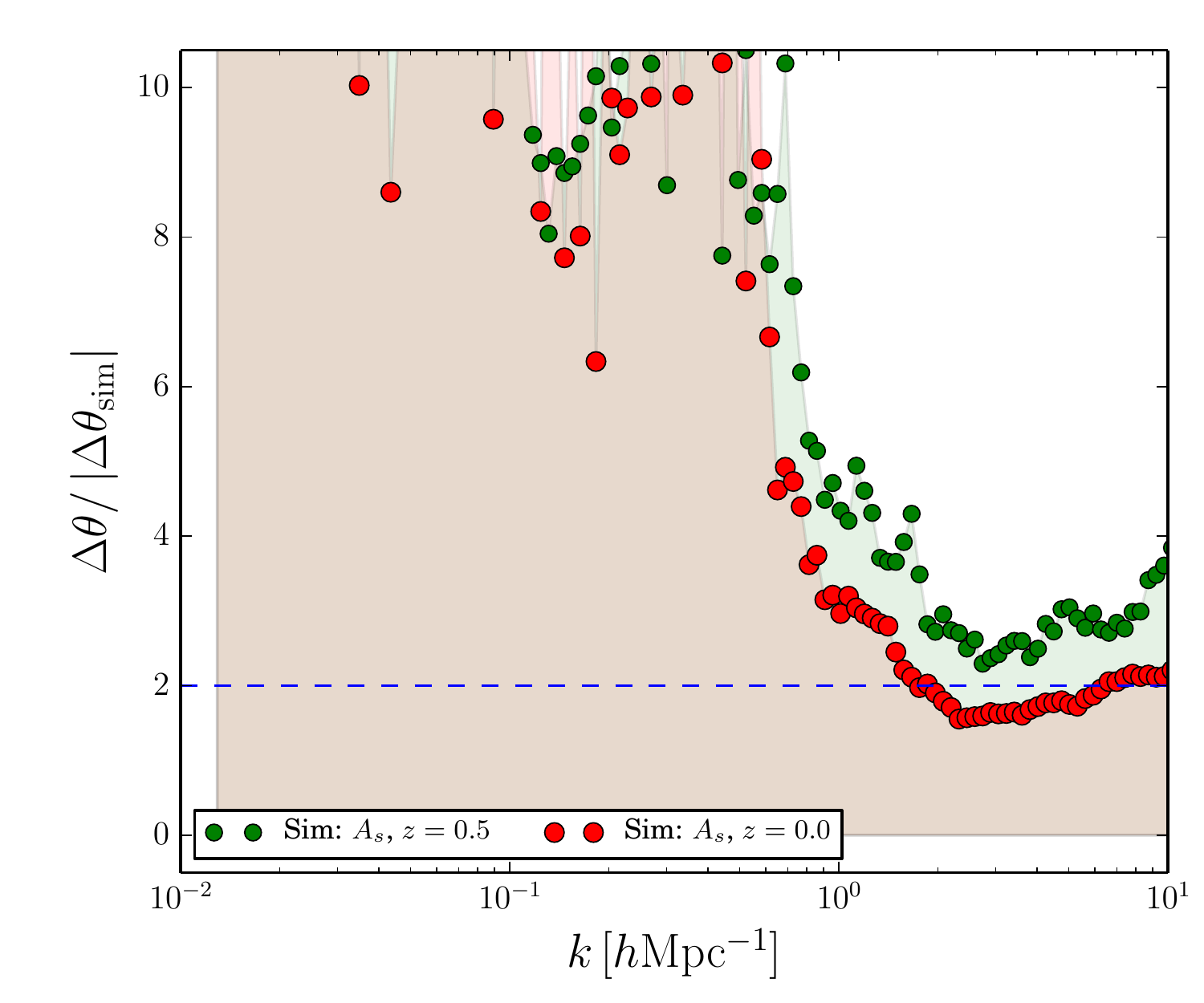}\hspace{0.3cm}
\includegraphics[width=6.5cm,angle=0]{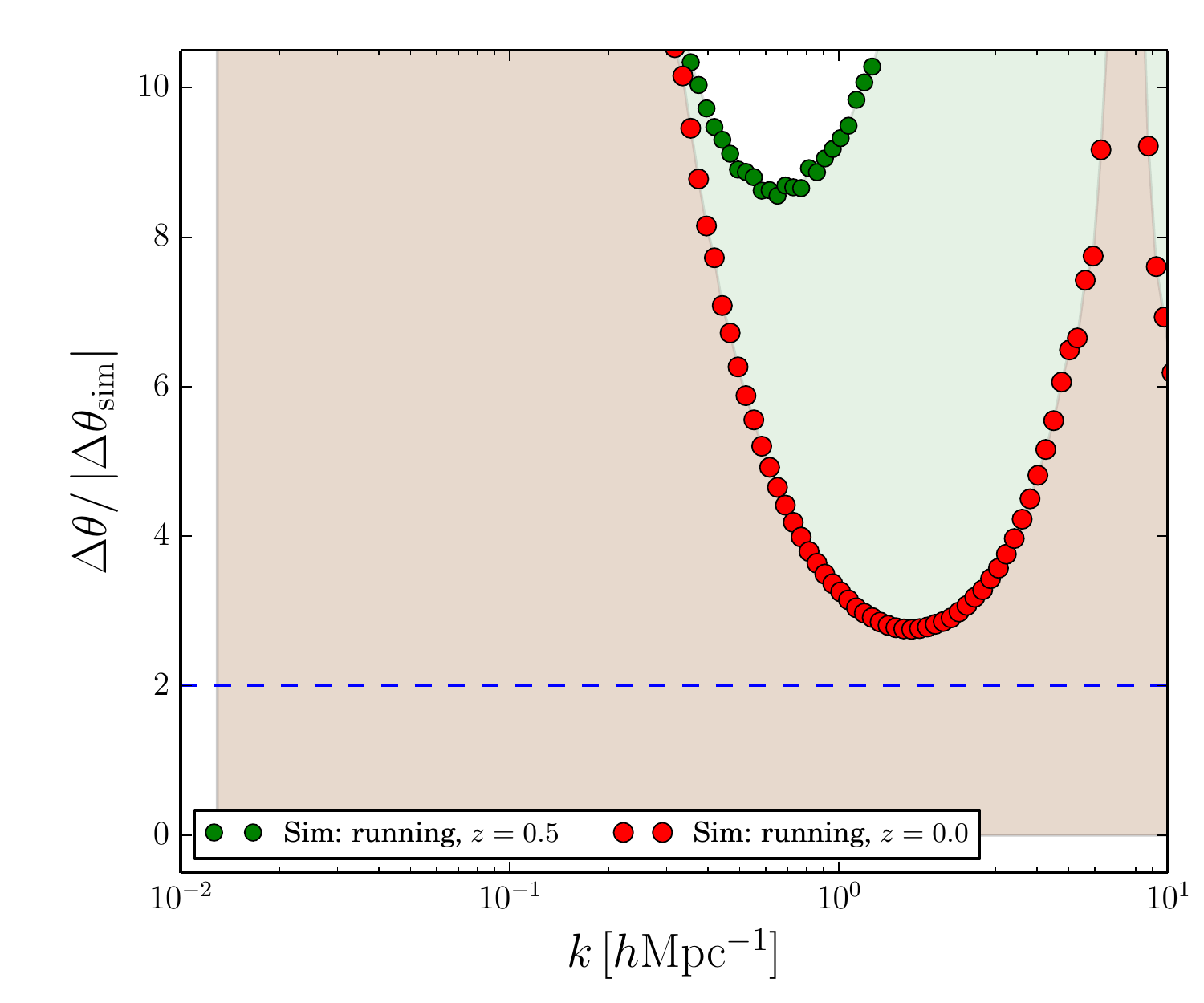}}
\caption{\small{Cosmological parameter step-size below which the
    Taylor expansion approach is precise to 3\%. $\Delta\theta$ is
    estimated from the simulations using \Eqn{eq:Delta} and we scale
    it in units of the variation step-sizes used in the D\"ammerung
    runs, and we show this as a function of wavenumber and
    redshift. The 8 panels show results for each of the cosmological
    parameter variations simulated. The large red points show results
    for $z=0$ and the small green points show results for $z=0.5$. In
    panels 2--7, the blue dashed lines indicate the line
    $|\Delta\theta| \lesssim 2 \left|\Delta\theta_{\rm sim}\right|$,
    and in panel 1 the line represents $|\Delta\theta| \lesssim 0.5
    \left|\Delta\theta_{\rm sim}\right|$. }
\label{fig:valid}}
\end{figure*}


\subsection{Estimate of overall precision and accuracy}\label{ssec:precision}

We now turn to the question of estimating the overall accuracy of the
Taylor expansion approach.  To do this recall \Eqn{eq:PNgen}, from
this we see that we need both:
\be
1 >  \sum_i{\mathcal R}_i^{(1)}(k,z|{\bm\theta}_0)\Delta\theta_i >
\frac{1}{2}  \sum_{i,j}{\mathcal R}_{ij}^{(2)}(k,z|{\bm\theta}_0)\Delta\theta_i\Delta\theta_j \ .
\ee
Let us focus on this relation for a single parameter, with all other
parameters held fixed at their fiducial values. Since we are only
including the diagonal terms of the Hessian one way to guarantee that
the linear order expansion is reasonably accurate would be to require
that the second order corrections are a small fraction compared to
unity, i.e., for the $i$th parameter, for second order corrections to
remain $\lesssim3\%$ we would require:
\be
0.03 \lesssim \frac{1}{2}{\mathcal R}_{ii}^{(2)}(k,z|{\bm\theta}_0)(\Delta\theta_i)^2
\ee
and this would imply that we should take steps in the parameter space
away from the fiducial that are given by the inequality:
\be \Longrightarrow \ \ \ \Delta\theta_i \lesssim 0.1
\sqrt{\frac{6}{{\mathcal R}_{ii}^{(2)}(k,z|{\bm\theta}_0)}}
\ . \label{eq:Delta} \ee
Since we do actually include the second order corrections (at least
for variation of a single parameter), this should provide us with a
conservative estimate of the precision for a single parameter. On the
other hand, since we do not have access to the full matrix ${\mathcal
  R}_{ij}^{(2)}$, it is the case that anything else may be viewed
circumspectly. We adopt the value of 3\% for the second order
correction, since in a recent study by \citet{Schneideretal2016} the
systematic error between various $N$-body codes on small scales was
shown to be of this order. Establishing which of the various $N$-body
codes is of higher accuracy will be a task for future work.

In figure \ref{fig:valid} we show the value of $\Delta\theta_i$ that
obtains from evaluating \Eqn{eq:Delta} and we scale this measured
value in units of the parameter step sizes used in the D\"ammerung
runs. The results are shown as a function of wavenumber and for two
redshifts. The plot clearly shows that for all of the parameters
considered a step size of $|\Delta\theta| \lesssim 2
\left|\Delta\theta_{\rm sim}\right|$ would guarantee that the second
order corrections would be below the required value (indicated as the
dashed blue line in each panel). The only exception is the case of
$w_0$, which appears to require the more restricted range of
$|\Delta\theta| \lesssim 0.5 \left|\Delta\theta_{\rm sim}\right|$.

This leads us to estimate that {\tt NGenHalofit} should be precise to
better than 3\% in the cubical region of parameter space given by the
range of values: $w_0\in\{-1.05,-0.95\}$, $w_a\in\{-0.4,0.4\}$,
$\Omega_{\rm m,0}\in\{0.21,0.4\}$, $\omega_{\rm c}\in\{0.1,0.13\}$,
$\omega_{\rm b}\in\{2.0,2.4\}$, $n_{\rm s}\in\{0.85,1.05\}$,
$A_s\in\{1.72\times 10^{-9},2.58\times 10^{-9}\}$,
$\alpha\in\{-0.2,0.2\}$. When the model under investigation is outside
of this region the code predictions smoothly revert back to {\tt
  halofit2012} and so the nominal values for that code apply.  Note
that we also set the transition speed to be,
%
\be \sigma^{\theta}_{{\rm cut},i}(k,z)=0.05\left[\theta_{{\rm cut},i}(k,z)-\theta_{0,i}\right]\ .\ee
%



\section{Conclusions \& Discussion}\label{sec:conclusions}

In this paper we have generated a suite of 27 high-resolution,
$N$-body simulations -- the D\"ammerung runs. These runs have enabled
us to explore the nonlinear evolution of the matter power spectrum in
the $w$CDM framework -- that is a cosmological model with a time
evolving dark energy equation of state. We chose as our fiducial model
the best fit parameters from the \citep{Planck2014XVI} data -- this we
covered with 11 simulations, 10 of which were of comparable resolution
to the Millennium Run1 simulation of \citet{Springeletal2005}. The
remaining 16 runs modelled the variations in 8 cosmological
parameters.  We have also performed the first detailed study of the
effects of a running primordial power spectral index $\alpha=dn/d\log
k$ on the nonlinear matter power spectrum. Our simulations were
described in \S\ref{sec:sims} and \S\ref{sec:spectraI}.

In \S\ref{sec:PT} and \S\ref{sec:semi} we compared our measured
nonlinear power spectra from the fiducial runs with various analytic
and semi-analytic methods.  We first explored the ability of MPT
theory at the 2-loop level to describe our data. We found that the
model reproduced the data to percent accuracy up to $k<0.15\kMpc$ at
$z=0$. On scales smaller than this, it deviated strongly from the
data. We compared our fiducial results with the upgraded {\tt halofit}
model of \citet{Takahashietal2012}, the {\tt EMU} code of
\citet{Heitmannetal2014} and the {\tt HMCode} of \citet{Meadetal2015}.
We found that all of these methods described the data on small scales
to a precision of 5\% -- with the {\tt EMU} code providing the best
description on small scales. However, somewhat surprisingly, this code
was not able to accurately capture the input large-scale power
spectrum, being in error at the level of $\sim10\%$ on scales of the
order $k\sim0.01\kMpc$. Considering all scales, we found that {\tt
  HMCode} best described our data overall, but it failed to accurately
capture the nonlinear processing of the acoustic oscillations.

In \S\ref{sec:newmodel} we showed how one could use a suite of
simulations to recalibrate a nonlinear power spectrum model to
accurately describe the 8-parameter cosmological model that we have
examined. The explicit case that we have developed utilises {\tt
  halofit2012} as the underlying base mode, and therefore we dubbed
our new improved model {\tt NGenHalofit}. We also showed that this
approach could work to high precision (<1\%) for the case of the
fiducial model for a range of redshifts and for scales $k<10\kMpc$.

In \S\ref{sec:cosmodep} we measured the ratio and logarithmic
derivatives of the nonlinear power spectrum with respect to the
cosmological parameters. We showed that {\tt halofit2012} captured well
the dependence on parameters $\{w_0,w_a,\Omega_{\rm
  DE},n_s,A_s\}$. However, in general, it showed variation of the
order several percent. For the parameters $\{\Omega_{\rm
  c}h^2,\Omega_{\rm b}h^2,\alpha_{s}\}$ it provided a poor
description.  We also measured a scaled logarithmic derivative and
also the diagonal terms entering the Hessian.  This enabled us to
build the cosmology dependent corrections for the {\tt NGenHalofit}
model.

We are also interested in constraining variations in the running of
the primordial power spectral index $\alpha$ from future large-scale
structure measurements, since placing constraints on this may help
constrain inflationary models \citep{Vieiraetal2017}. Currently, the
only models to do that is the {\tt Halofit2012} method.  We
demonstrated that this model was unable to describe such variations
accurately, but that our updated approach enables significantly
improved modelling of the impact of spectral running.

Finally, in \S\ref{sec:cosmodep} we used {\tt NGenHalofit} to predict
all of the measured nonlinear spectra ratios. We found that our new
approach was able to capture the dependence of the spectrum on
cosmology at a level of the order 1\% precision.

\vspace{0.2cm}\noindent {\bf Future work:} The fact that ${\mathcal
  R}^{(1)}_i(k,a)\lesssim4$ for the parameters $\{\Omega_{\rm
  c}h^2,\Omega_{\rm b}h^2,\alpha\}$
suggests that the underlying model {\tt halofit2012} did not describe
the variations in these parameters as well as the other parameters.
One obvious way to improve this implementation would be to include the
off-diagonal components of the Hessian matrix. In order to compute
these terms one needs to construct the following quantity:
\ba \widehat{\frac{\partial^2 P({\bm\theta})}{\partial\theta_i\partial\theta_j}}
& \approx &
\frac{1}{2\Delta\theta_i\Delta\theta_j}
\left[\widehat{P}_{+i,+j}+\widehat{P}_{-i,-j}-2\widehat{P}_{0,0}\right.  \nn\\
  & & \hspace{-0.5cm}-\left(\widehat{P}_{+i,0}+\widehat{P}_{-i,0}-2\widehat{P}_{0,0}\right) \nn \\
  & & \hspace{-0.5cm}\left.-\left(\widehat{P}_{0,+j}+\widehat{P}_{0,-j}-2\widehat{P}_{0,0}\right)
  \right]+ {\mathcal O}(\Delta\theta_i\Delta\theta_j) \ ,
\ea
where $(i\ne j)$ and where we have introduced the notation:
\ba
\widehat{P}_{+i,+j} & \equiv &
\widehat{P}(\theta_{0,1},\dots,\theta_{0,i}+\Delta\theta_i,\dots,\theta_{0,j}+\Delta\theta_j,\dots,\theta_{0,N}) \nn \\
\widehat{P}_{+i,-j} & \equiv &
\widehat{P}(\theta_{0,1},\dots,\theta_{0,i}+\Delta\theta_i,\dots,\theta_{0,j}-\Delta\theta_j,\dots,\theta_{0,N}) \nn \\
\widehat{P}_{+i,0} & \equiv &
\widehat{P}(\theta_{0,1},\dots,\theta_{0,i}+\Delta\theta_i,\dots,\theta_{0,j},\dots,\theta_{0,N}) \nn   \\
\widehat{P}_{0,+j} & \equiv &
\widehat{P}(\theta_{0,1},\dots,\theta_{0,i},\dots,\theta_{0,j}+\Delta\theta_j,\dots,\theta_{0,N}) \nn   \\
\widehat{P}_{0,0} & \equiv &
\widehat{P}(\theta_{0,1},\dots,\theta_{0,i},\dots,\theta_{0,j},\dots,\theta_{0,N}) \nn  
\ea
etc. and where the fiducial point in our $N$-dimensional parameter
space is $\bm\theta_0=\{\theta_{0,1},\dots,\theta_{0,N}\}$. Thus in
order to obtain the power spectrum Hessian with respect to the
cosmological parameters, we would need to run an additional two
simulations for every element of the matrix in the upper half matrix.
That is for $N$ cosmological parameters we would need to compute
$N(N-1)/2$ elements of the Hessian, each requiring an extra 2
simulations. For the 8 parameters that we are simulating that would
mean an extra $8(8-1)/2\times 2=56$ simulations. If we limit this to
the terms that depend on the 3 parameters that we have identified then
a good approximation would be to add $3(3-1)/2\times 2 = 6$ runs, and
set all other off-diagonal elements to zero. We will explore this in
future work.

Another interesting avenue would be to increase parameter space
coverage by stitching together Taylor expansion at various points in
parameter space. This approach may have advantages over the alternate
approach of building emulators using Latin hyper-cube sampling of
parameter space, since if the validity of the Taylor expansions
encompasses a larger range of parameter space at a given accuracy with
fewer simulations than the emulator, then one may need fewer
simulations overall to construct solutions over the entirety of the
parameter space. We expect that this will be important for future Dark
Energy missions like DESI, 4MOST, Euclid, LSST, and WFIRST. We shall
leave further discussion of this topic for future work.

Finally, we recognise that in this work we have made no attempt to
account for the impact of nonlinear late-time baryonic physics effects
on the evolution of matter perturbations. A number of works have
established, through theoretical analysis and detailed hydrodynamic
simulations, various results
\citep{ZhanKnox2004,Jingetal2006,SomogyiSmith2010,vanDaalenetal2011,Schneideretal2018}.
In most cases the inclusion of hot baryons can lead to a supression of
the power by $\sim$10\%. However, the precise value depends explicitly
on the details of the feedback model, resolution of the simulations,
the method for simulating the gas physics, and the physics
treated. For instance the work of \citep{Jingetal2006}, besides
showing a supression, showed that if radiative heating/cooling effects
are turned-off, the matter power spectrum will in fact be enhanced on
smaller scales. Similar results were noted in the work of
\citet{vanDaalenetal2011} and \citet{Sembolonietal2011}, where gas
cooling led to a boost in clustering on small scales. Owing to the
fact that this is a very complex problem to disentangle, we shall
defer a detailed examination of this area for future works, but note
that in the meant time, one may follow the phenomenological approach
that was advocated by \citet{Sembolonietal2011} to build a
parameterised baryonic physics template whose physical effects can be
marginalised over. We also note that we have neglected to take account
the effects of massive neutrinos. We expect that this will also
require careful corrections, but note that the work of
\citet{Birdetal2012} should still be a viable extension to our model.


\section*{Acknowledgements}
We would like to thank Martin Crocce, Roman Scoccimarro, Volker
Springel, and Simon White for useful discussions. We would like to
thank Volker Springle for providing access to the {\tt Gadget-3}
code. We would like to thank Martin Crocce, Roman Scoccimarro and
Francis Bernardeau for making public their {\tt MPTbreeze} code and
Roman Scoccimarro and Marc Manera for making public their {\tt 2LPT}
code.  RES thanks Cameron Brown for comments. RES acknowledges support
from the Science and Technology Facilities Council (grant number
ST/P000525/1).  REA acknowledges support of the European Research
Council through grant number ERC-StG/716151, and of the Spanish
Ministerio de Economía and Competitividad (MINECO) through grant umber
AYA2015-66211-C2-2. We acknowledge that the results of this research
have been achieved using the PRACE Tier-0 Research Infrastructure
resource SuperMuc based in Garching Germany at the Leibniz
Supercomputing Centre (LRZ) under project number 2012071360.



\bibliographystyle{mnras}

\vspace{5mm}
 

\onecolumn

\appendix

\section{Linear growth of density modes}\label{app:one}

\subsection{4th order Runge-Kutta solution to linear growth equations}

Following \S\ref{ssec:growth}, we now describe how to solve
\Eqn{eq:g2} for the evolution of the linear growth factor. To begin,
we introduce a new variable $s=D'(a)$. One can now rewrite this second
order ODE as a pair of coupled, linear, first order, ordinary
differential equations:
\ba 
D'(a) & = & s(a) \ ; \label{eq:D'} \\
s'(a) & = & -\Gamma_1(a)s(a) -\Gamma_2(a)D(a) \ . \label{eq:s'}
\ea
This system is evolved from the initial values \mbox{$D(a_i)=a_i$} and
\mbox{$s(a_i)=1$}, which owes to the fact that at early times in the
evolution we know that \mbox{$D(a)\propto a$}, deep in the matter
dominated era.  We will solve \Eqns{eq:D'}{eq:s'} using a 4th order
Runge-Kutta method.  First let us rewrite the above equations more
generally as
\ba 
\frac{dq_1}{da} & = & A(q_1,q_2,a) \ ; \label{eq:q1'} \\
\frac{dq_2}{da} & = & B(q_1,q_2,a) \ ; \label{eq:q2'} 
\ea
where it is the evolution of the variables $q_1$ and $q_2$ that we
wish to solve for and where $A$ and $B$ may be general functions of
$q_1$, $q_2$ and the scale factor $a$. In our case we have
$q_1(a)=D(a)$ and $q_2(a)=s(a)$ and also $A(q_1,q_2,a) = q_2$ and
$B(q_1,q_2,a) = -\Gamma_1(a)q_2 -\Gamma_2(a)q_1$.  The 4th order
Runge-Kutta solution proceeds from time step $a^{(0)}$ to time step
$a^{(1)}=a^{(0)}+\Delta a$ through the following algorithm:
\begin{align}
k_1^{(1)} & =  A(q_1^{(0)},q_2^{(0)},a^{(0)})            \ ; &  q_1^{(1)}  & =  q_1^{(0)} + k_1^{(1)} (\Delta a/2) \ ; \\
k_2^{(1)} & =  B(q_1^{(0)},q_2^{(0)},a^{(0)})            \ ; &  q_2^{(1)}  & =  q_2^{(0)} + k_2^{(1)} (\Delta a/2) \ ; \\
k_1^{(2)} & =  A(q_1^{(1)},q_2^{(1)},a^{(0)}+\Delta a/2) \ ; &  q_1^{(2)}  & =  q_1^{(0)} + k_1^{(2)} (\Delta a/2) \ ; \\
k_2^{(2)} & =  B(q_1^{(1)},q_2^{(1)},a^{(0)}+\Delta a/2) \ ; &  q_2^{(2)}  & =  q_2^{(0)} + k_2^{(2)} (\Delta a/2) \ ; \\
k_1^{(3)} & =  A(q_1^{(2)},q_2^{(2)},a^{(0)}+\Delta a/2) \ ; &  q_1^{(3)}  & =  q_1^{(0)} + k_1^{(3)} \Delta a \ ; \\
k_2^{(3)} & =  B(q_1^{(2)},q_2^{(2)},a^{(0)}+\Delta a/2) \ ; &  q_2^{(3)}  & =  q_2^{(0)} + k_2^{(3)} \Delta a \ ; \\
k_1^{(4)} & =  A(q_1^{(3)},q_2^{(3)},a^{(0)}+\Delta a)   \ ; &  q_1^{(4)}  & =  q_1^{(0)} + k_1^{(4)} \Delta a \ ; \\
k_2^{(4)} & =  B(q_1^{(3)},q_2^{(3)},a^{(0)}+\Delta a)   \ ; &  q_2^{(4)}  & =  q_2^{(0)} + k_2^{(4)} \Delta a \ ,
\end{align}
where the final estimate of the functions propagated to the next
time step is given by:
\ba 
\hat{D}(a^{(0)}+\Delta a) & = & \hat{D}(a^{(0)}) + \left(k_1^{(1)}+2k_1^{(2)}+2k_1^{(3)}+k_1^{(4)}\right) \frac{\Delta a}{6} \ ; \\
\hat{s}(a^{(0)}+\Delta a) & = & \hat{s}(a^{(0)}) + \left(k_2^{(1)}+2k_2^{(2)}+2k_2^{(3)}+k_2^{(4)}\right) \frac{\Delta a}{6}\ ;
\ea
We take $a_i=0.001$ and employ 1000 timesteps to reach $a=1.0$. It is
useful to note that the above solution for $D(a)$ also provides a
solution for the logarithmic growth rate $f(a)\equiv d\log D(a)/d\log
a$. Since:
\be \hat{f}(a) \equiv \frac{d\log \hat{D}}{d\log a} =  a\frac{\hat{s}(a)}{\hat{D}(a)}  \ .\ee

Owing to the fact that the fiducial model is $\Lambda$CDM, we can
compare our Runge-Kutta solution with the celebrated integral
expression of \citet{Heath1977} for presureless Friedmann-Lemaitre
models
\be D_+(a)=\frac{5}{2}\Omega_{\rm m,0} H(a) \int_{0}^a
\frac{da'}{\left(a'H(a')\right)^3} \  \ .\ee
Also we can compare this with the well known fitting formula given by
\citet{Carrolletal1992}:
\be D_{\rm CPT}(a) = a g_{\rm CPT}(a,\Omega_{\rm m}(a),\Omega_{\Lambda}(a)) \ ,\ee
where 
\be g_{\rm CPT}(a,\Omega_{\rm m}(a),\Omega_{\Lambda}(a))  \equiv \frac{5}{2}\Omega_{\rm m}(a)
\left[\Omega_{\rm m}^{4/7}(a)-\Omega_{\Lambda}(a)+\left(1+\frac{\Omega_{\rm m}(a)}{2}\right)\left(1+\frac{\Omega_{\Lambda}(a)}{70}\right)\right]^{-1} \ .\ee
We can also compare to the approximate growth factor expression from
\citet{Linder2005} given in \Eqn{eq:growlinder}.

\begin{figure*}
\centerline{
  \includegraphics[angle=0,width=8.4cm]{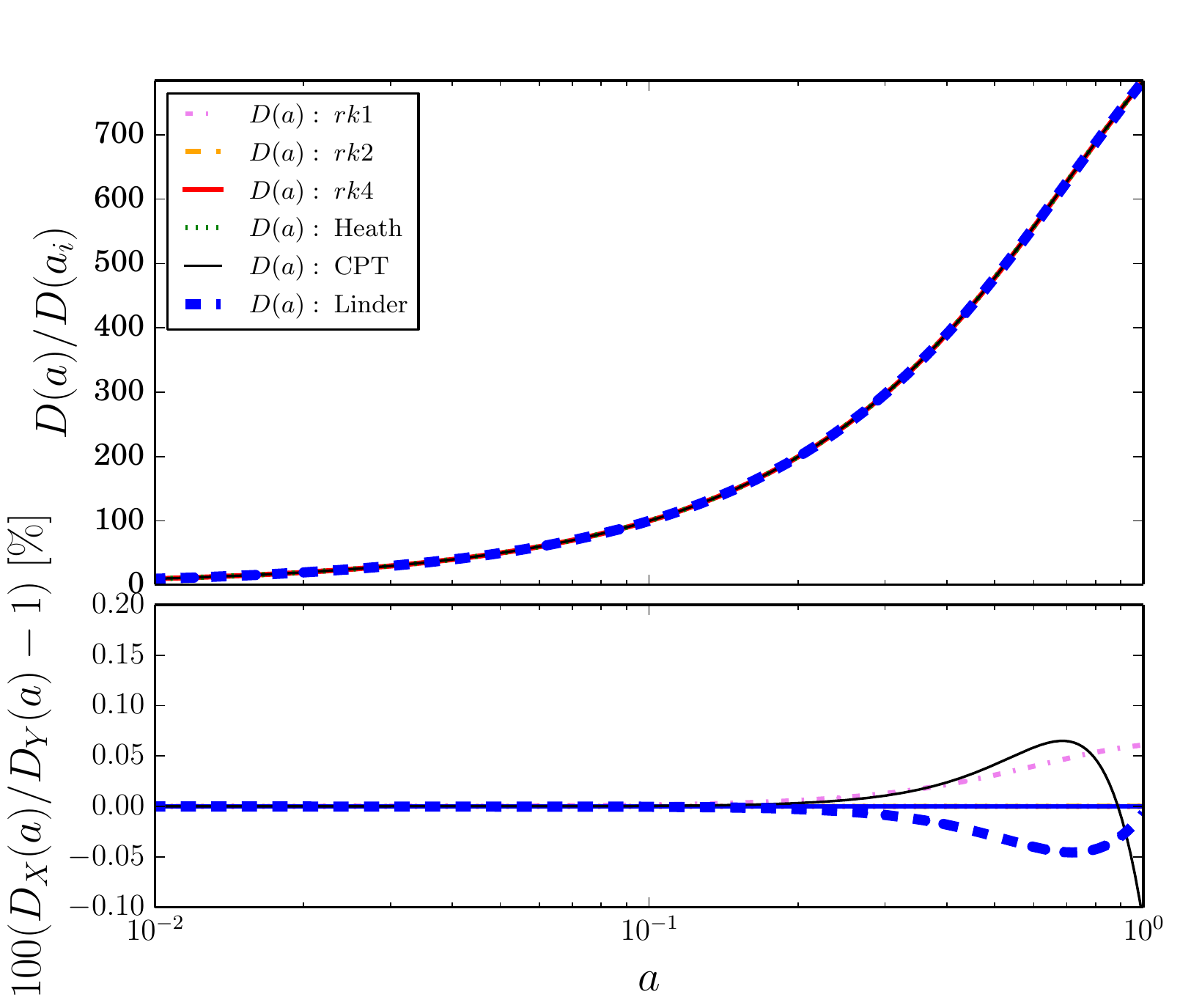}\hspace{0.3cm}
  \includegraphics[angle=0,width=8.4cm]{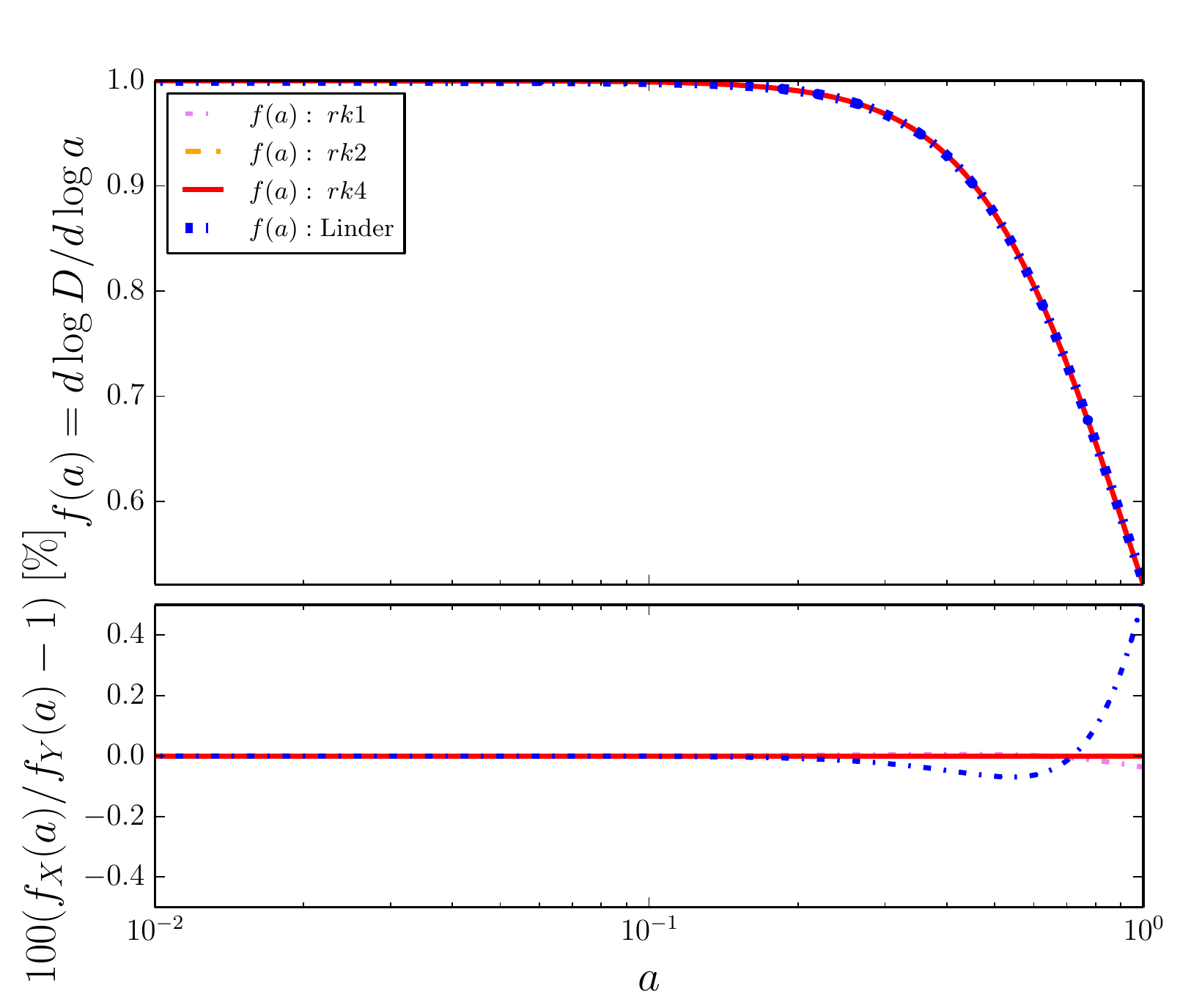}
}
\caption{\small{{\bf Left:} Upper panel shows the linear growth
    factor normalised to unity at the present day as a function of the
    cosmic expansion factor. The red dotted line shows the results for
    our 4th order Runge-Kutta solution. The solid blue and orange
    dashed show the 1st and 2nd order Runge-Kutta solutions,
    respectively. The green dot-dashed lines shows the
    \citet{Heath1977} integral solution, and the solid black line
    shows \citet{Carrolletal1992}. The lower panel shows the ratio of
    the various results with respect to the 4th order Runge-Kutta
    solution, expressed as a percentage difference. {\bf Right:}
    evolution of the linear growth rate as a function of expansion
    factor. Again we show the various Runge-Kutta solutions and
    approximation from \citet{Linder2005}. The line styles are as in
    the left panel.}
\label{fig:growthTEST}}
\end{figure*}

The left panel of Figure~\ref{fig:growthTEST} shows the evolution of
$D(a)$ as a function of expansion factor $a$ for our Fiducial
cosmological model. It compares the results from our 4th order
Runge-Kutta solution (red solid line) with results from a 1st
(dot-dashed violet), 2nd order (orange dashed) Runge-Kutta solution,
and the \citet{Heath1977} integral solution (green dotted) and the
\citet{Carrolletal1992} (thin black solid), and the \citet{Linder2005}
approximation (thick dashed blue). We find the solutions are accurate
to better than 0.1\%. The right panel of panel of Figure~\ref{fig:growthTEST} shows the
evolution of the logarithmic growth rate $f(a)$ as a function of
expansion factor $a$ for our Fiducial cosmological model. It again
compares the results from our 1st, 2nd and 4th order Runge-Kutta
solution, and the approximate solution of \citet{Linder2005}. The
figure shows that the Runge-Kutta solutions are consistent and that
the Linder approximation is accurate to better than 0.5\%.


\begin{figure*}
\centerline{
  \includegraphics[angle=0,width=8.4cm]{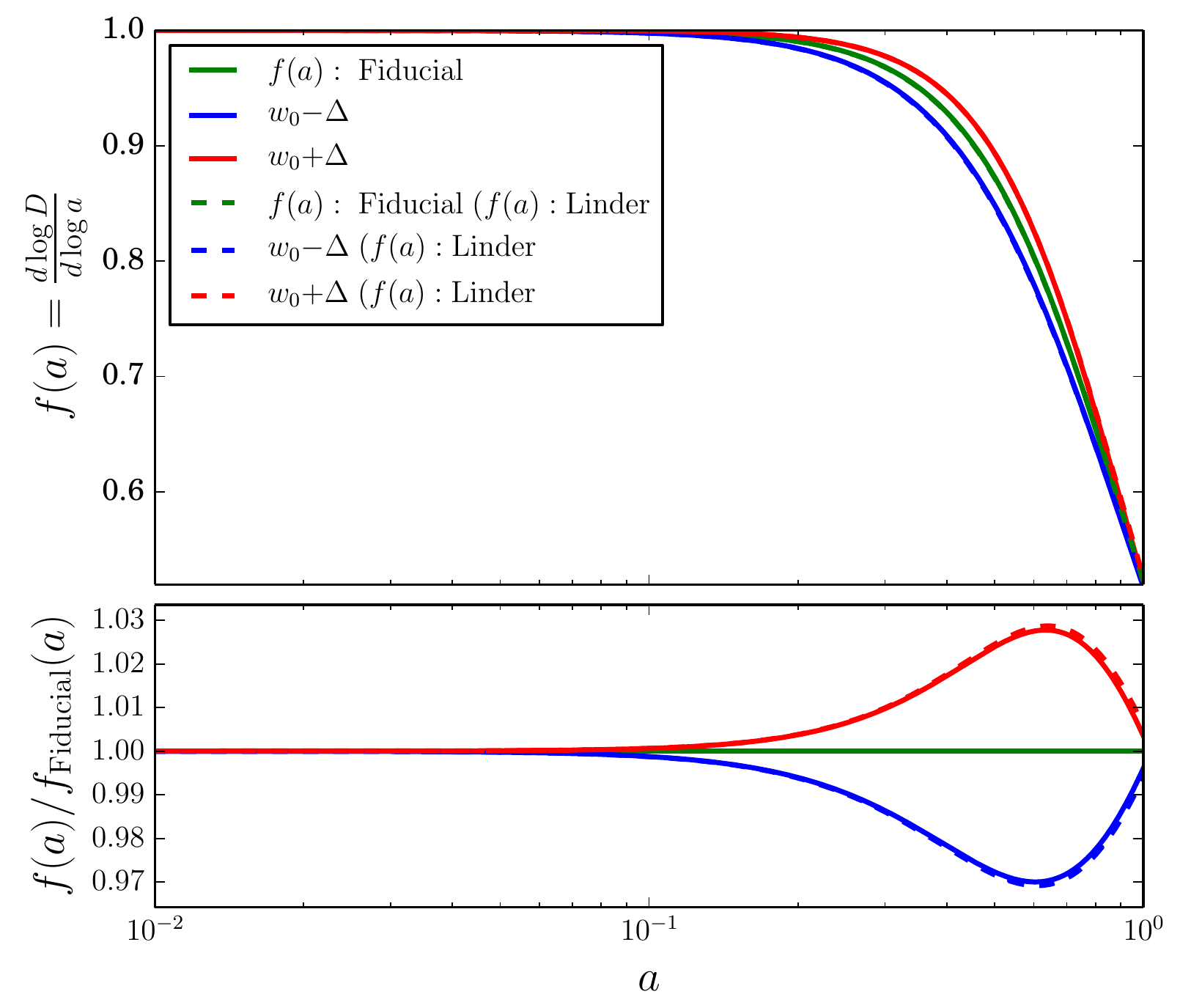}\hspace{0.3cm} 
  \includegraphics[angle=0,width=8.4cm]{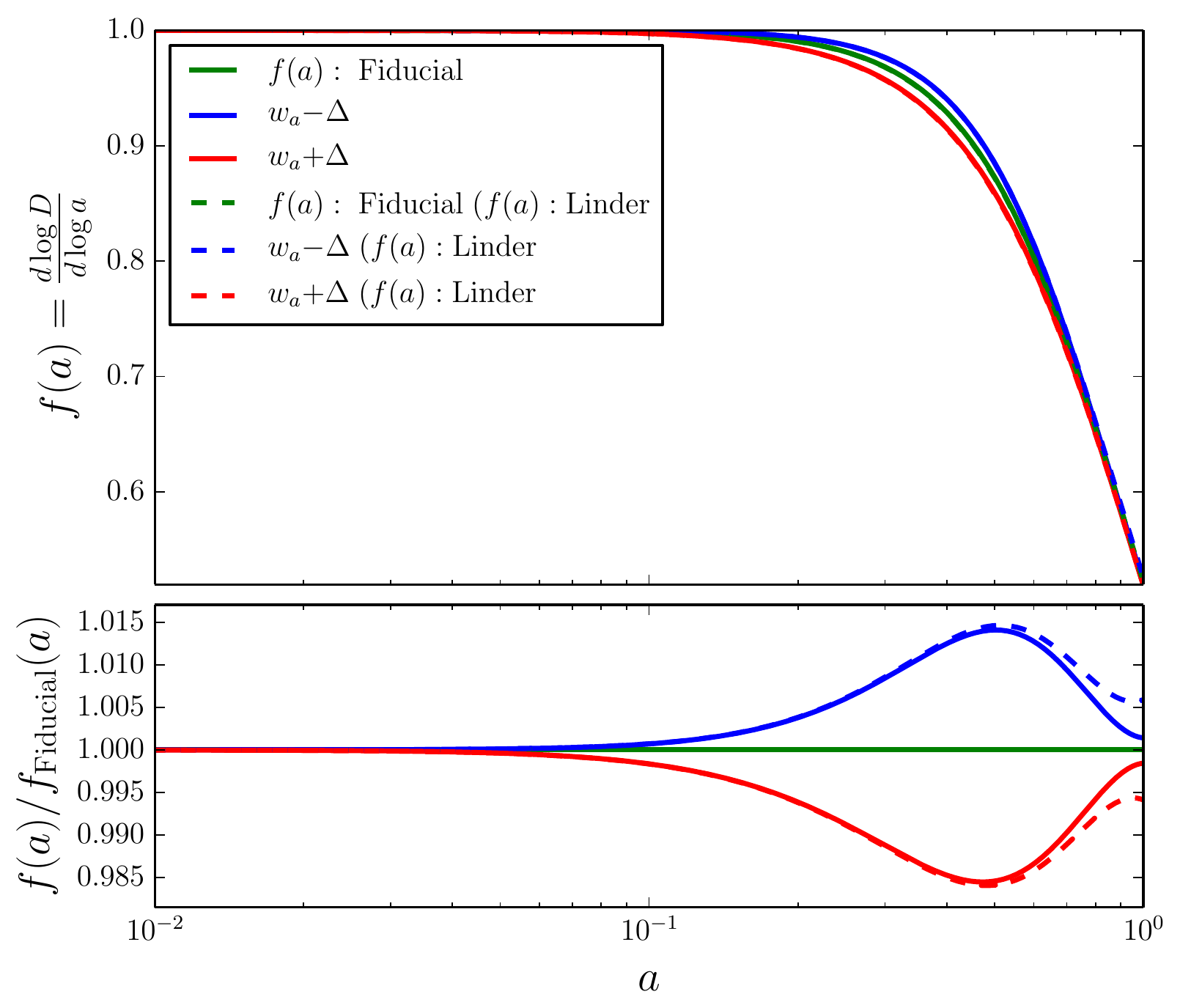} }
\caption{\small{Logarithmic linear growth rate factor as a function of
    the cosmic expansion factor $a$. Left and right panels show the
    results for the four dark energy models listed in
    Table~\ref{tab:cospar}.}
\label{fig:growthrate}}
\end{figure*}

Figure~\ref{fig:growthrate} shows the evolution of the logarithmic
growth rate as function of expansion factor for the dark energy models
considered in this paper.  The figure shows that the approximate
method of \citet{Linder2005} is an excellent description to our 4th
order Runge-Kutta solution, being better than 0.5\% for all the times
of interest.


\begin{figure*}
\centerline{
\includegraphics[width=5.5cm,angle=0]{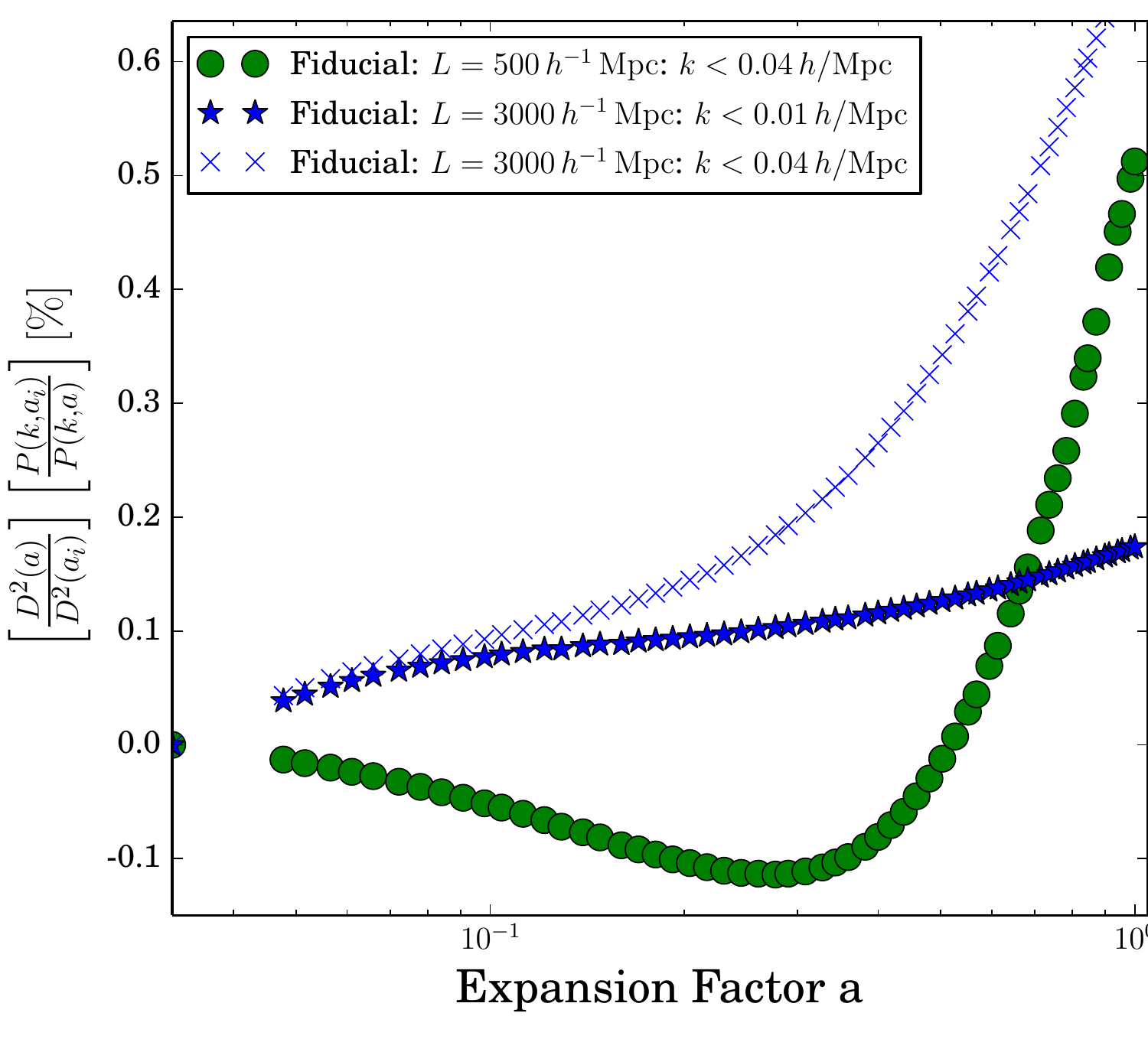}\hspace{0.1cm}
\includegraphics[width=5.5cm,angle=0]{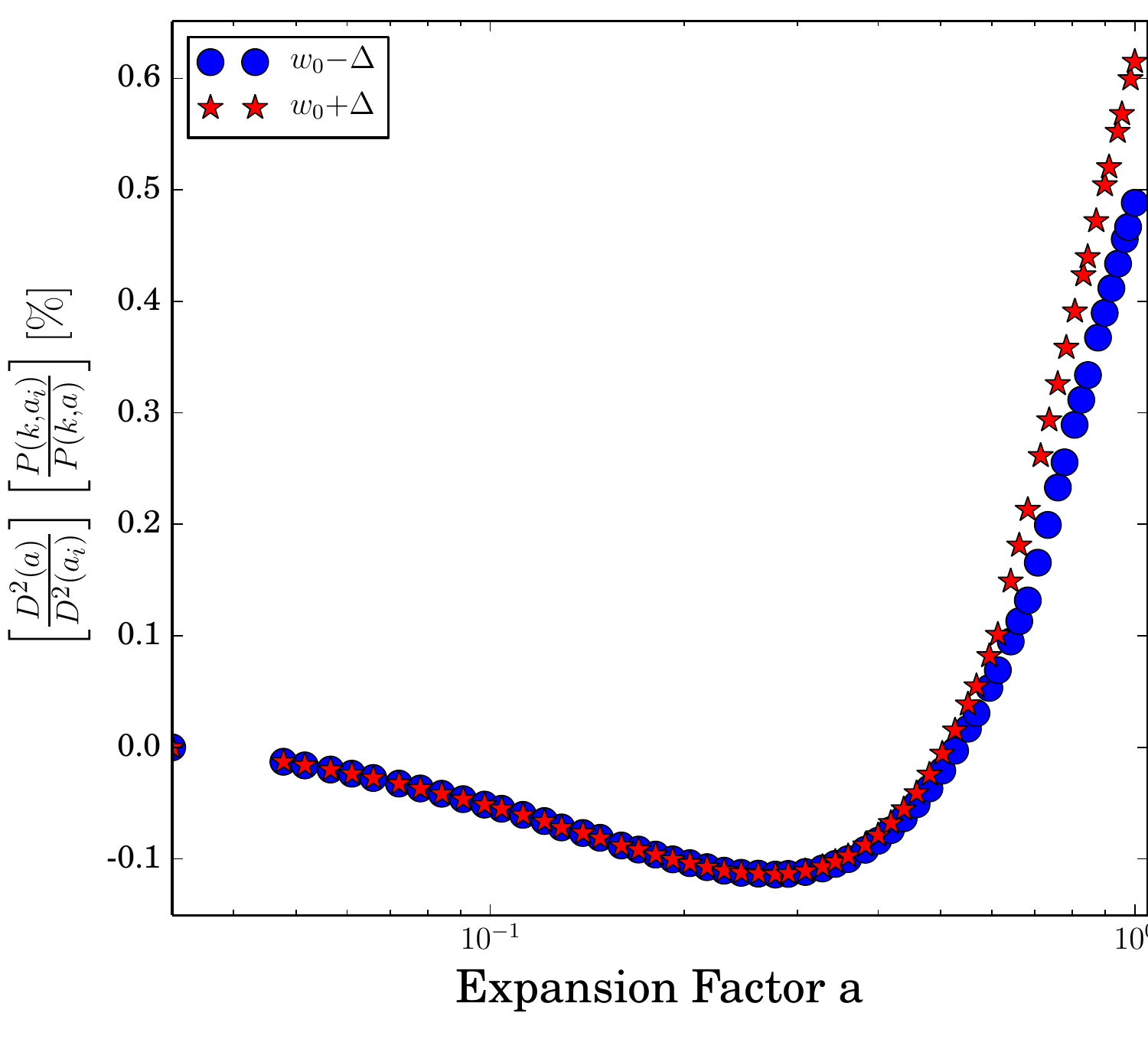}\hspace{0.1cm}
\includegraphics[width=5.5cm,angle=0]{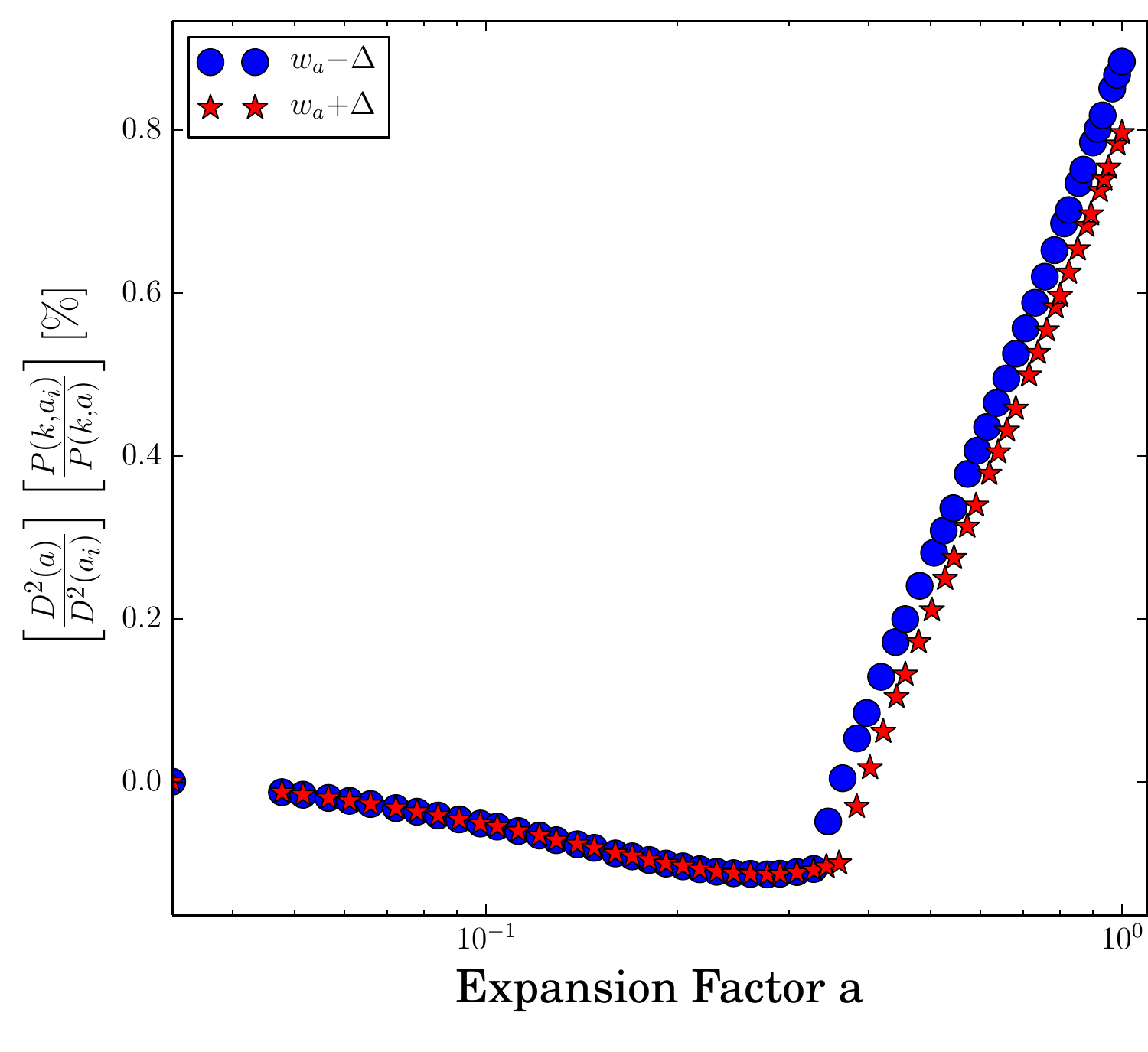}}
\vspace{0.1cm}
\centerline{
\includegraphics[width=5.5cm,angle=0]{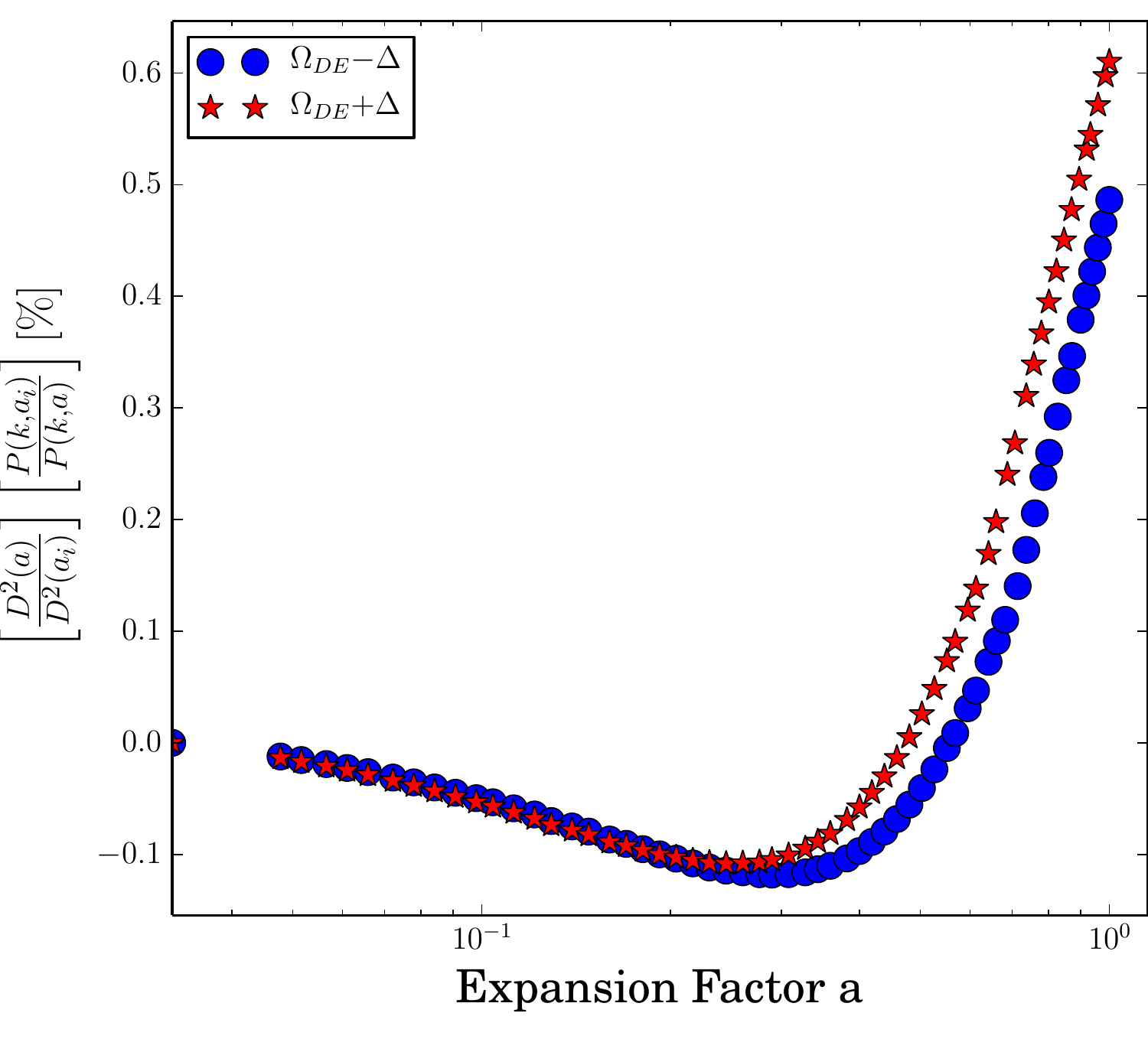}\hspace{0.1cm}
\includegraphics[width=5.5cm,angle=0]{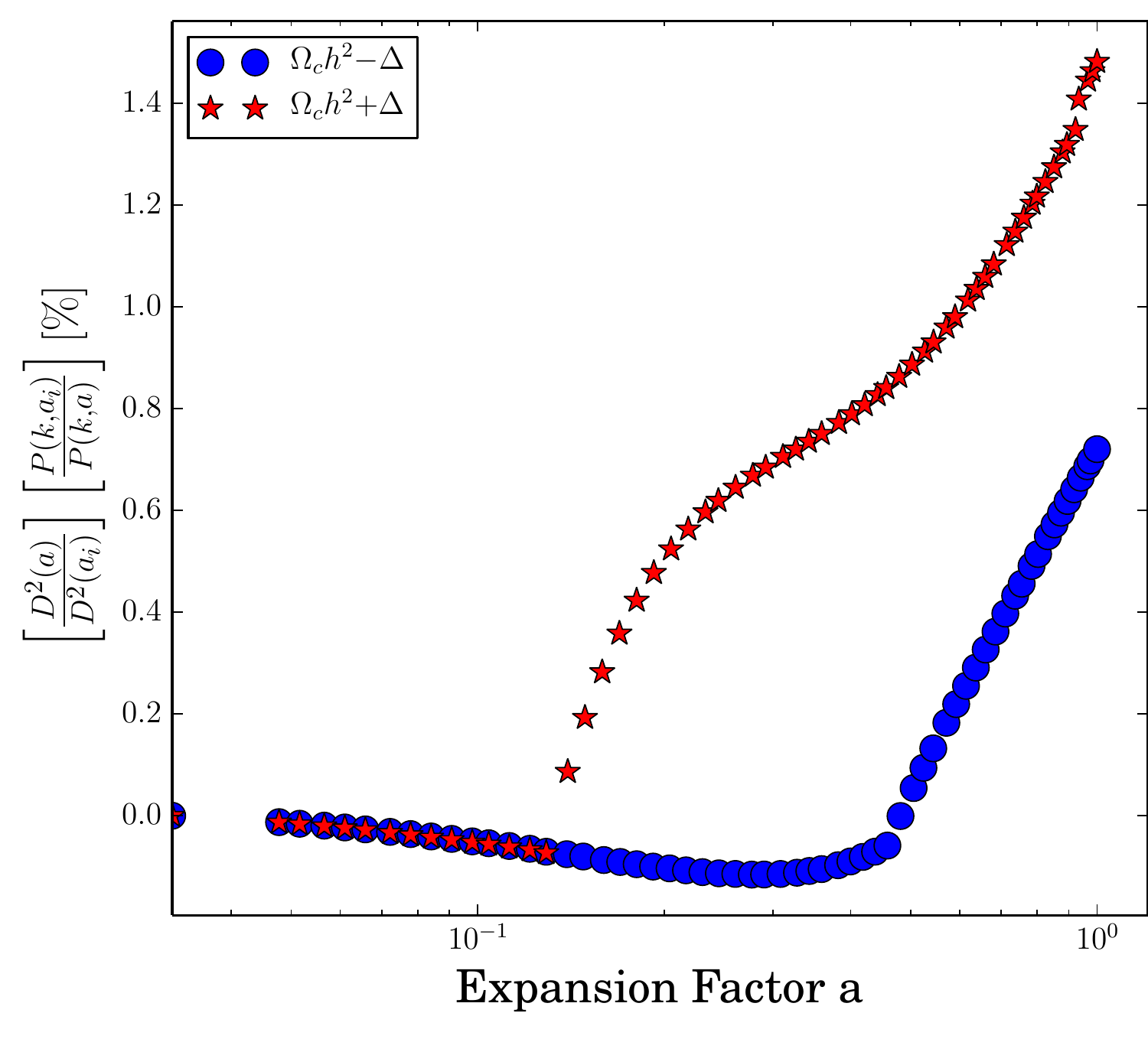}\hspace{0.1cm}
\includegraphics[width=5.5cm,angle=0]{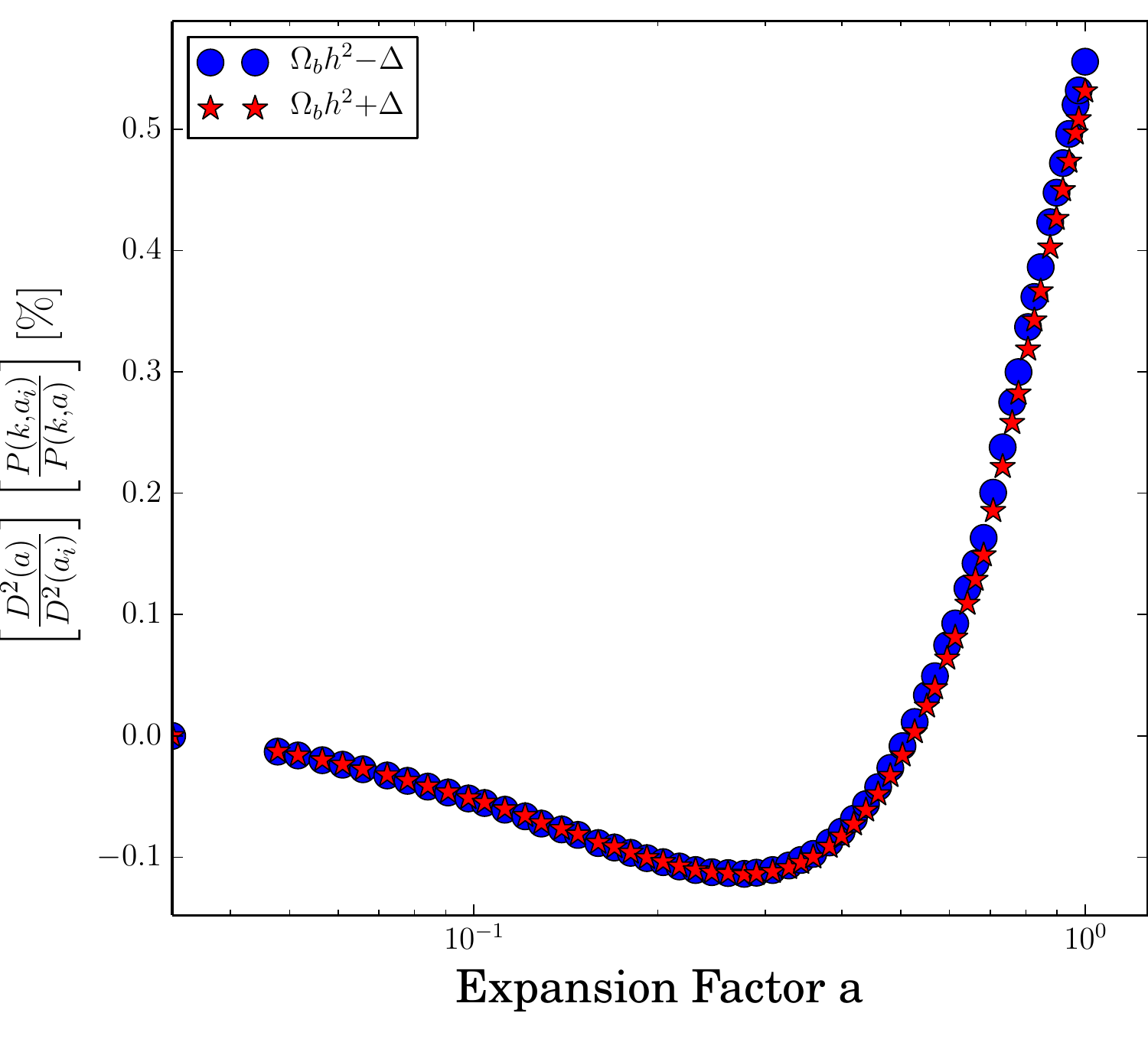}}
\vspace{0.1cm}
\centerline{
\includegraphics[width=5.5cm,angle=0]{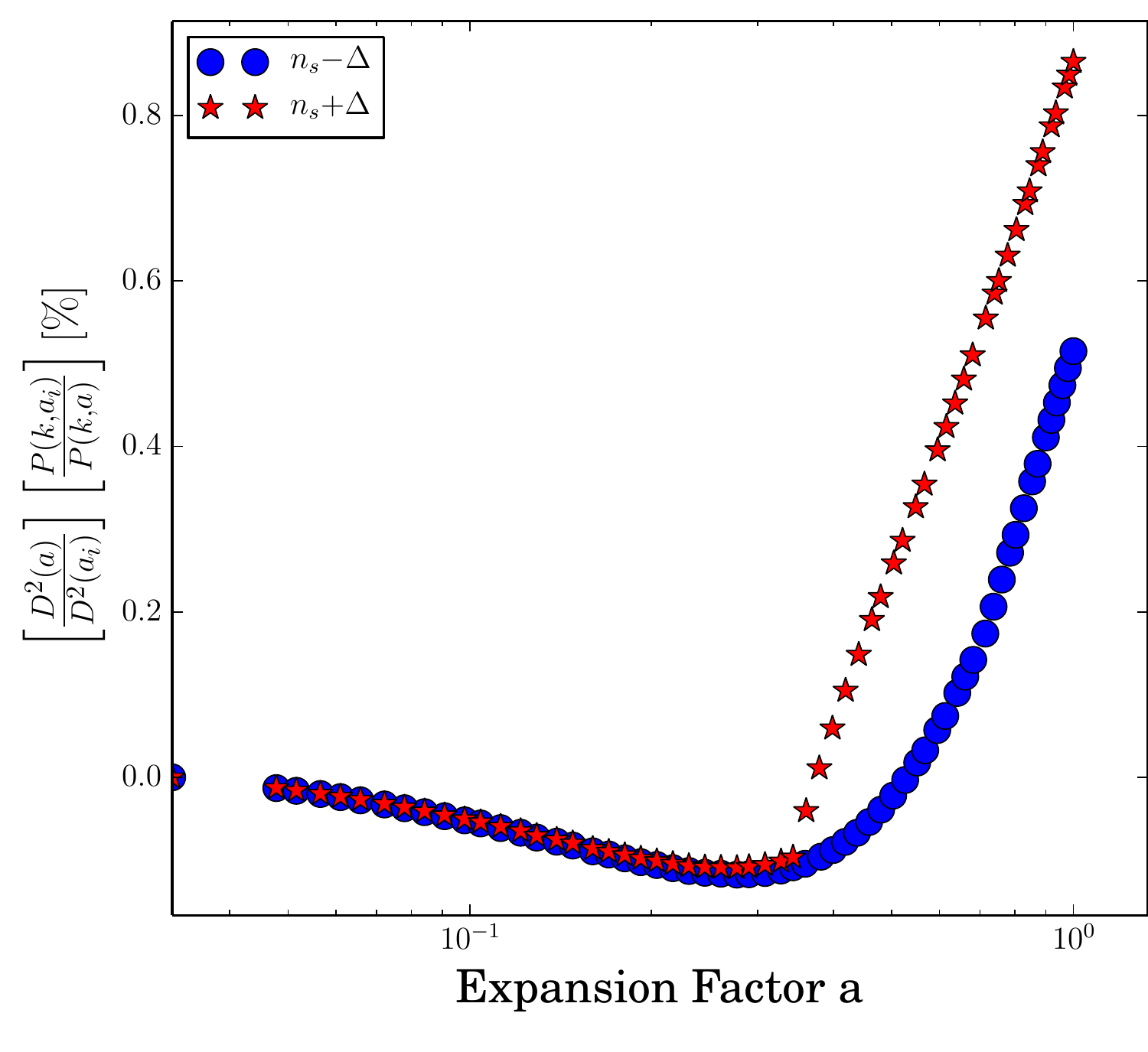}\hspace{0.1cm}
\includegraphics[width=5.5cm,angle=0]{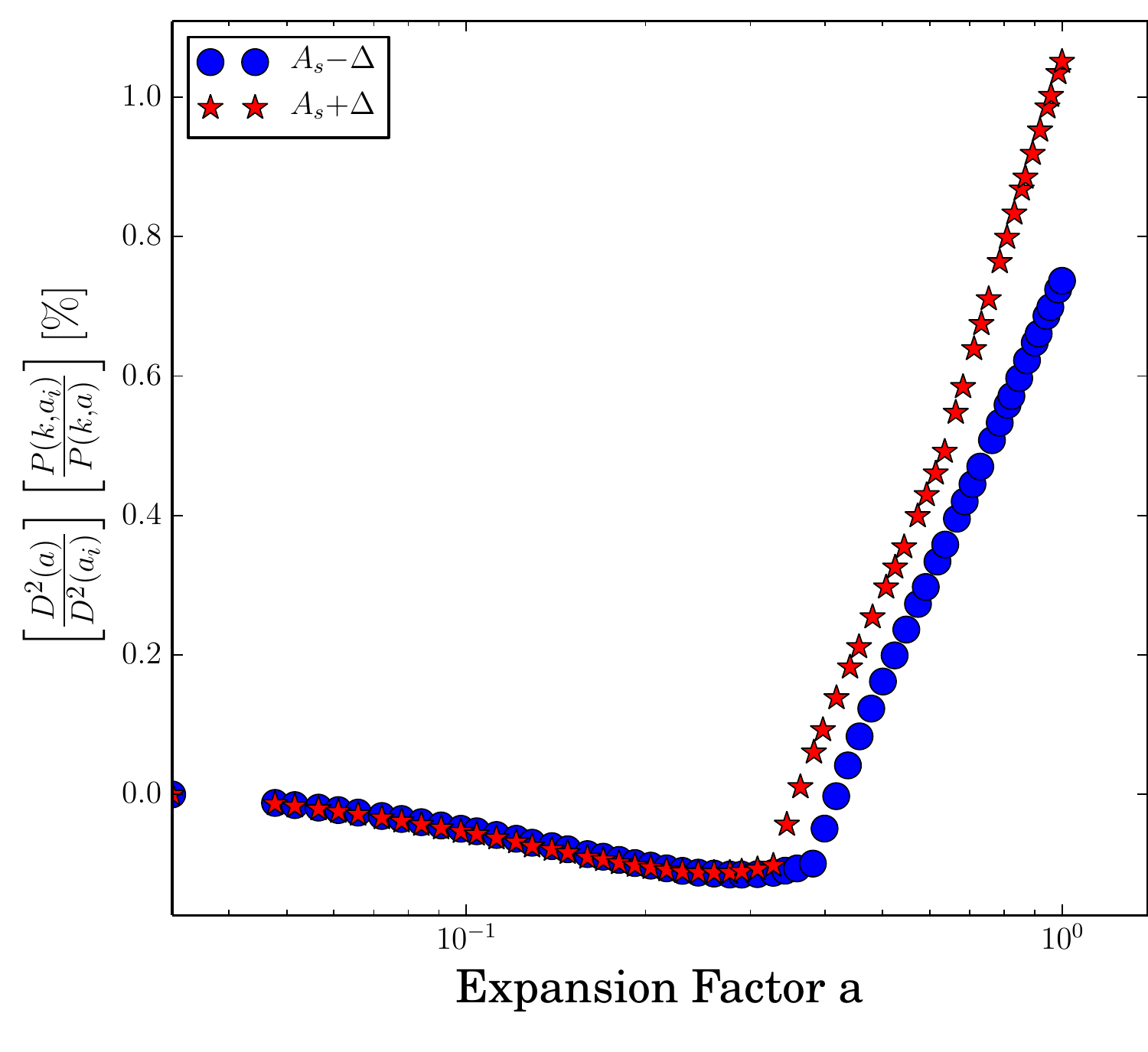}\hspace{0.1cm}
\includegraphics[width=5.5cm,angle=0]{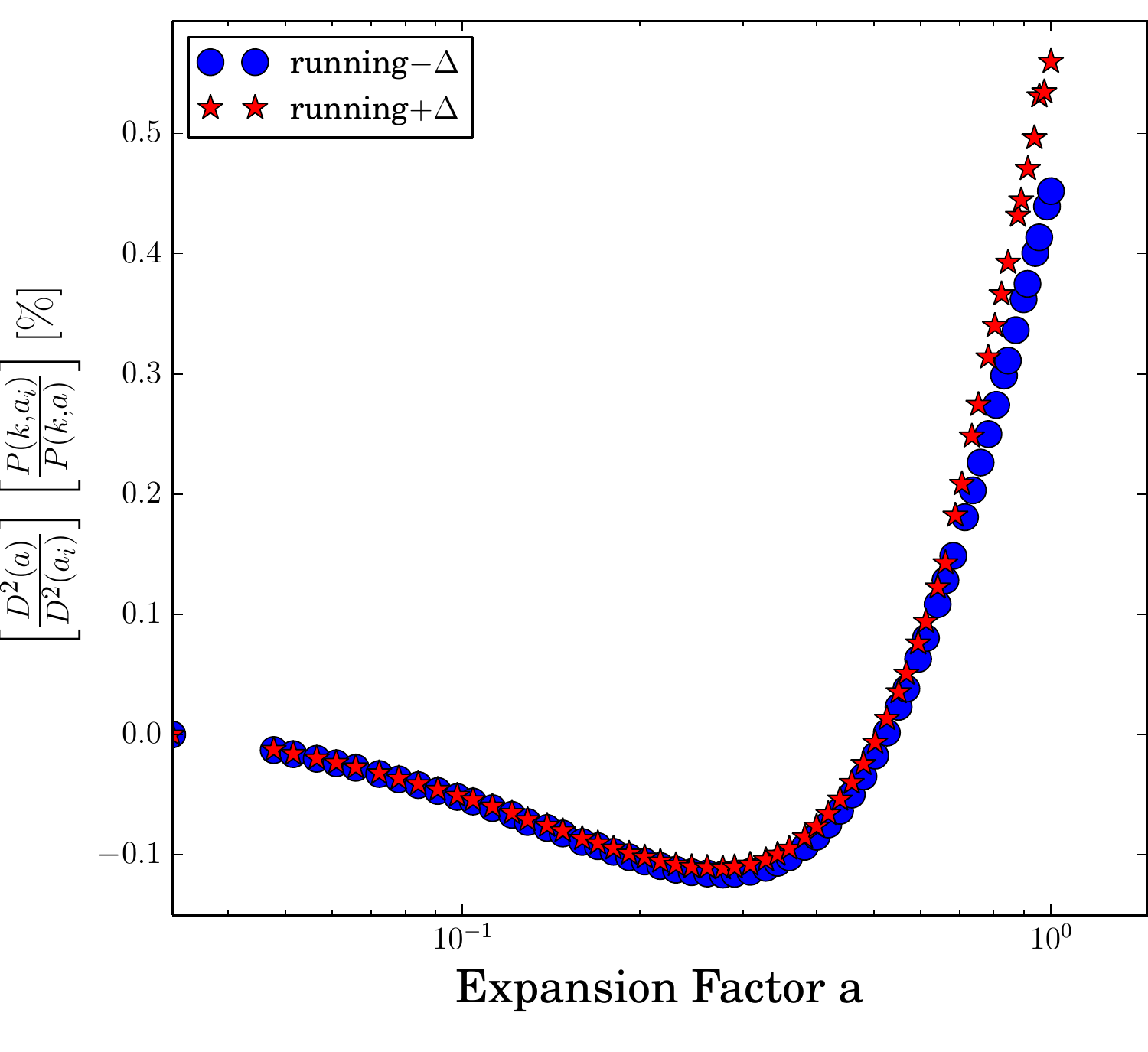}}
\caption{\small{Testing the linear growth of the $N$-body
    simulations. The plot shows how the square of the growth function,
    normalised in terms of its amplitude at the initial time, divided
    by the ratio of the power spectrum normalised in terms of its
    initial value scales as a function of expansion factor.} The
  top-left panel shows the results for the fiducial model: green
  circles shows results for the $L=500\Mpc$ box and for all scales
  $k<0.04\kMpc$; the blue points are for the $L=3000\Mpc$ run, with
  the crosses and stars denoting the results for all scales
  $k<0.04\kMpc$ and $k<0.01\kMpc$, respectively. The other 8 panels
  show the same but for the variational runs in the $L=500\Mpc$ boxes
  and with blue and red points denoting the lower and upper
  cosmological variation. All of these measurements were based on
  using all scales $k<0.04\kMpc$.
\label{fig:PkGrow}}
\end{figure*}


\subsection{Linear growth and the D\"ammerung simulations}

Figure \ref{fig:PkGrow} demonstrates how well linear growth is
preserved for the largest scale modes in the D\"ammerung simulations.
The y-axis of the plots shows:
\be r(a) = \frac{1}{N_<}\sum_{j\in N<}\frac{D^2(a)}{D^2(a_i)}\frac{P(k_j,a_i)}{P(k_j,a)}-1 \ .\ee
where $N_<$ are all modes with $k<0.04\kMpc$. We plot this ratio as a
function of the expansion factor $a$ for all of our runs.

First, looking at the small-box fiducial runs (top left panel) we see
that the ratio is less than 0.5\% for the full range of epochs
considered. For the larger box this measure has a higher
amplitude. This suggests that there may be some nonlinear
mode-coupling at these scales affecting $r$. Changing the cut-off
scale from $k=0.04\kMpc$ to $k=0.01\kMpc$ we see that the linear theory
growth is better preserved.

Considering some of the variations we see that same exact trend as for
the fiducial run. There are however some outliers where the ratio $r$
deviates to 1.5\% -- see for example the plot showing the $\Omega_{\rm
  c}h^2+\Delta$ variation. There appears to be something of a
discontinuity around $a=0.2$, where we see a sharp upturn in the
deviations from linear growth. This seems somewhat nonphysical. 

We speculate that this may arise in the following way. Owing to the
fact that most super-computer facilities impose a maximum time policy
on any one job it is likely that $N$-body runs will need to be
restarted and this may happen more than once depending on the strength
of clustering. With the code {\tt Gadget-3} it is possible to resume a
run in one of two ways. The first is from a restartfile, and the
second is to restart from a snapshot.  In the latter case, position
and velocities are synchronised, whereas in the former positions and
velocities are different by a half time step, which owes to the
structure of the leapfrog time integration.  In addition, in the
latter case the domain decomposition and tree-construction are done a
new, whereas in the former case the exact particle load is preserved
as if the simulation was not interrupted. In some cases, the
simulations were resumed by starting from snapshot and so potentially
this could lead to an error. However, without further investigation,
which is beyond the scope of the paper a definitive answer is still to
be found. We take these errors into account using the method described
in \Eqn{eq:corr}.


\section{Auxiliary multi-point propagator theory functions}\label{app:MPT}

In order to implement the MPT 2-loop calculation the following results
are required.


\subsection{Standard perturbation theory kernels}

Following \citet[][and references therein]{Bernardeauetal2002} the
unsymmetrised SPT kernels that we require are:
\ba
F_2(\bq_1,\bq_2) & = & \frac{5}{7}\alpha(\bq_1,\bq_2)+\frac{2}{7}\beta(\bq_1,\bq_2)\ ;\\
G_2(\bq_1,\bq_2) & = & \frac{3}{7}\alpha(\bq_1,\bq_2)+\frac{4}{7}\beta(\bq_1,\bq_2)\ ;\\
F_3(\bq_1,\bq_2,\bq_3) & = &
\frac{1}{18} \left[\frac{}{}
  7\alpha(\bq_1,\bq_2+\bq_3)F_2(\bq_2,\bq_3)
  +2\beta(\bq_1,\bq_2+\bq_3)G_2(\bq_2,\bq_3) \right.\nn \\
& & \left. \frac{}{} +7\alpha(\bq_1+\bq_2,\bq_3)G_2(\bq_1,\bq_2)
  +2\beta(\bq_1+\bq_2,\bq_3)G_2(\bq_1,\bq_2)\right] \ ; \label{eq:F3}
\ea
where the mode coupling kernels are:
\be
\alpha(\bq_1,\bq_2) \equiv \frac{(\bq_1+\bq_2)\cdot\bq_1}{q_1^2} \ \ \ ; \ \ \ 
\beta(\bq_1,\bq_2) \equiv \frac{\left[(\bq_1+\bq_2)\cdot(\bq_1+\bq_2)\right](\bq_1\cdot\bq_2)}{2q_1^2q_2^2} \ .
\ee
The symmetrised kernels $F_n^{(s)}$ are obtained from unsymmetrised
kernels by summing over all possible permutations of the arguments and
dividing through by the number of permutations:
\ba F_2^{(\rm s)}(\bq_1,\bq_2) & = &
\frac{5}{7}+\frac{1}{2}\frac{\bq_1\cdot\bq_2}{q_1q_2}
\left(\frac{q_1}{q_2}+\frac{q_2}{q_1}\right) +\frac{2}{7}
\left[\frac{\bq_1\cdot\bq_2}{q_1q_2}\right]^2 \ ;\\
G_2^{(\rm  s)}(\bq_1,\bq_2) & = &
\frac{3}{7}+\frac{1}{2}\frac{\bq_1\cdot\bq_2}{q_1q_2}
\left(\frac{q_1}{q_2}+\frac{q_2}{q_1}\right) +\frac{4}{7}
\left[\frac{\bq_1\cdot\bq_2}{q_1q_2}\right]^2 \ ;\\
F^{(\rm s)}_3(\bq_1,\bq_2,\bq_3)
& = & \frac{1}{6}\left[F_3(\bq_1,\bq_2,\bq_3)+F_3(\bq_2,\bq_1,\bq_3)+F_3(\bq_3,\bq_1,\bq_2) \right.\nn \\
& & \left.+ F_3(\bq_1,\bq_3,\bq_2)+F_3(\bq_2,\bq_3,\bq_1)+F_3(\bq_3,\bq_2,\bq_1) \right]\ . \label{eq:F3s}
\ea
The symmetrised version of the $F_3$ kernel can be developed further
by repeated substitution of \Eqn{eq:F3} into \Eqn{eq:F3s}, to obtain:
\ba
F^{(\rm s)}_3(\bq_1,\bq_2,\bq_3)
& = & \frac{7}{54}\alpha(\bq_1,\bq_{23})F_{2}^{(s)}(\bq_2,\bq_3)
+\frac{1}{27}\left[7\alpha(\bq_{23},\bq_1)+4\beta(\bq_{23},\bq_1)\right] G_{2}^{(\rm s)}(\bq_2,\bq_3)
+ 2\cyc \ .
\ea
%


\subsection{Auxiliary function for the MPT propagators}

In order to compute the MPT propagator damping factors given by
\Eqnss{eq:gam1}{eq:gam3} we need to evaluate the 1D integral:
\ba
f(k)&=&\int \frac{d^3q}{(2\pi)^3} P_0(q) \, \frac{1}{504 k^3 q^5}
\left[6k^7q-79k^5q^3+50q^5k^3-21kq^7 +\frac{3}{4}(k^2-q^2)^3(2k^2+7q^2)\ln \frac{|k-q|^2}{|k+q|^2} \, \right] \ . 
\label{eq:fk}\ea
We do this using the Gaussian quadrature routine {\tt
  gsl\_integration\_qag} provided in the {\tt GSL} standard library.

Our {\em ad hoc} correction to the $f(a)$ function is:
\ba
f_{\rm NL}(k)&=&\int_{0}^{q=1\kMpc}\!\!\! \frac{dq q^2}{2\pi^2}  \, \frac{P_{\rm halofit2012}(q)}{504 k^3 q^5}
\left[6k^7q-79k^5q^3+50q^5k^3-21kq^7 +\frac{3}{4}(k^2-q^2)^3(2k^2+7q^2)\ln \frac{|k-q|^2}{|k+q|^2} \, \right] \ . 
\label{eq:fkNL}
\ea

\label{lastpage}

\end{document}